\documentclass[twoside,fleqn]{ActaStyle}
\usepackage{times,cite}
\usepackage{graphicx}
\usepackage{psfrag}

\def\no{N_{\rm o}}
\def\ns{N_{\rm stat}}
\def\neff{N_{\rm eff}}
\def\veps{\epsilon}
\def\eeps{\varepsilon}

\def\bc{\begin{center}}
\def\ec{\end{center}}
\def\be{\begin{equation}}
\def\ee{\end{equation}}
\def\ef{\end{figure}}

\def\Rleft{R_{\rm left}}
\def\Rright{R_{\rm right}}
\def\Lleft{L_{\rm left}}
\def\Lright{L_{\rm right}}

\def\authorheading{P. Marko\v{s}}
\def\shorttitle{Numerical Analysis of the Anderson Localization}

\def\be{\begin{equation}}
\def\ee{\end{equation}}

\def\ds#1{\displaystyle{#1}}

\def\ba{\begin{array}}
\def\ea{\end{array}}

\def\bQ{\textbf{Q}}

\def\m2#1#2#3#4{%
\left(
\begin{array}{lr}
\ds{#1}  &  \ds{#2}  \\
\ds{#3}  &  \ds{#4}
\end{array}
\right)
}

\def\mv#1#2{%
\left(
\begin{array}{l}
#1    \\
#2  
\end{array}
\right)
}

\def\mvr#1#2{%
\left(
\begin{array}{r}
#1    \\
#2  
\end{array}
\right)
}

\begin{document}
\pagerange{1}{123}   

\title{NUMERICAL ANALYSIS OF THE ANDERSON LOCALIZATION}

\author{P.~Marko\v{s}\email{peter.markos@savba.sk} 
       }
       {Institute of Physics, Slovak Academy of Sciences, 845 11 Bratislava, Slovakia}

\abstract{%
The aim of this paper is to demonstrate, by simple numerical simulations, the main
transport  properties of disordered electron systems. These systems undergo
the metal insulator transition when either Fermi energy crosses the mobility edge
or the strength of the disorder increases over critical value.  We study how disorder
affects the energy spectrum and spatial distribution of electronic eigenstates
in the diffusive and insulating regime, as well as in the critical region
of the metal-insulator transition.
Then, we introduce the transfer matrix and  conductance, and we discuss how the quantum character
of the electron propagation influences the transport properties of disordered samples.
In the weakly disordered systems, the weak localization and anti-localization  as well as
the 
universal conductance fluctuation are numerically simulated  and discussed. The localization
in the one dimensional system 
is described and interpreted as a purely quantum effect.
Statistical properties of the conductance in the critical and localized regimes
are demonstrated.
Special attention is given to the numerical study of the transport
properties of the critical regime and to the numerical 
verification of the  single parameter scaling theory of localization. Numerical data for
the critical exponent in the  orthogonal models  
in dimension $2< d\le 5$  are compared with theoretical predictions.
We argue that the discrepancy between the theory and numerical data 
is due to the absence of the self-averaging of transmission quantities.
This  complicates the  analytical analysis of the disordered systems.
Finally, theoretical methods of description of weakly disordered systems
are explained and their possible generalization to the localized regime is discussed.
Since we concentrate on the one-electron propagation at zero temperature, no
 effects of electron-electron interaction and incoherent scattering 
are discussed in the paper.
}

\setcounter{tocdepth}{1}

%
	\section{Introduction}

Localization of electrons in disordered system has fascinated scientists for almost  fifty years.
In 1958, Anderson \cite{Anderson-58} predicted that randomly distributed
impurities in the crystal lattice can localize
an electron in a certain spatial region. The localization is given by the   quantum character
of electron propagation:
electron wave
function is scattered on randomly distributed impurities, and mutual  interference of scattered
components cancel  the wave function   on large distances. 
Localization of the electron is responsible for a new kind of insulators - the Anderson insulator,
which possesses zero electric conductivity, $\sigma$,
in a part of the energy bands, where the  density
of states, $\rho(E)$, is non-zero. 
Propagation of the  electron in disordered systems is therefore qualitatively different
from that in periodic structures. Although the 
electron wave function is 
reflected also by a periodic potential, the  interference of the reflected and the transmitted waves 
in a regular lattice gives rise
to bands and gaps in the electron energy spectrum. 
In the bands, where the density of states is non-zero, 
the electron propagates freely throughout the structure.

Of course, disorder is always present in the real world, and it influences the electric
transport. Scattering on weak impurities causes diffusive propagation of
electrons. The  electrical conductivity can be expressed through the diffusive coefficient,
$D(E)$, and the density of states, $\rho(E)$,
\be\label{v-1}
\sigma=e^2 D(E)\rho(E).
\ee
In the derivation of the expression (\ref{v-1}), it is assumed that
an electron on its travel through the sample is scattered on individual impurities. 
This assumption is correct only if 
the electron wavelength, $\lambda_F$, which is determined by the wave vector, $k_F$
on the Fermi energy, is smaller than the electron mean free path due to the  coherent scattering
on impurities, $\ell$,
\be\label{lif}
\frac{\ell}{\lambda_F}\gg 1.
\ee
\cite{Lifshitz}.
The mean free path represents, in the first approximation, the mean distance between two 
impurities. Clearly, $\ell$ is large 
in the limit of weak disorder, and decreases when disorder increases.
Therefore, condition (\ref{lif})
is violated for strong disorder and localization is expected when 
 $\ell\sim \lambda_F$. The last relation   is known as the  Lifshitz criterion for localization.

Increase of disorder changes the transport regime considerably.
In the limit of very strong disorder, all electronic states become  localized. 
The wave function of localized electron decreases exponentially  as a function of
distance from the localization center, $\vec{r_0}$,
\be\label{i-2}
\Psi(\vec{r})\sim \exp -\frac{|\vec{r}-\vec{r_0}|}{\lambda},
\ee
where  $\lambda$ is the \textsl{localization length}.

Anderson showed in his pioneering work \cite{Anderson-58} 
that all electronic  states become insulating when disorder increases above the  critical value.
The transition from metal to insulator due to the increase in disorder is called the
\textsl{Anderson transition}.  Similar to the theory of phase transitions,
it is believed that Anderson transition is universal and  can be described 
by the one-parameter \textsl{scaling theory}  \cite{AALR}. 
The key parameter of the scaling theory  is the conductance, $g$. Introduced by 
Landauer \cite{Landauer}, the conductance measures the transmission properties 
of disordered systems both in the metallic and the localized regime.

The scaling theory of localization  analyzes the size dependence of the conductance
in the limit of large system size.  Only three transport regimes exist in this limit:
the system is either in  the metallic, localized or critical regime.
In the metallic regime, the conductance increases with the system size, and the system
possesses non-zero electric conductance. In the localized regime, all electronic states
are localized, and $g$ decreases exponentially due to further increase of the system size.
The system is in the critical regime only at the critical value of the disorder.

While the localized regime exists in any system, provided  the disorder is sufficiently strong,
the existence of the metallic regime is not guaranteed, especially for systems
with lower spatial dimension. It is well-known that all electronic states are localized
in any one dimensional (1D) disordered system \cite{Mott-61}.
Spin-less electrons are always localized already in the  two dimensional (2D) systems,
with the only exception caused by an external magnetic field. 

The scaling theory of localization predicts that the metal-insulator transition is universal.
The size and disorder dependence of the conductance in the neighborhood of the critical point
is governed only by universal critical exponents. All the parameters, which define the microscopic
structure of the model, become irrelevant when the size of the system is sufficiently large.
Verification of the universality of the metal-insulator transition  and the calculation of the critical
parameters  
- the critical disorder and critical exponents -
for systems of various dimension and physical symmetry
are the main theoretical and
numerical problems of the theory of localization. 

\smallskip

Contrary to the scaling theory of localization which discusses the transport regimes
in the limit of $L\to\infty$,
in everyday life we must deal with systems of finite
size. Here, the transport regime depends on the relation of the system size, $L$, to
the characteristic lengths.  For instance, if
\be\label{c-diff}
\ell \ll L\ll\lambda,
\ee
then the system
exhibits metallic behavior with the conductivity given by Eq. (\ref{v-1}). 
This happens in 2D systems, where the localization length $\lambda$ is extremely huge
for weak disorder \cite{McKK-1983}.  
Since the electron diffuses through the sample, we called the transport regime,
defined by inequalities (\ref{c-diff}), the \textsl{diffusive regime}.  Of course,
an increase of the system size over the localization length  causes localization 
of all electronic states and the conductivity decreases to zero.  

There are  small quantum corrections to the
conductivity (\ref{v-1}) when the conditions (\ref{c-diff}) are fulfilled.
For instance, if the size of the two dimensional (2D) sample is much larger than the mean free path,
$L\gg \ell$, 
the mean value of the conductance decreases logarithmically when the  size 
of the system increases. 
This effect - weak localization corrections to the conductance -
is the first manifestation of the quantum character of the electron propagation.
Similar weak localization corrections to the conductivity exist 
also in one and three dimensions and in quasi-one dimensional (quasi-1d)  geometry.

In the  opposite limit of very small systems, $L\ll \ell$  we observe the \textsl{ballistic} regime.
In this regime, the electron, in its travel through the sample, is scattered only on a few impurities.
Clearly, the sample is almost transparent and the conductance might be large already in 1D systems.

\medskip

While the weakly disordered systems can be described analytically,  for instance
by the Dorokhov-Mello-Pereyra-Kumar (DMPK)  equation \cite{DMPK}, or perturbation Green's function
methods \cite{Langer,LSF}, the  theoretical description of
the critical regime is still not complete.   The main problem is the
absence of small parameter,  since the critical disorder is of the order or even 
larger than the bandwidth in three
dimensional systems.   
The critical disorder is  small only when the dimension of the system, $d$, decreases
to the lower critical dimension, $d_c=2$ \cite{Wegner-76}. Analytical theories in dimension
$d=2+\veps$
\cite{Wegner-76,Hikami,AKL} calculate critical parameters of the model in powers
of $\veps$. 

Since each sample contains
randomly distributed scatterers,  the measured quantities, like the conductance $g$,
fluctuate from sample to sample and  must
be averaged over random disorder. One possibility is to average  the conductance over 
the statistical ensemble
of macroscopically equivalent samples, which differ only in the microscopic realization
of the disorder. The ergodic hypothesis \cite{EH} states that the same  
statistical ensemble can be constructed with the use of the single sample by varying  the Fermi energy
or the  magnetic field. This enables us to compare the theoretical or numerical results,
 which use the ensemble
averaging, with  experiments, where usually only a few samples are analyzed.

The key parameter of the scaling theory,  the conductance, $g$, is not a 
self-averaged quantity.
Already in the metallic regime,  the conductance becomes
sample-dependent. Measured values of the conductance \cite{WW}
fluctuate  around the mean value.
These fluctuations, of the order of $e^2/h$,  depend neither on the size
of the system, nor on the mean value \cite{LSF}
and they lead to the \textsl{absence of self-averaging} of the 
conductance. In the localized regime, the fluctuations of the conductance \cite{Fowler}
are so strong that
the mean value, $\langle g\rangle$, is not a representative quantity \cite{ATAF}.

The absence of self-averaging
and  huge fluctuations of the conductance which do not disappear 
in the limit of infinite size of the system must be included in the theoretical
description of the localization.  First, it is not clear how 
the averaging over the disorder should  be performed. Moreover, not only the mean values,
but also the higher cummulants of the conductance  must be calculated. Contrary
to the classical systems, the higher cummulants do not diminish in the limit of large system size.
Also, the averaged quantity must be carefully chosen. In the localized regime, it is more suitable
to average the \textsl{logarithm} of the conductance than the conductance itself.
The average procedure is easier in the numerical simulations  than in the analytical theory.
This is the  reason why numerical methods, based on the finite size scaling
\cite{McKK} provides us with the most reliable information about the Anderson transition
in higher dimensions.

\smallskip

In this Paper, we discuss the basic ideas of the localization theory. 
We concentrate on numerical methods of investigation of disordered electronic
systems.  Numerical simulations enable us to describe the metallic, the critical and the 
strongly localized regime. They verify theoretical predictions, and, last  but not least,
they provide us with data necessary for the analytical description of 
the localization.

As already mentioned, both the weak localization and the localization are quantum effects,
caused by the mutual interference of the scattered components of the electron wave function.
The quantum coherence of the electron propagation is destroyed by the incoherent scattering.  
Since the incoherent  mean free path, $L_\phi$, increases when the  temperature decreases
\cite{Thouless-PRL}, we  expect that 
the best experimental conditions for the analysis of the localization effects are  those 
in the limit of small temperature, when $L_\phi\gg L$.
In this paper, we restrict our discussion only to the limit of zero temperature, when $L_\phi\to\infty$.
For the case of non-zero temperature, when $L_\phi>L$,  we can 
obtain a  qualitative estimation of the transport 
in the  diffusive regime if  the size of the system is replaced by $L_\phi$.  
The transport in the strongly localized regime, $L_\phi>\lambda$
 requires more detailed analysis \cite{serota,Kramer-AP}
which is above the scope of this paper.

We will  consider only the one-electron problem. 
Although the electron-electron interaction
might play important role in the localization \cite{Kravchenko}, numerical analysis of systems of disordered interacting electrons is too difficult to be able at present
to answer the main questions of the scaling theory.

Since the single electron localization is caused by interference of electron wave function,
the theory can be easily applied also to propagation of classical waves in disordered media.
Of special interest is propagation of electromagnetic waves through random
dielectrics \cite{SEGC,RN,Genack}. 

\medskip

The Paper is organized as follows.

In Sect. \ref{sec:loc} we describe  the Anderson localization and demonstrate the
localization of electron in two dimensional (2D) strongly disordered system.
In Sect. \ref{sect:model} we introduce several models,
used in numerical simulations.  For numerical convenience,
all the models are based on the propagation of electrons on the
$d$ - dimensional lattice, with disorder represented by the random on site energies.
The Anderson transition is explained briefly in Sect. \ref{sect:AT}. 
Spatial structure of the electron wave functions and spectra of eigenenergies is shown
in Sect. \ref{WF}. 
Transfer matrix and the conductance are introduced in Sect. \ref{sect:g}. 

In the most simple case
of the one dimensional (1D) disordered chain, it was proved \cite{Mott-61}
that all electronic states are localized even for weak disorder
provided that  the system is sufficiently long (longer than localization length, $\lambda$).
We discuss the transport in the 1D systems in Sect. \ref{sect:1D}. The one dimensional
systems are not only easy to simulate numerically, but allow also 
to derive exact analytical results. The statistics of the conductance and  the resistance
is discussed in details, and the quantum origin of the  localization is explained.

Section \ref{sect-diff} analyzes the electronic transport in the diffusive regime.
The scaling theory of localization, formulated in terms of the conductance, is introduced
in Sect. \ref{sect:scaling}.  It is argued that the metal - insulator transition is
a universal phenomenon, and the conductance, $g$, is a function of only one parameter
in the critical regime. This scaling hypothesis is verified numerically. First, in 
Sect. \ref{sect:crit} and \ref{sect:loc} we discuss statistical properties of
the conductance in the critical and localized regime. The critical conductance
distribution is presented and discussed in detail. Then, in Sect. \ref{section:nsa},
we review the numerical scaling analysis of the Anderson transition in the Anderson model.
Sect. \ref{sect:dd} presents 
the critical exponent, obtained by numerical scaling analysis of systems
with dimension $2< d \le 5$. Obtained numerical data are compared with theoretical
predictions. The two dimensional critical regime is discussed in Sect. \ref{sect:2d}.
Finally, Sect. \ref{sect:poss} discusses two possibilities of the analytical description
of the localized regime.  

The Paper contains five Appendices, which present some technical details useful
for the understanding of discussed  numerical data. Properties of the transfer
matrix are reviewed in Appendix \ref{app-a}. The next two Appendices 
introduce two successful theories of 
the transport in diffusive regime: the DMPK equation in App. \ref{app-dmpk} and
random matrix theory in App. \ref{app:rmt}. Lyapunov exponents of the product
of transfer matrices are introduced in Appendix \ref{app-le}. The last Appendix
\ref{app-b} discusses numerical algorithms for calculation of the conductance.

\smallskip

Various aspects of the electronic transport in disordered systems were reviewed recently.
The weak localization effects are discussed in 
\cite{Bergman}. Refs. \cite{Thouless-2,LR,Janssen} review the main ideas of the localization
theory. 
Experimental and theoretical aspects of the localization are reviewed  in \cite{McKK-93}.
The Quantum Hall effect is discussed in refs. \cite{Hall,KOK}. Supersymmetric field theory and its application
to electron transport is explained  in \cite{Efetov}.
Statistical properties of spectra and wave functions are reviewed in
\cite{Mirlin}.  Random matrix theory and its application to
electronic transport is reviewed in
\cite{Beenakker}.  The development of research is in conference proceedings
\cite{KramerSchoen,Kramer,Brandes}.

Wave transport in disordered media is a subject of many textbooks and monographs. 
We want to mention the classical book of Economou \cite{Economou} and Mott and Davis \cite{MD}.
Transport of electrons in mesoscopic systems is discussed in \cite{Imrybook,Datta,ping}.
The book of Mehta \cite{Mehta} presents the theory of random matrices.

\section{Localization}\label{sec:loc}

Localization of an electron in disordered system was predicted by Anderson \cite{Anderson-58}.
Anderson calculated the probability $p$ that an electron, being at time $t=0$ at point 
$\vec{r_0}$  with the wave function $\Psi(\vec{r},t=0)=\delta(\vec{r}-\vec{r_0})$
returns to the  same point in time $t>0$. It is evident that  $p=0$ for regular lattice,
since the  electron propagates freely, provided that its energy lies in the allowed energy band
The question is whether  there are  such  disordered systems, 
where $p\to 1$ in the limit $t\to\infty$. If yes, then the electron is localized
in the space  around $\vec{r_0}$.
If not, then electron can propagate through the sample,
in spite of the disorder.

\begin{figure}[t!]
\begin{center}
\includegraphics[clip,width=0.45\textheight,height=0.45\textheight]{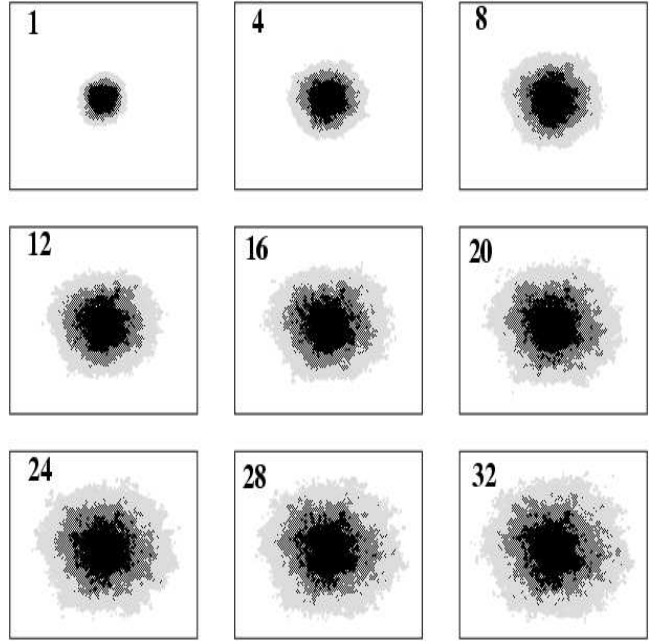}
\end{center}
\caption{The time dependence of the wave packet in the two dimensional disordered system
defined by Hamiltonian (\ref{ham}) with 
Box disorder and  $W=6$. This disorder corresponds to the
localization length $\lambda\approx 50$ \cite{SAD}.
Different colors  distinguish points where the wave function
 $|\Psi(\vec{r})|>0.0001$ (gray), $|\Psi(\vec{r})|>0.0010$ (dark gray),  and
 $|\Psi(\vec{r})|>0.0050$ (black) at different times shown in legend.
The time is measured in units of $100\hbar/V$ with hopping term $V=1$. The  
size of the lattice is $512\times 512$ (in units of lattice spacing $a$). 
The initial wave function, $\Psi(\vec{r},t=0)$, is  the eigenfunction of the
sub-lattice of the size $N_h\times N_h$ ($N_h=24$) of the system,
located in the middle of the lattice. The eigenfunction  belongs to
the eigenenergy  closest to the band center, $E=0$.
From the time development of the wave packet,
one can see that the electron is indeed  localized in the center of the lattice.
}
\label{fig-sche-yy}
\end{figure}

To demonstrate Anderson's idea,  
we simulated  numerically the time evolution of a single
electron wave function in two-dimensional disordered lattice.  
At time $t=0$, we add the electron to the center of the lattice and 
solve the time-dependent Schr\"odinger equation,
\be\label{time-sche}
\displaystyle{i\hbar\frac{\partial \Psi(\vec{r},t)}{\partial t}}={\cal H}\Psi(\vec{r},t)
\ee
where ${\cal H}$ is the \textsl{Anderson Hamiltonian}, 
\be\label{ham}
{\cal{H}}=W\sum_r \eeps_r c_r^\dag c_r+V \sum_{[rr']}c_r^\dag c_{r'}.
\ee
Following Eq. (\ref{ham}),
the electron moves on the $d$-dimensional lattice of size $L^d$.
The first sum is a local term, where $\vec{r}$ counts
sites on the lattice 
and the disorder is given by random on-site energies
$\eeps_r$.
The second term in Hamiltonian enables the hopping of electron between the nearest neighbor sites.
Parameter $V$ is given by the overlap of the wave functions localized on neighboring sites, and $W$
defines the strength of the disorder. The distance between two neighboring sites,
$a$, is used as a unit length, $a=1$.

\begin{figure}[t!]
\begin{center}
\includegraphics[clip,width=0.45\textheight,height=0.45\textheight]{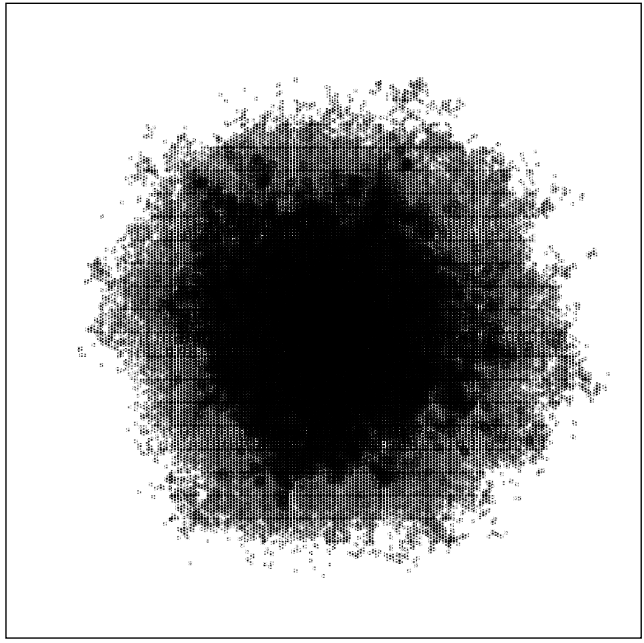}
\end{center}
\caption{The wave function of the electron in time $t=2800~\hbar/V$. 
The size of the lattice is $512\times 512$ (in units of lattice spacing, $a$).
 Note that the radius  of occupied region is much larger
than the localized length, which is in this case  $\lambda\approx 50$.
The localization length determines only the exponential decrease
of the wave function at large distances, not the size of the region in which the electron
is localized.}
\label{fig-sche-y1}
\end{figure}

Figure  \ref{fig-sche-yy} shows the time development of the electron wave function
in the strongly disordered two dimensional (2D) lattice.
At the beginning, 
the electron wave function  broadens with time. However, later 
  this  broadening  becomes slower
and finally stops. The electron is
localized in the  central part of the lattice and  its  wave function far 
away from the localization center is  negligibly small. 
Figure \ref{fig-sche-y1} shows the detailed  spatial distribution of the wave function
at time $t=2800V/\hbar$. 

The presented data show that indeed the electron might be localized
for sufficiently strong disorder. Anderson calculated the
critical strength of the disorder, $W_c$, such that the electron is delocalized 
(and the system is a metal) when $W<W_c$, and the electron is localized  and the system becomes an
insulator when $W>W_c$.
The transition of the system from the metallic to insulating regime due to increase
of the disorder is called metal-insulator transition.

We will not reproduce Anderson's analysis here, only present his main result:
there is indeed the critical disorder which separates localized and delocalized 
electron states. In the most simple approximation, critical disorder can be
found as
\be\label{wc-anderson}
\displaystyle{\frac{W_c}{V}}\approx 2eK\ln(eK)
\ee
where $K$ is a connectivity of the lattice. A detailed discussion of Anderson analysis is given 
in Refs. \cite{Ziman}.

Note, relation (\ref{wc-anderson}) does not contain the dimension $d$  of the system.
Later \cite{Wegner-76} it became clear that, similarly  to critical phenomena in statistical physics, the
dimension of the system is crucial for the existence of the Anderson transition. 
The dimension $d=2$ is a lower critical dimension; there is no metallic state for
$d<2$.  That means, all electron states are localized in space when $d< 2$.
However, localization can be observed only when  the system is large enough,
\be\label{lockrit}
 L\gg\lambda.
\ee
In 2D, the localization length is extremely large for weak disorder \cite{McKK-1983}.
This is the  reason  why good metallic behavior  is observed in
numerical simulations performed on 2D weakly disordered samples of finite size, $L$.
(Sect. \ref{sect-diff}).

Analytical  estimation of the critical disorder  is possible only when the dimension
of the system is close to 2, $d=2+\veps$ with $\veps\ll 1$.
In realistic three dimensional systems, $W_c$ can be obtained only numerically.
Various methods of calculation of $W_c$ and of other critical parameters
will be explained in Sect. \ref{section:nsa}.

\section{Models and symmetries}\label{sect:model}

The most suitable model for the numerical analysis  of the localization
is Anderson's model, defined by the Hamiltonian (\ref{ham}).
It  represents an isotropic model, in which
the hopping term, $V$, is the same in all directions.
In the isotropic models, we consider $V=1$. 
This also defines the scale of the energy.

It is often suitable to consider anisotropic  models, with different hopping
terms in different directions. For instance, we will use
the three dimensional  (3D) model 
\be\label{hama}
{\cal{H}}=W\sum_r \eeps_r c_r^\dag c_r+
t \sum_{[rr']}c_{xyz}^\dag c_{x'yz}+
t \sum_{[rr']}c_{xyz}^\dag c_{xy'z}+
V \sum_{[rr']}c_{xyz}^\dag c_{xyz'}
\ee
where $t<V$. The hopping term, $V$, defines the scale of the energy. We often use $V=1$.

Random energies $\eeps_n$ are distributed either with the Box distribution, 
\be\label{dis-box}
P_B(\eeps)=(2/W)\Theta(W/2-|\epsilon|)
\ee
or with the Gaussian distribution, 
\be\label{dis-gauss}
P_G(\eeps)=(2\pi W^2)\exp -(\epsilon^2/2W^2).
\ee
The parameter $W$ measures the strength of the disorder. Other distributions
of random energies (binary, Lorentzian) are often 
used in the literature too.

The Anderson model defined by Eq. (\ref{ham}) will be used in the present paper to demonstrate
numerically the basic ideas of the localization theory. 
For zero disorder, $\eeps_r\equiv 0$, we easily find all the eigenenergies
of Hamiltonian (\ref{ham}).
For instance, for the 3D anisotropic model we have
\be
E=2t\cos k_x+2t\cos k_y + 2V\cos k_z.
\ee
The energy band spans between $-V-2t$ and $V+2t$. We define the bandwidth,
\be
B=2V+4t.
\ee

Anderson's model, given by Hamiltonian (\ref{ham}), belongs to models with
time reversal symmetry. It describes the propagation of
a single \textsl{spin-less} particle in the disordered lattice.  
Such models are called \textsl{orthogonal}.
Systems with \textsl{orthogonal} symmetry do not exhibit the metal-insulator transition
in  dimension $d_c=2$.
This changes when the hopping between neighboring sites becomes  dependent
on the orientation of the spin of the electron
\cite{Wegner-76}. 
Evangelou and Ziman \cite{EZ}, and Ando \cite{Ando-89} showed numerically 
that the two dimensional (2D)  systems with spin dependent hopping
(called \textsl{symplectic} models) 
exhibit the  metal-insulator transition already at $d_c=2$.
In this paper we will discuss 
the Ando model which  is defined  by the Hamiltonian
\be\label{hamando}
{\cal{H}}=W\sum_r \eeps_r c_r^\dag c_r+
V_x \sum_{[rr']}c_{xy}^\dag c_{x'y}+
V_y \sum_{[rr']}c_{xy}^\dag c_{xy'}
\ee
with spin dependent hopping terms given by Ando \cite{Ando-89},
\begin{equation}\label{hamando-1}
V_x=\left(
\begin{array}{rr}
 c  & s\\
-s & c 
\end{array}
\right),
~~\textrm{and}~~~V_y=\left(
\begin{array}{rr}
 c  & -is\\
-is & c 
\end{array}
\right),
\end{equation}
with $c^2+s^2=1$.
Note how hopping depends on the direction of propagation.
Also, the wave function, $c_r$, has  two components,
\be
c_r=\mv{c_{r\uparrow}}{c_{r\downarrow}},
\ee
for different orientations of the spin of electron.
Another 2D model with \textsl{symplectic} symmetry is the 
Evangelou-Ziman model. In this model, the  hopping term has the form
\be
V=\m2{1+ivt_z}{-v(t_y-it_x)}{v(t_y+it_x)}{1-ivt_z},
\ee
where $t_x$, $t_y$ and $t_z$ are real  random variables chosen from box distribution,
and  $v$ measures the strength of (random) spin-orbit hopping.
In Refs. \cite{ASD}, SU(2) model is analyzed,   with the  random hopping term
\be\label{su2}
V=\m2{e^{i\alpha}\cos\beta}{e^{i\gamma}\sin\beta}{e^{-i\gamma}\sin\beta}{e^{-i\alpha}\cos\beta},
\ee
where random phases $\alpha$ and $\gamma$ are uniformly distributed in the interval
$(0,2\pi)$ and the distribution of random phase $\beta$ is $p(\beta)d\beta=\sin(2\beta)d\beta$
for $0\le \beta\le \pi/2$.

An external magnetic field breaks the time reversal inversion of the electron
propagation. Such systems  
 belong to the \textsl{unitary} universality class. 
Magnetic field can be  added to the two dimensional
Hamiltonian (\ref{ham}) by Peierls factor,
\be\label{ham-peierls}
\begin{array}{lcl}
{\cal H}&=&\sum_r \eeps_r c^\dag_r c_r+
\sum_{xy} c^\dag_{x,y+a}x_{xy}+
\sum_{xy} c^\dag_{x,y-a}x_{xy}\\
 & &+\sum_{xy} e^{+i\phi y} c^\dag_{x+a,y}c_{xy}+
\sum_{xy} e^{-i\phi y} c^\dag_{x-a,y}c_{xy}
\end{array}
\ee
where 
$\phi=B(ea)^2/\hbar$ 
is a  magnetic flux through the elementary
plaquette of the size $a^2$. 

Hamiltonian (\ref{ham-peierls}) is used for the numerical analysis of the
critical quantum Hall regime of 2D disordered system in a strong magnetic field.
Although this model  does not exhibit metal-insulator transition, there is, in each Landau band, 
the critical energy, $E_c$, at which the electron is delocalized \cite{Hall}. The existence
of the critical - delocalized - state inside  
the energy band is crucial for the transmission of the 
system between two neighboring quantum Hall plateaus \cite{Huck}. Another model, often used in the
numerical analysis of the critical quantum Hall regime is the model of Chalker and
Coddington \cite{CCh,KOK}.

\section{Anderson transition}\label{sect:AT}

The Anderson transition is the transition from the metallic (extended) regime 
to the localized regime.  Before describing the scenario of the Anderson
transition, we must understand how the disorder influences the density
of electronic states in the conducting band. Then, we introduce
the mobility edge, describe qualitatively the transition
from metal to insulator, and define critical exponents.

\subsection{The density of states.}\label{sec:density}

\begin{figure}[t!]
\begin{center}
\psfrag{re}{$\!\!\!\!\!\!\rho(E)$}
\psfrag{rE}{$\!\!\!\!\!\!\rho(E)$}
\includegraphics[clip,width=0.3\textheight,height=0.25\textheight]{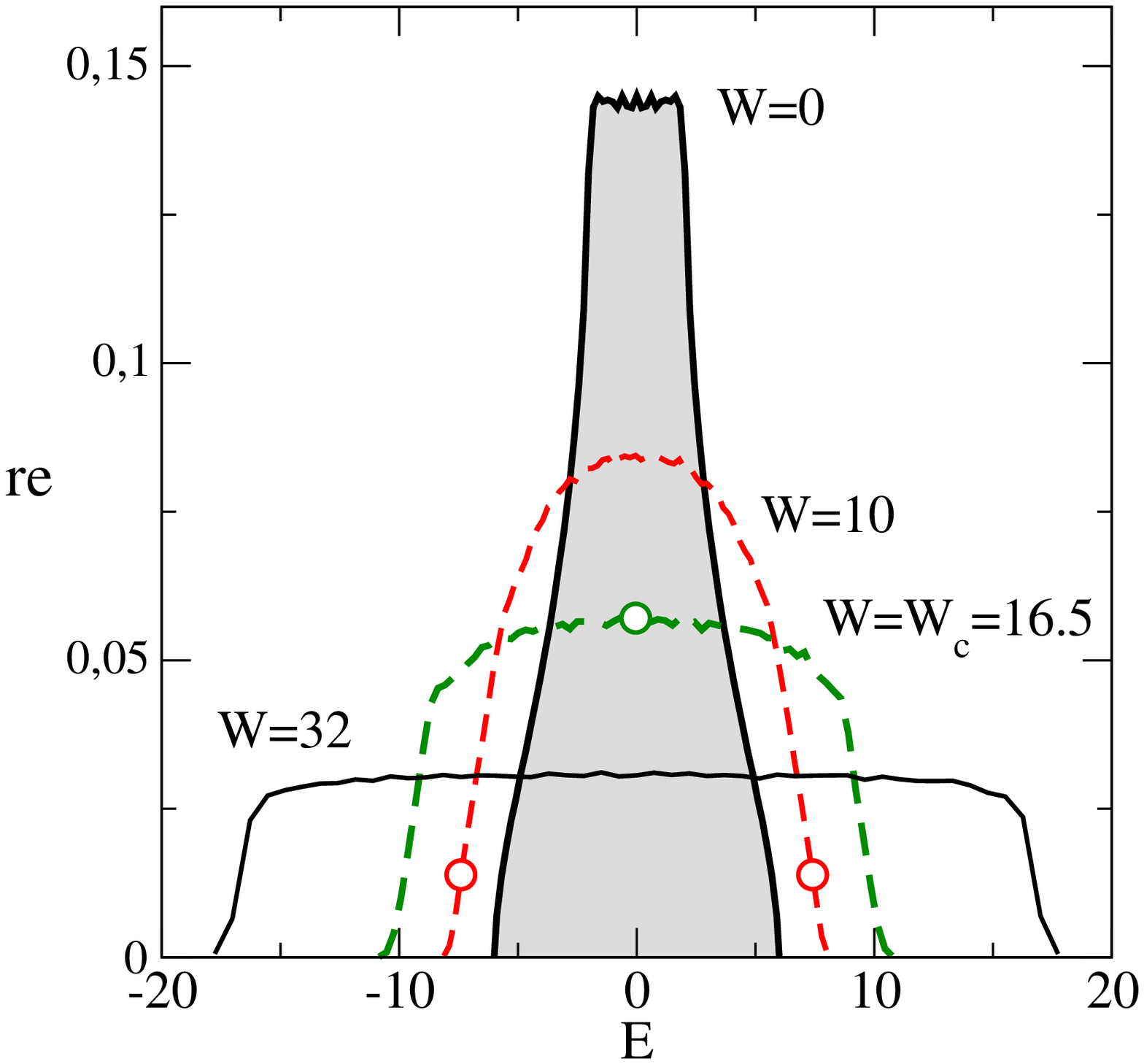}~~~~~
\includegraphics[clip,width=0.3\textheight,height=0.25\textheight]{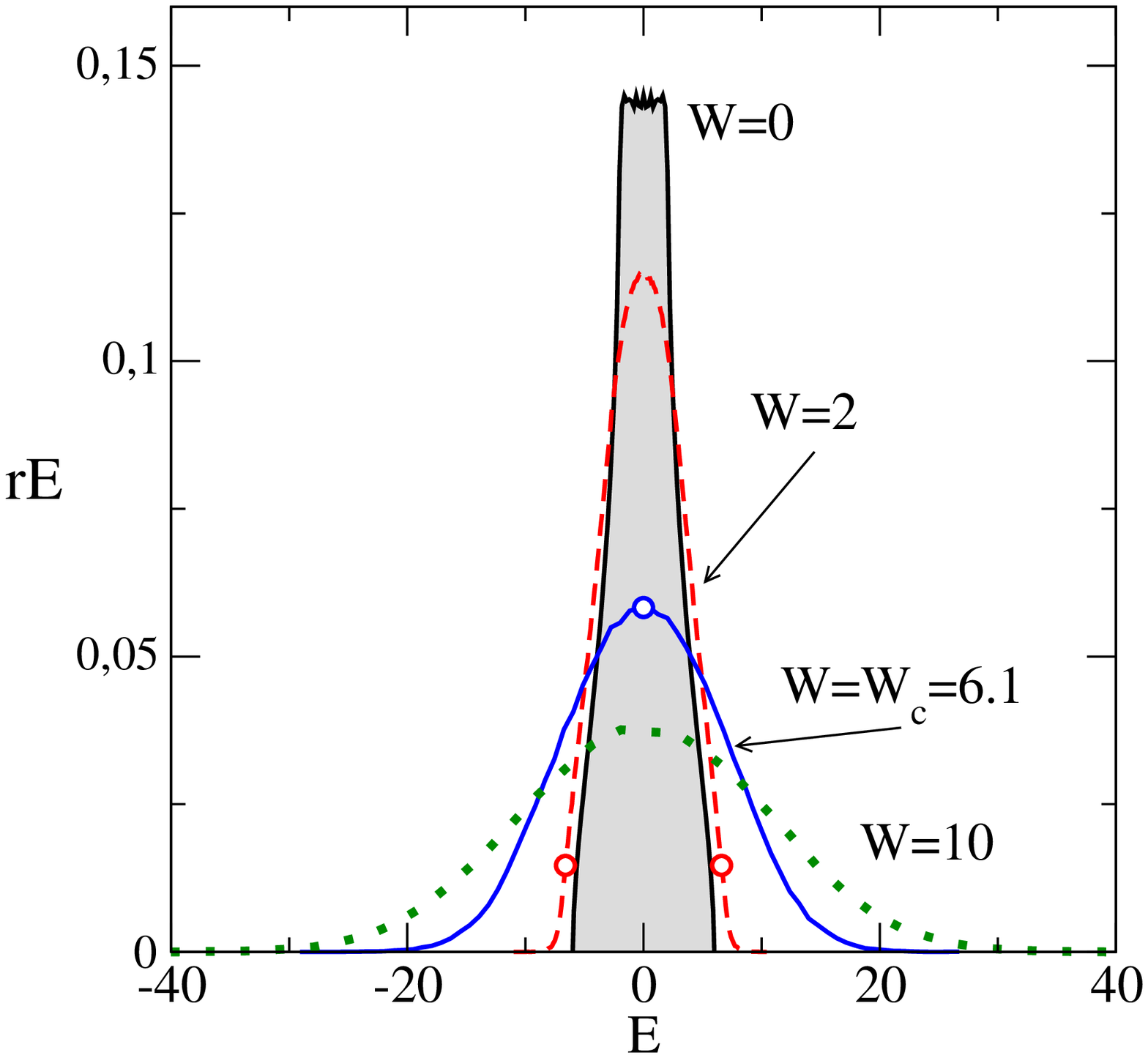}
\end{center}
\caption{The density of states, $\rho(E)$,
of  the three-dimensional Anderson model with box (left) and Gauss (right) 
distribution of random energies $\eeps_n$,
defined by Eqs. (\ref{dis-box}) and (\ref{dis-gauss}), respectively.
As expected, the energy band becomes broader when disorder 
increases. In the tails of the band, 
the electronic states become localized (Fig. \ref{fig-edge}).
Open circles show the position of three critical points, one
for $E=0$ and $W=W_c$, and two other for fixed disorder $W<W_c$ 
and the mobility edge $E_{c1}=-E_{c2}$. Note, for weak disorder ($W=10$ in the left figure)
 the mobility edges lie outside the unperturbed energy band, $|E_{c}|>B$.
}
\label{fig-rho}
\end{figure}

For zero disorder, $W=0$, the energy of the electron in the Anderson model is given by
the dispersion relation
\be\label{density-1}
E=2V\cos k_x+2V\cos k_y+2V\cos k_z,
\ee
where $\vec{k}=(k_x,k_y,k_z)$ is the wave vector of electron.
The density of states,
\be\label{eq-rho}
\rho(E)=\displaystyle{\frac{1}{2\pi}\frac{\partial k}{\partial E}}.
\ee
can be  calculated analytically \cite{Economou}, and is shown in  Fig. \ref{fig-rho}.
The energy band spans from $-6V\le E\le +6V$ and is symmetric, $\rho(E)=\rho(-E)$.
This symmetry is typical for tight-binding Hamiltonians given by Eq. (\ref{ham}).

Note that the density of states, given by Eq. (\ref{eq-rho}) does not include the degeneracy
of the electron system due to the two possible orientations of the spin.
When spin of the electron is included, $\rho(E)$ must be multiplied by a factor of 2.

Figure \ref{fig-rho} demonstrates the effect of disorder on the energy spectrum of the
three dimensional Anderson model for Box and Gaussian disorder.
As expected, the band becomes broader when  disorder increases.  This is more pronounced for
the Gaussian disorder which allows larger fluctuations of random energies.
 Open circles indicate the position of the mobility edge,
discussed in the next Section.

\begin{figure}[t!]
\psfrag{rE}{$\!\!\!\!\!\!\rho(E)$}
\begin{center}
\includegraphics[clip,width=0.3\textheight]{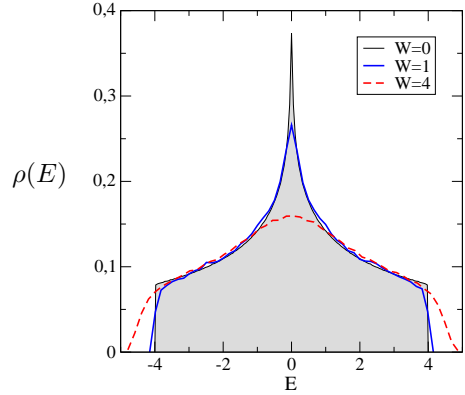}
\end{center}
\caption{The density of states for the disordered two dimensional Anderson model.
Note the typical singularity of the density of states 
at the band center for the unperturbed tight binding
Hamiltonian ($W=0$) \cite{Economou}.
}
\label{rho-2d}
\end{figure}

For completeness, we also show in Fig. \ref{rho-2d} the  density of states of the disordered 
two dimensional Anderson model.

\subsection{Mobility edge}\label{me}

\begin{figure}[t!]
\begin{center}
\includegraphics[clip,width=0.25\textheight]{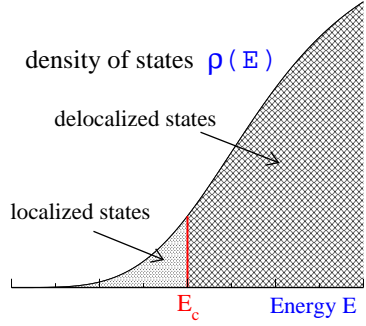}
\end{center}
\caption{The mobility edge, $E_c$, separates the localized states in the band tail from
the conducting states. By changing the Fermi energy, $E_F$, the system exhibits a transition
from the metallic regime ($E_F>E_c$) into the localized regime ($E_F<E_c$). 
}
\label{fig-edge}
\end{figure}

Intuitively, we expect that the  localized states appear first 
in tails of the energy band. These states are namely created by large random energies, $\eeps_r$.
One expects that the states with the energy close 
to the band center are less affected by randomness.
Another argument for the localization in tails was given by Lifshitz   \cite{Lifshitz}.
The classical scattering of electrons on impurities, which led to an expression for the conductance,
given by Eq. (\ref{v-1}), is only possible when the mean free path, $\ell$, is much larger than
the wavelength of the electron, $\lambda_F$, given as the  inverse of the Fermi wave vector, $k_F$.
When both lengths become comparable,
\be\label{lifshitz}
\lambda_F\sim \ell,
\ee
then the spatial extent of the electron wave function is comparable with the
typical distance between two impurities, so that we cannot discuss the propagation
of electron in terms of individual scattering procedures.  The electron wavelength, $\lambda_F$,
 is small, comparable with the lattice distance, $a$, at the band center, but increases with the
distance of the Fermi energy from the band center. Consequently, the Lifshitz criterion for the localization,
given by Eq. (\ref{lifshitz}) is fulfilled first in the band tails.

Separation of the energy spectra into localized and delocalized intervals is schematically shown 
in Fig. \ref{fig-edge}. For a given strength of the disorder, $W$,
there is an energy, $E_c$, called the \textsl{mobility edge}, which 
separates the localized states from the delocalized states.
When $E_F<E_c$, the system is an insulator, since all states at the Fermi energy are localized 
(we remind the reader that the temperature $T=0$).  
We call this insulator \textsl{the Anderson insulator} to emphasize the fact,
that the system possesses 
the zero electric conductance, although   the density of states is non-zero.
When the Fermi energy crosses the mobility edge, the system undergoes the transition
from insulator to metal. For $E_F>E_c$, the system possesses a finite electric conductivity.

Note that there are two mobility edges, $E_{c1}$ and $E_{c2}$ 
which separate localized states in the upper and lower 
band tails from the delocalized states in the central part of the band. 
For a particular disorder, we show the mobility edges in    Fig. \ref{fig-rho}.
Of course, the position of the mobility edges depends on the strength of the disorder.
With increasing disorder, both mobility edges  move to the band center. 
There is a critical value of the disorder, $W_c$, for which $E_{c1}=E_{c2}$ approach the band 
center. For $W>W_c$ there are no propagating states in the spectrum.
The function  $E_c=E(W)$ defines the critical line, which separates 
the metallic and the localized states
in the phase diagram.
An interesting feature of the Anderson transition is that  the disorder 
creates propagating  states \textsl{outside} the unperturbed energy band.  
We see that for non-zero disorder, the metallic regime
exists also for energies $|E|>6V$.

\begin{figure}[t!]
\begin{center}
\includegraphics[clip,width=0.4\textheight]{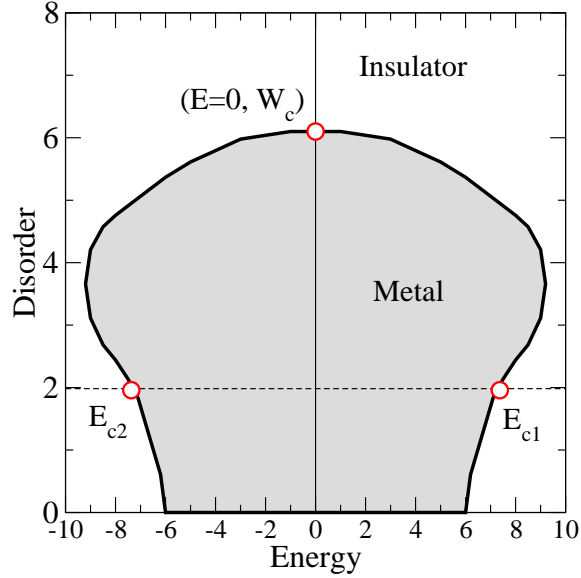}
\end{center}
\caption{Schematic phase diagram for the 3D Anderson model with the Gaussian disorder.
Metallic (conductive) states are separated from the localized states by the critical line,
which is an envelope of the shadow region.
Open circles show the position of three critical points, one
for $E=0$ and $W=W_c=6.1$, and two other for fixed disorder $W=2<W_c$ 
and the mobility edge $E_{c1}=-E_{c2}=6.58$. Note, these mobility edges lie
outside the unperturbed energy band. When disorder reaches its critical value, $W=W_c$,
two mobility edges reach the band center, $E_{c1}=E_{c2}$. No metallic states exist
when $W>W_c$.
}
\label{fig-schema-pom}
\end{figure}

The phase diagram of the three dimensional Anderson model is shown 
schematically in Fig. \ref{fig-schema-pom}.
For $W=0$, all the electron states inside the energy band, $|E|<6V$
are delocalized. Increasing disorder
broadens the density of states, and the interval $E_{c1},~E_{c2}$ of the metallic states
becomes broader, too.  Only for rather strong disorder, close to 
to the critical disorder, $W_c$,
the mobility edge starts to converge towards the band center.
For the disorder $W>W_c$, all electron states are localized.
The \textsl{phase diagram} 
for the 3D Anderson model was calculated numerically in Ref.  \cite{bulka}.

\subsection{Critical exponents}

The metallic states are separated from the localized ones by the critical line.
Any crossing of the critical line is accompanied by the metal insulator 
transition. The common belief,
supported by the scaling theory of localization \cite{AALR}
is that this transition is universal. This means that
the transition from metal to insulator should not depend on
microscopic details of the model,
and on the position of the critical point, lying at the critical line.  
However, it depends on the dimension and on the physical symmetry of the model.
We can, 
similarly to the theory of phase transitions,  formulate the theory
of the metal insulator transition in terms of the order parameter and of the critical 
exponents \cite{Wegner-76}. The critical exponents are defined in terms of the energy
or disorder dependence of the metallic conductivity and localization length.

Metallic conductivity, $\sigma$, characterizes the transport properties
of the metallic regime. We have non-zero conductivity
$\sigma>0$ in metal, and $\sigma$
decreases when the system  approaches the critical point. For many years it was
not clear whether  the conductivity
\textsl{at the critical point}, $\sigma_c$, is zero or not.
Mott \cite{Mott,MD}  suggested that the conductance possesses
non-zero value at the critical point and drops discontinuously to zero
on the insulating side of the transition. At present, the common belief is that
the critical conductivity $\sigma_c=0$ in 3D orthogonal systems. 
The decrease of the conductivity in the neighborhood of the critical
point is determined  by the critical exponent $s$,
\be\label{mit-1}
\sigma\sim (W_c-W)^s
\ee

\begin{figure}[t!]
\begin{center}
\includegraphics[clip,width=0.25\textheight]{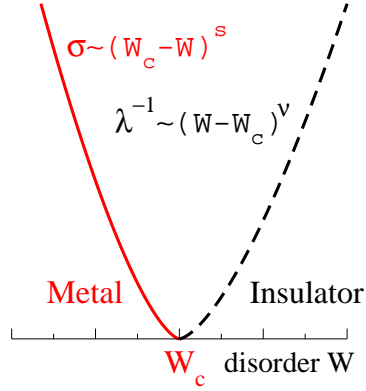}
\end{center}
\caption{Definition of the critical exponents. At the metallic side of the critical point,
the conductance $\sigma$ decreases to zero as $\sigma\propto (W_c-W)^s$, whereas
at the insulating side the  localization length diverges as $\lambda\propto (W-W_c)^{-\nu}$.
}
\label{fig-wc}
\end{figure}

On the insulating side of the transition, we can characterize the
electronic states by the localization length, $\lambda$.
By definition, the localization length which characterizes
the exponential decrease of the localized wave function far away
from the localization center $\vec{r_0}$, is given by Eq. (\ref{i-2}),
\be
\Psi(\vec{r})\propto \exp -\ds{\frac{|\vec{r}-\vec{r}_0|}{\lambda}}.
\ee
It is intuitively clear that $\lambda$ should increase when system approaches
critical point, and become infinite at the critical point,
\be\label{mit-11}
\lambda\sim(W-W_c)^{-\nu}.
\ee
As discussed in the Sect. \ref{sec:density}, the
Anderson transition is induced either by a change of the Fermi energy
with fixed disorder strength, or by an increase in the disorder, $W$, for fixed
energy ($E=0$).  Therefore, we  expect also that
\be\label{mit-2}
\sigma\sim (E_c-E)^s~~~~~~~\lambda\sim(E-E_c)^{-\nu}
\ee
for the case when the Fermi energy crosses the mobility edge.

Scaling theory of localization \cite{AALR} assumes that
the metal-insulator transition is universal.  
That means that critical exponents,
$s$ and $\nu$, are universal, depending only on the dimension of the system, 
and on the physical universality class. 
They do not depend on the microscopic details of the model. 
The scaling analysis gives  \cite{Wegner-76}
\be
s=(d-2)\nu.
\ee
Proof of the universality of critical exponents and their dependence on the
dimension of the system represents the main problem of the theory of 
Anderson localization.

\section{Wave functions and energy spectrum of disordered systems}\label{WF}

For the case of a regular lattice ($W=0$), all  the  eigenenergies and eigenfunctions
of the Hamiltonian can be found analytically. Of course, disorder influences the
spectrum of eigenenergies, and also the form of the corresponding wave functions. 
In this Section, we describe briefly the main qualitative and quantitative
properties of the wave functions and of the energy spectra both in the limit
of weak and strong disorder, and discuss the non-homogeneous spatial
distribution of electrons in the critical regime.

\subsection{The wave function}\label{wavefunction}

Following  Anderson \cite{Anderson-58},  
we expect that an electron is delocalized if  disorder is small,
and localized if disorder is strong. To verify this assumption, we calculate
numerically all the eigenenergies and eigenvectors of the 2D Ando model, defined
by Eqs. (\ref{hamando},\ref{hamando-1}).
We will see later that the Ando model exhibits MIT for critical disorder 
$W_c=5.838$. Thus, for $W\ll W_c$ the system is in the metallic regime 
and we expect that the electronic wave function
is almost homogeneously distributed throughout the sample.
On the other hand,  the  electron should be localized in some small region.
if $W\gg W_c$.

\begin{figure}[t!]
\begin{center}
\includegraphics[clip,width=0.35\textheight]{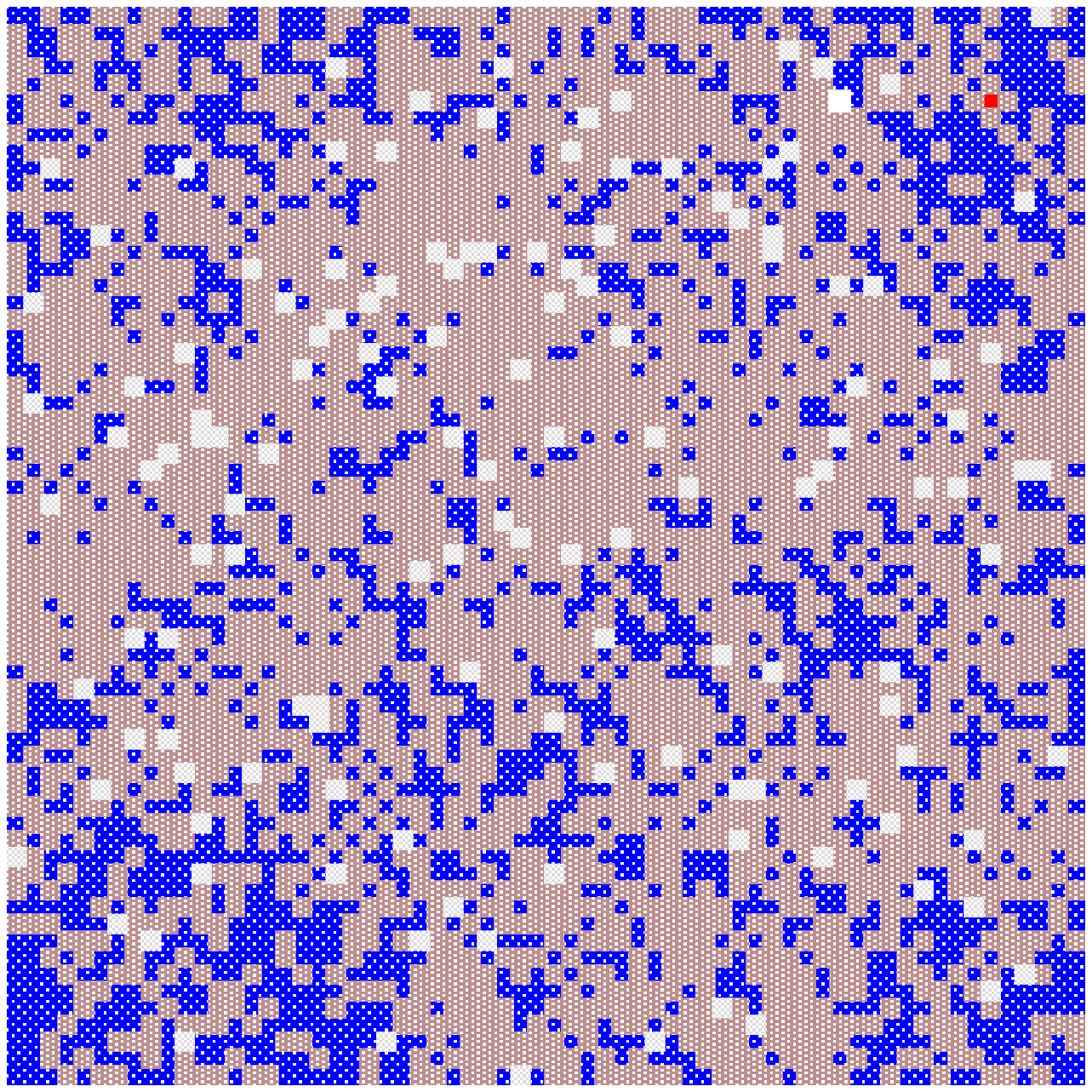}~~~~~
\includegraphics[clip,width=0.35\textheight]{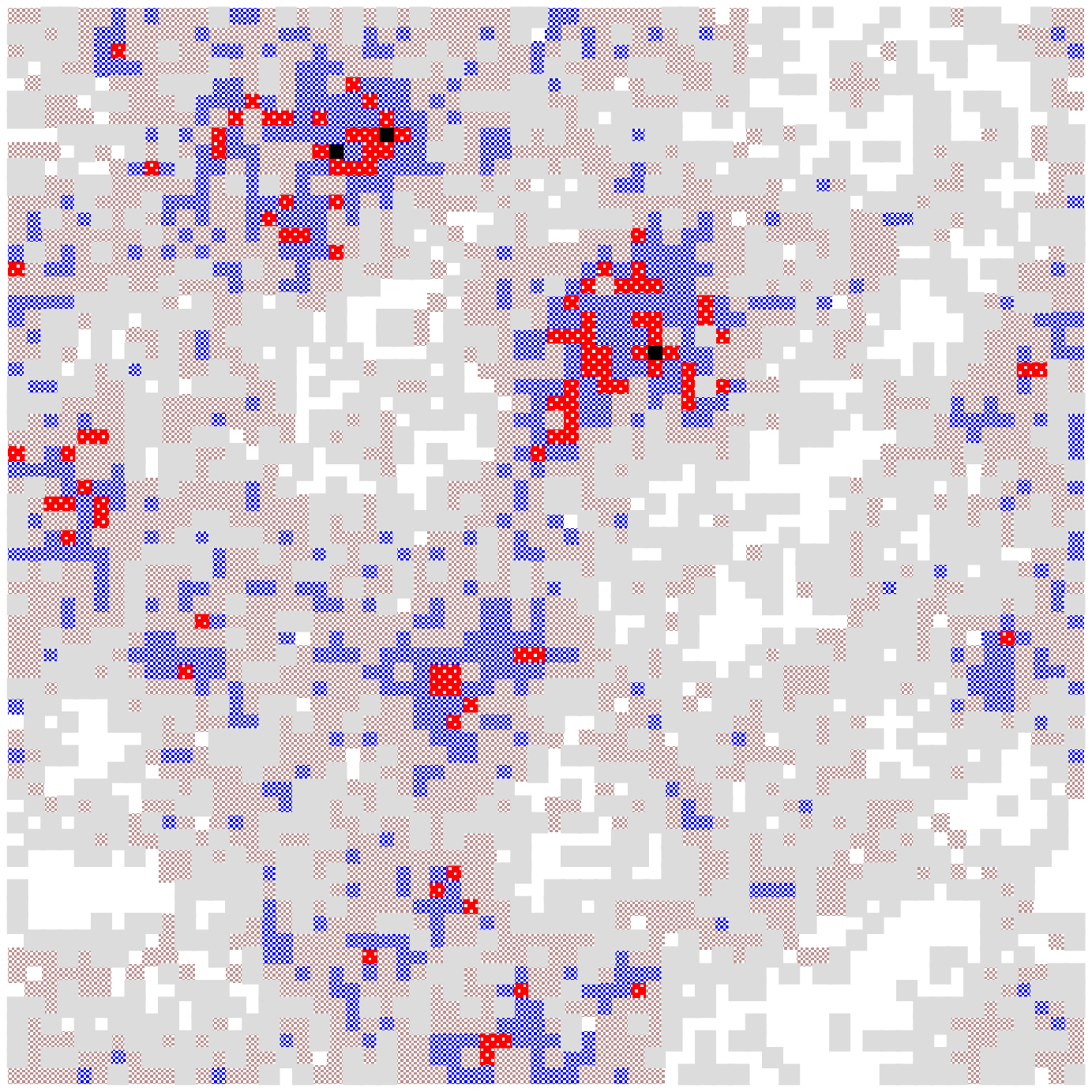}
\end{center}
\caption{The spatial distribution of the electron, given by the absolute 
value of the eigenvector, $|\Psi(\vec{r})|^2$, 
with the eigenenergy closest to the band center $E=0$
for  the two-dimensional Ando model. In the metallic regime (left, $W=2$), 
the electron is homogeneously
distributed in the system, while in the localized regime (right, $W=8$), 
the electron is localized
in a small part of the lattice and its wave function is very small elsewhere.
The size of the system is $64 \times 64$ lattice sites. The critical disorder
of the Ando model is $W_c=5.838$. 
The data was obtained by numerical diagonalization of the random 2D Hamiltonian (\ref{ham})
using the LAPACK subroutines.
}
\label{fig-ipr-ando}
\end{figure}

These expectations are confirmed by  numerical  results  presented in Fig.
\ref{fig-ipr-ando}  where we compare the spatial distributions of two electron 
eigenstates. The ``metallic'' eigenstate ($W=2\ll W_c$) is almost homogeneously 
distributed within the sample\footnote{Note, similar result can be obtained also in the 2D 
orthogonal systems, if the localized length is larger than 
the size of the system, $L\ll\lambda$.}. 
Also,  the localized eigenstate, calculated
numerically for the system with disorder $W=8\gg W_c$, is spatially localized in a certain
region of the sample, and is  orders of magnitude smaller elsewhere.

\begin{figure}[t!]
\begin{center}
\includegraphics[clip,width=0.35\textheight]{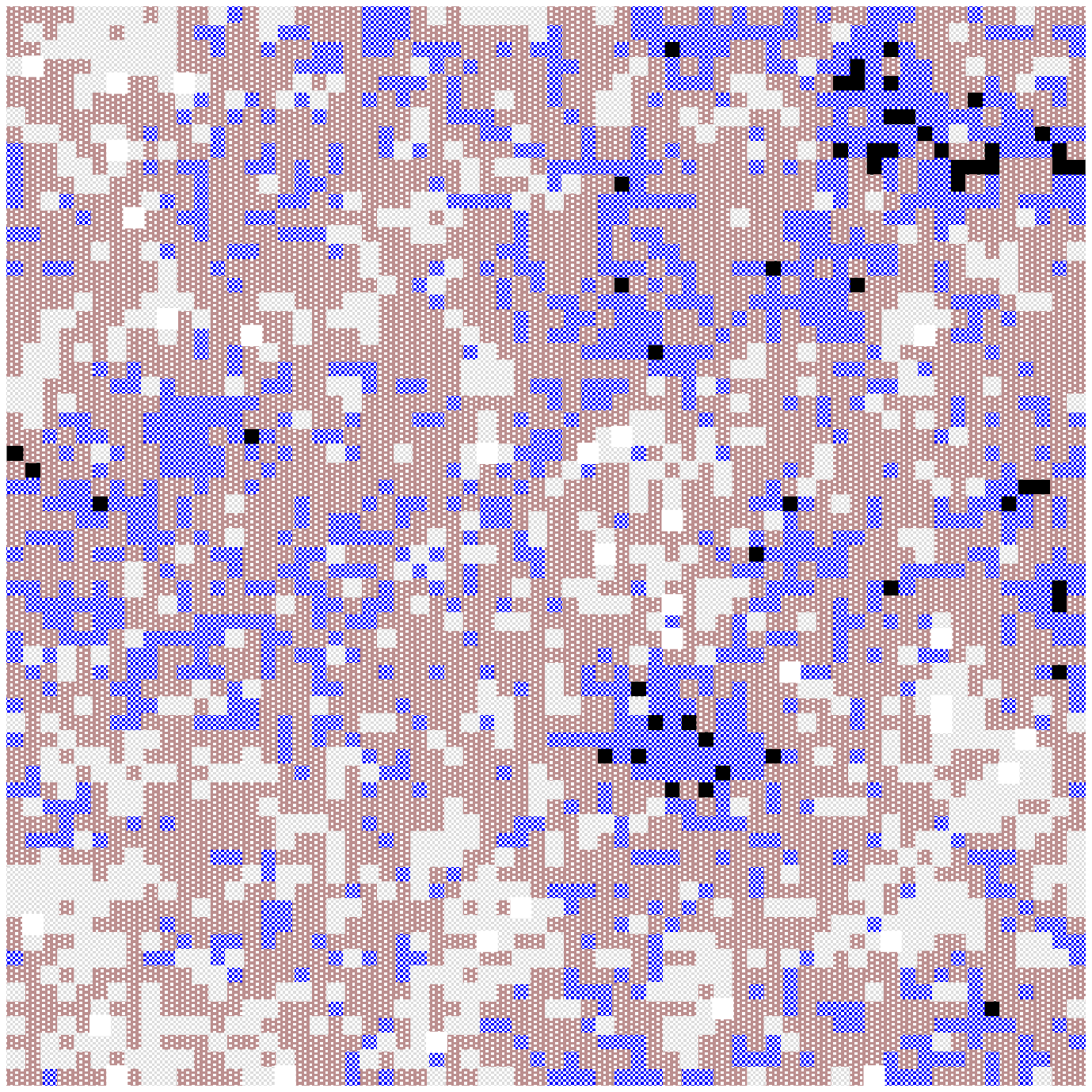}~~~~~
\includegraphics[clip,width=0.35\textheight]{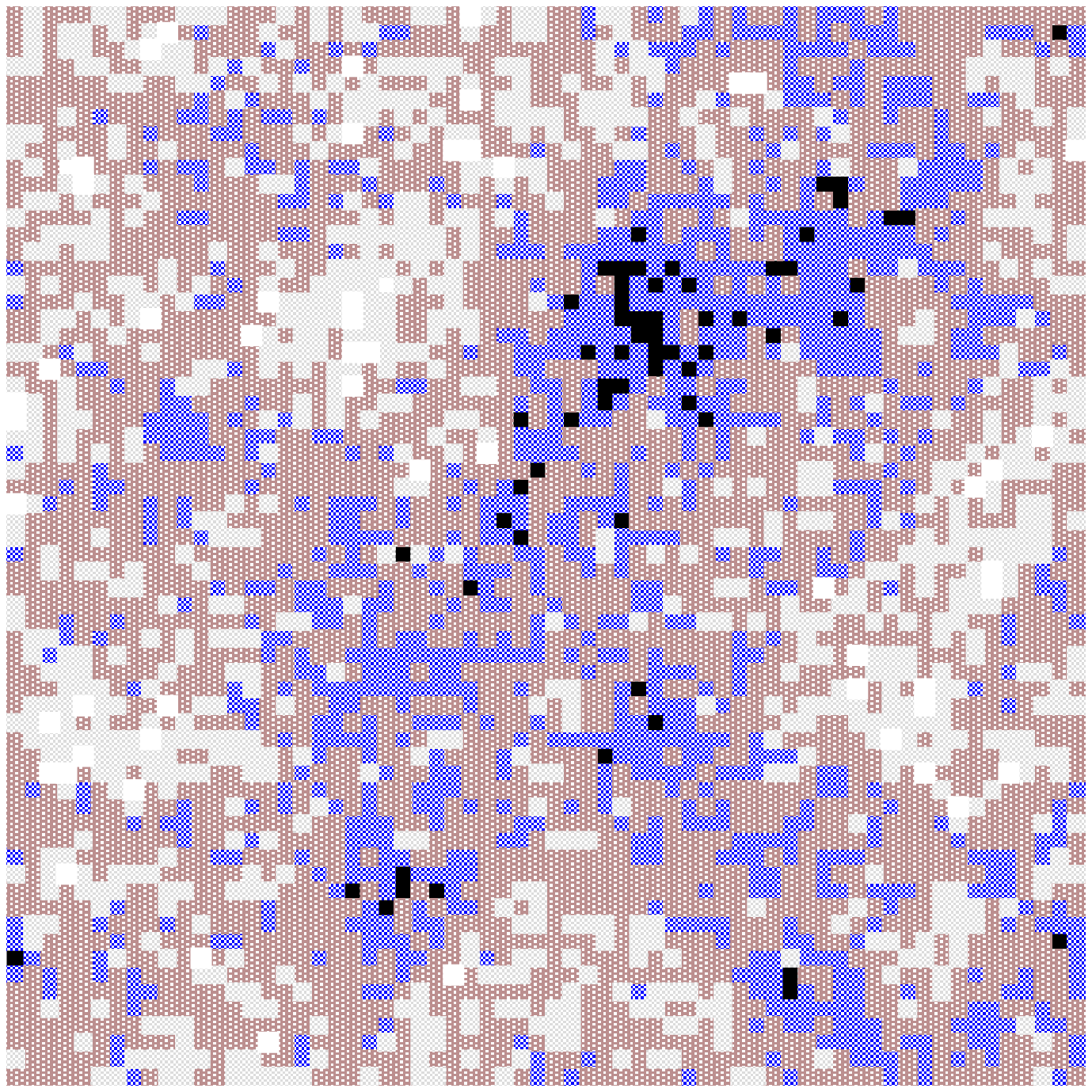}
\end{center}
\caption{Two electron eigenstates, $|\Psi(\vec{r})|^2$, 
in the critical regime of the  two-dimensional Ando model.
The size of the system is $64\times 64$, and $W=W_c=5.838$ for the electron energy $E=0$
\cite{Ludwig-1}.
For a given realization of disorder, we calculate two eigenvalues closest to the
band center ($E=-0.028$ and $E=-0.01476$ for the left and right figure, respectively).
The spatial distribution of the electron is more complicated
than in the metallic or localized regime. Detailed  analysis  of the wave function
(discussed in Sect. \ref{sect:2d}) shows that
the wave function  $\Psi(\vec{r})$ possesses  the \textsl{multi-fractal}
spatial structure.
}
\label{fig-ipr-ando-wc}
\end{figure}

\begin{figure}[t!]
\begin{center}
\includegraphics[clip,width=0.3\textheight]{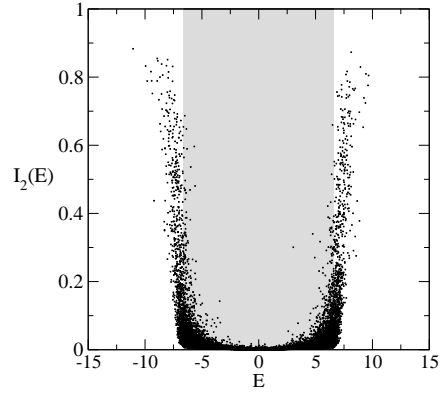}~~~~~~
\end{center}
\caption{The inverse participation ratio, $I_2$, as a function of energy for 
the 3D Anderson model with Gaussian disorder $W=2$. The size of the lattice is $L=16$.
The shaded area indicates propagating states, $|E|<|E_c|=6.58$ \cite{brndiar},
where   $I_2$ is very small. In the tails
of the energy band, the electronic states are localized and $I_2$ is close to 1.
Note, $I_2$ wildly fluctuates and see Fig. \ref{fig-ipr} for the probability distribution
of $I_2$.
}
\label{ipr-2}
\end{figure}

The critical point deserves special attention.
Figure \ref{fig-ipr-ando-wc} shows the electron wave function for two eigenstates
close to the band center for the Ando model with $W=W_c$.
Detailed analysis shows that the wave function at the critical point 
 possesses the \textsl{multi-fractal} spatial structure.  To define multifractality, we
introduce the quantities \cite{Mirlin}
\be\label{ipr}
I_q(E_n)=\sum_{\vec{r}} |\Psi_n(\vec{r})|^{2q}.
\ee
In Eq. (\ref{ipr}), $E_n$ is the eigenenergy of the Hamiltonian, and $\Psi_n(\vec{r})
$ is 
the corresponding eigenfunction.
It is evident that $I_1(E_n)\equiv 1$ for all eigenstates.  The quantity $I_2$ is the
\textsl{inverse participation ratio}, 
 often used for the characterization
of the eigenvectors in disordered systems (Fig. \ref{ipr-2}).

We can easily estimate the size dependence of the parameters $I_q$.
If the eigenstate $n$ belongs to the metallic part of the 
energy spectra, then the wave function is 
homogeneously distributed throughout the sample. Assuming that
$|\Psi_n(\vec{r})|\sim L^{-d/2}$ for all lattice sites $\vec{r}$,
we immediately see that in the metallic regime
\be\label{ipr-kov}
I_q\sim L^d\times \left[L^{-d/2}\right]^{2q}=L^{-(q-1)d}.
\ee

On the other hand, in the localized regime, we expect, in agreement with the right panel of
Fig. \ref{fig-ipr-ando},  that the wave function is
localized in a certain small region and is almost zero in the rest of the system.
Then, the wave function $|\Psi_n(\vec{r})|\sim 1$ in
the localization region and is almost zero 
otherwise. Inserting this Ansatz into the definition
(\ref{ipr}) we obtain that $I_q\sim 1$ for all $q>1$.

At the critical point,  the wave function is multifractal \cite{spiros-jpa}.
That means,  by definition, that
\be\label{ipr-crit}
I_q(E_n)\sim L^{-(q-1)d_q}
\ee
\cite{Mirlin}. In Eq. (\ref{ipr-crit}),  $d_q<d$ are 
\textsl{fractal dimensions}. Note, $d_q$ depends on $q$.

To understand the meaning of multifractality of the wave function,
note that  the wave function $|\Psi(\vec{r})|<1$, for all $\vec{r}$.
Therefore, the value of $|\Psi(\vec{r})|^{2q}$ decreases when $q$ increases.
Hence, the higher $q$ projects out such sites on the lattice, where $|\Psi(\vec{r})|$ is large.
For each value of $q$, these sites create a  fractal structure. Since different  $q$ 
projects different 
fractal structures, the  complete description
of the spatial distribution of the wave function requires 
the knowledge of $d_q$ for all values of $q$.

\begin{figure}[t!]
\begin{center}
\includegraphics[clip,width=0.3\textheight]{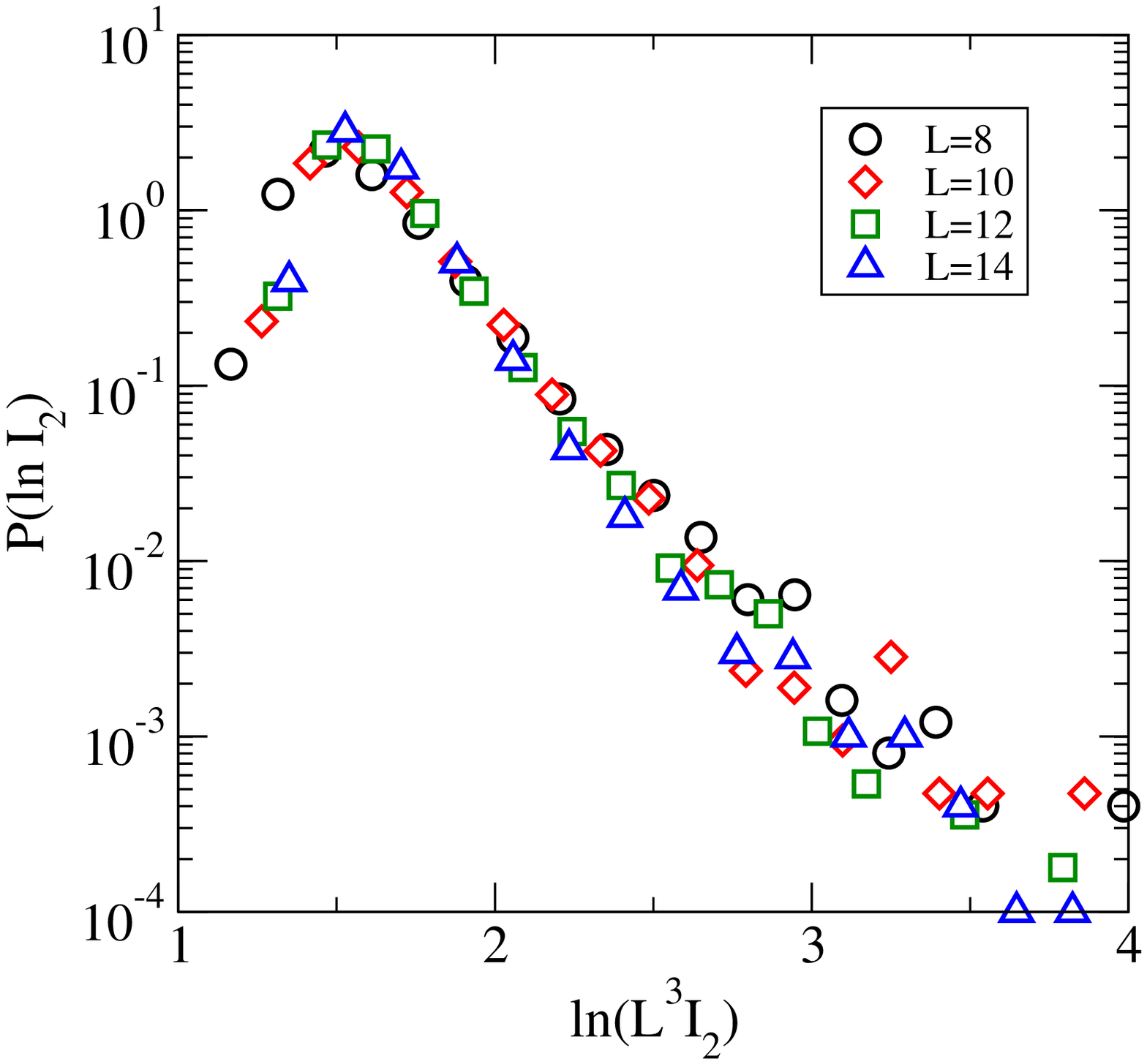}~~~~~~
\includegraphics[clip,width=0.3\textheight]{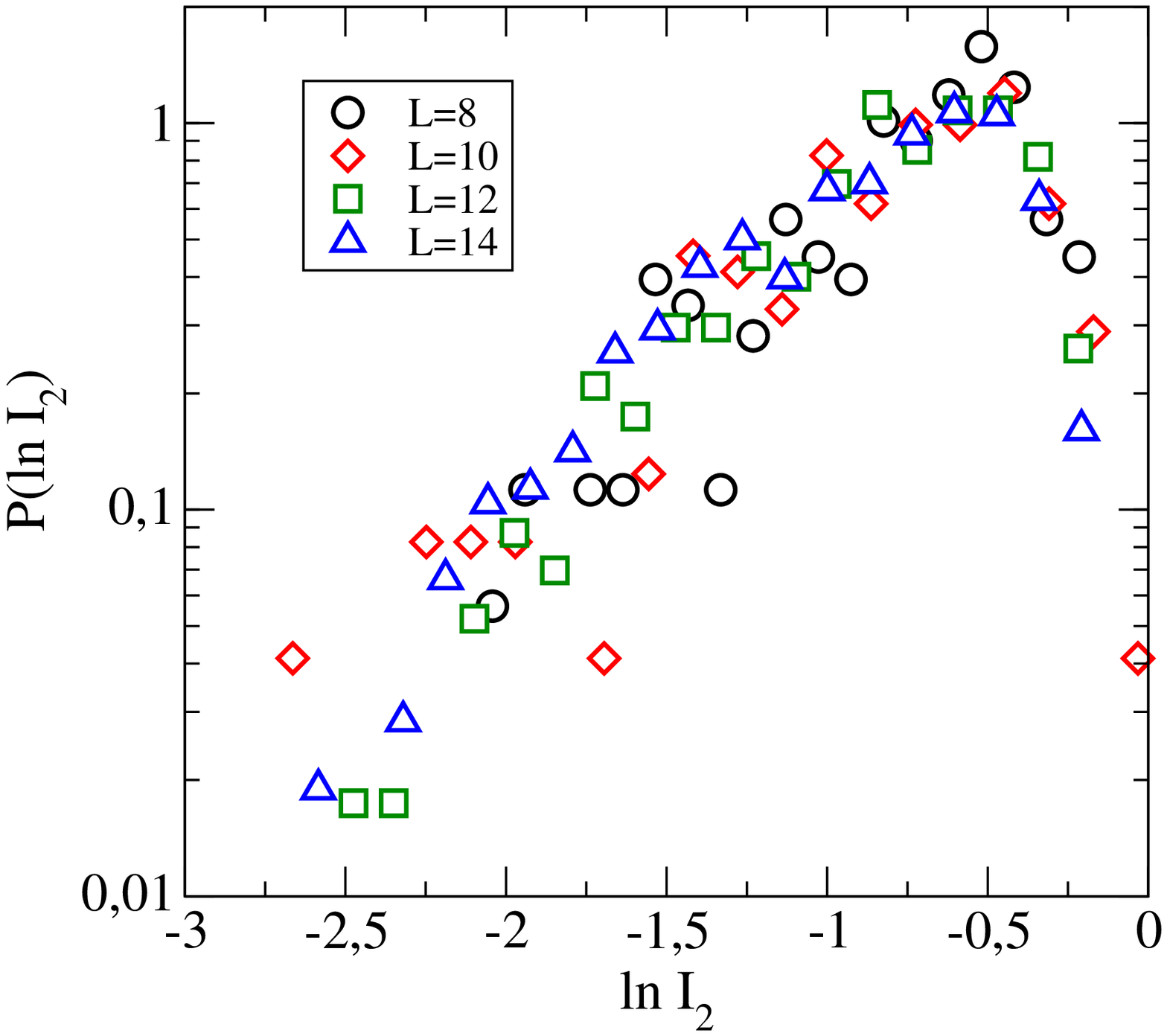}~~~~~~
\end{center}
\caption{The probability distribution of the \textsl{logarithm} of the 
inverse participation ratio, $\ln I_2$,
defined by Eq. (\ref{ipr}) 
for the 3D  Anderson model with Gaussian disorder $W=2$
and for system size given in the legend.
Left:  data for all eigenstates with energies $|E_n|<0.2$ (band center). 
From the phase diagram, shown in Fig. \ref{fig-schema-pom}, we know that
these eigenstates are metallic. We indeed see that the typical values of $I_2$
decrease as $L^{-3}$, in agreement with Eq. (\ref{ipr-kov}). 
The distribution decreases exponentially for large values of $I_2\gg L^{-3}$
indicating that there are no insulating states in the metallic phase.
Right: the
data for the eigenstates with $7.5<|E_n|<7.7$. These states are localized
(see Fig. \ref{fig-schema-pom}).  The typical values of $I_2$ are of the order of 1,
as expected.
}
\label{fig-ipr}
\end{figure}

Figure \ref{ipr-2} shows values of $I_2$ calculated for all eigenstates
of the 3D Anderson model with Gaussian disorder $W=2$.  Data confirm our 
qualitative estimations, namely that
$I_2$ is small in the metallic part of the spectra, $|E|\ll |E_c|=6.58$,
but possesses values of order of 1 in the  band tails, where localization is expected.
We also see that  
$I_2$  is a statistical variable, which might fluctuate from one eigenstate to another.
It is therefore  useful to calculate
the mean value over all eigenvalues lying in a given narrow interval of energies,
$\delta E$. Also, since  $I_2$  might fluctuate in many orders of magnitude, we
analyze $\ln I_2$. This is useful especially in the metallic regime.
Figure \ref{fig-ipr} shows the probability distribution
of $\ln I_2$ for the three dimensional Anderson model. 
In the metallic regime, the maximum of the  distribution $p(\ln I_2)$ 
is around the values $\sim L^{-3}$, in agreement with Eq. (\ref{ipr-kov}). 
Also, the distribution decreases exponentially on both sides of the maxima, indicating,
that the probability to find $I_2\sim 1$ rapidly decreases  when $L$ increases.
This agrees with the commonly accepted paradigm, that there are no localized states inside
 the metallic phase.
In the localized regime, $I_2$ possess values of $\sim 1$, as expected. Again,
the probability to find delocalized state in the tail is very small.

\subsection{Level statistics}
\label{levelstatistics}

\begin{figure}[t!]
\begin{center}
\includegraphics[clip,width=0.45\textheight]{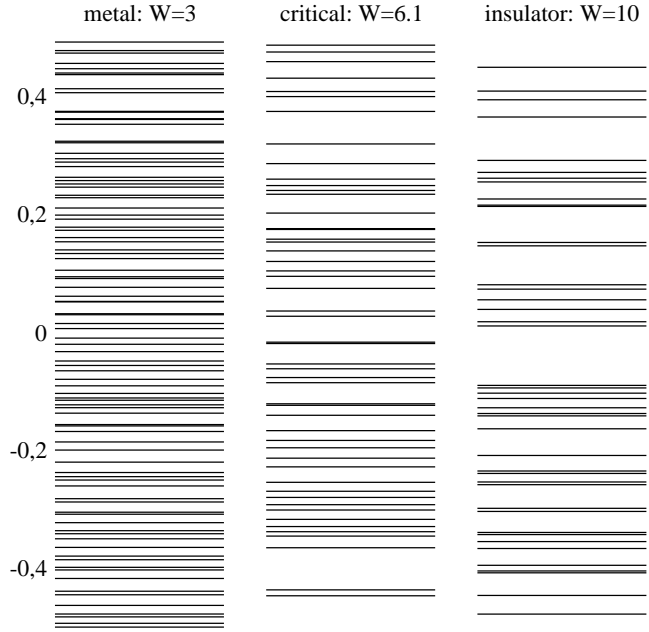}
\end{center}
\caption{Eigenenergies of disordered  three dimensional Anderson model with Gaussian
disorder. The size of the system $L=10$. Left: metallic regime. The spectrum is almost
equidistant, there are no degenerate energies. Right: localized regime. The degeneracy of
the energy spectra is more probable than in the metallic regime. Middle: spectrum at the critical point.
}
\label{fig-spectrum}
\end{figure}

Other important information about the transport regime can be obtained 
from the analysis of the spectrum of the eigenenergies of the disordered Hamiltonian.
In Fig. \ref{fig-spectrum} 
we show a typical spectrum of the eigenenergies
for the metallic, critical and insulating regime.  
One sees that the three spectra are qualitatively
different. This difference can be quantitatively 
measured by the statistical distribution of
\textsl{differences}
\be\label{eq-diff}
s_n=
{E_{n+1}-E_n}
\ee
of two neighboring eigenenergies of the random Hamiltonian (\ref{ham}).

Because of the randomness of the Hamiltonian,  $s$ is a statistical  variable.
It turns out that the probability distribution $p(s)$ converges in the limit
$L\to\infty$ to three characteristic universal functions, depending on whether
the system is in the  metallic,  critical or localized regime.

\begin{figure}[t!]
\psfrag{dE}{$s$}
\psfrag{pdE}{$\!\!\!\!\!p(s)$}
\begin{center}
\includegraphics[clip,width=0.3\textheight]{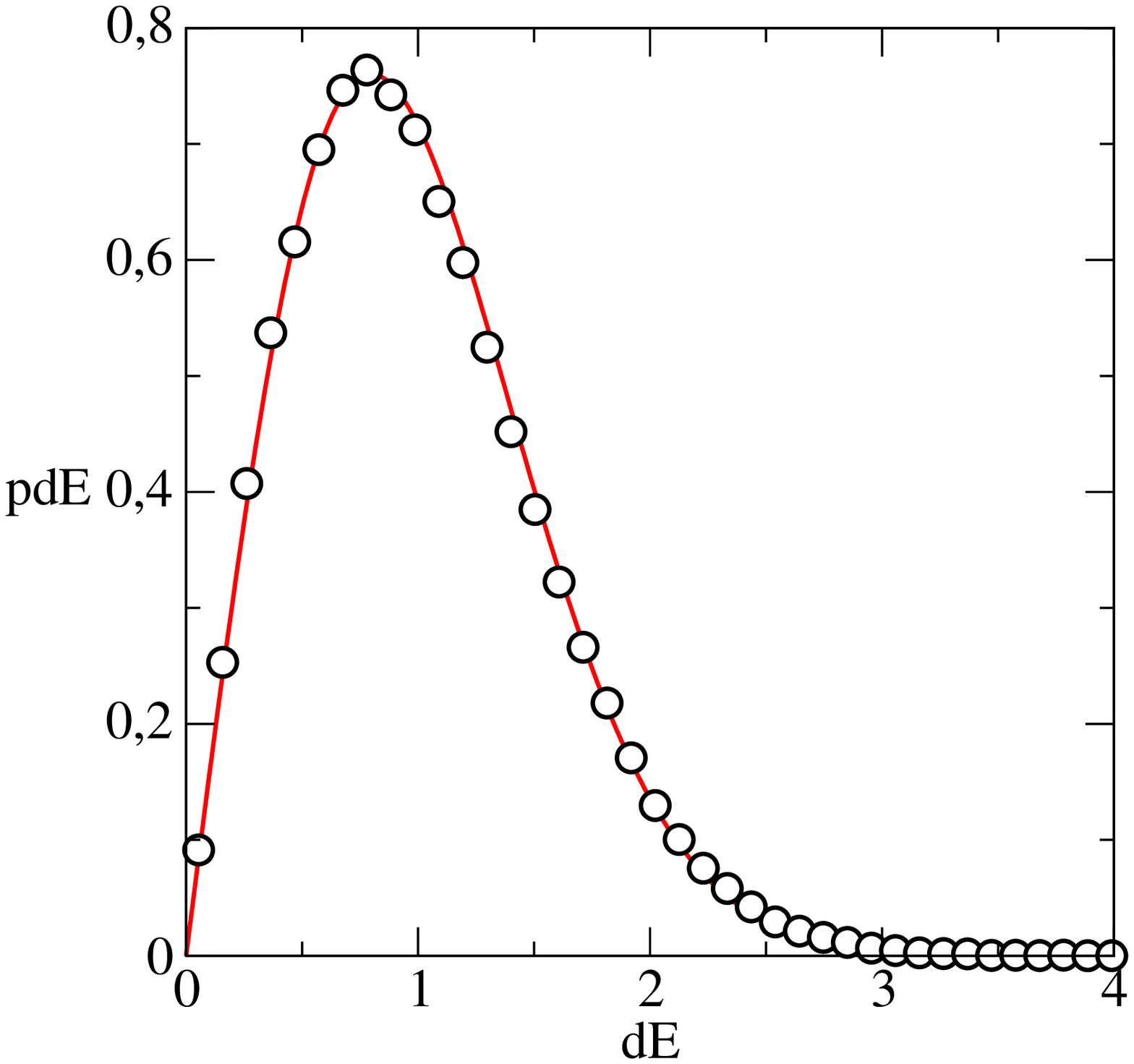}
\includegraphics[clip,width=0.3\textheight]{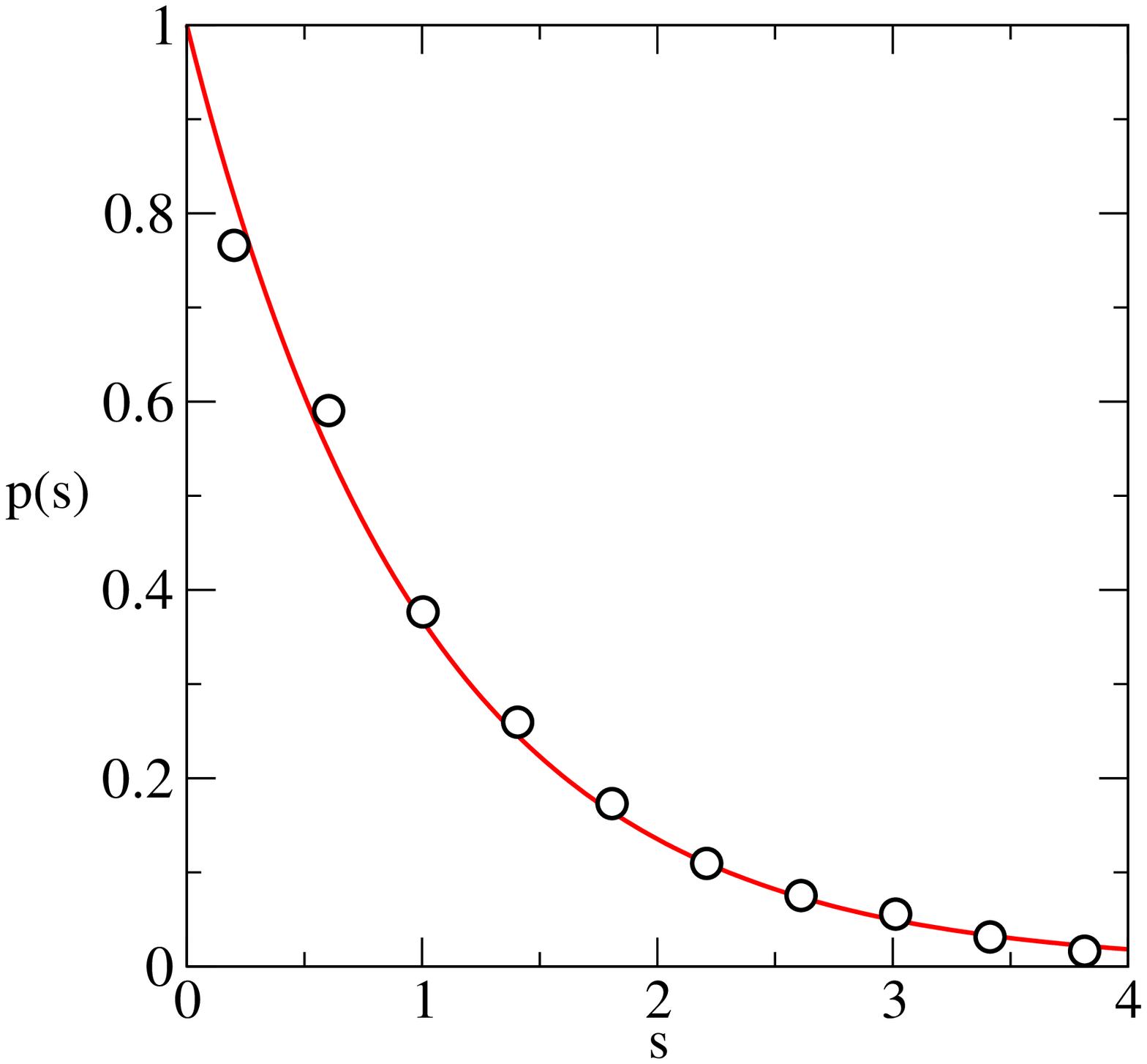}
\end{center}
\caption{The 
probability distribution $p(s)$ of the \textsl{normalized} differences, $s$,
between two neighboring eigenenergies
in the 3D Anderson model. Only eigenstates close to the band center, $|E_n|<0.5$, were considered. 
Left panel shows the distribution for the model  with
Gaussian disorder $W=2$, which is smaller than the critical disorder, $W_c\approx 6.1$. 
The system is
in the metallic regime and the distribution $p(s)$ is close to the Wigner distribution $p_1(s)$,
given by Eq. (\ref{wigner-1}).
(solid line).
The size of the system $L=14$ and  $N_{\rm stat}=450$. 
Right panel shows data for the
box disorder $W=32$, $L=10$. System is in the localized regime, and $p(s)$ is close to the
Poison distribution, $p(s)=\exp -s$.
}
\label{fig-wigner}
\end{figure}

If the system is in the metallic regime, then  the hopping term 
between sites causes level repulsion,
typical for random matrix theory. (Random matrix theory is discussed in Appendix
\ref{app:rmt}). 
Random matrix theory states that the distribution 
$p(s)$  of \textsl{normalized} differences $s$, defined by Eq. (\ref{eq-diff})
is universal, and depends on the physical symmetry. 
For orthogonal systems, we have
\be\label{wigner-1}
p_1(s)=\ds{\frac{\pi}{2}}~s~\exp -\ds{\frac{\pi}{4}s^2}.
\ee
For unitary and symplectic  systems, we have the Wigner distributions  given by
\be\label{wigner-2}
p_2(s)=\ds{\frac{32}{\pi^2}}~s^2~\exp -\ds{\frac{4}{\pi}s^2},
\ee
and
\be\label{wigner-4}
p_4(s)=\ds{\frac{2^{18}}{3^6\pi^3}}~s^4~\exp -\ds{\frac{64}{9\pi}s^2},
\ee
respectively.  Note,  all three distributions can be obtained analytically if we
assume that
\be
p_\beta(s)=c_0~s^\beta~\exp -c_1s^2,
\ee
where $\beta=1$, 2 and 4 for orthogonal, unitary and symplectic symmetry, respectively,
and  coefficients $c_0$ and  $c_1$ are given by the requirement of normalization,
\be
\int_0^\infty d~s~p_\beta(s)=1,
\ee
and  by the normalization,
\be
\langle s\rangle=\int_0^\infty d~s~sp_\beta(s)=1.
\ee
A characteristic property of the Wigner distribution is level repulsion. 
We indeed see in left panel of Fig.
\ref{fig-spectrum} that the spectrum in the metallic phase is almost 
equidistant, and the probability
to find two degenerate energies is very small. 
The absence of the degeneracy can be understood already
in the simple case of the $2\times 2$ matrix,
\be
\m2{\Lambda}{v}{v}{\Lambda},
\ee
which can represent the two site Hamiltonian. Without the overlap of the wave functions ($v=0$),
the spectrum is degenerate, $E_1=E_2=\Lambda$. 
However, any non-zero overlap removes the degeneracy,
and  the eigenenergies become $E_{1,2}=\Lambda \pm v$. Similarly, in more complicated systems, 
we expect that the overlap 
of wave functions prevents the eigenenergies being degenerate.
In random matrix theory, the absence of
degeneracy follows from the probability distribution of the eigenvalues of random matrices,
which is derived in Appendix \ref{app:rmt}.

The energy spectrum of the insulator  is different. 
The simplest model of localized systems is the Hamiltonian given by Eq. (\ref{ham}) with $V=0$. 
In this limit, the  energy spectrum consists of  random energies $\eeps$. 
It is well known \cite{Mehta} that the differences of
random \textsl{uncorrelated} numbers are distributed with the  Poisson distribution,
\be\label{poisson}
p_{\rm loc}(s)=e^{-s}.
\ee
We therefore expect that in the localized regime, the distribution $p(s)$ will be close to
the Poisson distribution. This is confirmed in the right panel of Fig. \ref{fig-wigner}.
Note, however, that the overlap $V\ne 0$ in any disordered system. Therefore, 
the level repulsion is always present in the energy spectra,
independent of the strength of the disorder, and
$p(s)\to 0$ when $s\to 0$ also in the strongly
localized models \cite{MSu}.

\begin{figure}[t!]
\psfrag{dE}{$s$}
\psfrag{pdE}{$p(s)$}
\begin{center}
\includegraphics[clip,width=0.3\textheight]{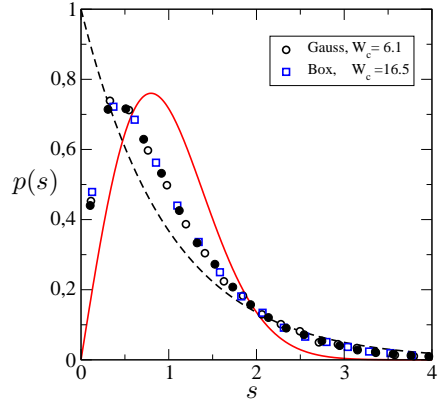}
\end{center}
\caption{The critical distribution $p_c(s)$ of the  normalized differences $s$.
Both the Gaussian and box disorder were considered.
The size of the system $L=10$. The solid circles show data  for the Gaussian disorder
and the system size  $L=14$.  
The data confirm that the distribution $p(s)$ does not depend
on the microscopic details of the model. Also, $p(s)$ is independent on the size of the 
system.
Only the eigenenergies $|E_n|<0.5$ were considered so that the density of
states can be considered as constant. For comparison, we show also the
Wigner distribution, 
given by Eq. (\ref{wigner-1}) (solid line)
and the Poisson distribution, $p_{\rm loc}(s)=\exp -s$ (dashed line).
}
\label{fig-merman}
\end{figure}

In numerical analysis of the level statistics, it is important to note that the 
mean  difference, $\langle s\rangle$,
depends also on the density of states. Indeed, $\langle s\rangle$ would posses smaller values
in that part of the  energy band  where the density of states, $\rho(E)$, is larger. Therefore, 
we restricted our  analysis of the statistics $p(s)$ to the energy interval
$E, E+\delta E$, where the density of states is approximately constant. The analysis of the 
entire energy band requires rescaling the differences $s$ by
$1/\rho(E)$.

When the  system undergoes the metal-insulator transition, the distribution   $p(s)$,
transforms from the Wigner distribution, given by Eq. (\ref{wigner-1}), to the Poisson distribution,
given by Eq. (\ref{poisson}).  
There is  the third universal distribution, characteristic for the
critical point  \cite{shkl}.
We do not know the analytical form of the critical distribution, but we can calculate it
numerically. 
In Fig.  \ref{fig-merman} we show the critical distribution
$p_c(s)$ for the  3D Anderson model.
Data confirms the theoretical expectation that $p_c(s)$
must decrease to zero when $s\to 0$. From 
the symmetry considerations, it follows that $p_c(s)\sim s^\beta$ for $s\ll 1$.
There is no agreement about the  form of exponential decrease for large $s\gg 1$.
Exponential decrease,  $p_c(s)\sim \exp(-s^\alpha)$, with $\alpha=1+1/(d\nu)$, was found in 
Ref. \cite{Aronov},  while 
the semi-Poisson distribution,
\be
p_c(s)=4se^{-2s},
\ee
was proposed for the orthogonal critical regime by other groups \cite{BMP}.
Numerical analysis \cite{ZK-e} did not distinguish between 
these two distributions. 

Figure \ref{fig-merman} also confirms the universality of the critical distribution
which does not depend on the size of the system and on the distribution function
of random energies. However, it was found that
$p_s$ depends on the choice of the boundary conditions \cite{BMP}.

\subsection{Boundary conditions}\label{sect:boundary}

The sensitivity of the electron eigenstates to the boundary conditions provides us with another
criterion for localization \cite{ET,LT}.
Consider a disordered $d$ dimensional system of size $L$. We can calculate all the eigenenergies
of the system with periodic  boundary conditions,
\be\label{bc-periodic}
\Psi(x+L)=\psi(x),
\ee
and then repeat the same calculation with the \textsl{anti-periodic} boundary conditions,
\be\label{bc-antiperiodic}
\Psi(x+L)=-\psi(x).
\ee
The change of the boundary conditions influences the spectrum of eigenenergies,
\be\label{th-de}
E_n\to E_n+\delta E_n,
\ee
where $\delta E_n$ is a change of the eigenenergy due to the change of the boundary conditions.
Clearly, $\delta E_n$ depends on the character of the electron eigenstate.

Suppose first that the electron in the $n$th eigenstate is
delocalized. Then its wave function is spread over the sample. We expect
therefore that the change of the boundary conditions influences strongly the position of the eigenenergy,
$E_n$,
so that  $\delta E_n$ might be comparable or even larger than the typical spacing between two neighboring 
eigenenergies. Since the wave function is delocalized for any system size,
$\delta E$ will not decrease when $L$ increases.
On the other hand,
If the electron is localized in a certain region inside  the sample, then its wave function 
decreases exponentially as a function of the distance from the center.  If the size of the system
is larger than the localization length $\lambda$,  the localized eigenstate is almost 
insensitive to the boundaries and th
localized electron does not know what happens on the boundary of the system,
so that $\delta E_n$ decreases 
exponentially when the size of the system increases,  $\delta E\sim\exp -L/\lambda$.


\begin{figure}[t!]
\begin{center}
\psfrag{theta}{$\theta/\pi$}
\psfrag{En}{$E_n(\theta)$}
\includegraphics[clip,width=0.6\textheight]{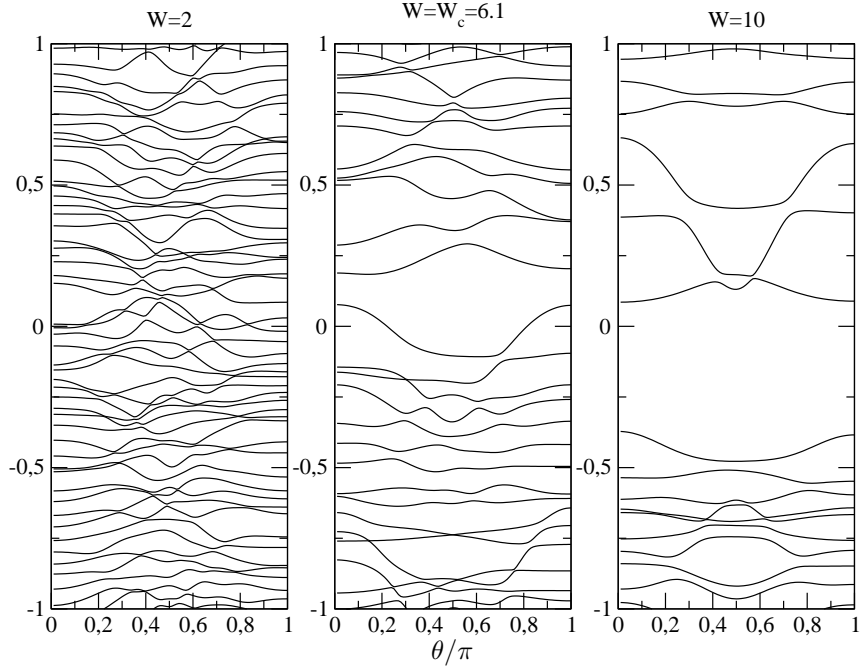}
\end{center}
\caption{Dependence of eigenvalues, $E_n(\theta)$ on the boundary conditions, given by Eq. (\ref{boundary}).
in the metallic (left), critical (middle) and the localized regime.
Only eigenvalues close to the band center are shown. The size of the system is $L=6$.}
\label{fig-theta}
\end{figure}

To measure the sensitivity to the change of the boundary conditions quantitatively, 
consider a small interval of energies, $E\pm \delta$, and calculate
the  parameter
\be\label{g-T}
g_T(E)=\frac{e^2}{h}\frac{\langle\delta E\rangle}{\Delta E},
\ee
where $\Delta E$ is a typical difference between two neighboring eigenenergies,
\be
\Delta E=\ds{\frac{1}{L^d\rho(E)}},
\ee 
and $\langle \delta E\rangle$ is a typical change of eigenenergies lying in
the interval $E\pm \delta$. 
Then, from the size dependence of $g$, we can distinguish between the metal and 
insulator. 

In Fig. \ref{fig-theta} we show how the eigenenergies $E_n$ of the disordered system depend
on the boundary conditions.
We consider the 3D  Anderson Hamiltonian, given by Eq. (\ref{ham}) with the random on-site energies,
$\eeps_n$, and the boundary conditions determined by the real parameter, $\theta$,
\be\label{boundary}
\begin{array}{lcl}
\Psi(L,y,z)&=&e^{i\theta}\Psi(0,y,z)\\
\Psi(x,L,z)&=&e^{i\theta}\Psi(x,0,z)\\
\Psi(x,y,L)&=&e^{i\theta}\Psi(x,y,0).
\end{array}
\ee
Clearly, $\theta=0$ corresponds to the periodic boundaries,  Eq. (\ref{bc-periodic}),
and $\theta=\pi$  gives us anti-periodic boundary conditions, (\ref{bc-antiperiodic}).

\medskip

It was  shown in Ref. \cite{ET} that if the energy 
$E$ belongs to the metallic part of the spectra, then
\be\label{gt-kov}
g_T= \sigma L^{d-2}~~~~~\textrm{metal}
\ee
where $\sigma$ is the conductivity of the sample. 
When the eigenenergies around the energy $E$ are localized,
$g_T$ decreases exponentially with the system size,
\be
g_T\sim e^{-L/\lambda}~~~~~~~\textrm{insulator}.
\ee
The quantity $g_T$ is called the Thouless conductance. It played an important role in the formulation of
the scaling theory of localization \cite{AALR}.

The mean value, $\langle\delta E\rangle$, is called the \textsl{Thouless energy}, $E_T$.
By comparison of the expressions (\ref{g-T},\ref{gt-kov}) with the  formula for the 
metallic conductivity, $\sigma=e^2D\rho$, we find that
\be
E_T=\frac{hD}{L^2}.
\ee
The corresponding Thouless time, 
\be
\tau_T=h/E_T,
\ee
represents the typical time, $L^2/D$, which the electron needs to diffuse from one side of the
sample to the opposite side.

To the best of our knowledge, the relation (\ref{g-T}) has so far not been  used for the numerical
analysis of the metal-insulator transition. The reason is that both 
$\delta E_n$ and $\Delta E$
fluctuate not only as a function of the microscopic realization of
disorder in a given sample, but also as a function of the energy
within one given sample. Also, it is not clear  which averaging procedure -   
arithmetic or geometrical - is more appropriate for the calculation of 
mean value, $\langle E_n\rangle$.
Nevertheless, an  introduction of the variable $g_T$  provided  the first step toward 
the scaling theory of localization. Since it is rather easy to estimate how both 
differences, $\delta E$ and $\Delta E$, depend on the system size, $L$, we can estimate
the size dependence of the Thouless conductance, $g_T$.

\section{Transfer matrix and conductance}\label{sect:g}

As discussed in the introduction, the electron localization  affects the transport properties
of disordered systems. Therefore, the theory of localization concentrates mostly on the 
electron propagation. A key role in this study is played by the conductance, a quantity defined
by Landauer \cite{Landauer} and Economou and Soukoulis \cite{ES}. The  conductance
is expressed  in terms of the transmission and reflection amplitudes of the electron.

\begin{figure}[b!]
\begin{center}
\includegraphics[clip,width=0.62\textheight]{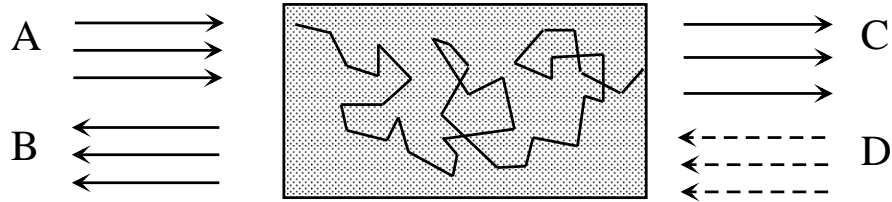}
\end{center}
\caption{%
Definition of amplitudes $A$ - $D$.}
\label{fig-scatt}
\end{figure}

Consider scattering experiment shown in Fig. \ref{fig-scatt}. 
The sample is connected on both sides to the  semi-infinite ideal leads. 
We are interested in the probability
that an electron, coming either  from the left or from the right side of the sample, 
can transfer to the opposite side.
Thus, we want to characterize the sample in terms of 
 four matrices: transmission of the wave
from the left to the right, $t^+$, from the right to the left, $t^-$,
and by the reflection coefficient from the right to the right, $r^+$, and from the left
to the left, $r^-$.  
In the most simple case of one-dimensional conductors, all the above parameters
are complex numbers.

Transmission and reflection amplitudes define the  \textsl{scattering matrix} $\textbf{S}$:
\be\label{one}
\textbf{S}=\left(\begin{array}{ll}
		t^+ &  r^-\\
		r^+ &  t^-
	\end{array}\right).
\ee
Figure  \ref{fig-scatt} shows that the scattering matrix,
$S$,  expresses  the wave functions of  outgoing waves $B$ and $C$,  in terms of 
the wave functions  of the incoming waves,  $A$ and $D$, 
\be\label{two}
\left( C\atop B\right)
=
\textbf{S}
\left( A\atop D\right)
\ee

Linear relation (\ref{two}) can be re-written into the form
\be\label{twox}
\left( C\atop D\right)
=
\textbf{T}
\left( A\atop B\right).
\ee
where $\textbf{T}$ is the transfer matrix, which determines the fields on one side
of the sample with the fields on the another side.
An explicit form of the transfer matrix, derived in Appendix\ref{app-a} reads
\be\label{TM}
\textbf{T}=\left(\begin{array}{ll}
t^+-r^-(t^-)^{-1}r^+&
 r^-(t^-)^{-1} \\
 -(t^-)^{-1}r^+  & 
(t^-)^{-1} 
\end{array}\right).
\ee
Some useful properties of the transfer matrix are given in Appendix \ref{app-a}.
If the system possesses time reversal symmetry, then the transmission amplitudes
$t^+$ and $t^-$ are related by
\be
t^+=(t^-)^T.
\ee
In the special case of 1D system, we have that $t^+=t^-$ Therefore, we omit the superscripts
in the transmission amplitudes in the following discussion.

\subsection{Conductance}

\begin{figure}[t!]
\begin{center}
\includegraphics[clip,width=0.42\textheight]{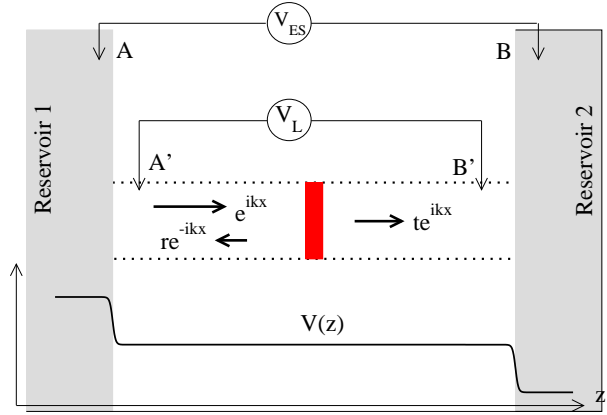}
\end{center}
\caption{%
Experimental setup for the measurement  of the  conductance $g$.
Two semi-infinite leads are attached to the sample. Electrons
are emitted from the left reservoir, propagate through the left lead and  scatter on the sample.
The transmitted electrons are absorbed in the reservoir on the opposite side,
the reflected electrons propagate back. There is no scattering in the leads. 
There are  two possibilities of the measurement of the voltage.
Voltmeter $V_L$ measures the voltage on leads; resulting  conductance is then
$g_L=j/V_L$, given by Eq. (\ref{gl}). The voltmeter $V_{ES}$ measures the voltage
on reservoirs. In this measurement, we obtain Economou-Soukoulis conductance $g_{ES}$,
given by Eq. (\ref{ges}). We show also the voltage for the case when the sample
is totally transparent to show the   voltage drops  due to the contact resistances between the
reservoirs and leads.
}
\label{land-1}
\end{figure}

Consider the one dimensional  experimental setup shown in Fig. \ref{land-1}. 
The sample is connected to two semi-infinite ideal leads, which 
transfer electrons from two reservoirs.  The current through the sample is
proportional to the voltage difference, $\Delta V$, and to the   conductance, $g$,
\be\label{g-j}
j=g\Delta V.
\ee
To find the conductance, we have to measure the voltage, $\Delta V$. Two
possibilities were considered. Landauer \cite{Landauer} proposed to measure the voltage
difference between the leads, and obtained the conductance
\be\label{gl}
g_L=\ds{\frac{e^2}{h}\frac{T}{1-T}}.
\ee
On the other hand, 
Economou and Soukoulis \cite{ES} considered the voltage difference on reservoirs
(Fig. \ref{land-1}), and  derived an alternative formula,
\be\label{ges}
g_{ES}=\frac{e^2}{h}T.
\ee
Both formulas are equivalent to each other only in the limit of small transmission, $T\to 0$.
However, they lead to different results 
when the sample becomes transparent ($T\to 1$).  Landauer formula
predicts that the conductance diverges (as it should be since the resistance
is zero). On the other hand, expression (\ref{ges}) converges to $e^2/h$.

The origin of the difference between two formulas lies in the presence of a
contact resistance between leads and reservoirs. We can write
\be\label{gdiff}
\ds{\frac{e^2}{h}\frac{1}{g_{ES}}=\frac{1}{T}=\frac{1-T}{T}+1=\frac{e^2}{h}\frac{1}{g_{L}}+1}
\ee
or, in terms of resistances,
\be\label{grho}
\rho_{ES}=\frac{1}{g_{ES}}=\rho_L+\frac{h}{e^2}.
\ee
Thus, in the measurement of Economou and Soukoulis, the total resistance
included in the measurement contains also contact resistance, $\rho_c=h/e^2$, measured on the 
contact of leads and reservoirs. 

In numerical simulations, we will calculate the conductance $g_{ES}$.
When necessary, we can extract the effect of contact resistance
with the help of Eq. (\ref{gdiff}).

Historically, both quantities, $g_L$ and $g_{ES}$, are called ``Landauer conductance'' in the literature.

\subsection{Conductance of multi-channel system}\label{sect:multi}

In the previous Section, we have introduced  the conductance of the  one dimensional systems. 
For a given energy,
we have only one possible value of the wave vector. This is the  reason why the maximum value
of the conductance is 1 (in units of $e^2/h$).
 In the real world, the leads might be quasi one dimensional, which means
that the electron can propagate also in directions perpendicular to the 
propagation direction. Since the cross section of the leads is finite, the wave vector 
in the transversal direction, $k_\perp$,  is quantized and possesses only  discrete 
values. Each value of the transversal wave vector defines one  channel.
It might happen that
transmission is not possible for some values of $k_\perp$. We call such channels evanescent, or closed.
For simplicity, we do not consider evanescent channels in the present discussion.
Evanescent waves are discussed in Appendix \ref{app-open}.
The number of channels  $\no$,  
defines the size of   the transmission matrices, $t^+$ and $t^-$.
The physical meaning of the matrix element is clear:
$t^+_{ab}$  is the  transmission amplitude of the 
electron from the channel $a$ to the channel $b$. Similarly, we define 
the matrix $r^+$  which contains reflection amplitudes from channel $a$ to channel $b$.
Then,
\be
T^+_{ab}=|t^+_{ab}|^2
\ee
is the probability that the electron, coming in channel $a$, is transmitted through
the sample into channel $b$, and
\be
R^+_{ab}=|r^+_{ab}|^2
\ee
is the probability that the electron is reflected back into channel $b$.

It is useful to introduce the probability
\be
T^+_a=\sum_b t^+_{ab}(t^+_{ab})^* = \sum_b T^+_{ab}
\ee
of the electron coming in the channel $a$ to be transmitted through sample,
and
\be
T^+=\sum_b T^+_{a} = \sum_{ab} T^+_{ab}
\ee
is the total probability that the electron transmits through the sample from 
left to right. Similarly,
\be
R^+=\sum_{ab}|r^+_{ab}|^2
\ee
is the probability that the electron is reflected back. Since electrons can not be
absorbed in the sample, we have
\be
T^++R^+=\no,
\ee
as is proved in Appendix  \ref{app-a}.
Then, generalization of expression (\ref{ges}) to the multi-channel system is straightforward,
\be\label{multichan}
g_{ES}=\displaystyle{\frac{e^2}{h}}\textrm{Tr}~t^\dag t,
\ee
\cite{Azbel,FL,LA,Land-1}.

\begin{figure}[t!]
\begin{center}
\includegraphics[clip,width=0.52\textheight]{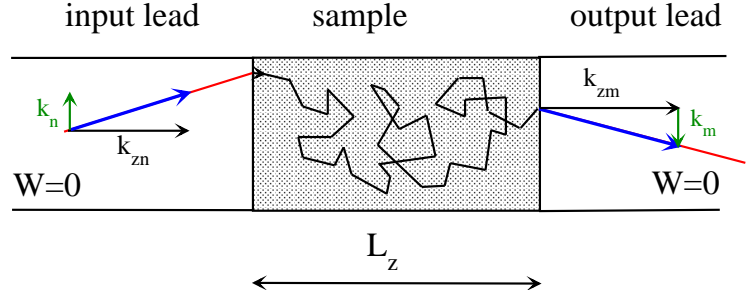}
\end{center}
\caption{%
Transmission of electron through disordered sample. The sample is connected to two semi-infinite
metallic leads. Channels are determined by transversal momentum, or, equivalently, 
by incident angle, $\alpha=\tan^{-1}(k_n/k_{zn})$.}
\label{fig-kanal}
\end{figure}

Multi-channel conductance is fully determined by eigenvalues of the matrix $t^\dag t$.
Using the parametrization of the transfer matrix, derived in appendix \ref{sect:param},
we obtain
\be\label{konduktancia}
g_{ES}=\displaystyle{\frac{e^2}{h}\sum_a^{\no}\frac{1}{1+\lambda_a}=\frac{e^2}{h}\sum_a^{\no}\frac{1}{\cosh^2 x_a/2}}
\ee
where $1/(1+\lambda_a)$ is the $a$th eigenvalue of the matrix $t^\dag t$ and
parameters $x_a$ are defined by the relation
\be
\lambda_a=\frac{1}{2}\left(\cosh x_a-1\right).
\ee
It was proved in Refs. \cite{FL,LA} that in the metallic regime,
the conductance $g$, given by Eq. (\ref{multichan})
is related to the conductivity $\sigma$, given by Eq. (\ref{v-1}), by
\be\label{g-v}
g_{ES}=\sigma L^{d-2}.
\ee

\subsection{Relation between the Thouless conductance and $g_{ES}$}\label{sect:rel}

In Sect. \ref{sect:boundary} we introduced the Thouless conductance, $g_T$,
which measures the sensitivity of the energy spectra  to the change of the boundary 
conditions.
In the diffusive regime, $g_T$  is given by
 Eq. (\ref{gt-kov}). 
 Comparison  with Eq. (\ref{g-v}), indicates  that two quantities, $g_T$ and $g_{ES}$
 are closely related to each other. This equivalence was studied numerically in Ref.
 \cite{vztah}.
Instead of changing the boundary conditions from periodic to anti-periodic,
the \textsl{level curvature},
\be\label{curvature}
c=\frac{\partial^2 E_n(\theta)}{\partial\theta^2}\Bigg|_{\theta=0}.
\ee
was studied numerically and compared with the  conductance $g_{ES}$.
The curvature, $c$, measures the sensitivity of the energy spectra 
to the change of the boundary conditions.
In the metallic regime, the curvature $c$ is related to $g_{ES}$ by the relation
\cite{vztah}
\be\label{bc-curv}
g_{ES}=\pi\rho(E) L^d\langle |c|\rangle.
\ee
Also, in the strongly localized regime  it was confirmed in Ref. \cite{vztah} that
\be\label{b-loc}
\langle \ln g_{ES}\rangle \propto \langle \ln |c|\rangle
\ee
Owing to  Eqs. (\ref{bc-curv}) and (\ref{b-loc}) we 
identify the conductance $g_T$ with $g_{ES}$ and 
conclude that 
the conductance $g_{ES}$  not only measures the transmission properties of the sample,
but also provides us with information about the sensitivity
of the wave functions to the change of boundary conditions. 
In what follows we discuss only the conductance $g_{ES}$, defined by Eqs. (\ref{ges}).
and use the notation
\be
g=g_{ES}.
\ee


\section{One dimensional systems}\label{sect:1D}

\begin{figure}[b!]
\begin{center}
\includegraphics[clip,width=0.25\textheight]{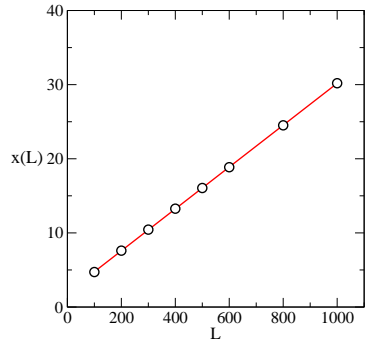}
\end{center}
\caption{%
The size dependence of parameter $x$ defined by Eq. (\ref{1d-es}).
Energy $E=1$ and disorder $W=1$. By combining 
the linear fit, $x(L_z)=1.94 + 0.0282 L_z$, 
shown by solid line,
with  Eq. (\ref{1d-ll}) 
we estimate the localization length, $\lambda=70.92$. This agrees with
the analytical estimation, $\lambda=[\textrm{Re}~\gamma]^{-1}$ for 
the real part of the Lyapunov exponent, 
$\textrm{Re}~\gamma(E)=1/24(4-E^2)=1/72$,
given by Eq. (\ref{9-result}).
}
\label{fig-1d-fit}
\end{figure}

It is instructive to analyze first the most simple problem, namely the 
one dimensional disordered (1D) chain.
The simplest system which exhibits localization is the  1D
Anderson model. If $\Psi_n$ is the wave function
of the electron on site $n$, then the Schr\"odinger equation reads
\be\label{1d-sche}
(\eeps_n-E)\Psi_n+\Psi_{n-1}+\Psi_{n+1}=0.
\ee

It is well known \cite{Mott-61} that all electron states 
in the disordered 1D system are localized.
Localization is characterized by the localization length, which
defines the exponential decrease of the wave function ,
\be
\Psi_n=\exp -L_z/\lambda
\ee
were $L_z=an$ is the length of the system.

We can calculate the conductance, given by Eq. (\ref{8-1}).
It is more   useful to start with the analysis of statistical properties
of the variable $x$, related to
the conductance by 
\be\label{1d-es}
g=\ds{\frac{1}{\cosh^2 (x/2)}}
\ee
(in units of $e^2/h$).

Since all states are localized in 1D, we expect that conductance decreases
exponentially when the length of the system increases, so that 
\be\label{1d-ll}
x(L_z)=2L_z/\lambda,~~~~~L_z\to\infty
\ee
as is shown in Fig. \ref{fig-1d-fit}.
This follows also  from the Oseledec's theorem, discussed in Appendix \ref{app-le}.

\begin{figure}[t!]
\begin{center}
\includegraphics[clip,width=0.42\textheight]{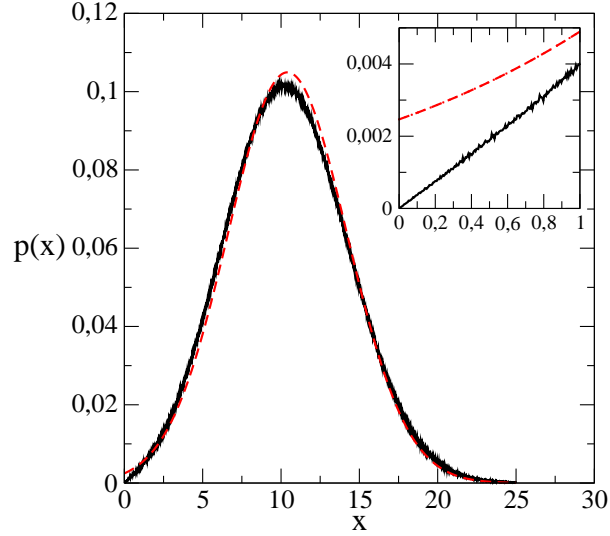}
\end{center}
\caption{%
The probability distribution, $p(x)$, of the parameter $x$ for the one-dimensional 
disordered chain of the length $L_z=300$, disorder $W=1$ and $E=1$. For comparison, we show 
also the Gaussian distribution
with the same mean value, $\langle x\rangle=10.41$ and variance, var $x=14.45$ (dashed line).
Inset shows how the two distributions differ for small $x$.
Note, $p(x)\sim x$ for small $x$.
To calculate the distribution, the statistical ensemble of
$\ns=10^9$ samples was  considered \cite{Kramer-AP}.
}
\label{fig-1d-x1}
\end{figure}

Conductance $g$ is a statistical
variable. To describe the transport properties of the systems with given 
disorder, we must study the statistical ensemble of samples, which differ
in the microscopic realization of random energies, and calculate the
mean, $\langle g\rangle$, and higher cummulants, or the entire
conductance distribution, $p(g)$.
We prefer to study the probability
distribution of the parameter $x$, shown in 
Fig. \ref{fig-1d-x1}
for a finite system length $L_z$. 
We see that the distribution $p(x)$ is similar to
Gaussian. 
However, $p(x)$ differs from Gaussian when $x$ is small. Indeed,
$p(x)\to 0$ when $x\to 0$ 
since no negative values of $x$ are allowed. Detailed numerical analysis confirmed 
(inset of Fig. \ref{fig-1d-x1}, \cite{Kramer-AP}) that
\be
p(x)\sim x~~~~ x\ll 1.
\ee
Note, no negative values of $x$ are allowed.
Thus, to first approximation, $p(x)$
can be written in the form
\be
p(x)=\propto  x~e^{-(x-a)^2/2b}
\ee
where    
\be
a=\langle x\rangle+O(1)
\ee
and
\be
b=\alpha\langle x\rangle.
\ee
Here,  $\alpha$ is a constant of order 1.  In the limit of $L_z\to\infty$,
$\alpha\to 2$  but $\alpha<2$ for finite $L_z$. For instance, $\alpha=1.388$ for system 
analyzed in Fig. \ref{fig-1d-x1}.

From known distribution of $p(x)$, we can calculate the distribution of conductance,
\be
p(g)=\int_0^\infty d~x~p(x)\delta\left[g-\ds{\frac{1}{\cosh^2 x/2}}\right],
\ee
or the mean values 
\be\label{1d-xx}
\langle g^n\rangle=\int_0^1~d~g~p(g)g^n=
\int_0^\infty d~x~p(x)\ds{\frac{1}{\cosh^{2n} x/2}}.
\ee
We evaluate the integral in the r.h.s of Eq. (\ref{1d-xx}) by the steepest descent method. 
The function $p(x)\cosh^{-2n} x/2$ possess a sharp maximum around $x_n$, which
solves the equation
\be
\frac{\partial}{\partial x}
\left\{ -\frac{(x-a)^2}{2b}+\ln x -2n\ln\cosh(x/2)\right\}\Bigg|_{x=x_n}=0.
\ee
It is easy to find that $x_n\sim O(1)$. Then, neglecting in Eq. (\ref{1d-xx})
$x_n$ with respect to $a$ which is $\propto L_z\gg 1$ 
we find that
\be\label{1d-gn}
\langle g^n\rangle = \frac{c_n}{L_z^{3/2}} e^{-a^2/2b}\propto e^{-\langle x\rangle/2\alpha}
\ee
with the constant $c_n$ independent of the length $L_z$\footnote{The factor $L_z^{-3/2}$ 
arises form the normalization constant of the distribution $p(x)$.}.

Equation (\ref{1d-gn}) agrees for $n=1$ with the 
analytical result derived in Refs.  \cite{abrikosov,Kirkman}.
For $n>1$, we recover the   universality of the moments of the conductance, derived in Refs. 
\cite{Kirkman,PMcKR},
namely that the ratio
\be
\ds{\frac{\langle g^n\rangle}{\langle g\rangle}=\frac{c_n}{c_1}}
\ee
does not depend on the system length.

\begin{figure}[t!]
\begin{center}
\includegraphics[clip,width=0.42\textheight]{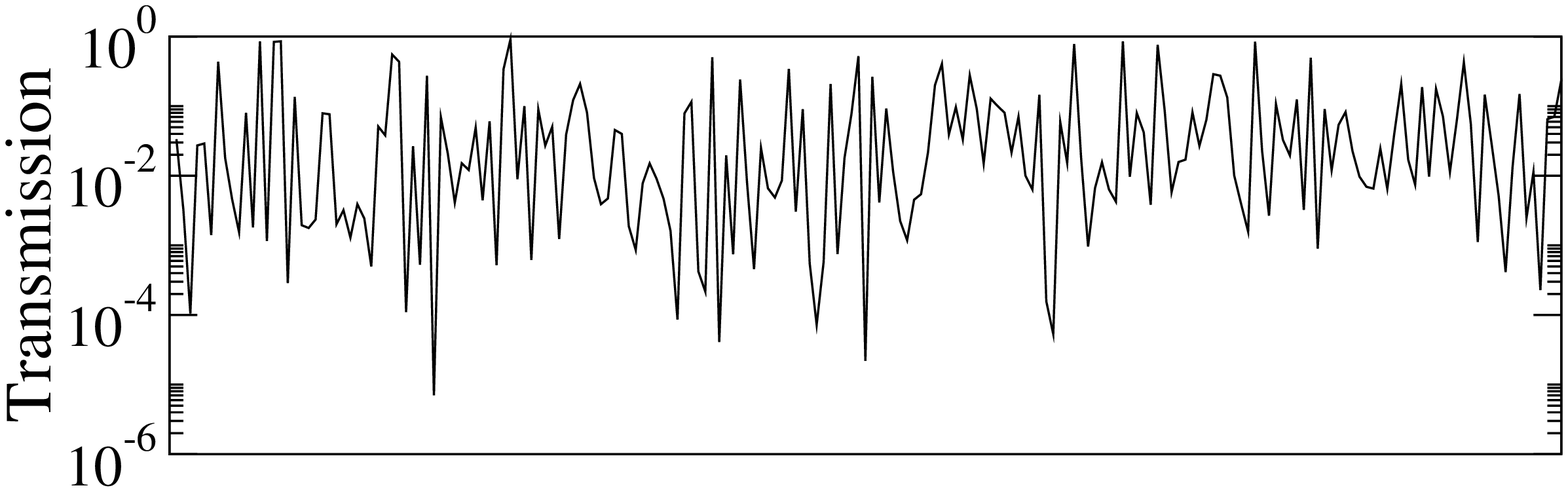}
\end{center}
\caption{%
Conductance calculated for $\ns=200$ realization of disorder on the 1D system 
of the length $L=200$ and box distribution of the disorder with $W=1$.
The localization length $\lambda\approx 100$.
}
\label{acta-x2}
\end{figure}

Also, note that 
since $\langle g^2\rangle \sim \langle g\rangle\gg \langle g\rangle^2$, we obtain that
the ratio
\be
\ds{\frac{\sqrt{\textrm{var}~g}}{\langle g\rangle}}=e^{+\langle x\rangle/4\alpha}\gg 1
\ee
increases exponentially when the length of the system increases.
Therefore, the mean value,  $\langle g\rangle$, is not a good representative of the
statistical ensemble. It follows from the derivation of Eq. (\ref{1d-gn} that
all the moments of the conductance are determined only by a vanishingly small number
of samples of the ensemble.  These samples with small $x$
are by no means representative. 
The same conclusion can be drawn out from
numerical data  in Fig. \ref{acta-x2}
which shows the conductance calculated for $\ns=200$ samples which differ from each
other only by the microscopic realization of disorder.  We see that the conductance fluctuates
from sample to sample in many  orders of magnitude. The mean value,
$\langle g\rangle$, is determined by a few samples with $g\sim 1$, while 
the most probable value of the conductance is in many order of magnitude
smaller. 

Since $x$ possesses a good probability distribution, and the transmission,
$T=\cosh^{-2} x/2$, it is evident that for long systems it is more convenient to
 analyze the distribution of the \textsl{logarithm}
of the transmission $\ln T$. 
We introduce a typical conductance,
\be\label{1d-gtyp}
g_{\rm typ}=e^{\langle\ln g\rangle}\propto e^{-a}\propto e^{-\langle x\rangle},
\ee
Comparing the typical conductance with the mean conductance,
\be\label{1d-gmean}
\ln\langle g\rangle \sim\langle x\rangle/(2\alpha)
\ee
we obtain that
\be
\ln g_{\rm typ}=\langle \ln g\rangle=\frac{1}{2\alpha}\ln\langle g\rangle.
\ee
The typical  conductance  is orders of magnitude smaller  than the
mean conductance. This is typical for the localized regime. The conductance,
$g$, is therefore not a good variable for description  of transport in the localized
regime. 

The same  holds for the resistance.  Using $\rho=R/T$, we have
\be
\rho=\sinh^2(x/2)
\ee
and 
\be
\langle\rho^n\rangle=\int dx p(x) \sinh^2(x/2)\approx e^{an+bn^2/2}.
\ee
In particular, for $n=1$ we obtain, 
\footnote{We remind the reader that $b=\alpha a$ and assume $\alpha=2$
in the limit of $L_z\to\infty$.}
\be\label{1d-odpor}
\langle\rho_L\rangle=\frac{1}{2}\left(e^{4L_z/\lambda}-1\right).
\ee
Using the composition law or transfer
matrices,  derived in the next Section,  the mean value of the Landauer 
resistance, $\langle\rho_L\rangle$,
increases exponentially with the length of the system,
but the \textsl{typical} resistance, 
\be
\rho_{\rm typ}=e^{\langle\ln \rho\rangle}
\ee
increases much slower than the mean resistance:
\be\label{1d-typ}
\rho_{\rm typ}=\frac{1}{2}\left(e^{2L_z/\lambda}-1\right).
\ee
Again, the reason for the difference between the mean and the typical resistance is that 
the value of $\langle \rho^n\rangle$ is determined by very specific samples which have extremely
huge resistance. 
Detailed analysis of the statistics of the resistance can be found in 
Ref. \cite{ATAF,D,Mello,BS,Vagner,book}

\subsection{Role of quantum coherence}

To understand the physical origin of the localization and huge fluctuations
of the conductance, we calculate the transmission
of the quantum particle through two barriers, shown in Fig. \ref{uk}. In Appendix
\ref{app:comp} we derived the transmission  \textsl{amplitude} for transmission
through two barriers, given by Eq. ({\ref{twobar}),
\be\label{2bar}
t_{12}^-=t_1^-\left[1-r_2^+r_1^-\right]^{-1}t_2^-.
\ee
From Eq. (\ref{2bar}) we obtain the transmission \textsl{probability},
\be\label{x-1}
T_{12}=|t_{12}^-|^2 = \ds{\frac{T_1T_2}{1+R_1R_2-2\sqrt{R_1R_2}\cos\phi}}
\ee
where we use $T_{1,2}=|t_{1,2}^-|^2$, $r_1^-=\sqrt{R_1}e^{i\phi_1}$,
$r_2^-=\sqrt{R_2}e^{i\phi_2}$ 
and $\phi=\phi_1+\phi_2$. 

Equation (\ref{x-1}) can be written in the form
\be\label{x-2}
\ln T_{12}=\ln T_1 + \ln T_2 - \ln (1+R_1R_2-2\sqrt{R_1R_2}\cos\phi).
\ee
We can introduce the disorder by considering the phase factor, $\phi$, being random
with uniform distribution in the interval $(0,2\pi)$  \cite{ATAF}.
Then, averaging over all realization of phase  gives the relation
\be\label{x-3}
\langle\ln T_{12}\rangle =  \ln T_1 +  \ln T_2,
\ee
since the integral of the last term on the r.h.s. of Eq. (\ref{x-2}) is zero \cite{ATAF}.

Relation (\ref{x-3}) can be generalized for the $N$ scattering centers. We obtain
\be\label{x-4}
\langle \ln T_N\rangle = N~ \overline {\ln T_1}
\ee
where $\overline{\ln T_1}$ is the average of the transmission probability through
one scatterer,
\be
\overline{\ln T_1}=\frac{1}{N}\sum_{a=1}^N \ln T_a.
\ee
Equation (\ref{x-4}) gives immediately the exponential decrease of the typical
transmission,
\be
e^{\langle T\rangle}=e^{-2L_z/\lambda}
\ee
where the localization length is given by
\be
\lambda^{-1}=\overline{\ln T_1}/2.
\ee
With the use of the expression $\rho_L+1=1/T$ (Eq. \ref{1d-odpor})
we recover the relation  (\ref{1d-typ}).

\medskip

On the other hand,
 Ohm's law  claims that the resistance of the
1D system increases linearly with the length of the system, or, equivalently, with the
number $N$ of scatterers.
To obtain this result, 
we have to assume that electron is a classical particle. Then, instead of
combination of transmission \textsl{amplitudes}, we combine  the 
transmission and reflection \textsl{probabilities}. We obtain that
\be\label{1d-clas}
T_{12}=\ds{\frac{T_1T_2}{1-R_1R_2}}.
\ee
With the use  of relations $T=1-R$ and $\rho=R/T$, we obtain from Eq. (\ref{1d-clas})
that
\be
\rho_{12}=\rho_1+\rho_2,
\ee
which is nothing but Ohm's law as we wanted.

\subsection{Distribution of the conductance}

Figure \ref{fig-1d-logg} presents numerical data for the distribution
$p(\ln g)$ for a strongly disordered 
1D system. The distribution is similar to a Gaussian, in agreement with 
theoretical expectation discussed in  Appendix \ref{app-le} 
and with our estimation of the distribution
$p(x)$.
The main difference from the Gaussian distribution
is that the distribution drops to zero  at $\ln g=0$ since the conductance never exceeds 1.

\begin{figure}[t!]
\begin{center}
\includegraphics[clip,width=0.25\textheight]{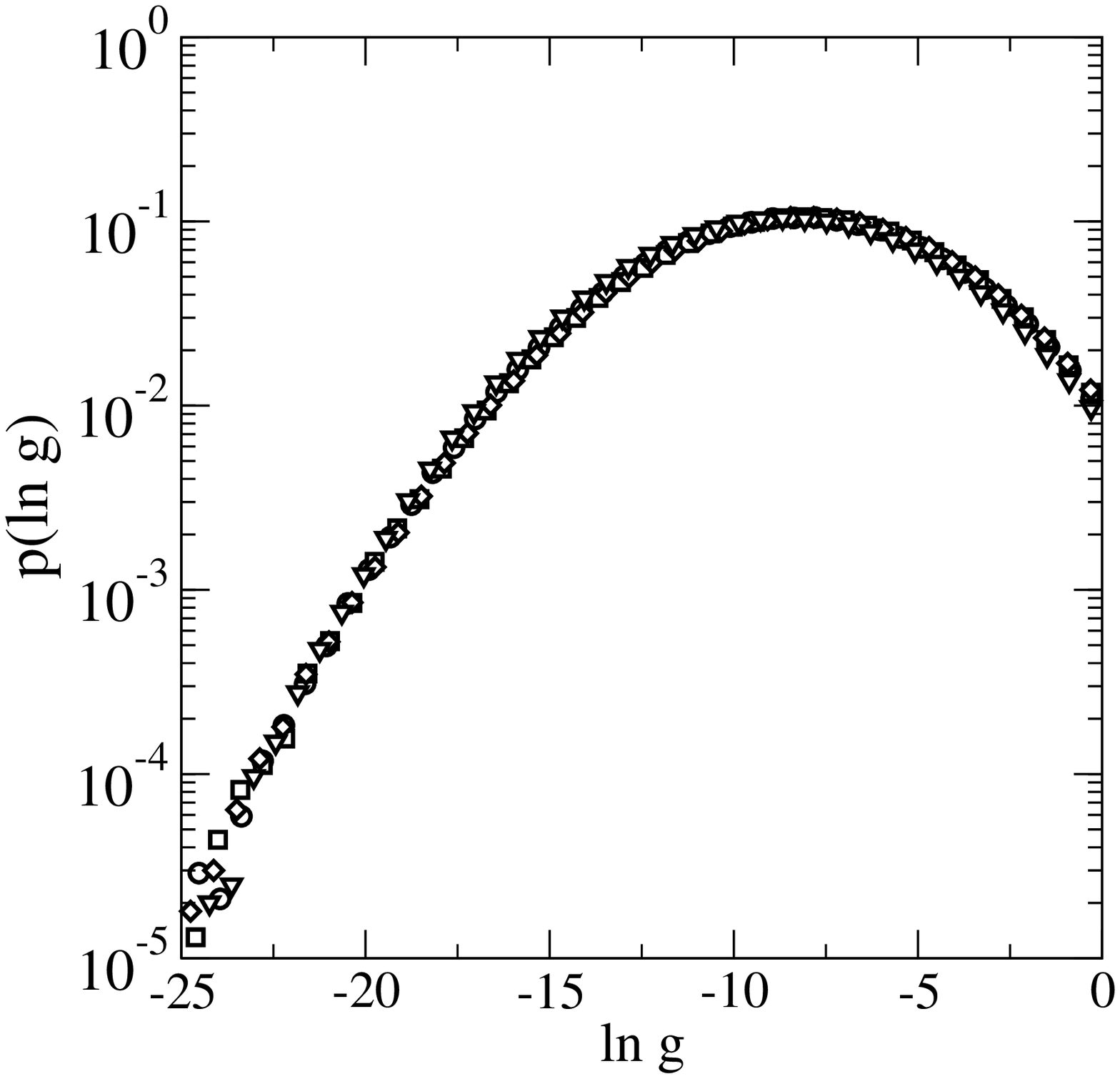}
\includegraphics[clip,width=0.25\textheight]{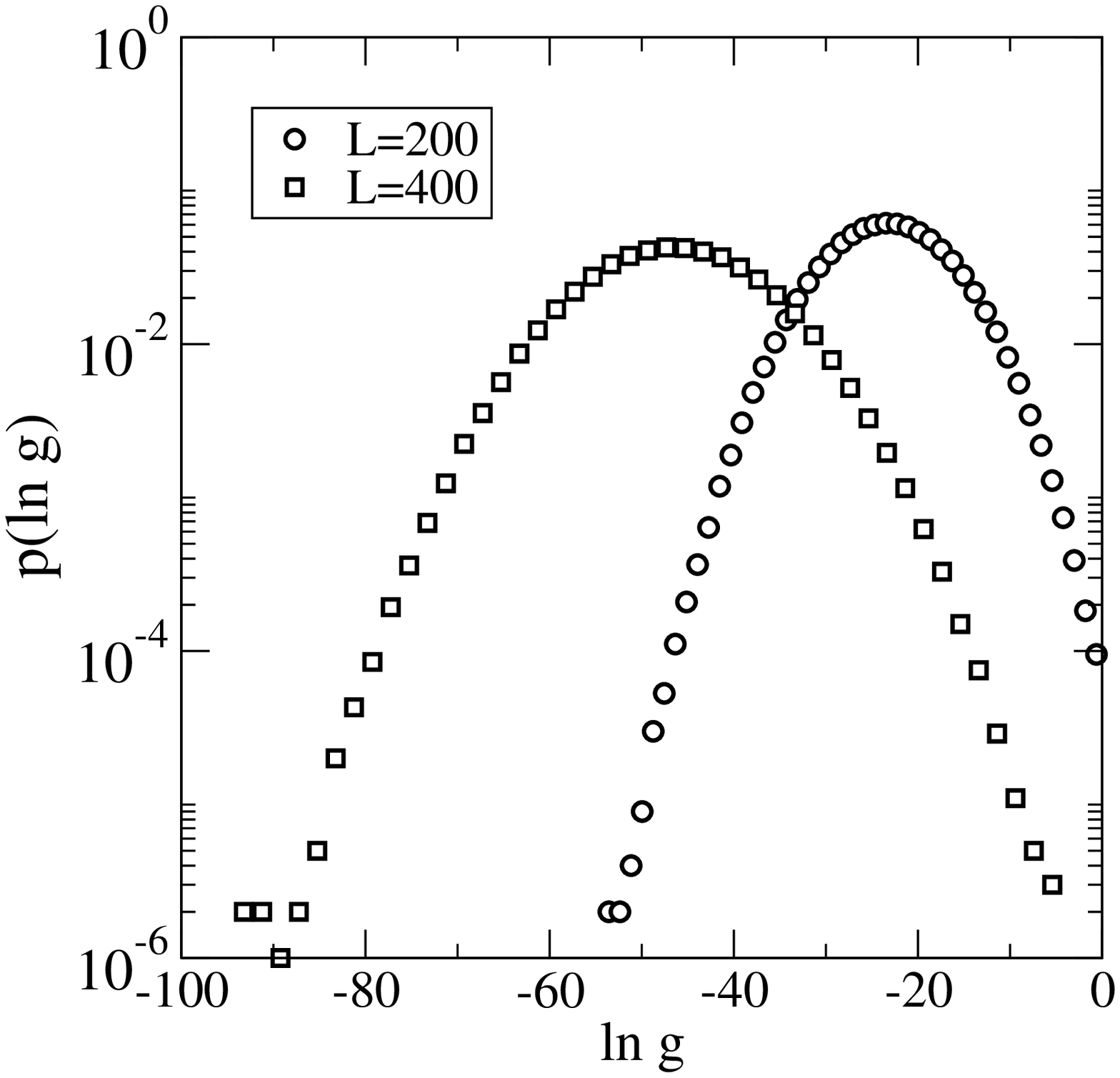}
\end{center}
\caption{%
Left: Distribution of the logarithm of the conductance, $p(\ln g)$ for
the 1D disordered Anderson model. The  energy of the electron is $E=1$, and the disorder
$W=0.25$, 0.5, 1 and 2. To keep the same value
of the ratio $L/\lambda$, we use the   analytical result,
$\lambda\sim W^{-2}$, 
for the localization length, discussed  in Appendix \ref{app-le},
and choose   the length of the system
$L_z=4800$, 1200, 300 and 75, respectively.
Note the drop of the distribution at $\ln g=0$. This is because
the conductance never exceeds 1 in the 1D systems. 
Right: $p(\ln g)$ for disorder $W=2$ and two lengths of the  system, $L_z=200$ and  $L_z=400$.
For a longer system, the probability to have $g$ close to 1 is negligible and $p(\ln g)$
is Gaussian.
}
\label{fig-1d-logg}
\end{figure}

Figure \ref{fig-1d-cond} shows the probability 
distribution, $p(g)$ for 1D chains shorter than the 
localization length. 
For very short systems, $L_z\ll\lambda$, the electron  ``almost always'' propagates
through the sample. The chance to be scattered is very small. 
Interestingly, the distribution is almost flat when $L_z/\lambda\approx 0.5$.

Numerical data shown in Figs. \ref{fig-1d-logg} and \ref{fig-1d-cond}
confirm that the distribution depends only on the ratio $L_z/\lambda$.
Although the system is defined by many parameters - energy, disorder,
length of the system - the only parameter which really determines the  transport
properties is the ratio $L_z/\lambda$. This observation simplifies considerably
the theoretical investigations of the localization, and it is the key assumption
in the scaling theory of localization.

\begin{figure}[t!]
\begin{center}
\psfrag{a1}{$L_z/\lambda=0.0694$}
\psfrag{a2}{$L_z/\lambda=0.139$}
\psfrag{a3}{$L_z/\lambda=0.278$}
\psfrag{a4}{$L_z/\lambda=0.444$}
\psfrag{a5}{$L_z/\lambda=0.555$}
\psfrag{a6}{$L_z/\lambda=0.833$}
\psfrag{g}{$g$}
\psfrag{pg}{$p(g)$}
\includegraphics[clip,width=0.62\textheight]{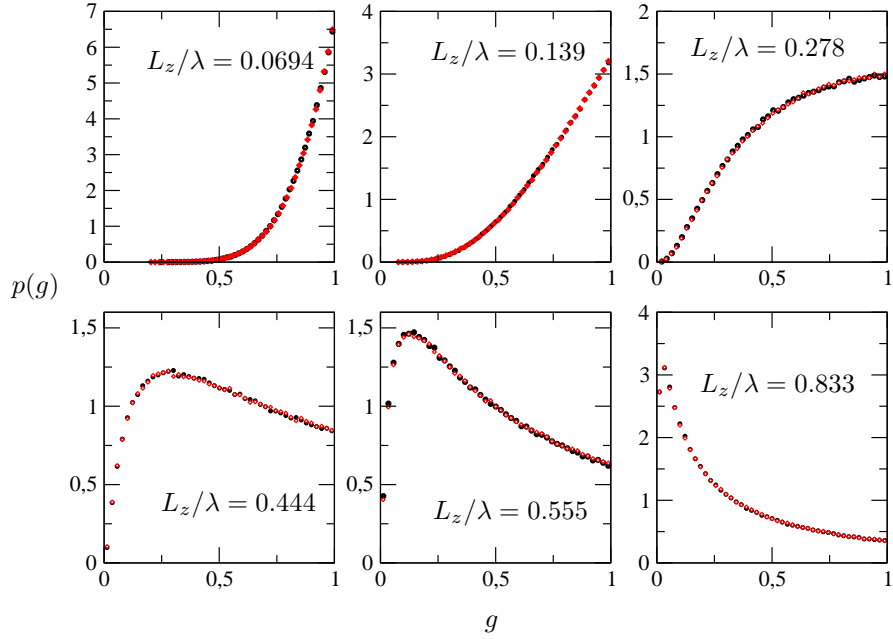}
\end{center}
\caption{%
Conductance distribution for the 1D Anderson model with short system length $L_z$. 
Figures contain data for
$W=0.5$ (the localization length $\lambda=360$) 
and $L=25$, 50, 100, 160, 200 and 300. These data are compared with
data for 
$W=0.25$ ($\lambda=1440$)  and a four times longer system, to show   that systems
with the same ratio $L/\lambda$ have the same distribution $p(g)$.
The energy of the electron is $E=0.5$.
}
\label{fig-1d-cond}
\end{figure}

\subsection{Ergodic hypothesis}

\begin{figure}[t!]
\begin{center}
\includegraphics[clip,width=0.42\textheight]{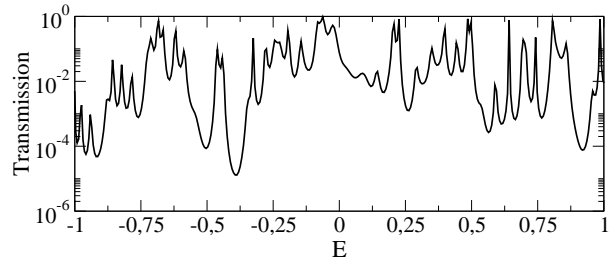}
\end{center}
\caption{%
The conductance of the single disordered one dimensional chain as a function of the
energy $E$ of the electron.
The length of the sample is $L=200$, disorder $W=1$.
Note, the conductance fluctuates within the same orders of magnitude as in Fig. \ref{acta-x2}.
The energy dependence of the conductance is studied in details in Ref. \cite{martin}
}
\label{acta-x1}
\end{figure}

As discussed above, the conductance of a 1D system  is a statistical quantity which wildly
fluctuates as a function of distribution of random energies, $\eeps_r$. This is demonstrated
in Fig. \ref{acta-x2}, which shows the conductance for $N_{\rm stat}=200$
realization of disorder. These data should be compared with those shown  
in Fig. \ref{acta-x1} which  plots the conductance of a given sample as a function
of energy $E$ of the electron. 
We observe that the conductance fluctuates similarly
as in Fig. \ref{acta-x2}. More detailed analysis proved that
the values $g(E)$ create the same statistical ensembles as the values
of the conductance calculated for different samples with fixed energy. 
This result is known as the \textsl{ergodic hypothesis} \cite{EH}.

The ergodic hypothesis plays a crucial role in the experimental studies of the 
transport in disordered systems. Contrary to numerical simulations,
it is impossible to study a huge number of microscopically different
samples in experiments.  Fortunately, due to the ergodic hypothesis, it is
sufficient  to use only one sample and vary the Fermi energy.
Similar equivalence holds also in the metallic regime. Here, not only Fermi energy,
but also the magnetic field can be varied \cite{Maily}. 
The ergodic hypothesis in higher dimensional systems was numerically tested in Ref.
\cite{Kramer-SSC}.


\section{Diffusive regime}\label{sect-diff}

By definition, the diffusive regime is characterized by the diffusive propagation of the
electron in the sample. This happens when the mean free path is sufficiently large
so that electron is scattered separately on two  successive impurities. Also,
the size of the system must be large enough,
\be\label{ucf-1} 
L\gg \ell,
\ee
so that the electron scatters many impurities  before it leaves the sample. 

The diffusive regime can be obtained in the 3D systems with disorder $W<W_c$.
Although no metallic phase exists in 2D systems,
 we can observe the diffusive regime if the size of the system 
 is smaller than the localization length, 
\be\label{ucf-2}
L \ll\lambda.
\ee
Similarly, in weakly disordered quasi one dimensional systems,
described by the DMPK equation (Appendix \ref{app-dmpk}), the diffusive regime exists
when the length $L_z$ of the system does not exceeds localization length.
Note, there is no diffusive regime in one dimensional samples
since  $\ell=\lambda$ \cite{Zdetsis}.

The diffusive regime is characterized by  the conductivity, $\sigma$, given by Eq. (\ref{v-1}), which is
 related to the conductance, $g$, by   Eq. (\ref{g-v}), which we now write in more accurate form,
\be\label{g-vodivost}
\langle g\rangle=\sigma L^{d-2}.
\ee
In Eq. (\ref{g-vodivost}), $\langle\dots\rangle$ means ensemble averaging.
The averaging is crucial in Eq. (\ref{g-vodivost}). Contrary to the conductivity, $\sigma$,
the conductance $g$, is a sample dependent statistical variable, which is not self-averaged.
In Sect. \ref{sect:ucf} we show that fluctuations of the conductance in the diffusive regime 
are universal, independent of the size of the system and the mean value, $\langle g\rangle$.

Owing to Eq. (\ref{ucf-1}),
the electrons are scattered many times inside the sample,
and their wave function becomes totally randomized. Thanks to this randomization,
electron transport  is totally universal in diffusive regime.
We discuss two main transport properties of the diffusive regime: the weak localization
correction to the mean conductivity, and the universal conductance fluctuations.

\subsection{Weak localization}\label{sect:weak}

\begin{figure}[t!]
\begin{center}
\includegraphics[clip,width=0.35\textheight]{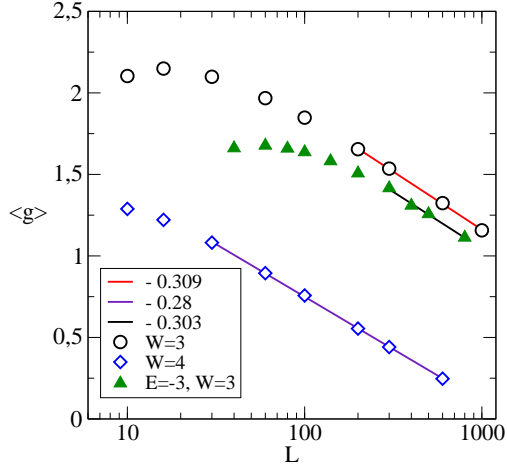}
\end{center}
\caption{%
The conductance (in units of $e^2/h$) in the 2D orthogonal system.
Weak localization corrections are given by a logarithmic decrease of $\langle g\rangle$
when the size of the system increases.
Symbols show numerical data
for the conductance, while the lines are logarithmic fits (note logarithmic scale on the horizontal axis). 
Corresponding slopes are given in the legend.
Theory predicts that the slope is universal, equal to $\pi^{-1}=-0.318$ (Eq. \ref{deltag-2d}).
}
\label{2D_w3_g}
\end{figure}

In the derivation of the conductivity, $\sigma$, given by Eq. (\ref{v-1}),  
no quantum interference processes were considered.  Experimental and
numerical results, however,  show that these processes play an important role
already in the diffusive regime, and lead to the so called \textsl{weak localization corrections}
to the conductance.
We demonstrate the weak localization effects in Fig. \ref{2D_w3_g} 
which shows  that the mean  conductance, $\langle g\rangle$,
of the 2D disordered Anderson model  depends logarithmically on the system size.
This is a consequence of coherent backscattering of quantum electrons.

To understand the origin of this correction, we  calculate  the
probability that the electron, propagating through  the sample,
is scattered back. Figure \ref{fig-back} shows the  typical
closed trajectory of the electron.   
The probability amplitude
associated with the $n$th closed trajectory is $A_n$.
Since the electron is a quantum particle,
the probability  to return back is given by   (in units of $e^2/h$)
\be\label{weak-1}
\delta g_q=\left|\sum_n A_n\right|^2.
\ee
where $n$ counts all possible   closed paths. 
The corresponding contribution for  the  \textsl{classical} particle
is
\be\label{weak-c1}
\delta g_c=\sum_n \left|A_n\right|^2.
\ee
Note, $\delta g_c$ is included already in the  conductance, $\sigma$, given by
Eq. (\ref{v-1}).
Therefore, the logarithmic corrections to the conductance, shown in Fig.
\ref{2D_w3_g} originate  form the difference
\be\label{weak-diff}
\delta g=\delta g_q - \delta g_c.
\ee

The r.h.s. of Eq. (\ref{weak-1}) can be written in the form
\be
\sum_n |A_n|^2+\sum_{n\ne n'} \left(A_n^*A_{n'}+A_nA_{n'}^*\right) =
\delta g_c+\sum_{n\ne n'} \left(A_n^*A_{n'}+A_nA_{n'}^*\right).
\ee
In random systems, the sum of the off-diagonal  terms is zero because of random phases
of complex amplitudes $A$. However, there is an important exception,
when $A_n$ and $A_{n'}$ correspond to the same closed path, but 
opposite in direction to the electron propagation. Consider
one loop, as shown in Fig. \ref{fig-back}. The electron can travel this path 
in both directions; both  possibilities give the same contribution, 
$|A|^2=|A_+|^2= |A_-|^2$. 
Moreover, if the system possesses time reversal symmetry (which is
the case of 2D Anderson model), then also $A_+=A_-$, and the off-diagonal terms,
\be\label{weak-2}
A_+A_-^*+A_+^*A_- = 2|A|^2
\ee
 are real, independent of phase. Then, in the quantum case, a
 given loop contributes to backscattering by 
\be\label{weak-3}
\delta g_q=|A_+|^2+|A_-|^2+ A_+A_-^*+A_+^*A_- = 4|A|^2.
\ee
In the classical case, this contribution is  only 
\be\label{weak-5}
\delta g_c=2|A|^2,
\ee
The difference between \textsl{classical} and \textsl{quantum} backscattering,
given by Eq. (\ref{weak-diff}),
gives rise to the \textsl{negative} weak localization corrections  to the conductance.

We do not calculate $\delta g$ here, only give a final result \cite{McKK-93},
\be\label{weak-int}
\delta g=\delta g_q-\delta g_c=-4\frac{e^2}{h}L^{d-2}\ds{\frac{1}{(2\pi)^d}\int \frac{d^d\vec{q}}{q^2}}.
\ee
Here, $q$ is the wave vector of the electron propagating on the loop. 
To avoid the divergences in integral (\ref{weak-int}), we introduce the
lower,  $q_{\rm min}=1/L$, and the upper, $q_{\rm max}=1/\ell$, 
integration  limits \cite{Kaveh}.
Then, integration gives for dimension $d=2$ the expression
\be\label{deltag-2d}
\delta g=-\frac{1}{\pi}\ln L/\ell,~~~~~~~~~d=2,
\ee
and for $d=1$ and $d=3$  the formulas
\be
\delta g= \left\{
\begin{array}{ll}
-\pi^{-2}(L/\ell -1)  & d=3\\
~~  & ~~~\\
\ell/L-1  &    d=1.
\end{array}
\right.
\ee

\begin{figure}[t!]
\begin{center}
\psfrag{r}{$\vec{r}$}
\includegraphics[clip,width=0.3\textheight]{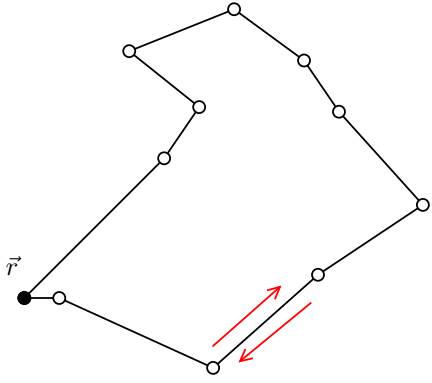}
\end{center}
\caption{The closed trajectory of the electron  in a disordered 
2D system. On the travel along the trajectory, the electron is  \textsl{coherently} scattered.
If the systems possess time reversal symmetry, then traveling in both
directions gives the same contribution to the conductance.
}
\label{fig-back}
\end{figure}

Since the backscattering is larger for the quantum particle, the correction
to the conductance must be negative.
Note, however, that the above results hold only in systems with time reversal symmetry.
When a strong magnetic field is applied to the system, 
the time reversal symmetry is broken
and $\delta g=0$.  Indeed, the electron acquires on the closed loop the phase
\be
\phi_n=\pm\int \vec{A}(\vec{r})d\vec{r}, 
\ee
where $\vec{A}$ is a vector potential and the sign $\pm$ depends on the propagation 
direction. 
Then, the product $A_nA^*_{n'}$ is not real,
but possesses the phase factor $e^{2i\phi_n}$.
Since the phase $\phi_n$ depends on the length of the loop,
in the summation over all loops, all complex contributions cancel.

In symplectic models, when the  hopping between two neighboring sites becomes spin
dependent, the quantum contribution to the backscattering changes the sign. The 
reason is that the wave function of the  particle with spin 1/2 transforms
into itself after rotation in  $4\pi$, contrary to spin-less particle
\cite{Bergman}. The weak anti-localization corrections to the conductance 
for the 2D Ando model is  shown
in Fig. \ref{2D-ando}.

\begin{figure}[t!]
\begin{center}
\includegraphics[clip,width=0.35\textheight]{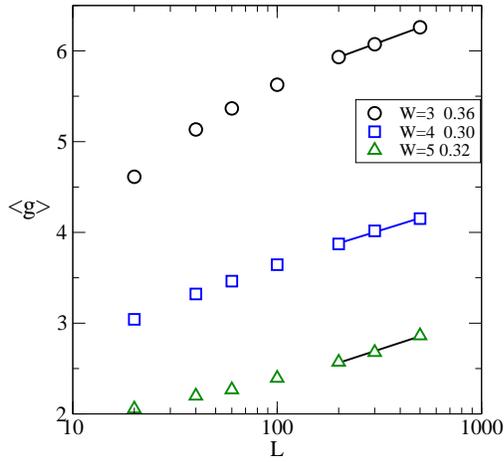}
\end{center}
\caption{The weak anti-localization corrections to the  conductance for
the 2D Ando model. The conductance is given in units $e^2/h$. 
The legend presents the calculated slopes. Theory predicts the slope $+\pi^{-1}=0.318$.
More accurate data for the weak anti-localization correction were obtained within the SU(2) model
in Ref. \cite{SOx}.
}
\label{2D-ando}
\end{figure}

\subsection{Weak localization in quasi-1d systems}

\begin{figure}[t!]
\begin{center}
\includegraphics[clip,width=0.32\textheight]{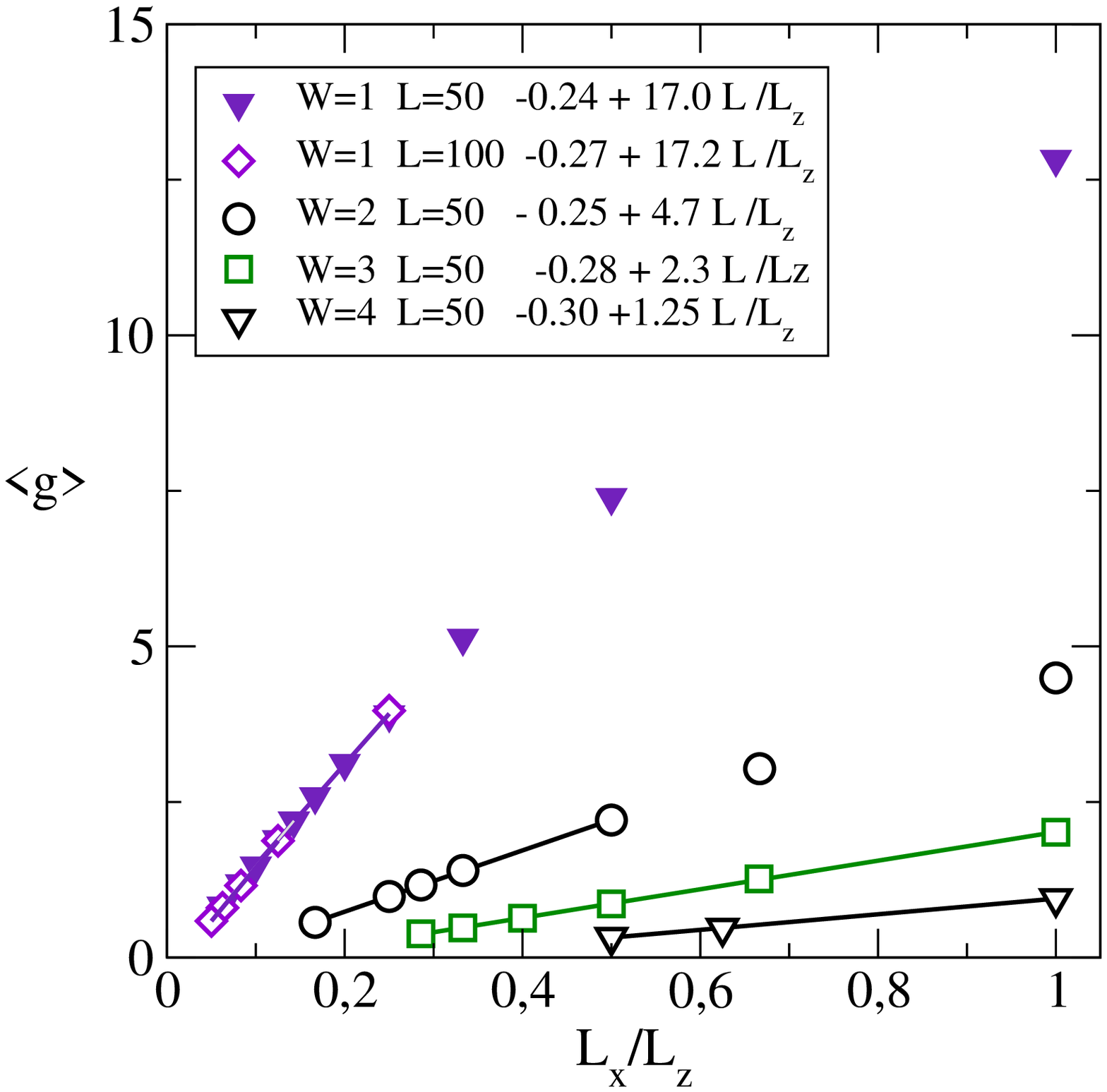}~~~~
\includegraphics[clip,width=0.32\textheight]{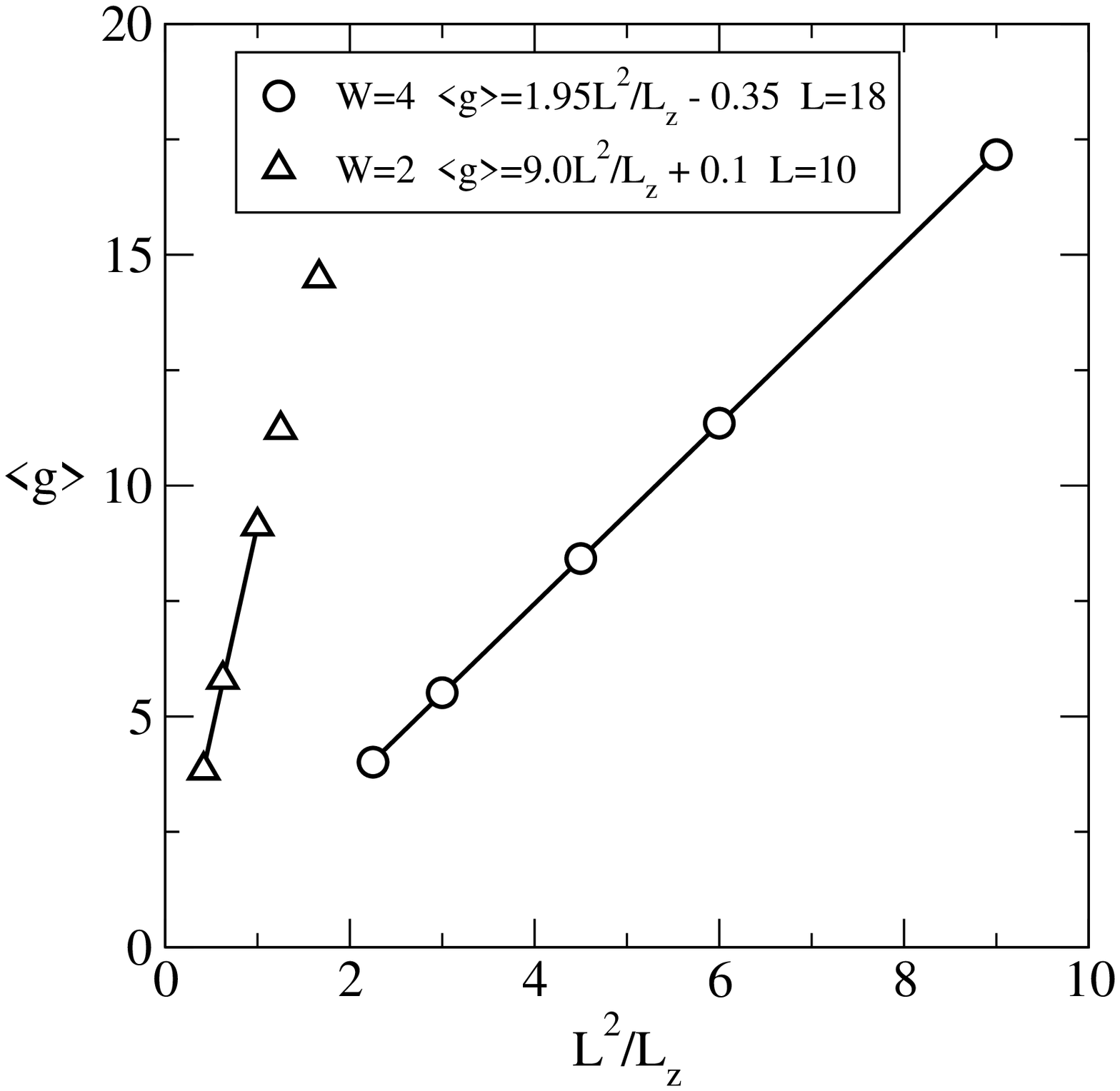}
\end{center}
\caption{The mean conductance, $\langle g\rangle$, as a function of $N/L_z$ for 
the 2D systems (left) and the  3D anisotropic model (\ref{hama}) with $t=0.4$ (right). 
The slope of the linear fit  determines, due to Eq. (\ref{weak-mfp}),
the mean free path, $\ell$. 
Left: The 2D system, $N=L=50$. We found $\ell=17.0$,
4.9, 2.30 and 1.25 for $W=1$, 2, 3 and 4, respectively.
For $W=1$, we add also data for $L=100$. 
Right: The 3D systems of the size $L\times L\times L_z$.  For $W=4$ (circles)
we found $\ell=1.92$, while $\ell\approx 9$ for $W=2$.
Note, the weak localization corrections in both systems 
are close to $-1/3$, as  predicted by
DMPK equation (Appendix \ref{app-dmpk}). 
The only exception is the 3D system with $W=2$, since $\ell\approx L$
in this case.
}
\label{fig-mfp}
\end{figure}

Of particular interest is transport in weakly
disordered quasi one dimensional systems. This problem is exactly
solvable. In Appendix \ref{app-dmpk} we introduced the DMPK equation for the
joint probability distribution of parameters $\lambda$ of the transfer matrix.
From the DMPK equation,  the following expression for the  mean conductance
can be derived \cite{Stone} (in units of $e^2/h$),
\be\label{weak-mfp}
\langle g\rangle =\frac{N\ell}{L_z}-\frac{1}{3},
\ee
(Eq. \ref{dmpk-g}).
The weak localization correction is constant, independent of the system length, provided that
$\ell\ll L\ll N\ell$. 

Note that  expression (\ref{weak-mfp}) was derived in the limit of infinite number
of transmission channels, $N\to\infty$. Therefore, it cannot be applied to the 1D
models, where $N=1$.

Figure \ref{fig-mfp} verifies both the $L_z$ dependence of the mean conductance and
the weak localization correction, $\delta g=1/3$ for  two different quasi-1d systems.
Note, relation (\ref{weak-mfp}) can be also used for the numerical
calculation of the mean free path, $\ell$.
Numerical data for the conductance in quasi-1d systems 
are presented in Fig. \ref{fig-mfp}.
The obtained values of the mean free path are  rather small, of the order of the lattice period.
Only when disorder is very weak ($W=1$ in 2D),
then $\ell\approx 17$. 
Note that in the Born approximation \cite{Zdetsis}, 
the mean free path decreases with the disorder as
\be
\ell\propto W^{-2}.
\ee
The numerical data 
for the mean free path of 2D systems, obtained in Fig. \ref{fig-mfp}
are in very good agreement with results of
Ref. \cite{Zdetsis}, shown in  Fig. \ref{mfp-data}.

\begin{figure}[t!]
\bc
\psfrag{ll}{$\ell$}
\psfrag{LL}{$\lambda$}
\includegraphics[clip,width=0.3\textheight,height=0.3\textheight]{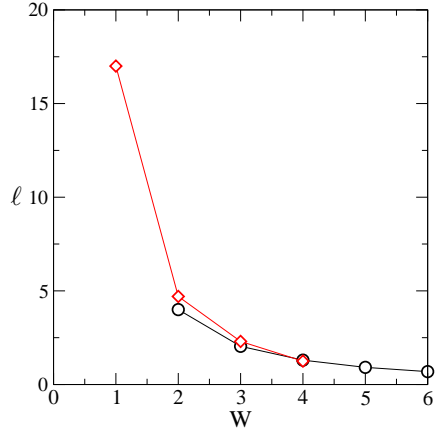}
\ec
\caption{The numerical data for the mean free path, $\ell$,
from Fig. \ref{fig-mfp} (diamonds), compared with results of Ref.
\cite{Zdetsis} (circles).
}
\label{mfp-data}
\end{figure}

\subsection{Conductance distribution. Universal conductance fluctuations}\label{sect:ucf}

Because of the randomness, the conductance is a statistical variable, and can vary
from sample to sample as a function of the  distribution of random energies 
$\eeps_n$. 
In the diffusive transport regime, the conductance distribution  is Gaussian, as 
shown in Fig. \ref{D-g}.

Since the disorder $W$ is weak, the statistical properties of the conductance
can be derived theoretically, either by the perturbation Green's function method \cite{LSF}
or within the framework of the DMPK equation, introduced in Appendix \ref{app-dmpk}.
The distribution of the conductance is Gaussian with universal width.
The variance,
\be
\textrm{var}~g=\langle g^2\rangle - \langle g\rangle^2
\ee
is a universal number, independent on the disorder strength (provided that
inequalities (\ref{ucf-1}, \ref{ucf-2}) hold). $\textrm{var}~g$ depends
only on the dimension of the system and on boundary conditions
\cite{LSF}. 
This phenomenon is known as universal conductance fluctuation. 

\begin{figure}[t!]
\begin{center}
\includegraphics[clip,width=0.25\textheight]{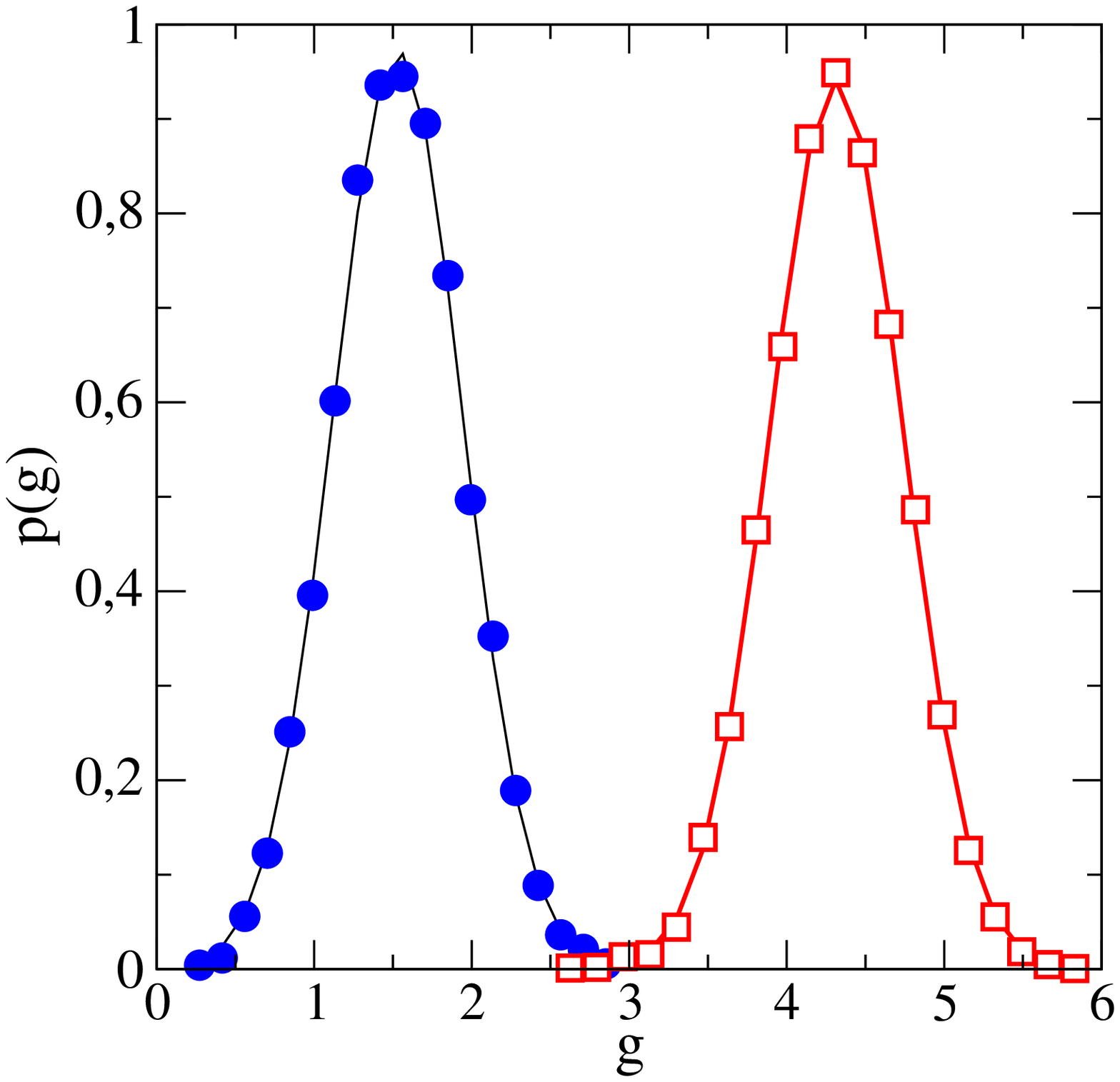}
\includegraphics[clip,width=0.25\textheight]{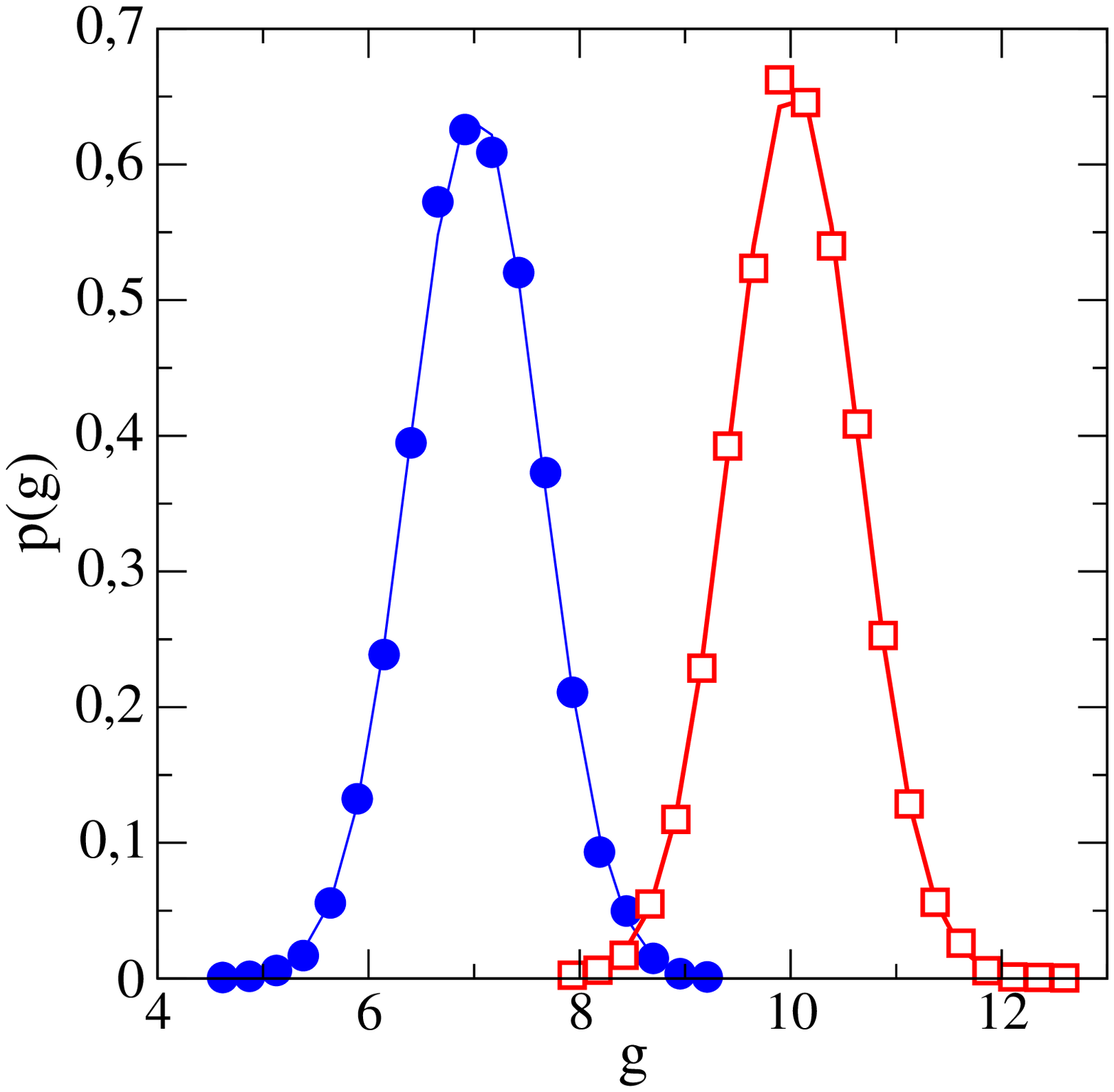}
\end{center}
\caption{The probability distribution $p(g)$ of the conductance for weakly 
disordered systems. Left: The 2D samples of the size $300\times 300$. Disorder
$W=3$ (circles) and $W=2$ (squares). Solid lines are Gaussian fits with variance
$0.168$ and $0.179$ for $W=3$ and $W=2$, respectively. This is close to the theoretical prediction, $0.185$,
obtained for hard wall boundary conditions in Ref. \cite{RMS-m}. 
Right: The 3D system with box distribution of random energies,
$W=6$ and $L=10$ (circles) and $L=14$ (squares). Solid lines are Gaussian fits
with mean value $\langle g\rangle =7$ ($L=10$) and $=10.03$ ($L=4$). These values
confirm the relation $\langle g\rangle =\sigma L$
with the conductance $\sigma=0.7$.
The width of distributions is $\textrm{var}~g=0.38$ and $0.366$ for $L=10$ and
$L=14$, respectively, are close to the theoretical prediction
$0.314$ \cite{RMS-m,igor}.  Data were obtained numerically with statistical ensembles of
$\ns=10^4$ samples.
}
\label{D-g}
\end{figure}

Also, $\textrm{var}~g$ depends on the physical symmetry of the model. It was proved
in Ref. \cite{LSF} that 
\be\label{ucf-beta}
\textrm{var}~g\propto\frac{1}{\beta}.
\ee
We remind the reader that $\beta=1,~2$ and 4 for the orthogonal, unitary and symplectic 
systems, respectively. For the quasi-1d systems, the same universality relation was
derived in Ref. \cite{Stone,Been}.

In systems with higher dimension, $\textrm{var}~g$ depends also on the boundary conditions
in the directions perpendicular to the propagation \cite{LSF}. This dependence
was theoretically investigated in Ref. \cite{BHMMcK} and numerically verified in \cite{RMS-m}.

\begin{table}[b!]
\bc
\begin{tabular}{|l|l|l|}
\hline
d    &    HW   &  periodic\\
\hline
1      &  2/15  &  2/15\\
2      &  0.185613   &  0.154078\\
3      &  0.314054   &  0.2194393\\
\hline
\end{tabular}
\ec
\caption{Universal values of the conductance fluctuations,
$\textrm{var}~g=\langle g^2\rangle-\langle g\rangle^2$ 
for the $d$-dimensional disordered systems with the hard wall
and periodic boundary conditions \cite{igor}. For other symmetry classes,
$\textrm{var}~g$ must be divided  by a factor of $\beta$, (Eq. \ref{ucf-beta}).
$\textrm{var}~g\sim \epsilon^{-1}$ in the dimension $d=4-\epsilon$.
}
\label{table-ucf}
\end{table}

The deviations  of $p(g)$ from Gaussian form can be measured by the third cummulant,
\be
\langle g^3\rangle_c=\langle g^3\rangle-3\langle g^2\rangle\langle g\rangle+2\langle g\rangle^3.
\ee
It was shown analytically \cite{vRAN} that $\langle g^3\rangle_c=0$
for the quasi-one dimensional systems and is $\sim 10^{-3}\langle g\rangle^{-1}$
for the 3D system. Such small values are not observable with today's computer facilities.

\begin{figure}[t!]
\begin{center}
\psfrag{LL}{$L/\ell$}
\psfrag{ll}{$\ell$}
\includegraphics[clip,width=0.3\textheight,height=0.3\textheight]{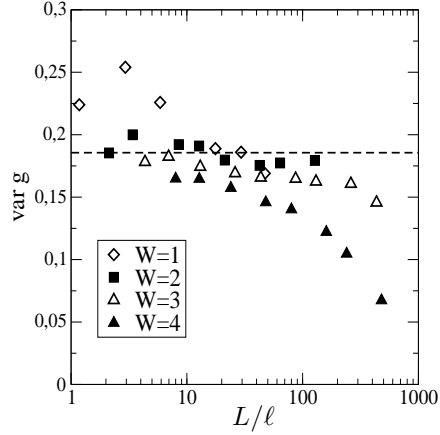}~~~~~~~
\end{center}
\caption{Conductance fluctuations for the 2D Anderson model.
Different symbols correspond to the disorder $W=1$, 2, 3, 4
and 5. The size of the system increases from $L=10$ to
$L=1000$. 
The horizontal dashed line is the theoretical prediction
$\textrm{var}~g\approx 0.1856$ of the universal conductance fluctuations 
for 2D orthogonal system with the  hard wall boundary conditions \cite{RMS-m}.
For $W=1$, the mean free path
$\ell\sim 17$ is already comparable with $L$ for some smaller samples. 
The system is in the ballistic regime and $\textrm{var}~g$  exceeds universal value.
A more detailed analysis of the universal conductance fluctuations has been done in
\cite{RMS-m}.
}
\label{2d-ucf}
\end{figure}

The conductance fluctuations for the 2D weakly disordered systems 
are  plotted in Fig. \ref{2d-ucf}. 
Note how the variance of the conductance is universal only when the size of the sample 
is much larger than the mean free path, $L> 10\times \ell$. 
However,
 $L$ must be much smaller than 
the localization length,  $L/\lambda\ll 1$.
Also, for small system size, when condition (\ref{ucf-1}) is not satisfied, the  system is
in the ballistic regime, in which the scattering of electrons is not sufficiently strong
to randomize the wave function sufficiently. In this regime, the conductance is large,
$g\sim\no$, and the fluctuations of the conductance increase. Of course, 
the variance decreases to zero in the limit
of a regular lattice ($W=0$).

\subsection{Other universal relations}

Universal conductance fluctuations are the consequence of the universality
of the diffusive transport. In Appendices \ref{app-dmpk} and \ref{app:rmt}
we discuss the statistical properties of the eigenvalues $\lambda_a$
of the matrix $\left[t^\dag t\right]^{-1}$.
It turns out, that in the diffusive regime the joint probability distribution
of the eigenvalues $\lambda$ is universal and depends only on the
length of the system and the number of the transmission channels,
$\no$. The disorder, $W$, influences the transmission parameters only in terms of the
ratio $L_z/\lambda$ of the system length and the mean free path.

From the universality of the probability distribution $p(\lambda)$ we can obtain other universal
relations for the transmission parameters.  In Appendix \ref{app:rmt}
we show that the spectrum of parameters $x_a$, defined by Eq. (\ref{konduktancia}) is linear,
\be
x_a\propto a,
\ee
as shown in Fig. \ref{fig-xa}.
Also, the differences, $x_{a+1}-x_a$  are distributed with the Wigner probability distribution
(Fig. \ref{fig-pdelta}).

\begin{figure}[t!]
\begin{center}
\includegraphics[clip,width=0.3\textheight]{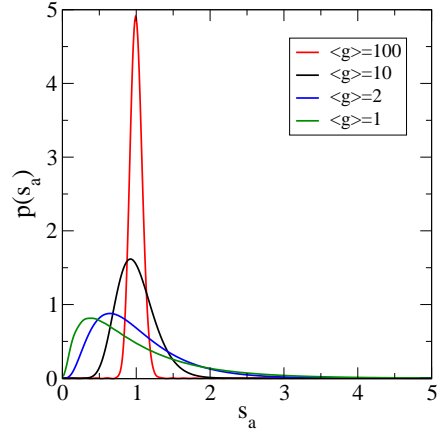}
\end{center}
\caption{Probability distribution $p(s_a)$, given by Eqs. (\ref{diff-ps1},\ref{diff-ps2})
for various values of the mean conductance,
$\langle g\rangle$.
}
\label{ps_ukazka}
\end{figure}

Here we want to show that not only the conductance, $g$, but also the 
 transmission probabilities, $T_{ab}$ and $T_{a}$, introduced in Sect. \ref{sect:multi} as
\be\label{diff-tab}
T_{ab}=|t_{ab}|^2
\ee
and
\be\label{diff-ta}
T_a=\sum_b|t_{ab}|^2= \sum_b T_{ab},
\ee
exhibit  universal statistical properties. We remind the reader that
$T_{ab}$ 
is the probability that the electron, coming in channel $a$ is transmitted through the
sample and leaves  the sample in channel $b$ and  $T_{a}$ is the probability
that the electron, coming in channel $a$, transfers through the sample.
It is clear that
\be\label{diff-g}
g=\sum_a T_a.
\ee
Note, $T_a$ is \textsl{not} the eigenvalue of the matrix $t^\dag t$.

\begin{figure}[t!]
\begin{center}
\includegraphics[clip,width=0.55\textheight]{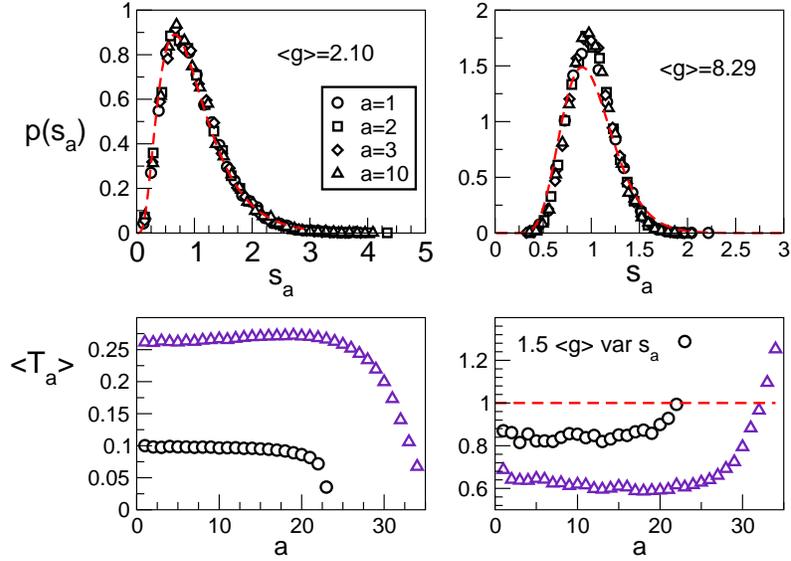}
\end{center}
\caption{Upper panels show the probability distribution of the parameters $s_a$,
defined by Eq. (\ref{eq-sa}) for the 
2D Anderson model,  with the random binary potential given by Eq.  (\ref{p-xx}).
The energy of the electron,  $E=0.14$ (left upper panel) and $E=0.31$ (right upper panel) is  
measured in this particular case from the bottom of the conductance band. 
Different symbols represent the data
for  channels $a=1$, 2, 3 and 10. The solid line is the theoretical prediction,
given by Eqs. (\ref{diff-ps1},\ref{diff-ps2}).
Two lower panels show the mean
transmission, $\langle T_a\rangle$ and  the variance,
$\textrm{var}~s_a$, given by Eq. (\ref{varsa}). Note that all channels are almost equivalent and 
they give the same contribution to the conductance. 
The number of open channels is
$\no=23$ and $\no=34$ for $E=0.14$ and $E=0.31$, respectively.
The size of the system is $192\times 192$ \cite{PRB2005}.
}
\label{s2c0}
\end{figure}

The universal probability distribution for the \textsl{normalized} transmission probability,
\be\label{eq-sa}
s_a=\frac{T_a}{\langle T_a\rangle},
\ee
was derived in Refs.  \cite{NR,Kogan}. 
The distribution $p(s_a)$ is determined  only by the mean conductance, and is given by the
following analytical formula,
\be\label{diff-ps1}
p(s_a)=\int_{-i\infty}^{+\infty}\frac{dx}{2\pi i}e^{xs_a-\Phi(x)}
\ee
where 
\be\label{diff-ps2}
\Phi(x)=\langle g\rangle \left[\ln\left(\sqrt{1+x/\langle g\rangle}+
\sqrt{x/\langle g\rangle}\right)\right]^2.
\ee
The probability distribution $p(s_a)$ 
is shown in Fig.  \ref{ps_ukazka}
for a few values of the mean conductance. 
Distribution $p(s_a)$ possesses universal 
variance,
\be\label{varsa}
\textrm{var}~s_a=\langle s_a^2\rangle-\langle s_a\rangle^2=\frac{2}{3\langle g\rangle}.
\ee

Universality of the distribution (\ref{diff-ps1}) was confirmed experimentally
in experiment with microwave electromagnetic waves \cite{Kogan,Genack}.
Numerically, it was studied in Ref. \cite{PRB2005}  for
the 2D system with correlated binary disorder
\be\label{p-xx}
p(\eeps)=(1-x)\delta(\eeps)+x\delta(\eeps+V_b).
\ee
Spatial correlations were introduced so that random energies create randomly
distributed potential wells of the size of $3\times 3$ lattice sites.

Figure \ref{s2c0} shows the numerically obtained data 
for the system of size $192\times 192$ and for two
energies of the electron. Data confirm that indeed 
the channels are equivalent to each other, 
and they give the same value of the transmission, $\langle T_a\rangle$. This is a typical
property of the diffusive regime: transmission in a given channel does not depend
on the incident angle, since electron, being many times scattered 
inside the sample, forgets the initial direction of propagation.

\begin{figure}[t!]
\begin{center}
\includegraphics[clip,width=0.45\textheight,height=0.45\textheight]{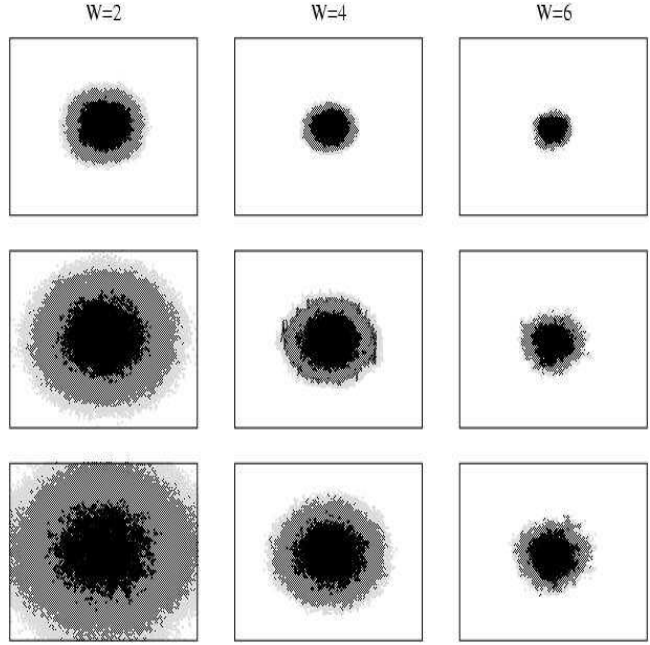}
\end{center}
\caption{The time dependence of the wave packet in the 2D disordered system,
defined by Hamiltonian (\ref{ham}). Shown are points where
 $|\Psi(\vec{r})|>0.0005$ (gray), $|\Psi(\vec{r})|>0.0010$ (dark gray),  and
 $|\Psi(\vec{r})|>0.0050$ (black) at time $t=100$, $500$ and $900$ 
 (from top to bottom).
The time is measured in units $\hbar/V$ with $V=1$ and 
the size of the lattice is $512\times 512$.   For weak disorder, 
the wave function of the electron diffuses and $\rangle r^2\langle =2Dt$
where $D$ is the diffusion coefficient. Clearly, $D$ decreases
when disorder increases. For $W=6$ (right column), diffusion already stops
and the electron is localized (see also Fig. \ref{fig-sche-y1}).
}
\label{fig-sche-w}
\end{figure}

\begin{figure}[t!]
\begin{center}
\psfrag{r2}{$\langle r^2\rangle$}
\includegraphics[clip,width=0.35\textheight]{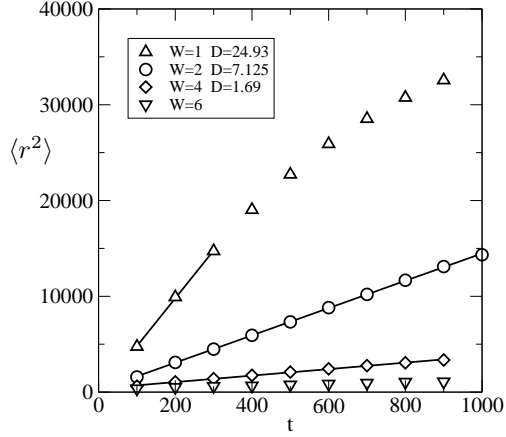}
\end{center}
\caption{Left: $\langle r^2\rangle$ \textsl{vs} time $t$ for the 2D Anderson model with box
disorder.
The slope of the linear dependence determines the diffusive coefficient $D$.
The data for disorder $W=2$ and $W=4$ corresponds to the wave function shown in Fig.
\ref{fig-sche-w}. For weak disorder, $W=1$, we see already the saturation 
of  $\langle r^2\rangle $ due to the finite size of the system and 
reflection of the wave packet from boundaries. Solid lines are linear fits,
$\langle r^2\rangle =2Dt$. 
Time is measured in units $\hbar/V$. The diffusion coefficient, $D$, measured in units
$a^2V/\hbar$, is given in legend.
}
\label{fig-sche-diff}
\end{figure}

\subsection{Diffusion}

Conductivity, $\sigma$, can be expressed through the diffusion  coefficient, $D$,
$\sigma=e^2D\rho$, given by Eq. \ref{v-1}. The density of states, $\rho(E)$ was calculated in section
\ref{sec:density}.
The diffusion coefficient  can be calculated numerically 
solving the time dependent Schr\"odinger equation,
Eq. (\ref{time-sche}). Figure \ref{fig-sche-w} shows the time evolution of the
single electron wave function in disordered 2D samples with different 
strength of the disorder. We use the same data to
calculate the width of the wave packet,
\be\label{diff-r2}
\langle r^2(t)\rangle=\int~d\vec{r}\Psi^*(\vec{r},t)r^2\Psi(\vec{r},t)
\ee
in time $t$.
When disorder is small, the  electron diffuses from the center, and we obtain that
\be\label{2dt}
\langle r^2\rangle = 2Dt.
\ee
In the case of strong disorder,  $\langle r^2\rangle$ should converge to 
the time-independent quantity, when the electron is localized. This was shown already 
in Fig.  \ref{fig-sche-yy} for the case of strong  disorder.

Figure \ref{fig-sche-diff} shows the time dependence of $\langle r^2\rangle$
for various strengths of the disorder.  The diffusion constant, $D$, is obtained 
as the slope of the linear dependence $\langle r^2\rangle$ \textsl{vs} $t$.
If the density of states, $\rho(E)$ is known, then   we can obtain the
conductance, 
\be\label{ggg}
g=\sigma=\frac{e^2}{h} h D\rho.
\ee
Using numerical data for the disorder $W=4$: $\rho(E)\approx 0.15/a^2V$ 
(Fig. \ref{rho-2d}),
and $D=1.69~a^2V/\hbar$ (Fig. \ref{fig-sche-diff}), 
we obtain that 
$g(W=4)=1.59(e^2/h)$.
This is consistent with numerical data shown in Fig. \ref{2D_w3_g}.

\begin{figure}[t!]
\begin{center}
\includegraphics[clip,width=0.28\textheight]{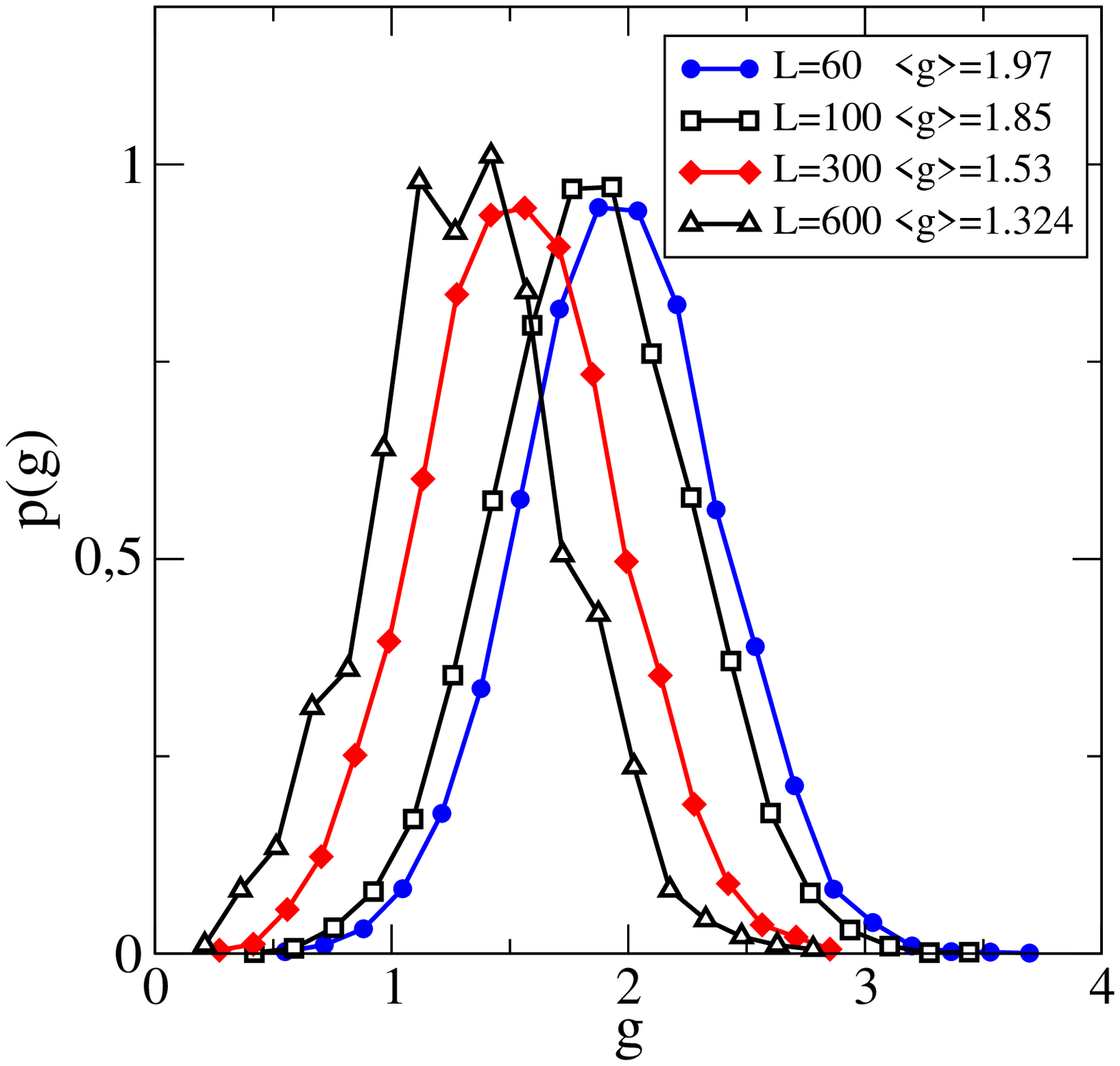}~~~~~~~~~
\includegraphics[clip,width=0.25\textheight]{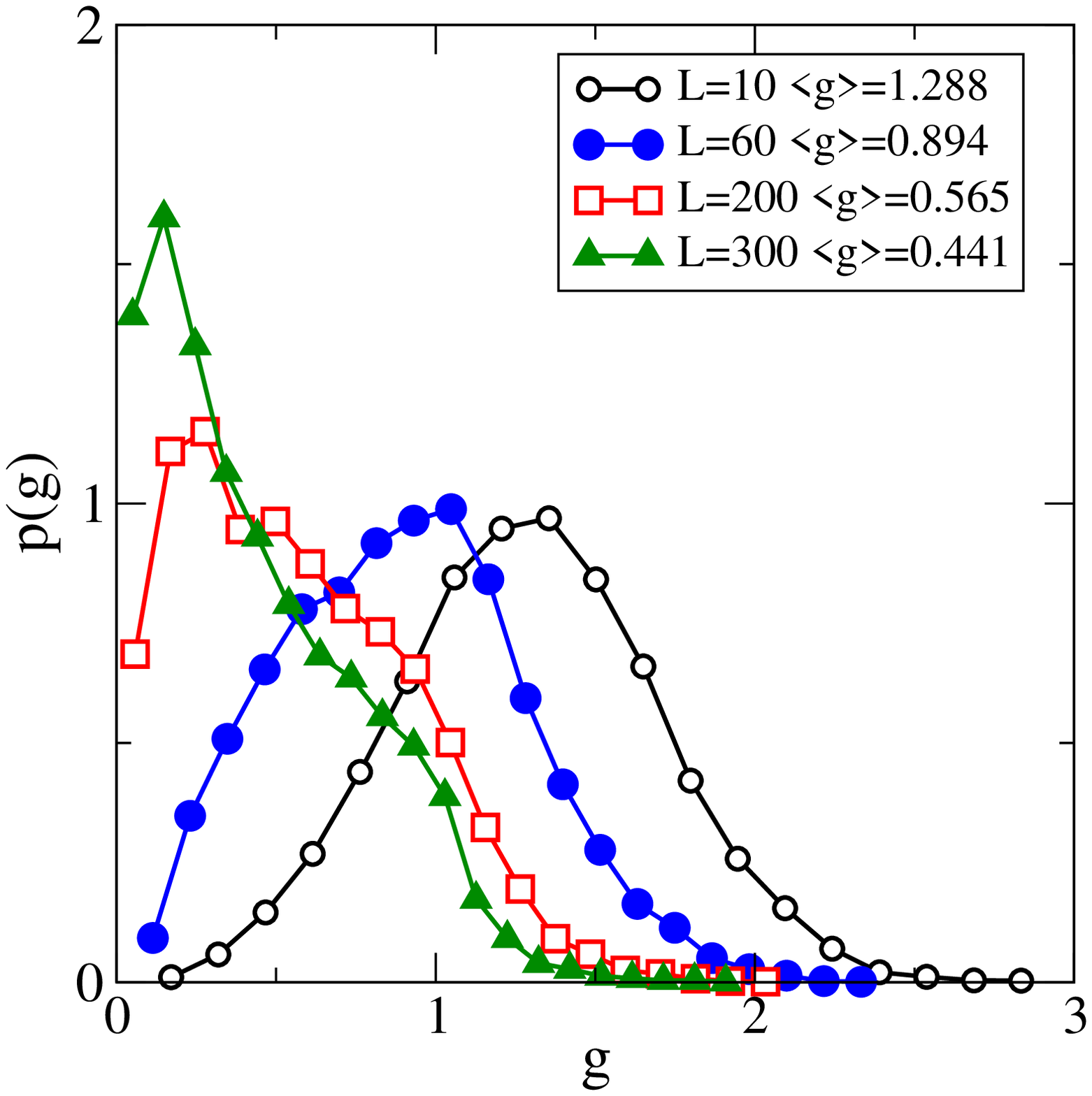}
\end{center}
\caption{The system size dependence of the conductance distribution, $p(g)$ for the  2D systems.
Left: $W=3$. The conductance is almost independent to the system size, since
system is in the diffusive regime ($\ell\approx 2.3$, and $\lambda=5046$). Right: 
$W=4$. The distribution $p(g)$ is Gaussian
only for very small systems ($L=10$), and it changes considerably when $L$ increases.
The mean free path, $\ell\approx 1.3$ and localization length, $\lambda=481$. 
The estimation of the localization length is taken from Ref. \cite{McKK-1983}.
}
\label{2D-g}
\end{figure}

\begin{figure}[b!]
\begin{center}
\includegraphics[clip,width=0.35\textheight]{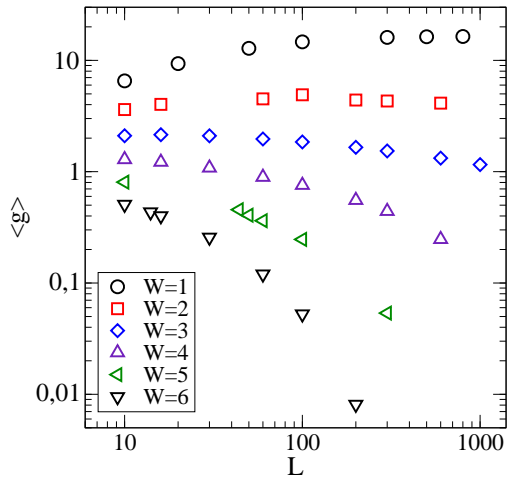}
\end{center}
\caption{%
The length dependence of the mean conductance, $\langle g\rangle$,
for the 2D disordered systems. 
When disorder is weak, $\langle g\rangle$ increases for small $L$ 
since the system is in the ballistic regime. Naively, this might be
interpreted as a metal-insulator transition:
since the mean  conductance increases for $W=1$ and 
decreases for $W=5$,  one might
conclude that the 2D system possesses the critical point, $W_c\le 1$, i. e.
 that disorder $W=1$  corresponds to the metallic regime. 
}
\label{2D-gW}
\end{figure}


\subsection{Beyond the diffusive regime}\label{sect:beyond}

As will be discussed in next Section, there is no true metallic regime in 2D
orthogonal systems.   What we have observed so far is the diffusive propagation
of the electron on a \textsl{finite} square lattice of size $L\times L$. 
When $L$ increases over the localization length, $\lambda=\lambda(W)$, the electron 
becomes localized and the $L$ dependence of the conductance changes from 
the logarithmic to the  exponential.

In Fig. \ref{2D-g} we demonstrate how 
the conductance distribution, $p(g)$, of the 2D disordered systems changes when the size
of the system, $L$, increases.
For disorder $W=3$, 
$p(g)$ is Gaussian even for $L=1000$. However, for the disorder $W=4$, the localization length
$\lambda\approx 50$ and the probability distribution changes its form 
considerably.

In the opposite limit of small $L$, the system is in the ballistic regime. 
In the limit of $L<\ell$, there is almost no scattering inside the sample,
so  the conductance $\langle g\rangle$ is  given by the number of open channels.
In 2D, this means that
$\langle g\rangle\propto L$. 
Increase of the conductance  with the system size  is shown in
  in Fig. \ref{2D-gW}.
Note, the ballistic regime can be observed also in the 1D systems. 
Fig. \ref{fig-1d-cond} shows that for sufficiently short systems, 
the conductance of 1D disordered system is close to 1 when $L\ll\xi$.

Since the electron   is scattered only on a few impurities during its travel 
through a ballistic sample,  the values of the conductance for a given sample depends
on the actual distribution of impurities. Consequently, fluctuations of the conductance increase
in the ballistic regime
\cite{ball}. This is confirmed  in Fig. \ref{2d-ucf}.


\section{Scaling theory of localization}\label{sect:scaling}
As discussed in the Introduction, the metal-insulator transition resembles the
phase transition in statistical physics. We would like to develop
the theory, similar to the renormalization group theory of second
order phase transitions. 

The first step in this direction is to
determine the relevant order parameter. 
It turns out that the most suitable candidate for such parameter is
 the Thouless conductance, $g_T$.
First, it can be defined both in the metallic and localized regime. Second, 
the analysis of the sensitivity to the boundary conditions tells us how the
conductance $g_T$ develops when the size $L$ of the system increases. Since $g_T$ is equivalent
to $g_{ES}$ \cite{vztah} both in the metallic and in the insulating regime,
we will consider the conductance $g_{ES}$ and use the simpler notation, $g$.

The main assumption we have to accept is that for the system size sufficiently
large,  the length dependence of the conductance is given only by the
conductance itself:
\be\label{beta}
\ds{\frac{\partial\ln g}{\partial \ln L}}=\beta(g)
\ee
The ``sufficiently large'' system size $L$ means that  $L$ exceeds all ``natural''
lengths which might determine the transport in the system. Some examples of such
scales are
the mean free path, $\ell$, which determines the coherent scattering on impurities,
the coherence length of the random potential, or magnetic length,
which is important if
the  magnetic field is present.
 If $L$ is much larger than all these scales, we expect that
microscopic details of the model become irrelevant.

Now, we want to derive some consequences from the Eq. (\ref{beta}). First of all,
we must admit that the form of the function $\beta(g)$ is unknown. However, we can
derive its limits for $g\to\infty$ and $g\to 0$. Consider for simplicity 
only the orthogonal symmetry, so that the symmetry parameter $\beta=1$.
In the limit of large conductance, we can use the $L$-dependence of the conductance,
derived in Sect. \ref{sect-diff}. The leading term of the conductance behaves as
$g=\sigma L^{d-2}$. Inserting into Eq. (\ref{beta}), we obtain that
\be\label{beta-lim1}
\lim_{g\to\infty}\beta(g)=d-2.
\ee
Since the first correction to the conductance, derived in Sect. \ref{sect:weak}
is \textsl{negative},
we immediately see that $\beta(\ln g)$ reaches the limit (\ref{beta-lim1}) from below.
In particular, for $d=2$, we obtain from Eq. (\ref{deltag-2d}) that 
\be
\beta(g)=\ds{\frac{\partial\ln g}{\partial\ln L}}=\ds{\frac{1}{g}\frac{\partial g}{\partial\ln L}}=
 -\frac{1}{\pi g}.
\ee
In the opposite limit, $g\ll 1$, we have exponential localization,
\be
g\sim e^{-2L/\lambda},
\ee
so that 
\be\label{beta-lim2}
\beta(g)=-\frac{2L}{\lambda}=\ln g.
\ee

Now, we can interpolate between  the  limits (\ref{beta-lim1}) and (\ref{beta-lim2}). This is 
straightforward if we \textsl{assume} that the function $\beta(g)$ is always continuous and monotonous.
Although we do not have a rigorous proof that $\beta(g)$ really fulfills these conditions,
there is no physical reason not to accept them. Then, connecting both limits, we obtain
that the function $\beta(g)$ behaves as shown in Fig. {\ref{fig-beta}.
\begin{figure}[b!]
\begin{center}
\psfrag{beta}{$\beta(g)$}
\psfrag{ln1}{$\ln g$}
\psfrag{ln2}{$\ln g_c$}
\includegraphics[clip,width=0.35\textheight]{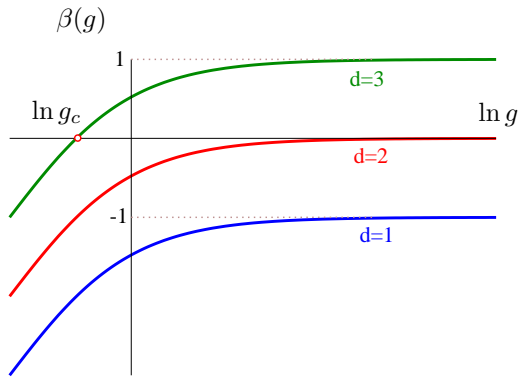}
\end{center}
\caption{The function $\beta(g)$ for the disordered \textsl{orthogonal} systems with
dimension $d=1$, 2 and 3. Notice the critical point, $g_c$ for $d=3$.}
\label{fig-beta}
\end{figure}

The form of the function $\beta(g)$ has some important consequences. First, we see that
for $d<2$, the function $\beta(g)$ is always negative.  So starting with a sample of finite size,
$L_0$, and conductance, $g_0$, the conductance, given by Eq. (\ref{beta})  decreases
and develops to a smaller conductance when the system size,
$L$, increases. Consequently, an infinite system with dimension $d\le 2$ does not
exhibit the Anderson transition  since all electronic states are localized, 
independent of the strength of the disorder.

For $d>2$ ($d=3$, for instance),  there is a critical point $g_c$ such that
\be\label{beta-crit}
\beta(g_c)=0.
\ee
When starting exactly  with $g=g_c$, the conductance remains constant, independent of the
system size even in the limit of $L\to\infty$. Thus, a disordered system in dimension
$d>2$ possesses a critical point. This critical point is unstable: when starting with
$g=g_c+\delta g$, we end up, in the limit of $L\to\infty$, in the metallic regime
with finite conductivity $\sigma$. Similarly, starting with $g=g_c-\delta g$,
the system will develop into the insulating regime with exponentially small conductance.

What remains is  the calculation of the function $\beta(g)$.
Analytically, this is possible only for systems close to the critical dimension: $d=2+\veps$,
($\veps\ll 1$) \cite{Wegner-89,hikami}. We will discuss these results
in Section \ref{sect:dd}.

The scaling theory of localization, formulated first in Ref. \cite{AALR}
represents the main milestone in our understanding of the Anderson transition. First,
it estimates the lower critical dimension,
\be
d_c=2.
\ee
There is no metallic regime in the disordered system with dimension $d\le 2$. 
However, 
this statement is true only for \textsl{orthogonal} systems, i.e. for systems with
time reversal symmetry. We know already that in symplectic systems
the first correction to the conductance in the diffusive regime is positive (Fig. \ref{2D-ando})
so that $\beta(\ln g)\to 0^+$ when $g\gg 1$ \cite{Wegner-76}. Consequently, 
the 2D systems with spin dependent hopping  exhibit the metal insulator transition.
This was for the first time predicted in Refs. \cite{EZ,Ando-89}  and confirmed numerically
by many authors \cite{Fastenrath,ludwig-isa,ASD,Ludwig-1}.

Soon after formulation of the scaling theory \cite{AALR} it became clear that 
 $g$ is a statistical quantity, which is not self-averaged in the limit of the infinite system size.
In the metallic regime, the probability distribution of the conductance, $p(g)$,
possesses an universal width, given by the universal conductance fluctuations.
In the insulating regime, $g$ does not represent the 
statistical ensemble, as discussed in  Section \ref{sect:1D}, and  we have to use 
its \textsl{logarithm}, $\ln g$. Even then, $p(\ln g)$ is Gaussian with the mean value
$\langle\ln g\rangle=-2L/\lambda$
and the variance, $\langle \ln^2 g\rangle-\langle\ln g\rangle^2$, being of  the same order as $-\langle\ln g\rangle$.
We do not know the analytical form of the critical distribution, $p_c(g)$,
but both theoretical \cite{BS,BS-1} and numerical \cite{Kramer-PM} results confirm that
$p_c(g)$ is independent on the size of the system. This is consistent with
the size independence of the critical conductance, given by Eq. (\ref{beta-crit}).
However, since the width of the critical distribution is non-zero,
the conductance is not the self-averaged quantity at the critical point. The statistical character
of the conductance opens new problems in  the scaling theory. First, we have to prove 
that both mean values, $\langle g\rangle$, and $\langle\ln g\rangle$ obey scaling relations
(\ref{beta}).
Then, it would be useful to prove the same for 
all cummulants of the conductance. This is, of course, an unsolvable task. 

The first step in the verification of the single parameter  scaling theory 
is to understand
the statistical properties of the conductance in the critical and localized regime.


\section{Statistical properties of the conductance in the critical regime}\label{sect:crit}

The statistical properties of the conductance at the critical point were discussed 
by Shapiro, \cite{BS,BS-1}. With the use of the Migdal-Kadanoff renormalization, he  studied
the size dependence of the  conductance distribution and he proved that
the critical conductance distribution, $p_c(g)$, is universal, independent of the system length.
The later works \cite{CRS} however
showed that the Migdal-Kadanoff renormalization overestimates the conductance fluctuations, and is
therefore not suitable for a quantitative description of the critical conductance distribution.

In dimension $d=2+\veps$, the 
 non-universality of higher order conductance cummulants has been found  analytically 
\cite{AKL} to be
\be\label{kumulanty}
\langle \delta g^n\rangle =\left\{\begin{array}{ll} \veps^{n-2} & n<n_0=\veps^{-1}\\

                                                   \sim L^{\veps n^2-n} & n>\veps^{-1}.
					\end{array}\right.
\ee
Expression (\ref{kumulanty}) states that
the higher order  cummulants, $n>n_0=\veps^{-1}$,
depend on the size of the system. This  
seems to be in contradiction with universality of the critical distribution. 
This discrepancy was explained in Ref. \cite{CS}. Starting from the known cummulants,
given by Eq. (\ref{kumulanty}), the critical conductance distribution was derived.
It was shown that $p_c(g)$ is  indeed universal in the limit of the infinite system size.
The non-universality of higher cummulants follows from  the form of the critical distribution.
 For small $\veps$, the bulk of $p_c(g)$ is approximately Gaussian near
 the mean value $\langle g\rangle$. The parameters of the Gaussian peak,
\be\label{edon}
\langle g\rangle\sim \veps^{-1}
\ee
and
\be\label{dvaa}
\textrm{var}~g\sim \veps^0,
\ee
but 
$p_c(g)$  possesses long tails for large values of $g$,
\be\label{chvost}
p_c(g)\sim g^{1-2/\veps}.
\ee
The power-law behavior of $p_c(g)$ explains the non-universality of higher cummulants, 
$\langle \delta g^n\rangle$, which are not defined for $n>2/\veps$.

\begin{figure}[t!]
\begin{center}
\includegraphics[clip,width=0.3\textheight]{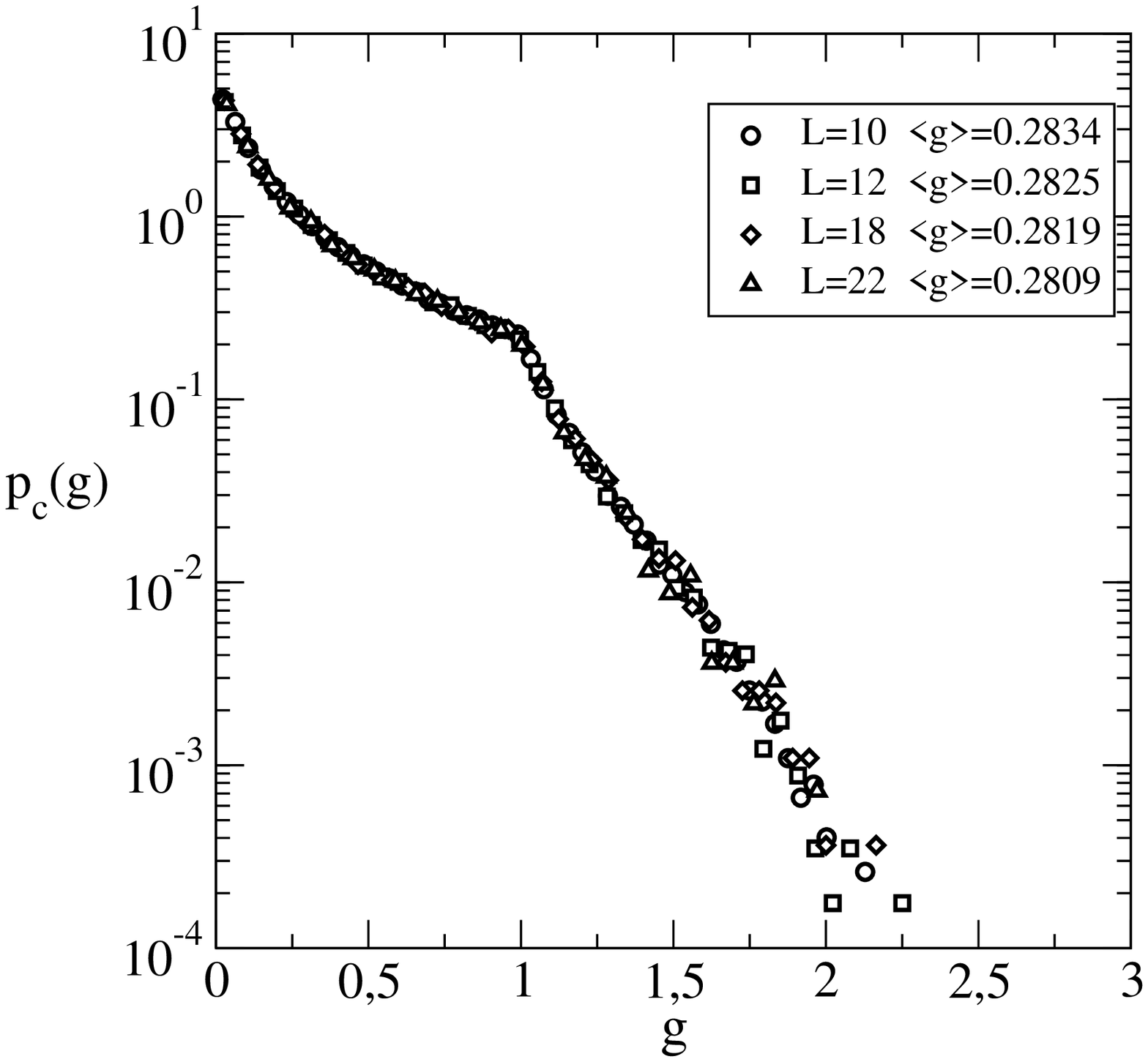}
\includegraphics[clip,width=0.3\textheight]{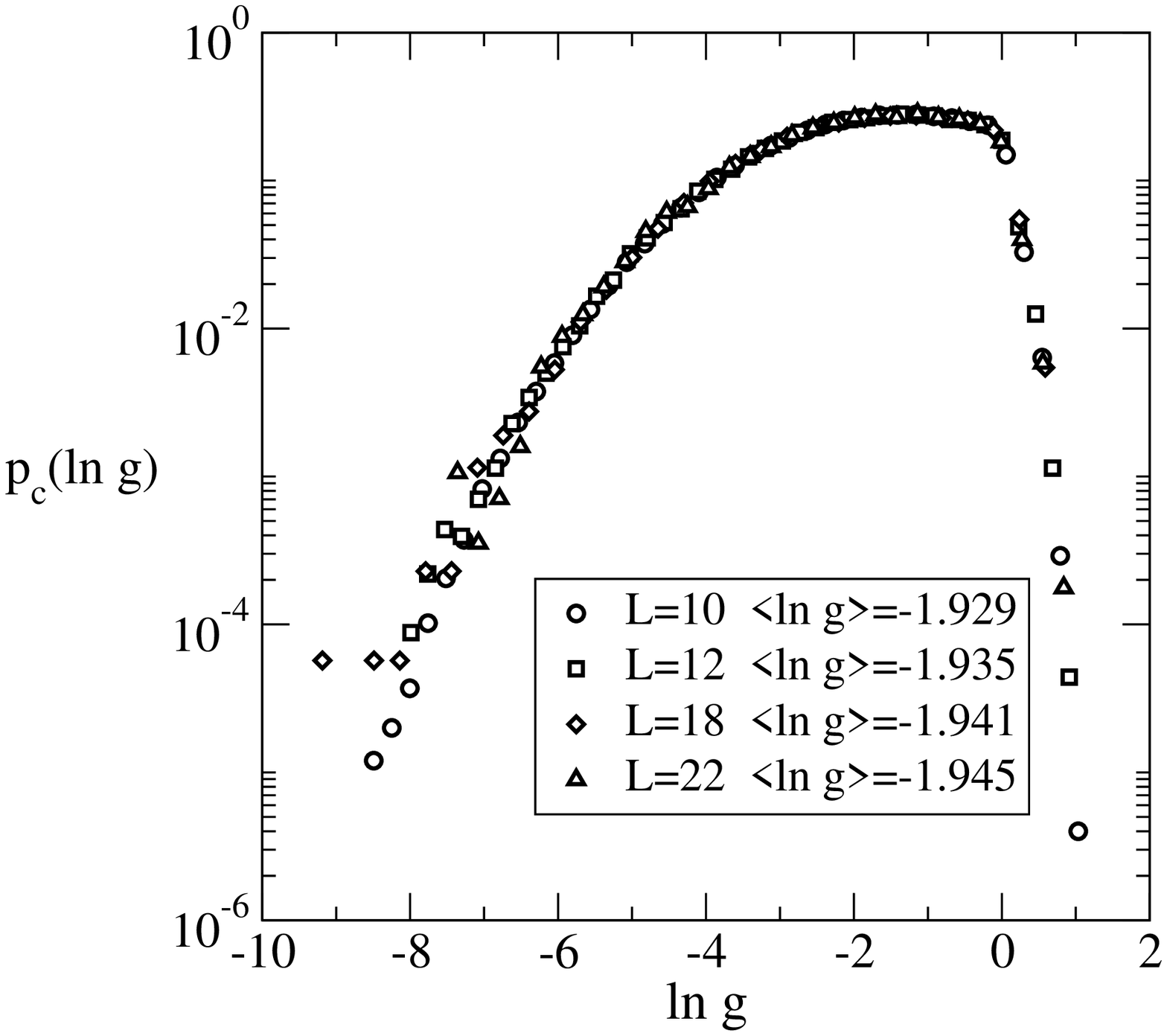}
\end{center}
\caption{The critical conductance distribution for the 3D Anderson model.
The data for the samples $L^3$ with  $10\le L \le 22$ are shown. Statistical ensembles of
 $\ns=10^6$ (20000) sample for $L=10$ ($L=22$), respectively were used to create the distribution.
 Left: the distribution of the conductance, right: the  distribution of the \textsl{logarithm}
of the conductance. The legends present the mean values, which should be independent
of the system size at the critical point. Note the non-analytical behavior of the 
distribution at $g=1$.
}
\label{3D-pcg}
\end{figure}

These  analytical results were derived only for dimensions close to 2, for $\veps\ll 1$.
Any attempt to apply them to the 3D system fails. For instance, Eq. (\ref{chvost}) predicts
that $p_c(g)^{(d=3)}\sim g^{-3}$ for large $g$, which is clearly not realistic. 
The most relevant information about the form of the critical conductance
distribution was therefore obtained from numerical  simulations
\cite{Kramer-PM,M-1999,RMS-pg,SO-97,C-7}
based on the  formula  
\be
g=\textrm{Tr}~t^\dag t=\sum_a\frac{1}{\cosh^2(x_a/2)},
\ee
already used in the diffusive regime.

In Fig. \ref{3D-pcg} we plot the critical conductance distribution 
for the 3D Anderson model. We see that 
$p_c(g)$ does not depend on the system size. This is what we expect if the scaling theory
really works. 

The form of $p_c(g)$ is rather unusual. Numerical data enables us to understand its main
properties. It turns out that it is useful to study also the
 distribution of the logarithm of the conductance, $p(\ln g)$, shown in the right panel of Fig. \ref{3D-pcg}.

\begin{figure}[t!]
\begin{center}
\includegraphics[clip,width=0.25\textheight]{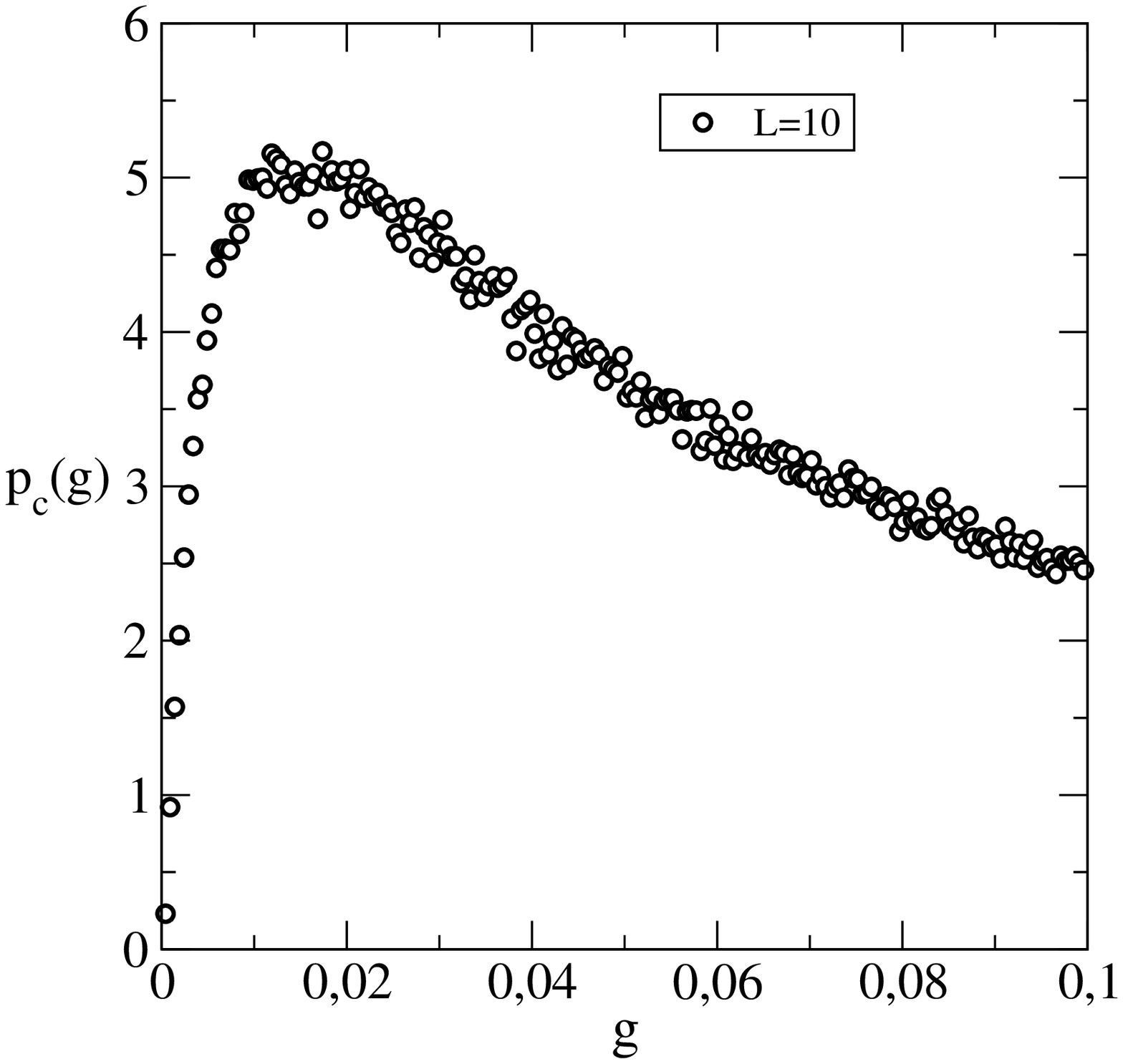}
\includegraphics[clip,width=0.25\textheight]{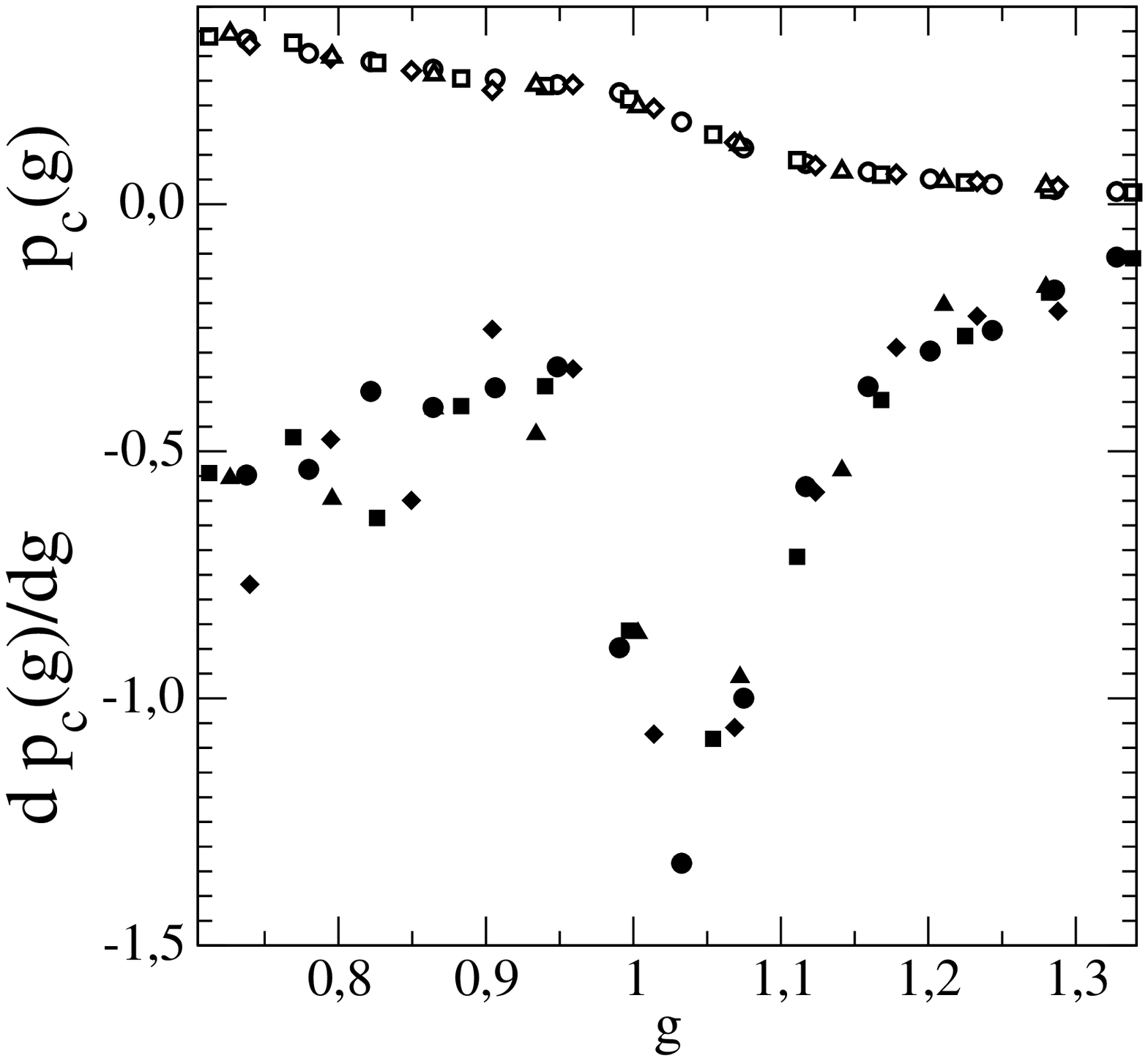}
\ec
\caption{Details of the critical probability distribution for the 3D Anderson model.
The left figure proves that $p_c(g)\to 0$ when $g\to 0$. The right panel shows 
both the distribution, $p_c(g)$ and its first derivative, $\partial p_c(g)/\partial g$
in the neighborhood of $g=1$. Numerical data indicate 
the
non-analytical behavior of the distribution at $g=1$.
}
\label{3D-pcg1}
\end{figure}

\begin{figure}[t!]
\begin{center}
\includegraphics[clip,width=0.42\textheight]{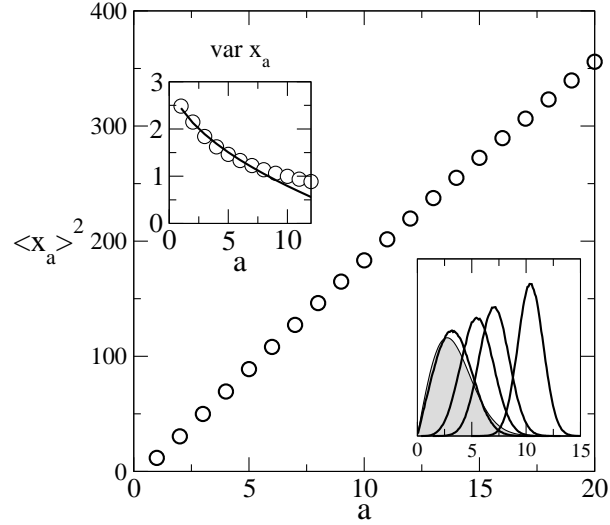}
\end{center}
\caption{%
The index dependence of the parameters $x_a$ at the critical point in the 3D Anderson model.
The data confirm that $\langle x_a\rangle^2\propto a$. 
Left inset   shows that the variance, $\textrm{var}~x_a$ decreases as a function of index
$a$.  The solid line is a fit $a_0+a_1x^\alpha$ with $\alpha=0.51$.
This gives an estimate of the variance, $\textrm{var}~x_a\approx a^{-1/2}$.
 Right inset shows the probability distributions $p(x_1)$, $p(x_2)$, $p(x_3)$ and $p(x_6)$.
Note, the distribution $p(x_1)$ is similar to 
the  Wigner distribution, $p_1(x)$, shown by the shaded area.
}
\label{3D-zi}
\end{figure}

First, we need to know the  form  of $p_c(g)$ for small $g$. 
The behavior of $p_c(g)$ for small $g$  can be easier obtained from the distribution of 
the $\ln g$. In the right panel of Fig. \ref{3D-pcg} we see that 
\be
\ln p_c(\ln g)\sim -(\ln g)^2~~~~~g\ll 1.
\ee
Since $p_c(g)dg = p_c(\ln g) d(\ln g)$, we immediately obtain that \cite{M-1999}
\be
p_c(g)\propto \exp \left[-(\ln g)^2-\ln g\right]~~~~~g\ll 1,
\ee
so that $p_c(g)\to 0$ when $g\to 0$.
The left panel of Fig. \ref{3D-pcg1} shows details of the critical distribution
for $g<0.1$. The data, collected from the  ensemble of the $\ns=10^7$ samples of the size $L=10$
confirm that $p_c(g)$ indeed decreases to zero when $g\to 0$.
However, the probability to obtain small values of $g$ is very small. 

\begin{figure}[b!]
\begin{center}
\includegraphics[clip,width=0.32\textheight]{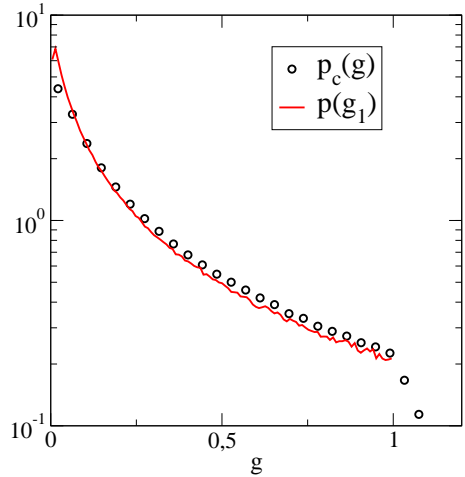}
\ec
\caption{Comparison of the critical conductance distribution with
distribution $p(g_1)$ of the contribution of the first channel. 
}
\label{px1-crit}
\end{figure}

Other properties of the critical distribution, namely  the form of the tail  for large values $g$
and the non-analytical behavior in the neighborhood of $g=1$,  
are  more convenient to analyze  in terms
of parameters $x_a$ \cite{Kramer-PM,M-1999}, introduced in Eq. (\ref{konduktancia}).
We remind that $x_a$ determine the eigenvalues of the matrix $t^\dag t$
and that the conductance is expressed in terms of $x_a$ by the following
formula
\be\label{konda}
g=\sum_a\ds{\frac{1}{\cosh^2 x_a/2}}.
\ee
Numerical data \cite{Kramer-PM,M-1995,M-2000} 
showed that the spectrum of $x_a$ 
consists of two qualitatively different parts. In the  lower part
of the spectra, for $a\le L$,
$\langle x_a\rangle$ are independent on the size of the system
\be
\langle x_a\rangle=\textrm{const}~~~~~~~~~~~~~~~a<L.
\ee
This is what we expect, since the mean conductance, $\langle g\rangle$,
is  size-independent  at the critical point.
The size independence of $x_a$ was confirmed numerically in Ref. \cite{M-2000} and will be
discussed in Sect. \ref{higher:LE}.  Contrary to the metallic regime, where 
$\langle x_a\rangle \propto a$ (Eq. \ref{univ-xa}), at the critical point 
we obtain  (Fig. \ref{3D-zi})  that
\be\label{xa-crit}
\langle x_a\rangle^2\propto a.
\ee
It is important to note that
Eq. (\ref{xa-crit}) is valid only for $a\le L$. The upper part of the spectra,
with $a>L$, is $L$ dependent with $\langle x_a\rangle\propto a$.  Since 
this upper  part of the spectra does not  contribute to the conductance,
we will concentrate only on the lower part, given by Eq. (\ref{xa-crit}).

Insets of   Fig. \ref{3D-zi} show that the distribution of $x_1$
is similar to the Wigner distribution $p_1$, and all higher parameters $x_a$, with  $a>2$
are distributed according to the  Gaussian distribution with decreasing variance,
\be\label{varxa}
\textrm{var}~ x_a \sim a^{-\alpha},
\ee
where the  exponent  $\alpha$ is close to 1/2 \cite{M-1995}.

We use the statistical properties of $x_a$ to estimate the
behavior of the critical  distribution for large $g$.
From the expression for the conductance, given  by Eq. (\ref{konda})
we see that we can obtain a large value of the conductance, 
\be
g\sim N\gg 1,
\ee
only when all $x_a$, $a=1,2,\dots, N$
 are small. However, since all parameters $x_a$ are distributed with the  Gaussian distribution,
we obtain that the probability to have $x_N\ll 1$ is
\be\label{3D-velkeg}
\exp - \frac{\langle x_N\rangle^2}{\textrm{var}~x_N}\sim \exp -g^{1+\alpha}.
\ee
In Eq. (\ref{3D-velkeg}) 
 we used that $\langle x_N\rangle^2\propto N= g$ 
and  we estimated the variance 
$\textrm{var}~x_N\sim N^{-\alpha}$ with the use of Eq. (\ref{varxa}).
Eq. (\ref{3D-velkeg}) confirms that $p_c(g)$ decreases \textsl{faster} than exponentially for large $g$.
This is confirmed by our numerical data. 
The probability to have  $g>1$  is very small (less than 3\%).  The higher values of $g$ appear
with marginal probability. For instance, 
in an  ensemble of $\ns=10^7$ samples,
we found only 470 samples with conductance $g>2$ and only one with $g>3$.

\begin{figure}[t!]
\begin{center}
\includegraphics[clip,width=0.35\textheight]{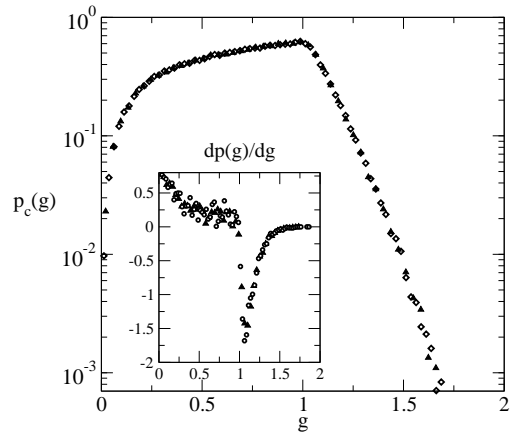}
\ec
\caption{The critical conductance distribution for the 2D Ando model. The
different symbols correspond
to the size of the system $L=82$ and $L=200$. The data prove that $p_c(g)$ is system
size invariant. Note the logarithmic scale on the vertical axis.
Inset describes the non-analyticity of the distribution at $g=1$.
The mean value of the conductance is $\langle g\rangle_c=0.71 e^2/h$.
\cite{Ludwig-1}
}
\label{pcg-ando}
\end{figure}

The most surprising property of the critical conductance distribution is that $p_c(g)$ seems to be
non-analytical at $g=1$. The right Fig. \ref{3D-pcg1}  shows that the first derivative,
$\partial p_c(g)/\partial g$, is discontinuous at  $g=1$.
The origin of the non-analyticity can be again explained  from the  spectra of the
parameters $x_a$.
The numerical data, for mean values,
$\langle x_1\rangle=3.42$, 
$\langle x_2\rangle=5.52$, 
and $\langle x_3\rangle=7.07$,
indicate that the conductance is given mostly by the first term, $g_1=\cosh^{-2} (x_1/2)$.
Indeed,  the distribution  $p(g_1)$ is a very good approximation to $p_c(g)$ for $g<1$,
as shown in  Fig. \ref{px1-crit}. 
Since $g_1\le 1$, the function $p(g_1)$ has a cutoff when $g\to 1^-$.  
It is difficult to estimate the contribution of  higher channels but it is reasonable to expect
that the non-analyticity of the distribution survives also when the contribution of
higher channels is included. This is supported by analytical calculations of
Muttalib et al \cite{MuttGW,MuttGW1}, who reported the non-analytical 
form of the conductance distribution
in the weakly disordered quasi-1d systems with the mean conductance $\langle g\rangle\sim 1$.

The above properties of the critical 
conductance distribution can be found also in other models, where
the critical conductance  is close to or less than to 1. 
For instance, Fig. \ref{pcg-ando} shows $p_c(g)$ for the 
2D Ando model. The  non-analyticity at $g=1$
is clearly visible, too.

\subsection{Properties of the critical conductance distribution}

The properties of the critical conductance distribution are important for the formulation of the
scaling theory. Besides the system size independence of $p_c(g)$, we need to understand
its universality, i.e. how the form of $p_c(g)$ depends on microscopic
details of the model, on the physical symmetry and on  the dimension of the system. Especially  the
dimension dependence of $p_c(g)$ 
is important for the comparison of the theoretical predictions with numerical data.

\begin{figure}[b!]
\begin{center}
\includegraphics[clip,width=0.35\textheight]{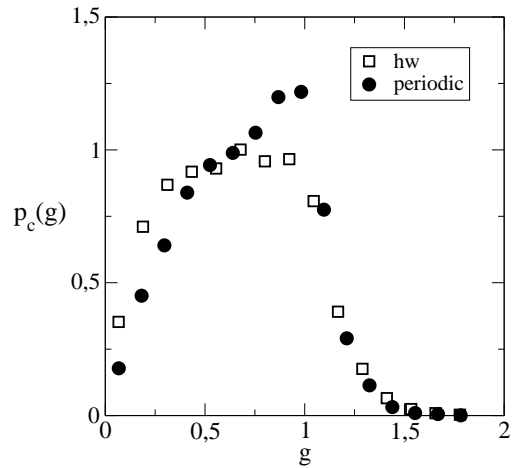}
\ec
\caption{The critical conductance distribution for the  2D Ando model with the
hard wall and periodic
boundary conditions in the direction perpendicular to the boundaries.
The critical conductance is $\langle g\rangle=0.655$ (0.71) $e^2/h$,
 and the variance $\textrm{var}~g=0.43$ (0.36) $(e^2/h)^2$
for the hard wall  (periodic) boundary conditions, respectively.
}
\label{pcg-boundary}
\end{figure}

As expected, the critical conductance distribution is independent of the size of the system.
This is confirmed by numerical data shown in Figs. \ref{3D-pcg} and \ref{pcg-ando}.
For a given universality class, it does not depend on microscopic details of the model.
This was confirmed in Refs. 
\cite{M-1994a,SO-01} where the critical distribution, $p_c(g)$ was calculated 
for the 2D and 3D models with various distributions of random disorder.
However, the form of $p_c(g)$ does depend on the symmetry class \cite{SO-97}. 
Also, the form of $p_c(g)$ depends   on the topology of the lattice.
For instance, $p_c(g)$ for the 3D Anderson model with triangular, 
hexagonal and rectangular lattice in the transversal direction possess different  
critical distributions $p_c(g)$
\cite{Travenec}.

Surprising   was the observation that $p_c(g)$ depends also on the boundary conditions
in the directions perpendicular to the propagation \cite{g0,SOK-00,BHMMcK}. This result is,
however, consistent with
our understanding of the conductance as a measure of the sensitivity of eigenenergies to the boundary
conditions \cite{BMP}. 
We remind the reader that also the values of the 
universal conductance fluctuations depend on the
boundary conditions.
Nevertheless, the difference in the form of $p_c(g)$ is surprisingly large
for the 3D Anderson model \cite{SOK-00},
Also, the
mean values $\langle g\rangle_c$,  are considerably different, being 0.445 (0.280) ($e^2/h$)
for the periodic and hard wall boundary conditions, respectively \cite{SOK-00}.
In Fig. \ref{pcg-boundary} we demonstrate the sensitivity to 
the boundary conditions for the  2D Ando model.

\subsection{Dimension dependence of the critical conductance distribution}\label{sect:pgd}

\begin{figure}[t!]
\bc
\includegraphics[clip,height=0.13\textheight]{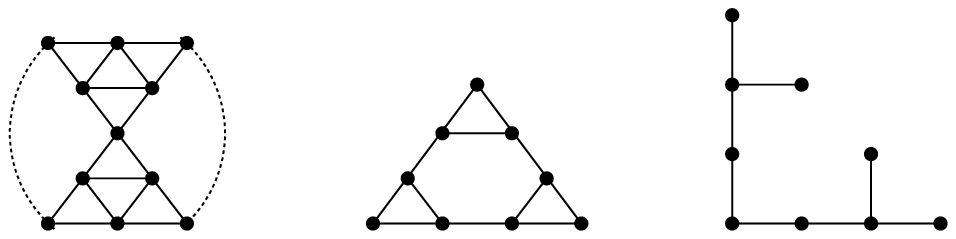}\\
\includegraphics[clip,height=0.19\textheight]{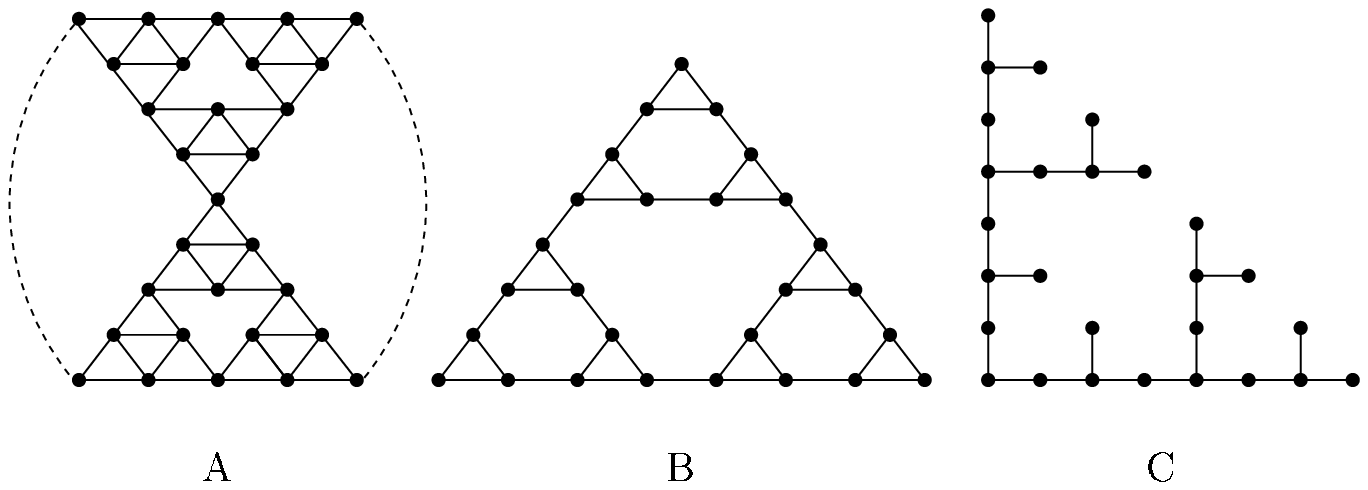}
\ec
\caption{Definition of three fractals used in the numerical  analysis of the 
conductance distribution of fractals. Shown are  the 2nd and the 3rd generations.
All three fractals have the same fractal dimension, $d_f=\ln 3/\ln 2\approx 1.58$.
Fractals A and B have \textsl{spectral} dimension $d'_s=1.365$, and fractal $C$ has $d'_s=1.226$.
Note the different number of nearest-neighbor lattice sites.
The bifractal lattice is created by combination of the fractal
with the linear chain in the perpendicular
direction. The spectral dimension of the bifractal is $d_s=d'_s+1$. 
The number of lattice sites on fractal is $3^n$ in the $n$th generation, 
and the length along the linear chains
is $2^n$ lattice sites.}
\label{fractals}
\ef

As discussed above, the analytical theories provide us only with the information
about the critical conductance in systems with dimension close to the lower critical dimension, $d=2+\veps$.
Since we do not expect that the analytical results could be applied also for $\veps=1$,
the numerical data for dimension $d=3$ are not suitable for verification of the theory.
Therefore,
it might be interesting to calculate the critical conductance distribution on lattices with fractal
dimension close to $d_c=2$.
This was done in Ref.  \cite{Travenec}.

By definition, the bifractal lattice \cite{SG} is linear in the direction of propagation and possesses the fractal 
transversal structure.  Three fractals, discussed in Ref. \cite{Travenec}
are shown in Fig. \ref{fractals}.  
For the analysis of critical phenomena on fractals, 
it is important to note that
 the ``dimension'' which is important for description of the 
critical phenomena is the \textsl{spectral}
dimension of the lattice, not the fractal dimension, $d_f$. 
We remind the reader that spectral dimension, $d_s$,  determines the
low-frequency behavior of the phonon density,
\be
\rho_{\rm phonon}(\omega)\sim \omega^{d_s}
\ee
\cite{ramal,SG},
while the fractal dimension, $d_f$, determines an increase $b^{d_f}$
 of the ``mass' (number of lattice points) when the scale of the system changes by a factor of $b$.
Note that the spectral dimension of a regular system equals to its dimension, $d_s=d$.

\begin{figure}[t!]
\bc
\includegraphics[clip,height=0.25\textheight]{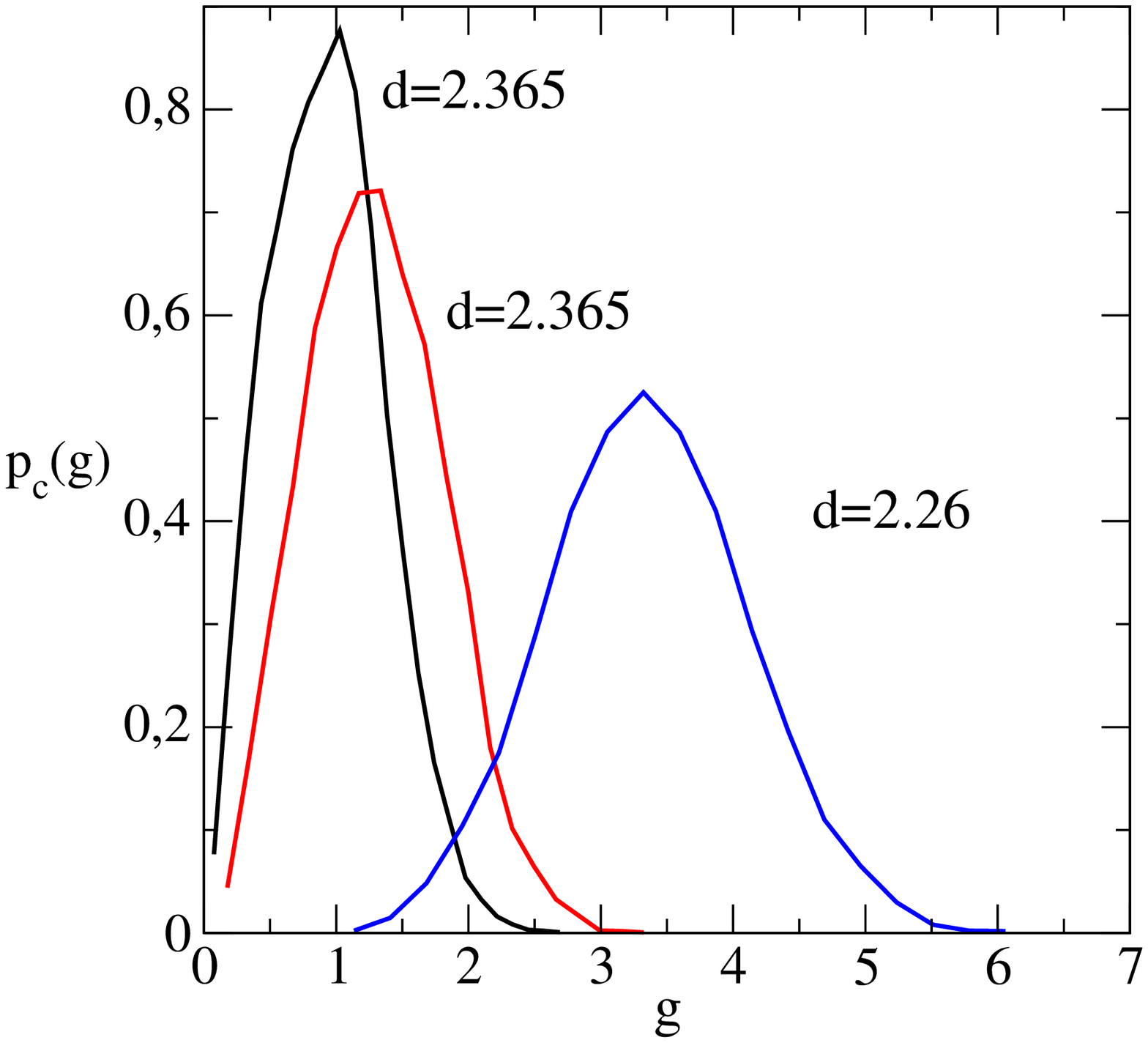}
\includegraphics[clip,height=0.25\textheight]{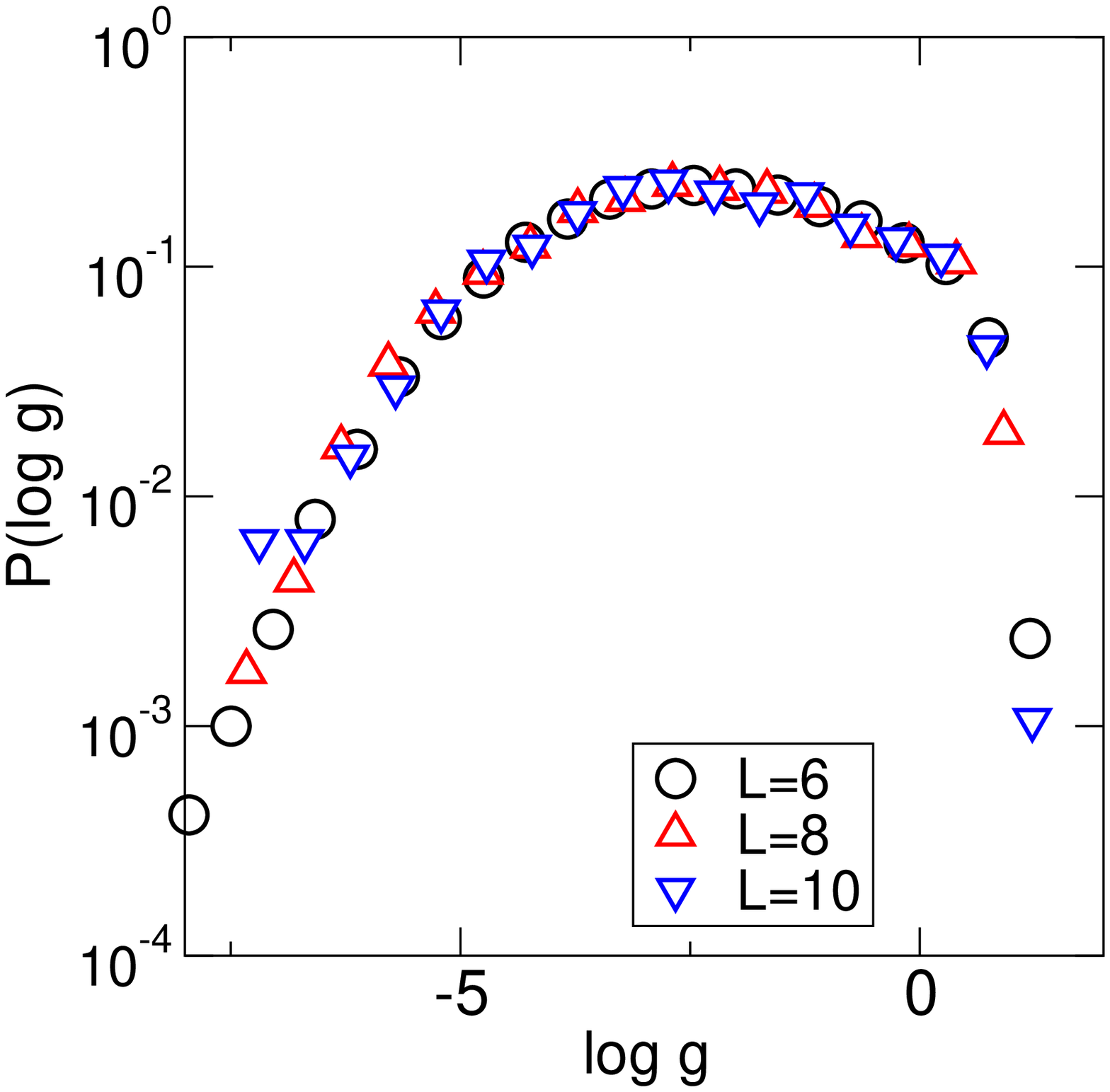}
\ec
\caption{Left: The critical conductance distribution for three bifractals,
 shown in Fig. \ref{fractals}. 
The legends give the spectral dimension, $d_s$, of the lattice.
Critical distribution
is close to Gaussian distribution when $d\to 2$. The power-law tail, predicted by 
the theory \cite{CS},
is not observable in numerical simulations. 
Note, although two bifractals, A and B, have the same spectral dimension,
the corresponding critical distributions differ from each other. This is because $p_c(g)$ depends
on the topology of the lattice.
 Note also, the non-analyticity of the
conductance distribution around $g=1$ disappears, since more than one propagating channel contribute to
the conductance in these models.
Right: Distribution of $p_c(\ln g)$ for the 4D Anderson model.
}
\label{fig-qq}
\end{figure}

All three fractals, shown in Fig. \ref{fractals} have the same fractal dimension, 
$d_f=\ln 3/\ln 2= 1.58$. Fractals A and B have the same spectral dimension, 
$d'_s=1.365$, and fractal $C$ has spectral dimension $d'_s=1.226$, so that the spectral dimension
is $d_s=d'_s=2.365$  ($2.226$) for bifractals A and B (C), respectively.
In Fig. \ref{nu-1} we will show that the critical exponent, $\nu$ is indeed a function
of the spectral dimension only.

Critical conductance distributions are shown in  Fig. \ref{fig-qq}. We see that
the mean conductance increases when $d_s\to 2^+$ and the distribution $p_c(g)$ is similar
to Gaussian, in agreement with theoretical prediction \cite{CS}. However, 
the width of the distribution, measured by the variance, $\textrm{var}~g$, increases
when $\veps\to 0$. This seems to be in contradiction with the analytical formula
(\ref{dvaa}).  The numerical data cannot confirm the power law decrease of the 
distribution, given by Eq. (\ref{chvost})  for large values of $g$.

For completeness, we show in the right panel of Fig. \ref{fig-qq} the critical distribution
$p_c(\ln g)$ for the 4D Anderson model. The distribution is similar to that for the 3D system,
including the non-linearity at $g=1$.
Similar to 3D models, the main contribution to the conductance is given by 
the first channel.


\section{Localized regime}\label{sect:loc}

It is commonly believed that the distribution of the logarithm of the conductance
is Gaussian, with the mean value 
\be\label{loc-1}
\langle \ln g\rangle =- 2L/\lambda.
\ee
The  variance,
\be\label{loc-2}
\textrm{var}~\ln g\propto -\langle\ln g\rangle
\ee
should be proportional to $-\langle\ln g\rangle$,
with the coefficient of proportionality close to 2.
This behavior is deduced from the analysis of the DMPK equation, discussed in Appendix  \ref{app-dmpk}.
It is argued that since all parameters $x_a\propto L_z$, the difference between
$x_2$ and $x_1$ becomes large enough so that the contributions $g_2$, $g_3$ \dots 
of higher channels become
negligible in comparison with $g_1=\cosh^{-2}(x_1/2)=\exp -x_1$. Since $x_1$ possesses
the Gaussian distribution in the limit of large $L_z$, it is expected that the logarithm 
of the conductance,  $\ln g\approx \ln g_1$ should posses a
Gaussian distribution, too.

\begin{figure}[b!]
\begin{center}
\psfrag{d}{$\delta_{21}$}
\psfrag{pd}{$\!\!\!\!\!p(\delta_{21})$}
\includegraphics[clip,width=0.35\textheight]{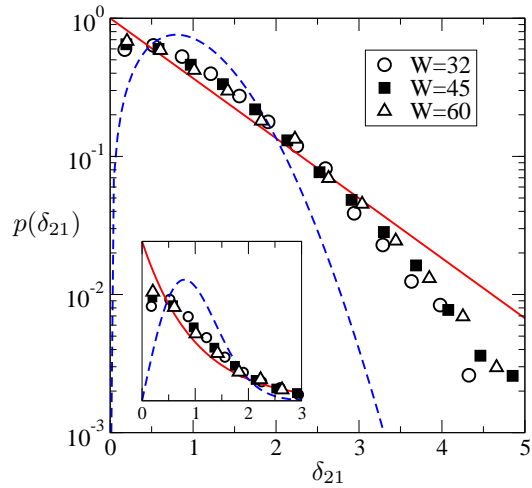}
\end{center}
\caption{%
The probability distribution of the \textsl{normalized} difference
$\delta_{21}=x_2-x_1$ for the 3D Anderson model ($L=10$)
compared with the Wigner distribution $p_1$ (dashed line) and Poisson distribution (solid line). 
Inset shows the same data in the linear scale \cite{Kramer-AP}.
}
\label{pd}
\end{figure}

This argument is supported by the numerical data for the \textsl{normalized}
difference, $\delta_{21}=x_2-x_1$, 
shown in Fig. \ref{pd}. 
As expected, $p(\delta_{21})$ differs considerably from
the Wigner distribution, and is more similar (although not identical)  to the  Poisson
distribution \cite{Kramer-AP}. 
Although the deviations  from the Poisson distribution indicate that $x_2$ and $x_1$ are not
statistically independent, we expect that  their correlation is  small. Then,
the contributions $g_1$ and $g_2$ can be assumed to be almost statistically independent
and we can split the spectrum of the transfer matrix to the first eigenvalue, $x_1$,
which determines the magnitude of the conductance, and the rest of the spectra, 
which have almost negligible contribution to $g$.

\begin{figure}[t!]
\begin{center}
\includegraphics[clip,width=0.3\textheight]{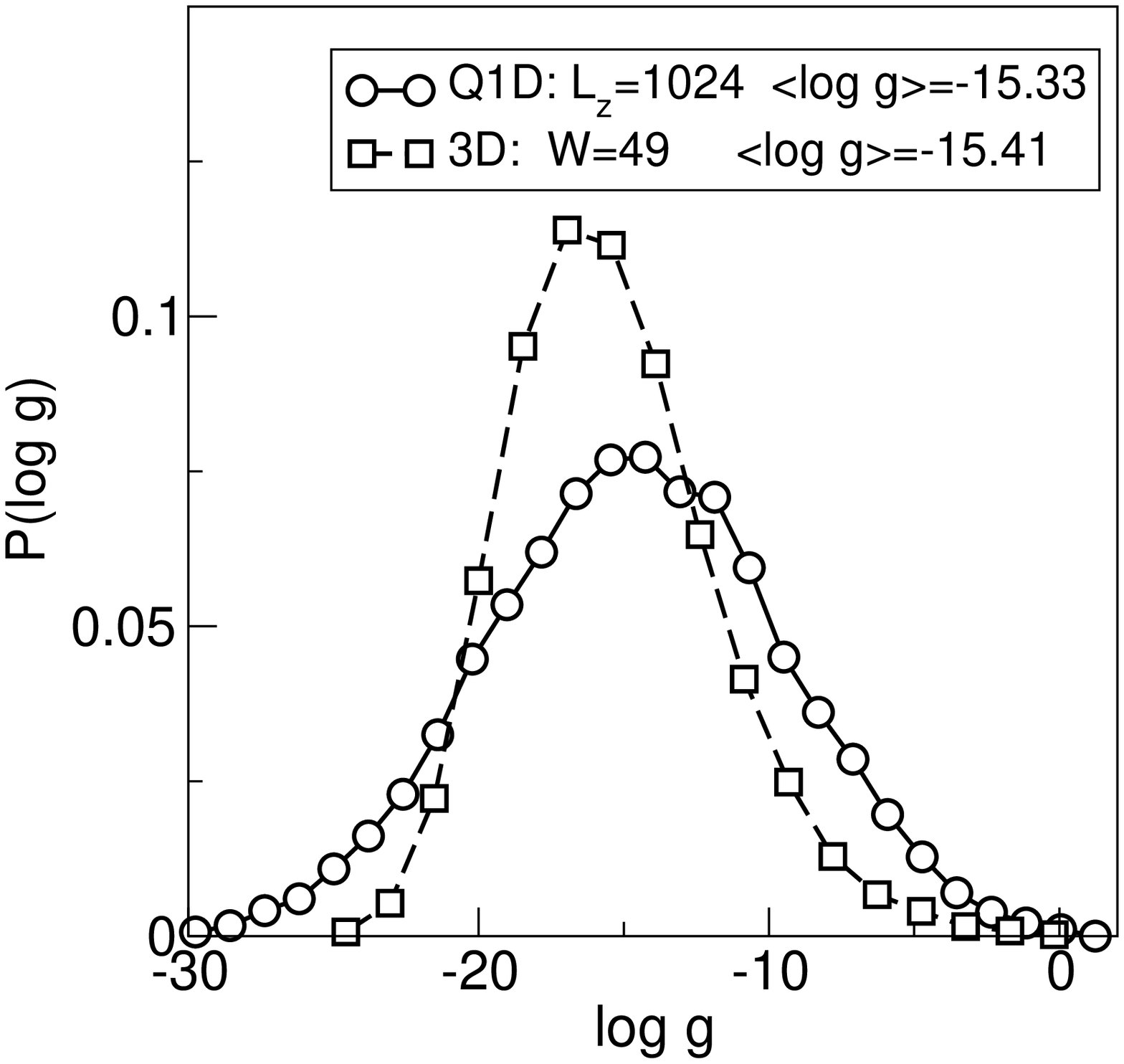}
\end{center}
\caption{%
Comparison of the conductance distribution for the quasi-1d   weakly disordered system
in the localized regime
and strongly disordered 3D system.
While the distribution of $p(\ln g)$ is Gaussian in  the quasi-1d system,
for the 3D system,
the distribution is asymmetric and narrower. 
The size of the quasi-1d system is $8\times 8\times L_z$ and disorder $W=4$. 
The size of the 3D system is $8^3$ and the strength of the disorder, $W$,
 is chosen to have the same value of the mean logarithm of the conductance
\cite{M-2002a}. 
}
\label{porovnaniex}
\end{figure}

However, the above arguments are  valid only in the quasi-1d weakly disordered systems
(see Appendix \ref{dmpk-ll}).
As shown in Fig. \ref{porovnaniex}, the probability distribution $p(\ln g)$ for the
strongly disordered samples is not Gaussian.
The origin of the difference between the strongly disordered 3D systems  and
weakly disordered quasi-1d systems
lies in the spectra of parameters $x_a$.
Contrary to the weakly disordered systems, parameters $\langle x_a\rangle$ do not increase
linearly with $a$, but fulfill the relation
\be\label{DDD}
\langle x_a\rangle = \langle x_1\rangle +\Delta_{a1}.
\ee
As is shown in  Fig.  \ref{delty},
$\Delta_{a1}$  depends neither on the system size nor on the disorder.

\begin{figure}[t!]
\begin{center}
\psfrag{z1}{$\langle x_1\rangle$}
\includegraphics[clip,width=0.3\textheight]{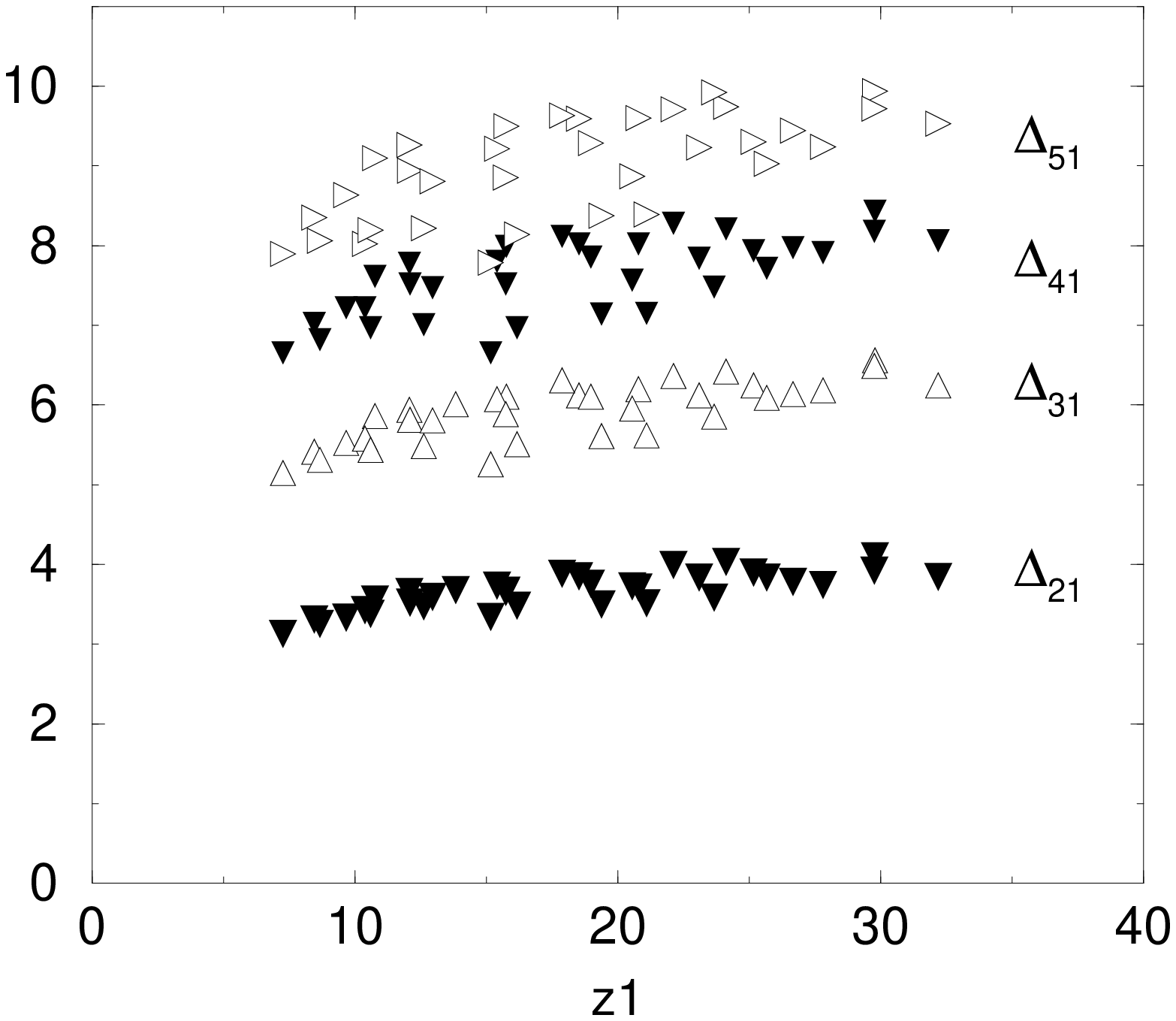}~~~~~
\includegraphics[clip,width=0.3\textheight]{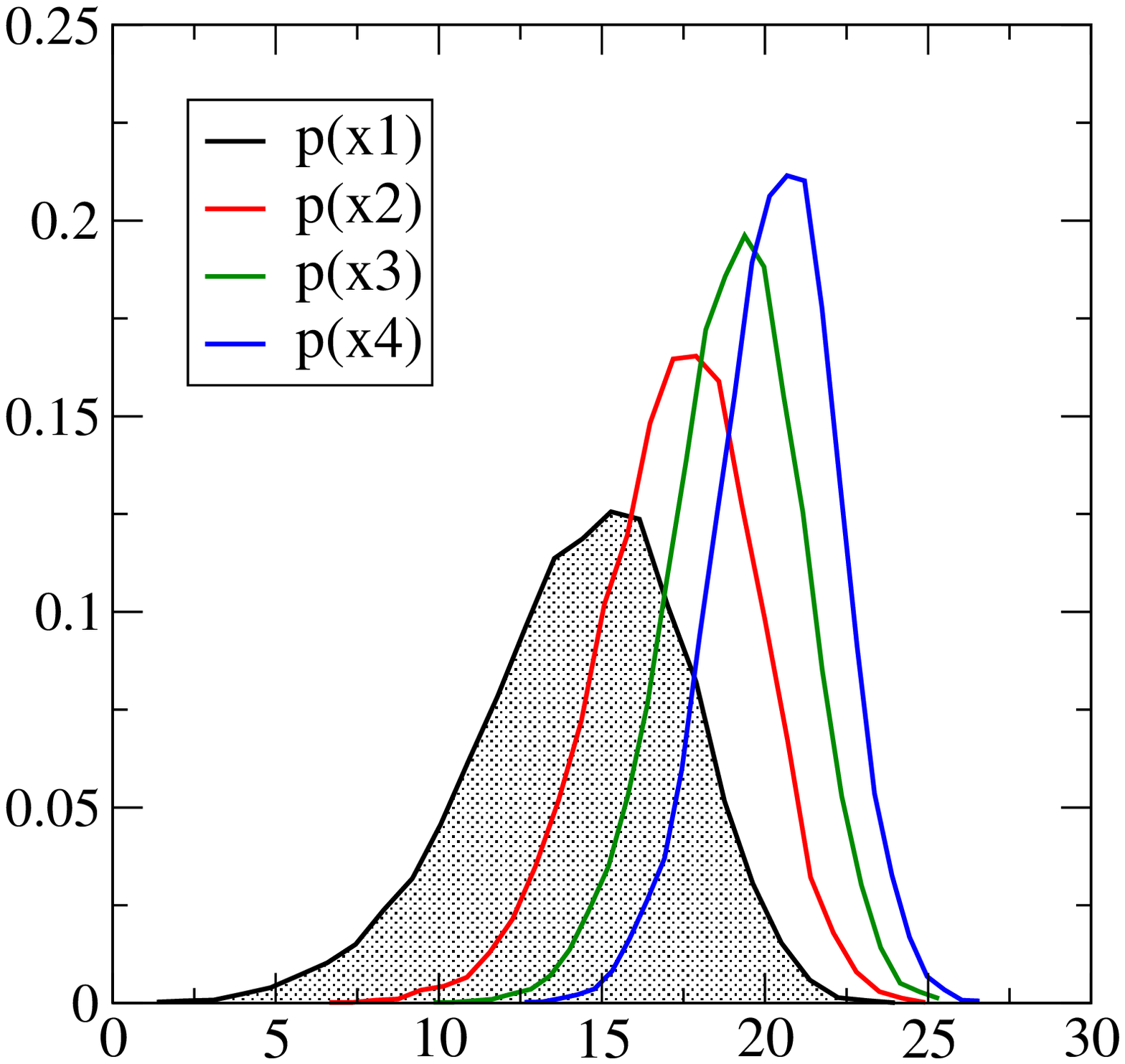}
\end{center}
\caption{%
Left: The differences $\Delta_{a1}=\langle x_a-x_1\rangle$ for the 3D Anderson model
\cite{M-2002a}.
Right: The probability distribution $p(x_a)$ for 
the strongly disordered cube ($W=20$, $L=18$). 
These data should be compared with  those in Fig. \ref{fig-z1} which shows the distribution
$p(x_a)$ for long weakly disordered quasi-1d systems.
Note, the mean values, $\langle x_a\rangle$ do not increase linearly.
Also, the width of the distribution $p(x_a)$ decreases when $a$ increases,
similar to the distributions of parameters $x$ at the critical point, shown in Fig. \ref{3D-zi}.
}
\label{delty}
\end{figure}

Apart from the deviation from the Gaussian form, $P(\ln g)$ possesses also the non-analyticity
at $\ln g=0$, shown in Fig. \ref{fig-logg}. 
The origin of this non-analyticity is the same as in the case of the critical
distribution, $p_c(g)$. The contribution to the conductance comes mostly from the first channel
and the chance to have $g>1$ is negligibly small.

\begin{figure}[t!]
\begin{center}
\includegraphics[clip,width=0.3\textheight]{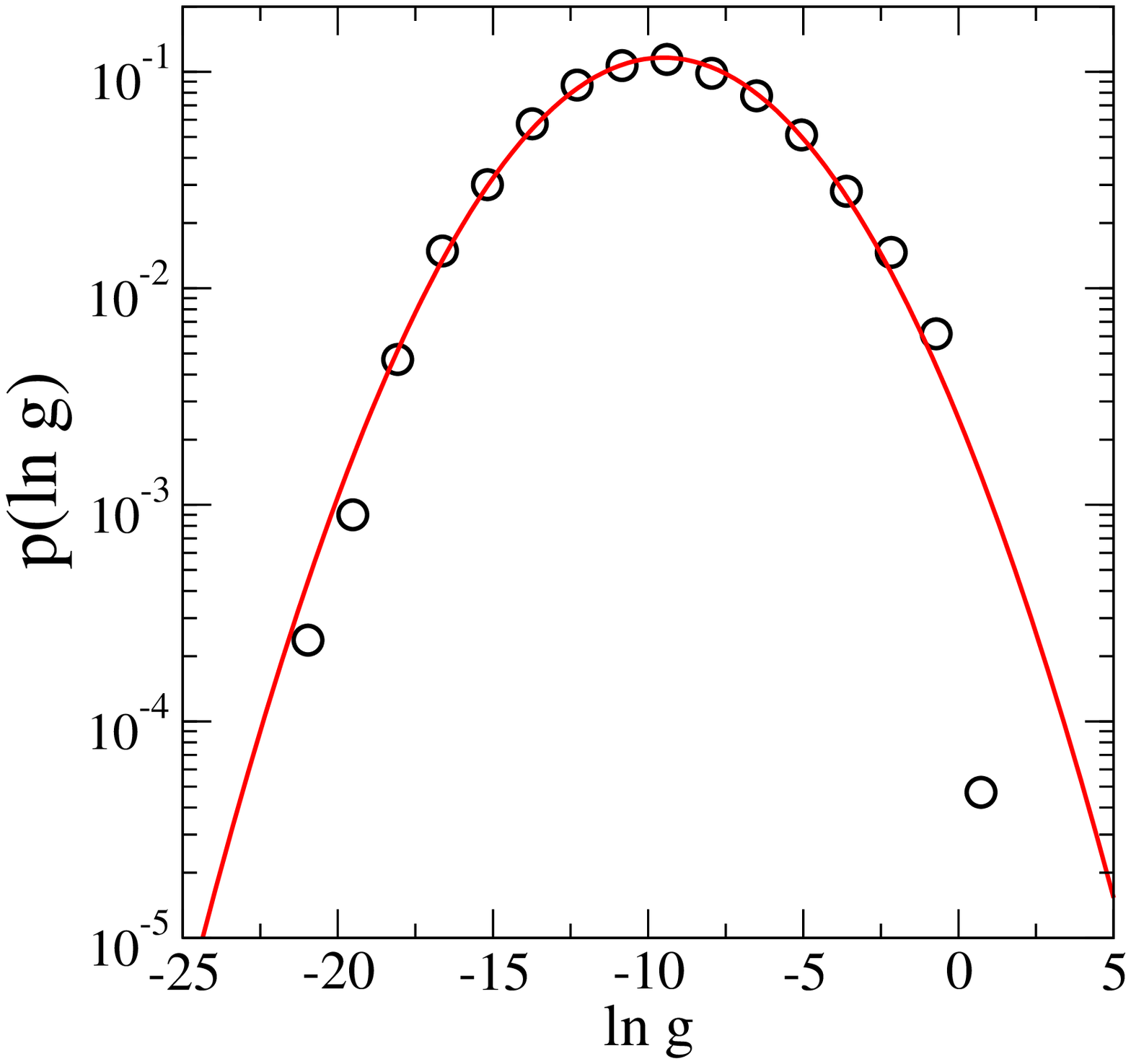}~~~~
\includegraphics[clip,width=0.3\textheight]{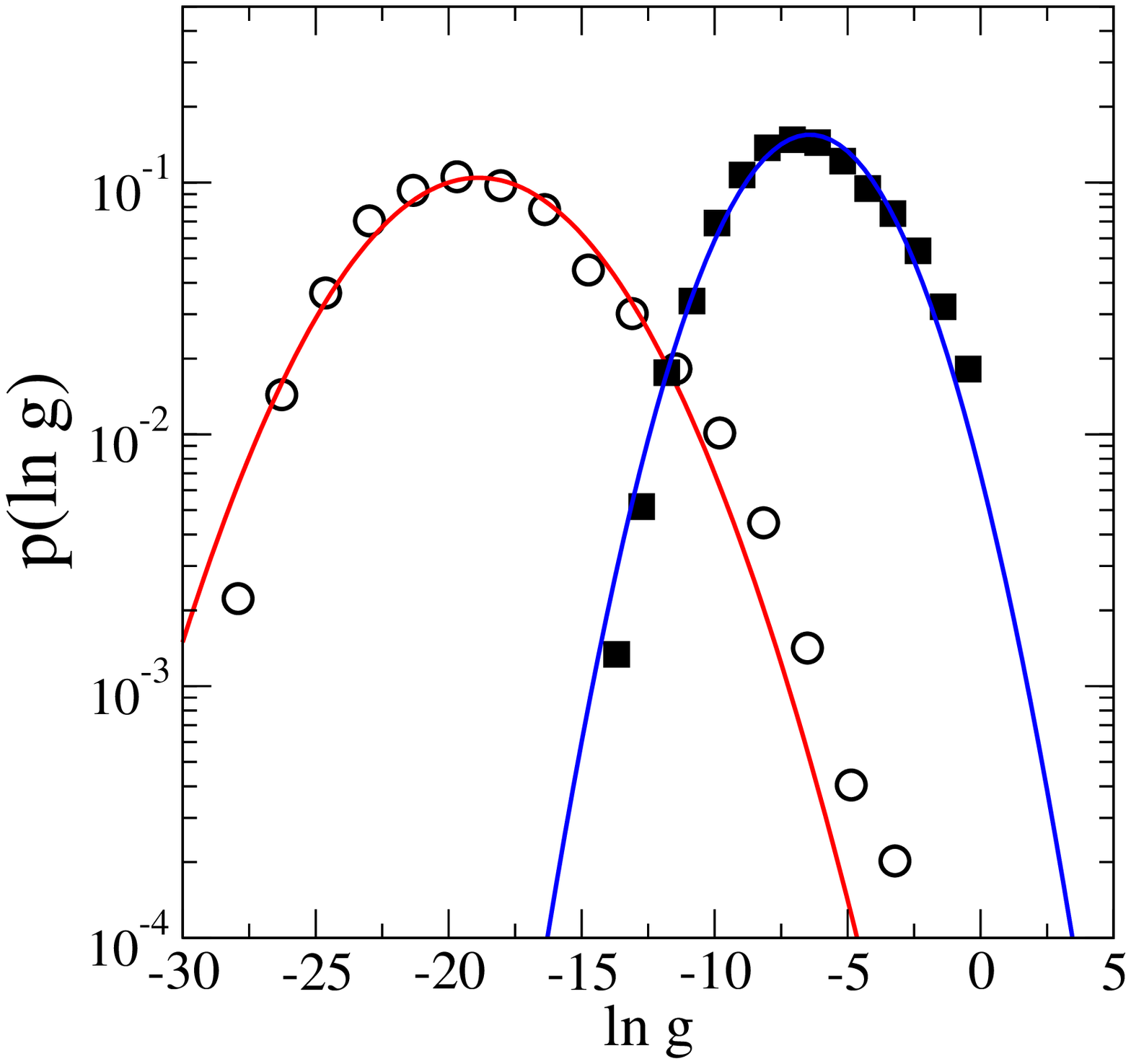}
\end{center}
\caption{%
The statistical distribution of the logarithm of the conductance in the strongly
disordered squares (left) and cubes (right). The 
solid lines are Gaussian fits with the same mean and variance.
Left: The 2D Anderson model, $W=6$, $L=200$, 
 $\langle\ln g\rangle=-9.51$, $\textrm{var}\ln g=11.78$.
Note the sharp decrease of the distribution at $\ln g=0$.
Right: The 3D Anderson model, $W=32$, $L=18$.
 $\langle\ln g\rangle=-18.88$, $\textrm{var}\ln g=14.57$
and  $W=14$ and $L=10$ (
 $\langle\ln g\rangle=-6.42$, $\textrm{var}\ln g=6.63$).
}
\label{fig-logg}
\end{figure}


\begin{figure}[t!]
\begin{center}
\includegraphics[clip,width=0.3\textheight]{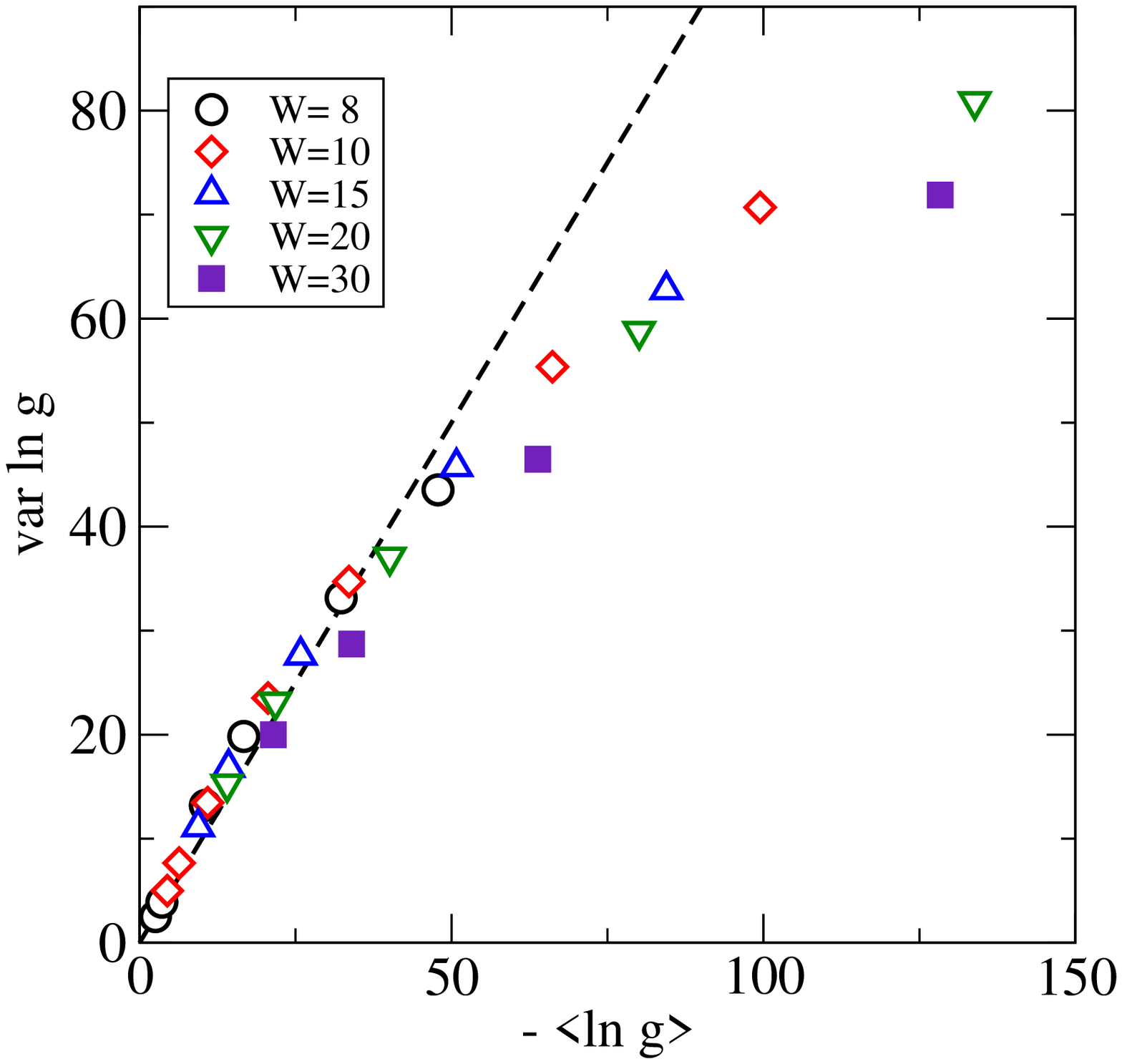}
\end{center}
\caption{%
The variance, $\textrm{var}~\ln g$ as a function of the mean value, $\langle\ln g\rangle$
for the 2D Anderson model. The expected linear relation, (\ref{loc-2}), is valid 
only for rather small values of $\langle\ln g\rangle$  \cite{M-2002a}.
}
\label{fig-2D-varlng}
\end{figure}

More important is the question whether the variance is an unambiguous function of the mean value.
This must be so if the one parameter scaling holds. The verification of the ambiguity
is difficult in 2D and 3D. In 1D, analytical calculations \cite{Roberts} showed that
there is no universal relation between the mean and the variance of $x$.
This was confirmed by numerical simulations \cite{SP,Kramer-AP}
of strongly disordered 1D systems,  indicating the existence of the second relevant
length scale in the strongly localized regime. 

The  second relevant length scale 
was indeed found in 1D systems \cite{deych} and estimated analytically by the formula
\be\label{ells}
\ell_s=\frac{1}{\sin~\pi\rho(E)},
\ee
where $\rho(E)$ is the density of states.
Single parameter scaling
is expected to be valid only when $\lambda>\ell_s$. This is not fulfilled 
in the band tail, where the density of states is small and, consequently, $\ell_s$ is large.
The second relevant length scale
was reported also in 2D systems in Ref. \cite{somoza1}.

Figure \ref{fig-2D-varlng} shows that the variance, $\textrm{var}~\ln g$ is \textsl{not} a linear
function of the mean value, $\langle \ln g\rangle$.
This was observed already in \cite{SP} for the
1D system. Very recently \cite{somoza1,somoza2}, the non-linear relation
between the mean value and the variance of $\ln g$,
\be
\textrm{var}~\ln g\propto\left\{
\begin{array}{ll}
\left(-\langle\ln g\rangle\right)^{2/3}   & d=2\\
\left(-\langle\ln g\rangle\right)^{2/5}   & d=3
\end{array}
\right.
\ee
was numerically observed.

Another test 
of the single parameter scaling was reported 
in Refs. \cite{bunde,deQ}, where the 
electron wave function of large 2D disordered system was calculated numerically. The statistical
distribution
of the logarithm of the wave function at the distance $r$ from the center of the lattice was discussed.
The numerical data was fitted to the distribution function
\be\label{jan}
H(-\ln|\Psi(\vec{r})|,r)=\frac{1}{\sqrt{2\pi\sigma}}\exp -\frac{(\ln |\Psi|+r/\lambda)^2}{2\sigma}.
\ee
The Gaussian form of the distribution $H$ seems to be natural, since 
we expect that the wave function decreases exponentially at large distance. This decrease is
controlled by the localization length, $\lambda$. Surprisingly, fitting the
numerical data to the distribution
(\ref{jan}) led to the $r$-dependent localization length, $\lambda(r)$. 
This result was interpreted as a 
failure of the scaling theory. 

However,  an assumption that the distribution function $H$, given by
Eq. (\ref{jan}) is exactly Gaussian,
is not correct. Indeed, the Gaussian distribution  possesses a long tail for both
negative and positive values of its argument, 
but the value of $-\ln|\Psi(\vec{r})|$ can not be negative if the wave function, $\Psi$, is normalized.
Therefore, the distribution $H(x,r)$ must converge to zero when $x=-\ln|\Psi(\vec{r})|\to 0^+$,
 in the same way as the distribution 
$p(x)$, shown in Fig. \ref{fig-1d-x1}.  We believe that the fit of numerical data
to the correct distribution function will recover the single parameter scaling.

\subsection{3D versus quasi-1d systems}

As discussed above, the statistical properties of the conductance in the 3D systems require
numerical simulations. On the other hand, similar transition from the metallic
to localized regime can be studied in
weakly disordered quasi-1d systems. Indeed, such a system exhibits metallic behavior
if the length of the system, $L_z$, fulfills the relations
\be
\ell\ll L_z\ll N\ell
\ee
where $\ell$ is the mean free path and $N$ is the number of channels. 
The conductance of such a system is $N\ell/L_z\gg 1$. By increasing $L_z$,
the conductance decreases. There is an interval of $L_z$, where $g\sim 1$.
Further increase of the system length draws  the system  into the localized regime,
where $g\sim \exp -2L/\lambda$, where $\lambda=n\ell$  is the localization length.

This scenario is very similar to the metal-insulator transition. Of course, we do not have
a critical behavior in quasi-1d system (there is no divergence of the correlation length),
but it is reasonable to expect that the conductance distributions, obtained in all three
regimes, $g\gg 1$, $g\sim 1$ and $g\ll 1$, might mimic the main properties
of the conductance distributions in the metallic, critical and localized regime.
The advantage of quasi-1d systems is that they might be, at least within some approximations,
solved analytically.

The above idea was developed by Muttalib \textsl{et al.} \cite{MuttW-99,MuttGW,MuttGW1}.
By solving the DMPK equation in the regime of $g\sim 1$, they indeed found that
the ``critical'' conductance distribution possesses non-analyticity close to 
$g=1$ \cite{MuttGW1}. Also, in the localized regime, they observed that the distribution
$p(\ln g)$ drops down at $\ln g=1$. Similar results were observed numerically in
Ref. \cite{froufe} where the
transport in the quasi-1d systems with a corrugated surface  was investigated
and the conductance distribution in the ``critical'' regime was studied.

However, these  results provide us only with a qualitative description
of the conductance distributions in the true critical regime and in the insulator. 
In the previous Section, we showed that the localized regime in the
3D system differs qualitatively from
the localization in
the  weakly disordered quasi-1d systems. The origin of this difference lies in the 
different form of the spectra of parameters $x_a$. 

To demonstrate the difference between the two insulating regimes, 
we showed in Fig. \ref{porovnaniex} the distribution $p(\ln g)$ for the 3D and
the weakly disordered quasi-1d systems.
Here, we compare in Fig. \ref{porovnanie}   the conductance distributions
for both systems in the metallic and critical regime (Fig. \ref{porovnanie}).
For a quantitative description of the critical and localized regimes, we need to
study the  systems with strong disorder.

\begin{figure}[t!]
\begin{center}
\includegraphics[clip,width=0.3\textheight]{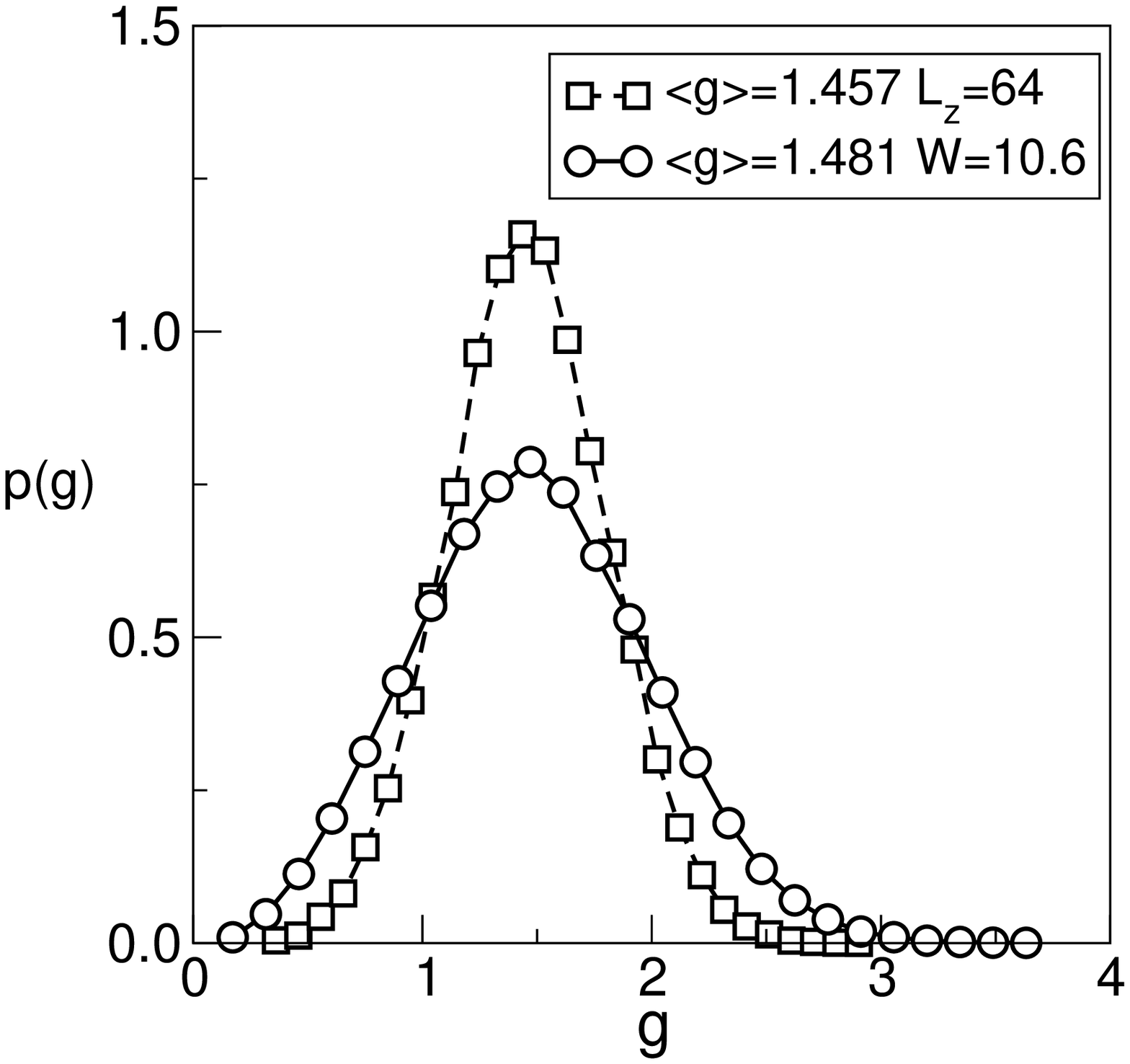}~~~~
\includegraphics[clip,width=0.3\textheight]{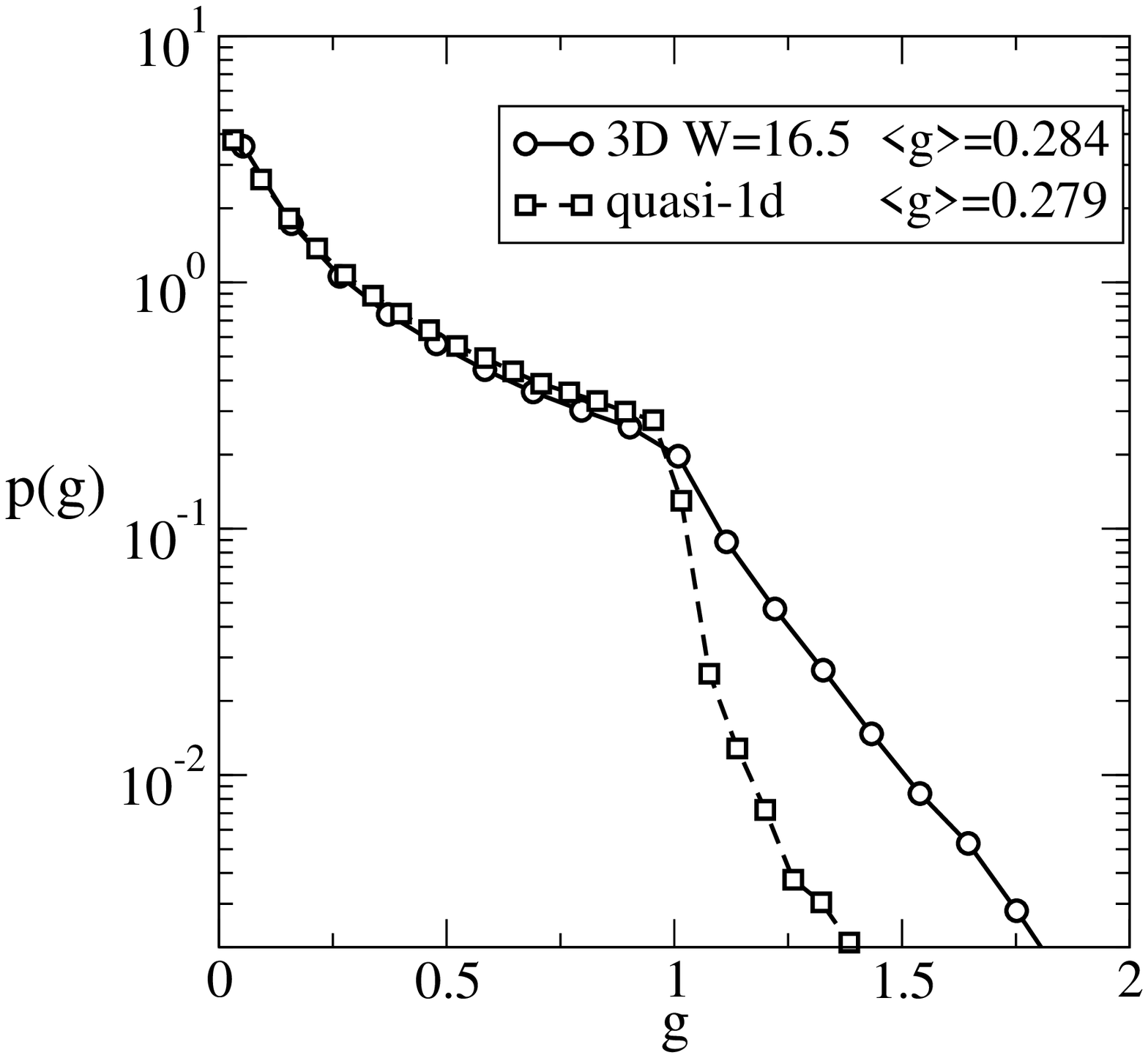}
\end{center}
\caption{%
Comparison of the conductance distribution for the 
quasi-1d systems (squares) and cubic systems (circles) \cite{M-2002a}. 
Left: The metallic regime:  $p(g)$ is Gaussian in both systems, but the widths of the
distributions differ. This is consistent  with our data for the universal conductance
fluctuations, given in Table \ref{table-ucf}, since $\textrm{var}~g$ of quasi-1d
systems differs from that of 3D. 
Right: The
critical conductance distribution  for the 3D Anderson model compared
with the distribution of the conductance for quasi-1d weakly disordered ($W=4$)
system of the size $8\times 8\times 210$. Although the mean conductance is the same, the 
distributions differ in the region $g>1$. The reason for this difference
is that the difference, $\langle x_2-x_1\rangle = \langle x_1\rangle$
in the quasi-1d system, while it  is much smaller in the 3D system, as shown in
Fig. \ref{delty}. The non-analyticity of the distribution for the quasi-1d system
was  explained and qualitatively described in Ref. \cite{MuttGW1}.
}
\label{porovnanie}
\end{figure}


\section{Numerical scaling analysis}\label{section:nsa}

Since the analytical calculations are possible only in the limit of the dimension $d=2+\veps$
($\veps\ll 1$),
numerical simulations provides us at present with the most relevant information about
the critical regime of the Anderson transition. 

The main problem of the numerical scaling analysis is the calculation of critical
exponent, $\nu$ in various physical models. Universality of the critical exponent
for a given symmetry class and dimension is interpreted as evidence of the
validity of the single parameter scaling. 

Historically, the first numerical simulations were performed for quasi-1d system
\cite{PS,McKK}.
The scaling of the smallest Lyapunov exponent was proved and the critical exponent,
$\nu$ was found. Later, scaling of other variables  was analyzed,
mostly with the motivation to treat the true 3D systems and 2D systems with
unitary and symplectic symmetry.  We will discuss in this Section the scaling
analysis of the conductance, conductance distribution and  level statistics.

\subsection{Scaling of the smallest Lyapunov exponent}\label{scal-LE}

Pichard and Sarma \cite{PS} were the first who proposed to use
the finite size scaling analysis for the calculation of the critical parameters of
the Anderson model.
They considered the quasi-1d system of the size $L^{d-1}\times L_z$ 
($d=2$ and 3) and calculate numerically
the \textsl{smallest} Lyapunov exponent, $\gamma_1$. 
As discussed in Appendix
\ref{le-q1d}, 
the Schr\"odinger equation for the Anderson model 
in the quasi-1d geometry can be 
written in the form
\be
\mv{\Psi_{n+1}}{\Psi_n}=M_n\mv{\Psi_n}{\Psi_{n-1}},
\ee
where $M_n$ is the transfer matrix for the $n$th slice of the system,
\be
M_n=\m2{E-{\cal H}_n}{-1}{1}{0}.
\ee
By multiplication of the transfer matrices, we obtain that the exponential decrease of the
wave function along $L_z$ is given by eigenvalues $\Lambda_a$ of the matrix
\be
M_{L_z}=\prod_{n=1}^{L_z} M_n
\ee
Oseledec's theorem states that in the limit of $L_z\to\infty$,
 all Lyapunov exponents, 
$\gamma_a$  of the matrix $M_{L_z}$
posses the Gaussian distribution with the variance proportional to the mean value.
The smallest (in absolute value)  Lyapunov exponent, $\gamma_1$, determines
the localization length in the $z$ direction.

Since $\gamma_1\propto L_z$, it is more convenient to use the parameter
\be
\Lambda=\frac{L_z}{L\gamma_1}.
\ee
In this paper, we  discuss the scaling behavior of parameter $z_1$,
\be
z_1=\frac{2}{\Lambda}=\frac{2 L\gamma_1}{L_z},
\ee
which is closely connected to parameter $x_1$, used in the previous Section for parametrization
of the contribution to the conductance.
From Oseledec's theorem it follows that 
$z_1$  converges in the limit of $L_z\to\infty$ to the mean value, and $\textrm{var}~z_1\propto
L/L_z$. Consequently, the difference between the numerically calculated
value of $z_1$ and the mean value is $\propto L_z^{-1/2}$
so that $z_1$ is free from any statistical problems provided that the system
is sufficiently long. 
Of course, $z_1$ is a function of the disorder, $W$, energy, $E$ and the system width, $L$.

\begin{figure}[b!]
\begin{center}
\psfrag{xx}{$\xi(W)$}
\psfrag{LLL}{$L/\xi(W)$}
\includegraphics[clip,width=0.32\textheight]{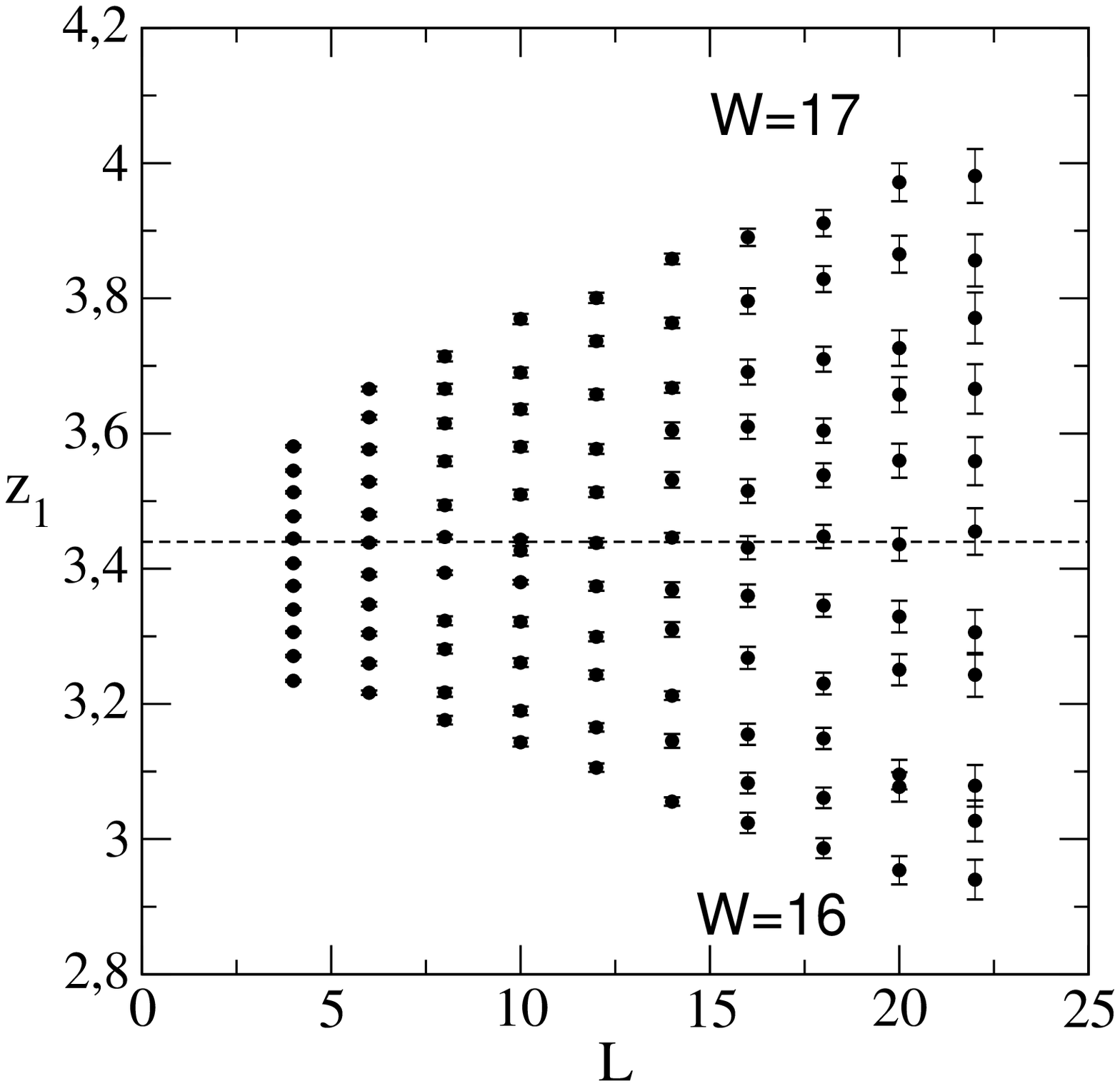}~~~~~~
\includegraphics[clip,width=0.32\textheight]{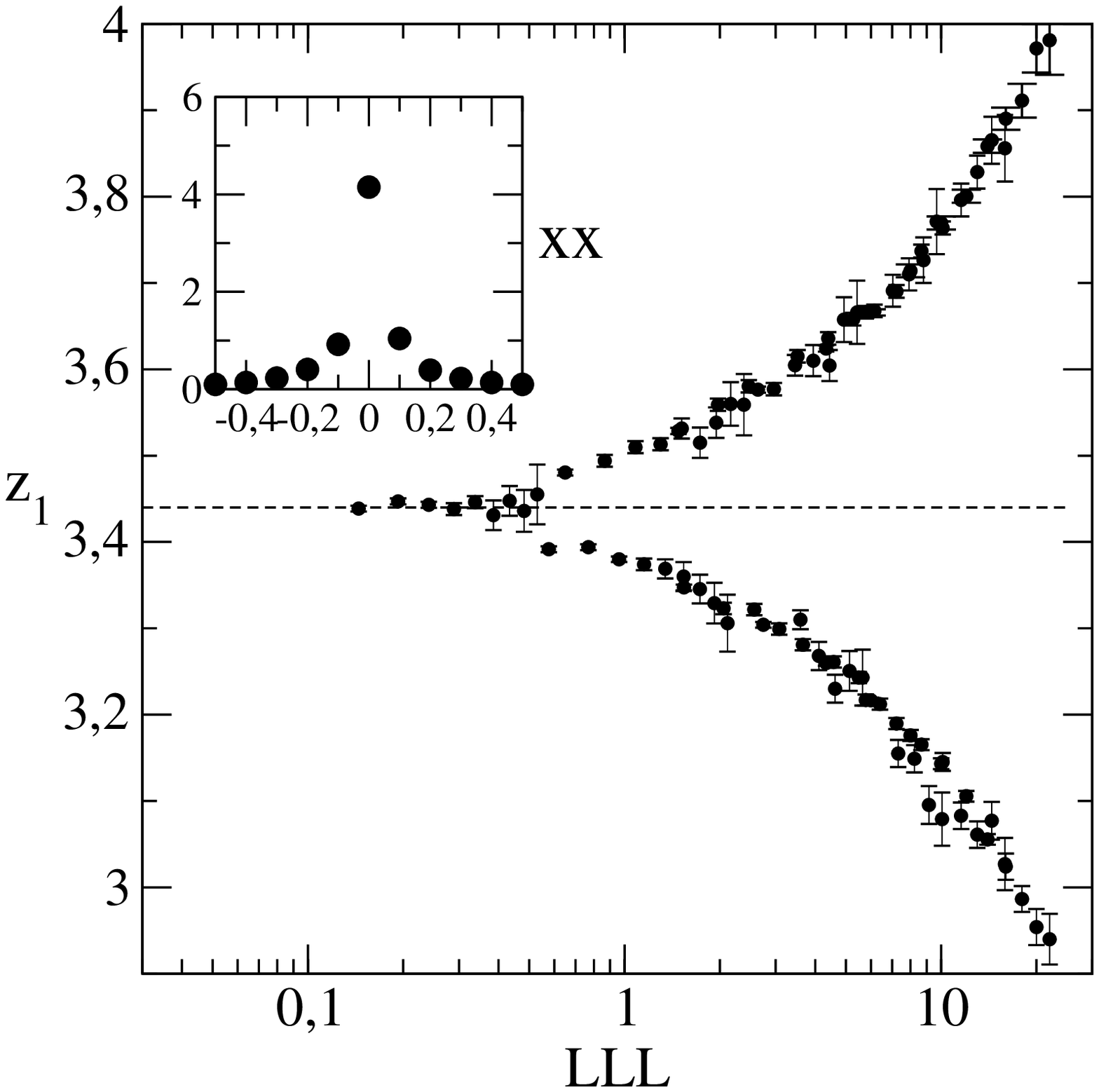}
\end{center}
\caption{Left: The $L$-dependence of the  variable $z_1=2L\gamma_1/L_z$ for the 3D Anderson model.
The data was calculated for the quasi-1d systems, $L^2\times L_z$
and for  disorder increasing from $W=16$ (bottom) to $W=17$ (top) with step
$\Delta W=0.1$.
The length $L_z$ is sufficiently
long to guarantee the relative accuracy of  
$0.2\%$ for small $L$, and of $1\%$ for $L=22$.
The dashed line is the fit of the critical value  $z_{1c}\approx 3.44$
\cite{M-2000}.
Right: The same data plotted as a function of $L/\xi(W)$, Eq. (\ref{z1-s2}).
Note the logarithmic scale on the horizontal axis. 
The function $F$ has two branches, the lower branch for the metallic regime, 
and upper one for the
insulating regime. The data with $L=4$ were excluded from the data sets,
because they do not lie on the universal curve due to the finite size effects.
Inset shows divergence of the correlation length, $\xi(W)$, at the critical point.
Note, $\xi$ is calculated up to a multiplicative factor.
}
\label{3D-z1}
\end{figure}

For a sufficiently large width $L$ of the system, we can estimate
the disorder and $L$ dependence of $z_1$ from basic physical considerations.
It is reasonable to expect that when  the system is in the  localized regime, $W\gg W_c$,
then $z_1$
converges to the ratio $2L/\lambda$. On the other hand,
in the metallic regime,  $W\ll W_c$, we have 
that the spectrum of $z_a$ is linear, as 
given by  Eq. (\ref{univ-xa}), 
of Appendix \ref{app:apl}.
Therefore, 
$z_1=\frac{2L}{N\ell}$, where $N$ is the number of channels.
Since $N\propto L^{2}$ in the case of the 3D Anderson model
we  obtain   $z_1\sim L^{-1}$ in the metallic regime.
Finally, at the critical point, we expect that  
$z_1=\textrm{const}$.

The above consideration can be summarized as follows:
\be\label{z1-scaling}
z_1=2\gamma_1\frac{L}{L_z}\sim\left\{
\begin{array}{ll}
L^{-1}  &  W<W_c\\
\textrm{const}  &  W=W_c\\
L/\lambda  &  W>W_c.
\end{array}
\right.
\ee
Relation (\ref{z1-scaling}) enables us to identify the critical point from the numerical data.
The left panel of Fig.  \ref{3D-z1}  presents the numerical data for the quasi-1d Anderson model
$L^2\times L_z$. We indeed see that for disorder $W>W_c$, $z_1$ increases when $L$
increases, while for $W<W_c$, $z_1$ decreases with $L$. From the $L$ dependence of $z_1$,
we estimate approximately the value of the critical disorder, $W_c\approx 16.5$
and the critical value, $z_{1c}=z_1(W=W_c)\approx 3.44$.

The next step in  the scaling analysis was done by MacKinnon and Kramer in Ref. 
\cite{McKK}. They assumed that
$z_1$ is a function of only one parameter,
\be\label{z1-s2}
z_1(L,W)=F(L/\xi(W))
\ee
where $\xi(W)$ is the \textsl{correlation length}. 
Relation (\ref{z1-s2}) follows from the assumption of the validity of single parameter
scaling. The right panel of Fig. \ref{3D-z1}  shows that indeed all the data $z_1(L,W)$
lie on the universal curve.

Since $z_{1c}=z_1(W=W_c)$ 
does not depend on the system size, $L$, 
the correlation length must diverge at the critical point,
\be\label{z1-s3}
\xi(W)\propto |W-W_c|^{-\nu}.
\ee
Figure \ref{3D-z1l} confirms that $z_1\propto (W-W_c)$ in the critical region.
Therefore, if we expand the function $F(x)$ in the Taylor series, and 
keep only the first two terms, $F(x)=F(0) + Ax^\alpha$,
we have that exponent $\alpha=1/\nu$. Consequently,  $z_1(L,W)$ 
is given in the critical region by the 
simple scaling equation,
\be\label{z1-s33}
z_1(L,W)=z_{1c}+A(W-W_c)L^{1/\nu}.
\ee
The fit of numerical data to Eq. (\ref{z1-s33}) enables us to calculate both the critical 
exponent, $\nu$, and critical disorder, $W_c$.

The easiest scaling analysis can be performed in the following
two steps \cite{spirosprivate}.
First, we can calculate the linear fit 
\be\label{z1-xx}
z_1(L,W) = z_1^{(0)}(L) + Wz_1^{(1)}(L)
\ee
Comparing the r.h.s. of Eq. (\ref{z1-xx}) with Eq. (\ref{z1-s33}), we have
\be\label{z1-xx1}
z_1^{(0)}=z_{1c}-AW_cL^{1/\nu}
\ee
and
\be\label{z1-xx2}
z_1^{(1)}(L)=AL^{1/\nu}.
\ee
Now, we can use  Eq. (\ref{z1-xx2}) to calculate the critical exponent.
Next, with the use of Eq. (\ref{z1-xx2}) we can write  Eq. (\ref{z1-xx1}) in the form
\be\label{z1-xx3}
z_1^{(0)}=z_{1c}-W_cz_1^{(1)},
\ee
so that 
the slope of the linear fit $z_1^{(0)}$  \textsl{vs.} $z_1^{(1)}$  determines 
critical disorder, $W_c$.

The physical meaning of the correlation length, $\xi(W)$  can be estimated
by comparing  Eq. (\ref{z1-s2}) with Eq. (\ref{z1-scaling}).
Clearly, $\xi(W)=\lambda$ in the insulating side of the transition.
Also, we find  that 
$F(x)\sim x$ in the insulator and $F(x)\sim x^{-1}$ in the metal. Then, from the expression
of $z_1=2L/(N\ell)=2\langle g\rangle^{-1}=1/(2L\sigma)$ ($\sigma$ is the conductance) 
we find  that in the metallic regime the correlation length
$\xi(W)\propto \sigma^{-1}$ \cite{McKK}.

\begin{figure}[t!]
\begin{center}
\includegraphics[clip,width=0.3\textheight]{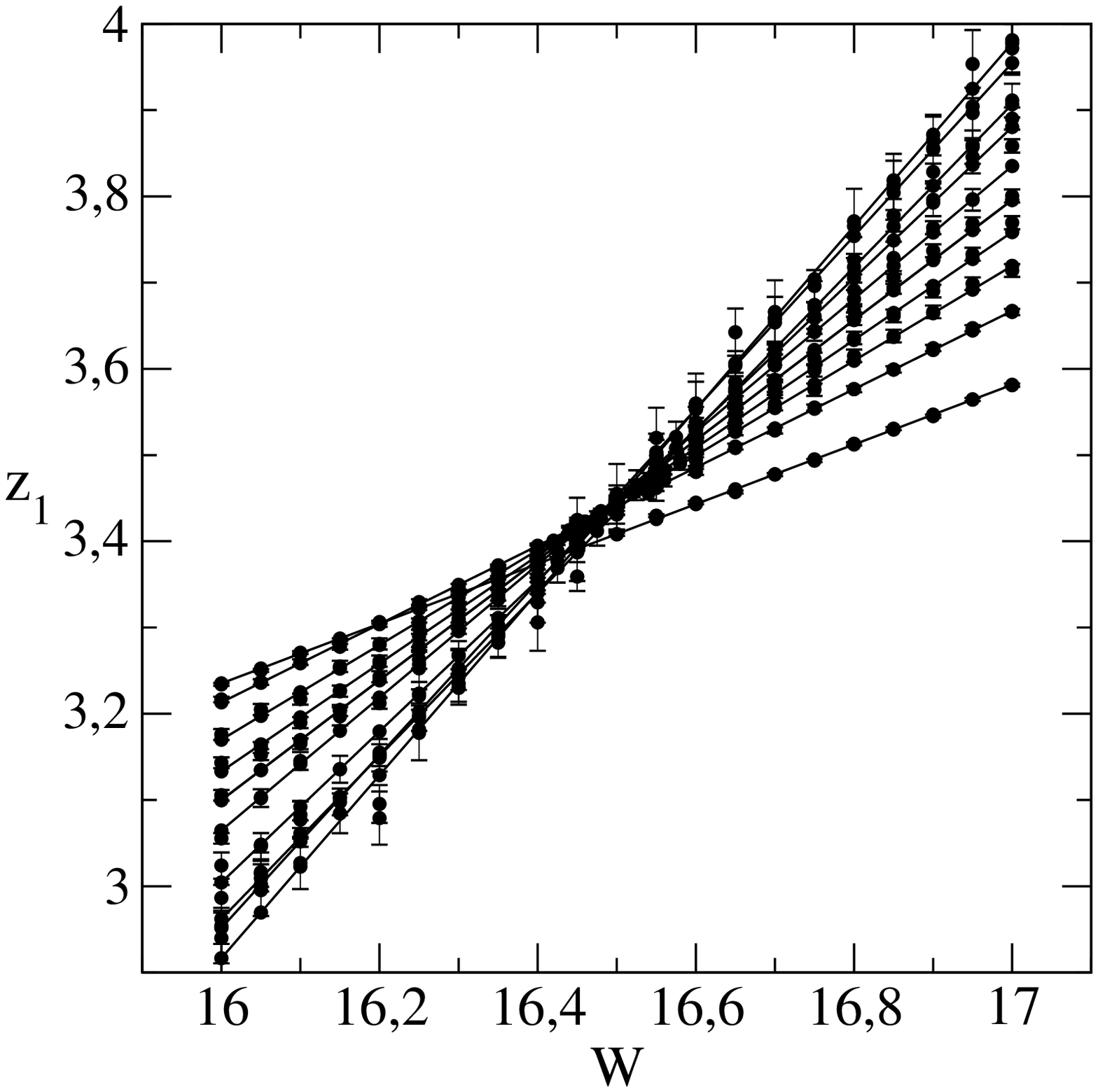}~~~~~~
\includegraphics[clip,width=0.3\textheight]{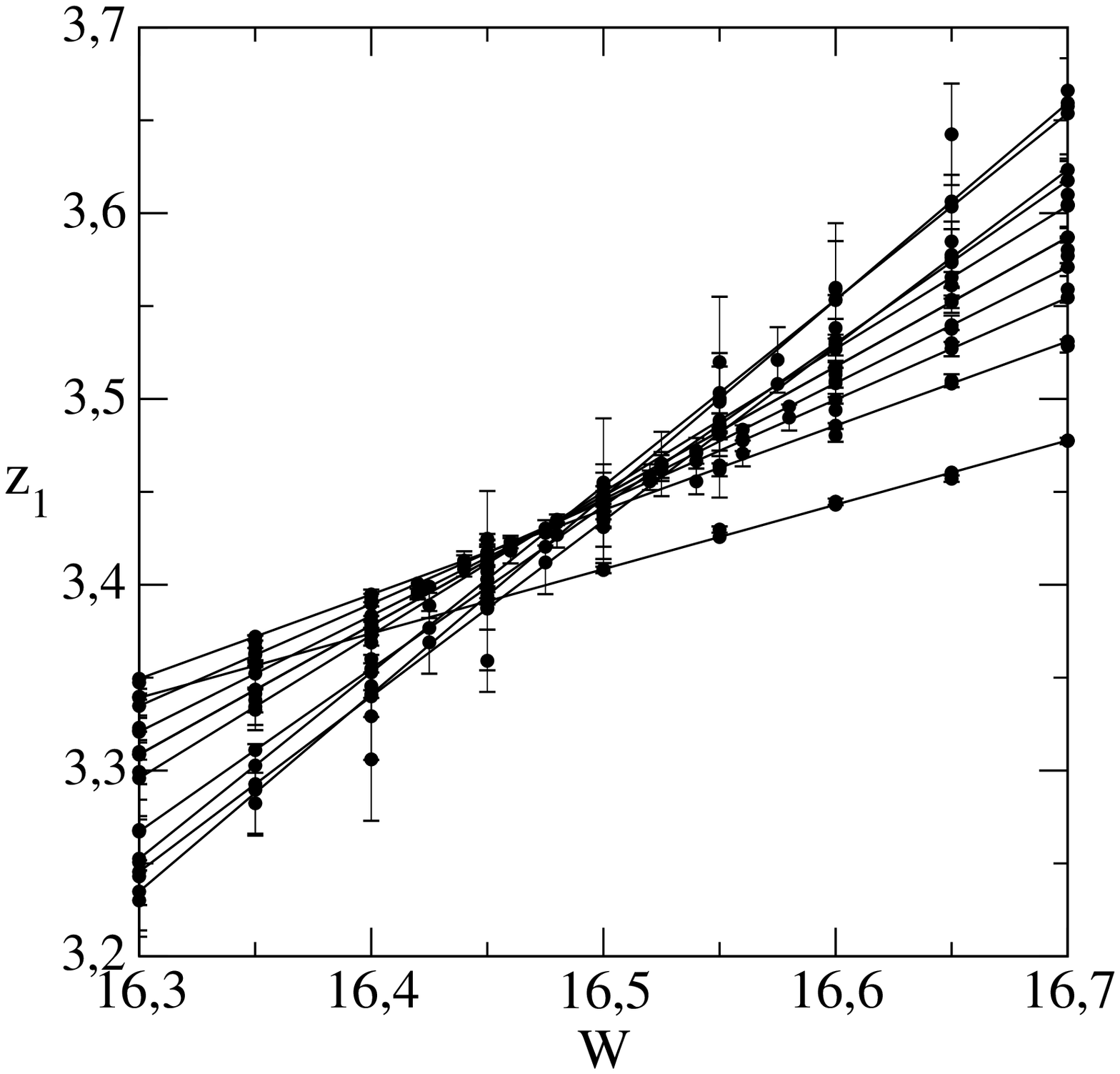}
\end{center}
\caption{Left: The same data as in left  panel of 
Fig. \ref{3D-z1} but plotted as a function of the disorder $W$.
Solid lines are linear fits, given by Eq. (\ref{z1-s33}) for $4\ll L\ll 22$.
In the ideal case, all lines have a common crossing point. This never happens for
numerical data, as shown in the
detailed plot in the right panel.  The deviations
from the universal scaling relation, given by Eq. (\ref{z1-s33})
are either due to 
the finite size effects,
discussed in Sect. \ref{sect:fse}, or due to the insufficient accuracy of numerical data.
}
\label{3D-z1l}
\end{figure}

The correlation length $\xi(W)$ was first calculated numerically 
in Ref. \cite{McKK-1983}. 
The critical disorder, $W_c\approx 16.5$, and the critical exponent, $\nu\approx 1.50$,
were calculated. These calculations were repeated by many other authors
\cite{ostatni,cain,schreiber}
with the use of various scaling analysis, and for increasing system size.
Surprisingly, these new analysis did not
bring any considerable corrections to the  critical parameters, obtained in the pioneering work, 
\cite{McKK}.
At present,  the most accurate estimation of the critical exponent is 
\be
\nu=1.57\pm 0.02,
\ee
obtained by very detailed scaling analysis of numerical data for $z_1$ in Ref. \cite{SO-99}.
In Ref. \cite{bulka},
the phase diagram in the energy-disorder phase space (shown schematically in Fig.
\ref{fig-schema-pom}) was calculated for the first time for various distribution
of random energies.

The scaling analysis \cite{McKK} solved also the problem of the existence of the 
Anderson transition in the 2D orthogonal systems.  It was shown, that there is no
metallic phase, but
the localization length, given by $\xi(W)$,
is extremely large for small disorder (of orders of $10^6$ lattice sites for $W=1$.
This result also explains why various  previous
works identify the Anderson transition also in 2D systems: these works analyzed only small 
systems \cite{McKKansw}.
The numerical data for the correlation length, $\xi(W)$ of the  2D Anderson 
model are given in Refs. \cite{McKK-1983,SAD}
 and for 3D Anderson model in Ref. \cite{McKK-1983}.

\subsection{Finite-size corrections}\label{sect:fse}

A more general formulation of Eq. (\ref{z1-s3}) is
\be\label{z1-irr}
z_1=F(\zeta_1,\eta_1,\eta_2,\dots),
\ee
where $\eta_i$ are further  $L$-dependent parameters. 
They   determine how $z_1$ depends on various microscopic
parameters of the model, for instance  distribution of the disorder, correlation length 
of the disorder, magnetic field. 
The one-parameter scaling is valid only if in the limit of $L\to\infty$
all 
$\eta_i\to 0$. Only under this assumption, the equation
(\ref{z1-s3}) can be recovered, with $\zeta_1=L/\xi(W)$. Thus, the assumption of
the single parameter scaling requires that
\be
\eta_i(L)\sim L^{-y_i}
\ee
where $y_i$ are irrelevant scaling exponents.

Although these parameters play no role
for sufficiently large systems, they might influence the scaling analysis
for smaller  $L$. For instance, the one parameter scaling, given by Eq. (\ref{z1-s33})
requires that all
linear lines in Fig. \ref{3D-z1l} cross in one common point for $W=W_c$ and
$z_1=z_{1c}$. This is evidently not true for the smallest $L=4$.

From Eq. (\ref{z1-irr}) it follows that when $L$ is not sufficiently large, 
one can generalize the scaling analysis 
by the inclusion of an additional term on the r.h.s. of Eq. (\ref{z1-s3}). In the most simple case,
when only one irrelevant parameter is considered, we obtain
\be\label{z1-s34}
z_1(L,W)=z_{1c}+A(W-W_c)L^{1/\nu} + BL^{-y}.
\ee
\cite{McK-jpa}.
We can estimate the critical parameters by fitting the numerical data to the function
(\ref{z1-s34}). 
A more detailed scaling analysis might consider also 
higher order terms in powers of $\zeta_1=(W-W_c)L^{1/\nu}$.
The most detailed scaling analysis 
was performed in Ref. \cite{SO-99}, where  
the  numerical data were fitted to a function of 11 parameters.

Another possibility to eliminate the finite size scaling effects is to perform
the scaling analysis with reduced data sets
by omitting data for the smallest system size \cite{M-2000,Ludwig-1}.
Using only the numerical data for $L>L_{\rm min}$, one can study the $L_{\rm min}$ dependence 
of the critical parameters.  We demonstrate this method in Fig. \ref{FSS}.
 
\begin{figure}[t!]
\begin{center}
\includegraphics[clip,width=0.43\textheight]{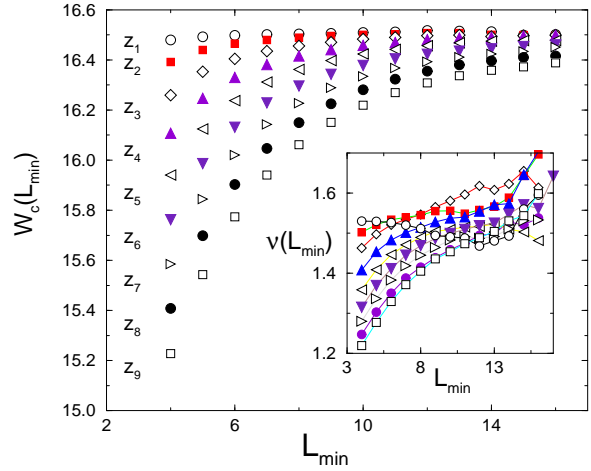}
\end{center}
\caption{The critical  disorder, $W_c$, and the critical exponent, $\nu$ (inset) calculated from the
finite size scaling analysis of the first 9 Lyapunov exponents \cite{M-2000}. 
The numerical data for
$4\le L \le 22$ were used.  The critical disorder,  $W_c$,
converges to the ``correct'' value, $W_c\approx 16.5$
when numerical data for smaller system size, $L<L_{\rm min}$ 
are not considered and $L_{\rm min}$ increases. 
The finite size effects are larger for the higher Lyapunov exponents,
and may lead to wrong estimation of the critical parameters, when too small systems are considered.
On the other hand, increase of $L_{\rm min}$ influences   the accuracy of the numerical fits,
especially in the case of  the power fit (\ref{z1-xx2}) for the  critical exponent.
Note, the finite size effects are very small in the  scaling analysis of the smallest 
Lyapunov exponent.
}
\label{FSS}
\end{figure}

\subsection{Scaling of higher Lyapunov exponents}\label{higher:LE}

\begin{figure}[t!]
\begin{center}
\psfrag{zz}{$\!\!\!\!\!\!\!\!\!\!\!\! z_a^{(1)}(L)$}
\includegraphics[clip,width=0.3\textheight]{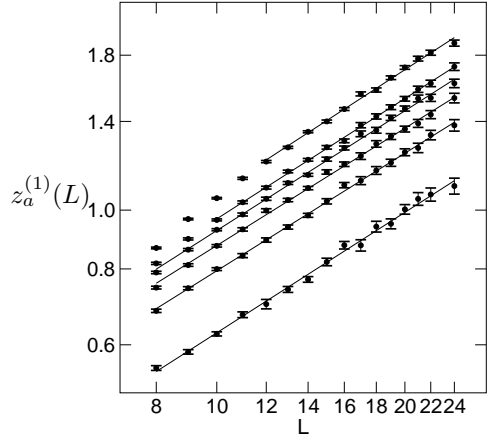}
\end{center}
\caption{Estimation of the critical exponent, $\nu$, from the power fit, given by Eq. 
(\ref{z1-xx2}). Data for the smallest Lyapunov exponent, $z_1$,  as well as for
$z_a$, $a=2-5$ and 9 are shown.
We remind the reader that $z_a^{(1)}$ is the slope of the $W$ dependence of $z_a$,
defined by Eq. (\ref{z1-xx1}). 
}
\label{higherLE}
\end{figure}

As discussed in Sect. \ref{sect:crit}, the critical conductance distribution is not a self
averaged quantity. Therefore,  the scaling theory might, at least in principle,
contain an infinite number of  independent \textsl{relevant} parameters. To verify this 
possibility, it is worth to verify whether the  higher Lyapunov 
exponents, $z_a$, scale in the same way as the $z_1$. This was done 
in Ref. \cite{M-2000} and reproduced in \cite{SO-01}.

As shown in Fig. \ref{higherLE}, already the most trivial scaling procedure,
given by Eqs. (\ref{z1-xx1}-\ref{z1-xx3}), provides us with reliable evidence that
the  nine smallest Lyapunov exponents indeed scale with the same
 critical exponent, $\nu$.   This proves that all Lyapunov exponents can be expressed
 as a functions of only one variable, $L/\xi(W)$, with the same correlation length,
 $\xi(W)$.

Scaling of the higher Lyapunov exponent serves also as a nice example
how the finite size effects influence the scaling analysis.
As was shown in Fig. \ref{FSS}, the finite size corrections are 
almost negligible in the case of $z_1$, but they increase when $a$ increases.
The reason lies in the properties of the spectra of parameters $z_a$, or,
equivalently, of $x_a$. In  Section \ref{sect:crit}, we found that 
for a given size of the system, $L$, only parameters $x_a$ with $a\le L$ are size
independent. The rest of the spectra depends on $L$. The same must hold for
$z_a$. Therefore, only the data with $L>a$ are relevant 
for the scaling analysis  of the higher Lyapunov exponent. In the case of $z_9$, only the 
data for $L\le 12$ are relevant for the scaling analysis.

\subsection{Scaling of the mean conductance}

\begin{figure}[b!]
\begin{center}
\psfrag{gm}{$\langle g\rangle$}
\psfrag{LL}{$L/\xi$}
\includegraphics[clip,width=0.3\textheight]{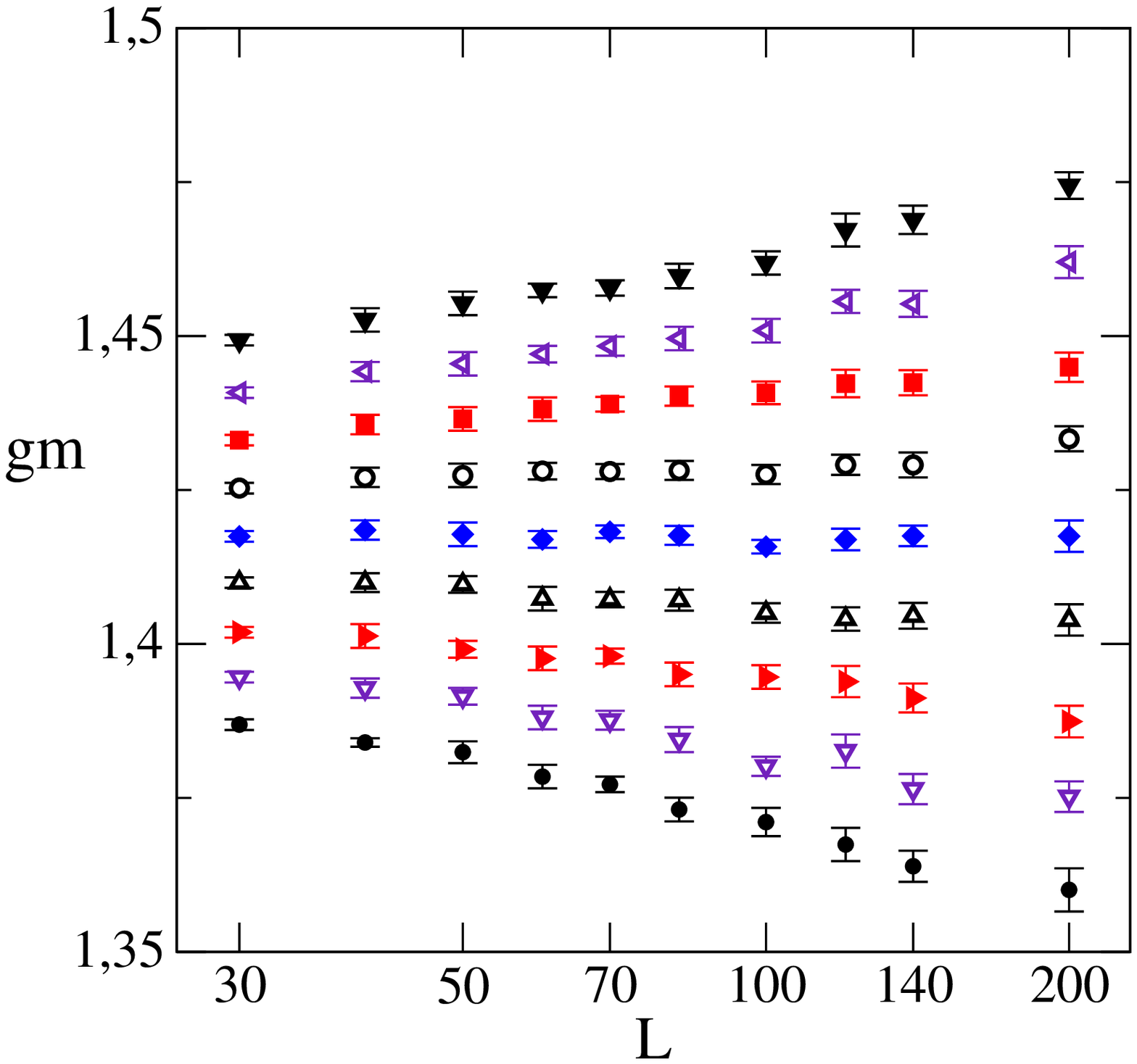}~~~~~~
\includegraphics[clip,width=0.3\textheight]{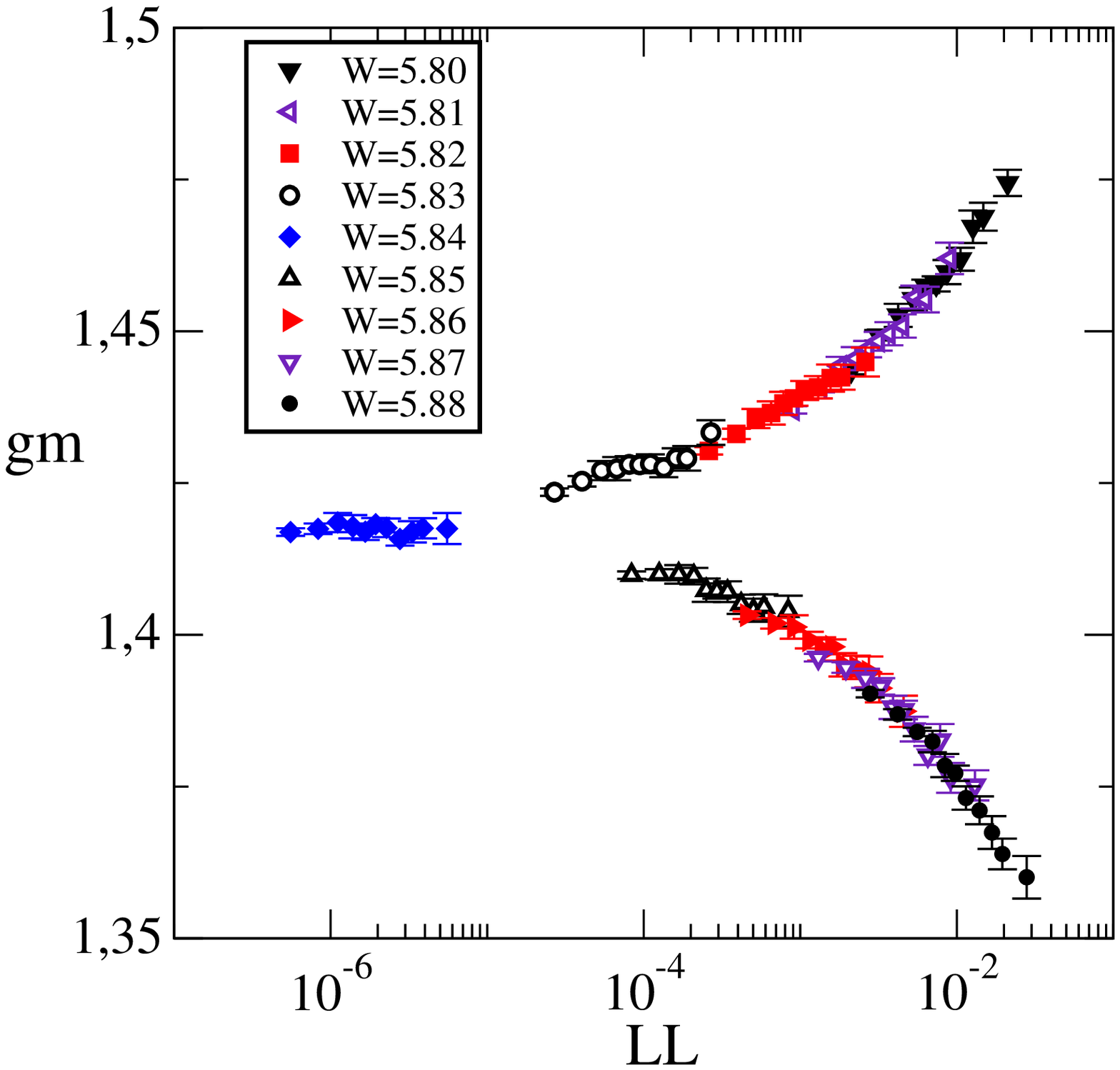}
\end{center}
\caption{Left: The $L$ dependence of the mean conductance of the 2D Ando model.
For disorder $W<W_c$, $\langle g\rangle$ increases with $L$,
and for $W>W_c$, $\langle g\rangle$ decreases with $L$, in agreement with
scaling hypothesis. Right panel shows the same data
plotted as a function of only one parameter, $L/\xi(W)$ with $\xi\propto |W-W_c|^{-\nu}$.
Appropriate choice of the correlation length, $\xi(W)$, for each value of the disorder, $W$
enables us to scale all numerical data to one universal curve. Note the logarithmic scale on the
horizontal axis.
}
\label{fig-ando-scaling}
\end{figure}

Numerical analysis of the quasi-1d systems  provides us with the most accurate estimation of
the critical parameters. Nevertheless, the use of quasi-1d geometry is rather artificial.
It would be more suitable to prove the scaling of the conductance of the true $d$ dimensional
systems.  This problem is rather complicated since, as we have seen in Sect. 
\ref{sect:crit}, 
 the conductance is not self-averaged quantity in the critical region.
 We need therefore to calculate the mean value,
$\langle g\rangle$ from the statistical ensemble of $\ns$ different cubes,
\be
\langle\ g\rangle=\ds{\frac{1}{\ns}}\sum_{i=1}^{\ns} g_i
\ee
where $g_i$ is the conductance calculated for the $i$th sample,
and to  estimate the accuracy of the mean value by using the relation
\be
\textrm{acc}~g=\ds{\sqrt{\frac{\textrm{var}~g}{\ns}}}.
\ee
Since both $\langle g\rangle$ and $\textrm{var}~g$ are  of order of unity, we need  
$\ns\approx 10^6$ to reach the relative accuracy $\sim 0.1\%$.

The scaling formulas, presented in previous Sections, 
are valid for the scaling behavior of the mean conductance as well.
Thus, we expect that the disorder and system size dependence
of the mean conductance  in the critical region is given by
\be\label{g-33}
\langle g\rangle = \langle g\rangle_c + A(W-W_c)L^{1/\nu}
\ee
where $\langle g\rangle_c$ is the critical conductance.
A similar equation can be constructed for $\langle\ln g\rangle$.

For the 3D Anderson model, the scaling of the mean conductance, $\langle g\rangle$,
and of the typical conductance, $\exp\langle\ln g\rangle$ was numerically confirmed
in Ref. \cite{SMO-1}.  The calculated critical exponent, $\nu=1.57$, agrees with the result
of the scaling analysis of the smallest Lyapunov exponent \cite{SO-99}.

As an example of the scaling analysis of the conductance,
we present  the most recent numerical data 
for the mean conductance of the 2D Ando model \cite{Ludwig-1}.
The left panel of Fig. \ref{fig-ando-scaling} shows the  $L$-dependence of the mean conductance
for fixed disorder. The metallic, localized and critical regime can be estimated
in the same way as for $z_1$, with the only difference, that  increasing
$\langle g\rangle$ indicates the metallic phase in this case.
The right panel of  Fig. \ref{fig-ando-scaling} shows the same data but as a function of $L/\xi$.
The data confirms that indeed $\langle g\rangle=g(L/\xi)$ is a function of only one variable.
Two branches of the function $g$ correspond to two different transport regimes, metallic
and localized.

The numerical proof of the scaling of the mean value is still not sufficient for
the verification of whether or not one parameter scaling really works. Namely, we 
cannot exclude that the higher cummulants of the conductance do not scale. 
Of course, it is impossible to verify the scaling of all cummulants. Instead, we
discuss in the next Section the scaling of conductance distribution, $p(g)$.

\subsection{Scaling of the   conductance distribution}\label{scal:pcg}

\begin{figure}[t!]
\bc
\includegraphics[clip,height=0.22\textheight]{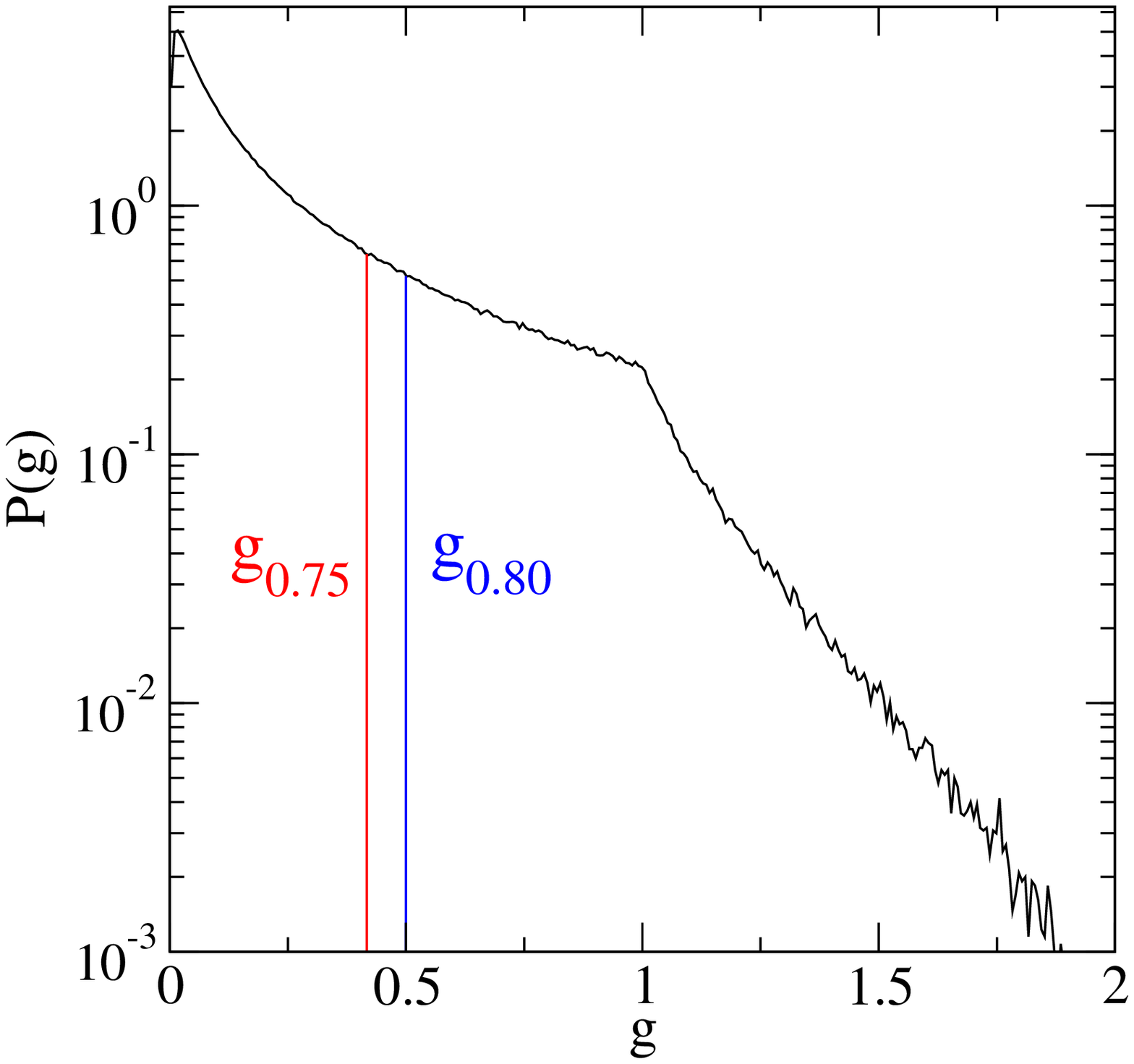}
\ec
\caption{Definition of percentiles. Note the logarithmic scale on the vertical axis.}
\label{perc-def}
\ef

In the previous Section we verified the scaling of the mean conductance,
$\langle g\rangle$. However, we still cannot exclude the possibility that
the length and disorder dependence
probability distribution, $p(g)$,
is determined by   an infinite number of parameters, for instance all  higher
cummulants of the conductance. To prove that the entire conductance distribution
scales as a function of only one parameter, 
the scaling  of the percentiles, $g_q$ was analyzed numerically in
Ref. \cite{SMO-2}. 

The percentile, $g_q$, is  defined by the relation
\be\label{eq-perc}
q=\int_0^{{g_q}}p_L(g)~\textrm{d}g.
\ee
By definition (\ref{eq-perc}), the probability that 
$g<g_q$, equals to $q$ (Fig. \ref{perc-def}).

In Ref. \cite{SMO-2}, the one parameter scaling of percentiles, $g_q$, was proved for four values of $q$, 
$q=0.025$, 0.17, 0.50 (median) and
0.83. All four variables obey the one parameter scaling with the critical disorder
$W_c$ close to 16.5 and with the critical exponent 
$1.56 < \nu <1.60$. 

Suppose now that  two percentiles, $g_\alpha$ and $g_\beta$,
($\alpha<\beta$) obey the single parameter scaling. Then, the percentile
$g_\gamma$ ($\alpha<\gamma<\beta$) must scale, too.
Also, if $g_\alpha$ and $g_\beta$ scale, 
then the difference  $g_\beta-g_\alpha$ scales. 
Therefore, for the proof of the one parameter scaling,
it is sufficient to prove the scaling of 
only a few percentiles, which was done in Ref. \cite{SMO-2}. We conclude that
the numerical verification of the single parameter scaling of a few percentiles 
provides us with the evidence that the entire probability distribution, $p(g)$,
obeys the single parameter scaling in the critical region.

\subsection{Scaling of the level statistics}

As discussed  in Sect. \ref{levelstatistics}, the distribution $p(s)$
of the differences between the neighboring eigenenergies depends on whether
the system is in the metallic, localized or critical regime.  
In Refs. \cite{Shklovskii,spiros,shkl}, the scaling analysis of the
level statistics was proposed and studied. 

The critical parameters were calculated for
the  2D Ando model \cite{ludwig-isa}, the 3D Anderson model \cite{batsch,ZK-e},
and for the problem of quantum percolation in 3D system \cite{tomi}.
Recently, the scaling analysis of the level statistics was applied  to the symplectic
model on the fractal lattice with the aim to prove that the lower critical dimension
for the symplectic systems is less than 2 \cite{SO-2006}.

\subsection{Scaling of the inverse participation ratios.}\label{scal:ipr}

\begin{figure}[b!]
\begin{center}
\psfrag{nn}{$\nu$}
\includegraphics[clip,width=0.32\textheight]{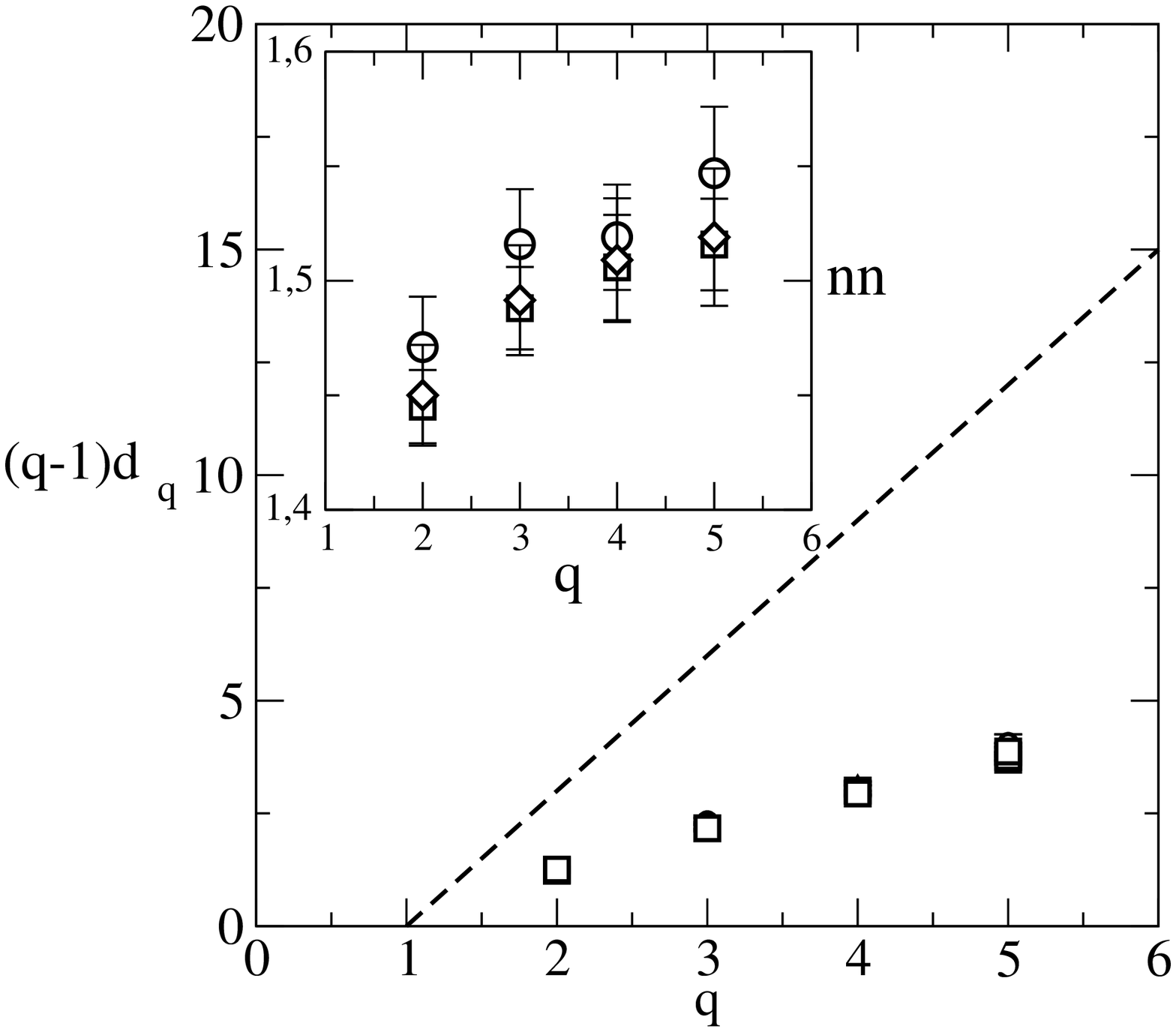}
\end{center}
\caption{The fractal dimension, $d_q$, obtained from the scaling analysis of 
the inverse participation ratios,
$I_q$, for the 3D Anderson model with the Gaussian and box disorder. The 
dashed line is
$(q-1)d$ for $d=3$. Inset shows the estimated critical exponent for three critical points.
At the band center with the Gaussian (circles) and box (boxes) distribution, and for
the Gaussian disorder $W=2$ with  critical energy $E_c\approx 6.58$ (triangles).
The electron eigenenergies and eigenfunctions were calculated 
numerically for cubes of the size $8\le L \le 54$ \cite{brndiar}.
}
\label{iqq}
\end{figure}

The scaling of inverse participation ratios, 
$I_q$, defined in Sect. \ref{WF}, was performed recently
in Ref. \cite{brndiar}. Contrary to the conductance, which is defined only for energies
inside the unperturbed energy band, $|E|<6V$, $I_q$ can be calculated for the 
entire energy spectrum
of the Hamiltonian, and can be used for the verification of the universality
of the metal-insulator transition along the critical line.  However, one has to keep in mind 
that $I_q$ is not size-invariant at the critical point but behaves as
\be
I_q\sim L^{-(q-1)d_q},
\ee
as discussed in Sect. \ref{WF}.

In the scaling analysis, the logarithm of $I_q$ was used, since the values of $I_q(E_n)$
might fluctuate in orders of magnitude  within a small energy interval, $\delta E$
(Figs. \ref{ipr-2}, \ref{fig-ipr}).
The quantity of interest is then
\be
\tilde{I}_q(L,W)=\langle\langle\ln  I_q(E_n)\rangle_{\delta E}\rangle,
\ee
where the averaging is performed within the energy interval, $\delta E$, and over
statistical ensemble of microscopically different samples. $\tilde{I}_q(L,W)$ is then
fit to the scaling equation,
\be
\tilde{I}_q(L,W)=A-(q-1)d_q\ln L + B(W-W_c)L^{1/\nu}
\ee
and critical parameters, $W_c$, $d_q$ and $\nu$ are calculated for the 3D Anderson model with
the Gaussian and box  distribution of random energies.

Figure \ref{iqq} shows that not only the critical exponent, $\nu$, but also the
fractal dimensions,
$d_q$, are universal, independent of the microscopic details of the model and on the position
of the critical point on the critical line. 
The fractal structure of the critical wave function was studied also in Refs.
\cite{evers,evers1,cuevas,cuevas2002a}.


\section{Scaling in the $d$-dimensional systems}\label{sect:dd}

The numerical scaling analysis provides us with rather accurate estimation of the critical
exponent for the 3D Anderson model. However, the obtained results are in disagreement with
expectations of the theory, which reports $\nu=1$ for $d=3$ \cite{Woelfle}.
It is therefore important,
for the detailed comparison of the theory and numerical data,
to calculate the dimension dependence 
of the critical exponent, and check, whether or not agreement
with theory is better for, $d\to 2^+$ or for $d>3$. We summarize here the very recent data 
for the critical exponent, calculated in Ref. \cite{Travenec} for $2<d\le 4$,
and we present also new data for $d=5$.

\begin{figure}[t!]
\begin{center}
\psfrag{eps}{$\veps^{-1}$}
\psfrag{nu}{$\nu$}
\includegraphics[clip,width=0.25\textheight]{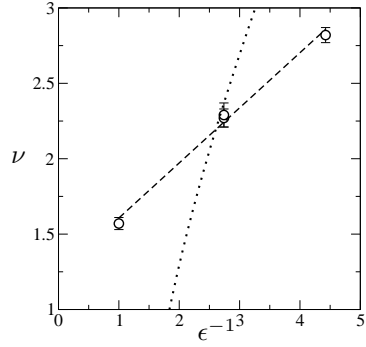}
\end{center}
\caption{The critical exponent, $\nu$, as a function of $\veps^{-1}$ ($\veps=d-2$).
Shown are the data for three bifractals, discussed in Sect. \ref{sect:pgd} (Fig. \ref{fractals}),
and for a cubic 3D system. The dashed line is the  linear fit, $1.24+0.365/\veps$. 
The dotted line is the analytical $\veps$ - expansion of the critical exponent,
given by Eq. (\ref{nu-eps}). Note, there is no agreement between the theoretical prediction
(dashed line) and the numerical data. 
Note also that  two systems with the same spectral dimension have the same critical exponent,
as expected.
}
\label{nu-1}
\end{figure}

\subsection{Dimension $d=2+\veps$}

Figure \ref{nu-1} shows the disorder dependence ($d=2+\veps$) of the critical exponent, $\nu$,
calculated from the finite size scaling of the smallest Lyapunov exponent
of quasi-1d bifractal lattices
with fractal structure of the cross-section, shown in Fig. \ref{fractals}.
These data are compared with the theoretical prediction, based on the
analytical calculation of the function $\beta(g)$.

In the limit of $\veps\ll 1$, the
critical disorder $W_c\sim\veps\ll 1$ 
and the critical conductance $g_c\sim \veps^{-1}\gg 1$.
The function $\beta(g)$ can be expanded 
in power series of $g^{-1}$.
It is more convenient to use, instead of the  conductance $g$, the parameter
\be
t=\frac{1}{2\pi g}.
\ee
The size dependence of the parameter $t$  is given by the equation
\be
\ds{\frac{\partial t}{\partial \ln L}}=\beta(t),
\ee
and the  critical exponent, $\nu$, is then given by the relation
\be\label{nunu}
\nu^{-1}=-\frac{\partial\beta(t)}{\partial t}\Bigg|_{t=t_c}.
\ee
For the orthogonal systems, the $t$-expansion of the function $\beta(t)$ reads \cite{Hikami}
\be\label{beta-eps}
\beta_{\rm O}(t)=\veps t-2t^2-12\zeta(3)t^5 +\frac{27}{2}\zeta(4)t^6+\dots .
\ee
In Eq. (\ref{beta-eps}),
$\zeta(3)= 1.202$ and $\zeta(4)=\pi^4/90$.

Since the expansion (\ref{beta-eps}) is known only up to the 6th power in $t$,
it is difficult to estimate the  accuracy of the obtained results for the critical 
conductance and  critical exponent, specially  when $\veps$ is not small.
For instance, 
for the 3D system, ($\veps=1$)  one finds, by solving the equation
$\beta_{\rm O}(t_c)=0$, that $t_c=0.395$.  This agrees qualitatively with
the estimation of critical conductance, $g_c=1/(2\pi t_c)\approx 0.40$
(we remind the reader that the numerically observed values of the critical conductance
are 0.445 and 0.280 for periodic and hard wall boundary conditions, respectively).
However, 
from   Eq. (\ref{nunu}) we obtain  $\nu=0.67$, which is far from 
the numerical result, $\nu\approx 1.57$. The agreement with the numerical data
is not better for small $\veps$, as is shown in 
Fig. \ref{nu-1} which  compares    the critical exponent, calculated from the 
$\veps$ expansion,
\be\label{nu-eps}
\nu=\frac{1}{\veps}-\frac{9}{4}\zeta(3)\veps^2
\ee
with our  numerical data. Clearly, there is no agreement between the theory and results of
numerical simulations.

For completeness, we  add the $\veps$ expansion of the 
$\beta$ function for
symplectic systems. It  can be obtained from expression (\ref{beta-eps})
with  the use of the symmetry relation \cite{Wegner-89}
\be
\beta_{\rm S}=-2\beta_{\rm O}(-t/2),
\ee
which gives
\be
\beta_{\rm S}(t)=\veps t +t^2-\ds{\frac{3\zeta(3)}{4}}t^5 -\ds{\frac{27\zeta(4)}{64}}t^6+\dots .
\ee
Note, $\beta_{\rm S}(t)$ is positive for $\veps\to 0$ which confirms the existence of
the critical point in the 2D symplectic systems.

\subsection{Dimension $d\ge 3$.}

\begin{figure}[b!]
\begin{center}
\psfrag{slope}{$z^{(1)}$}
\includegraphics[clip,width=0.35\textheight]{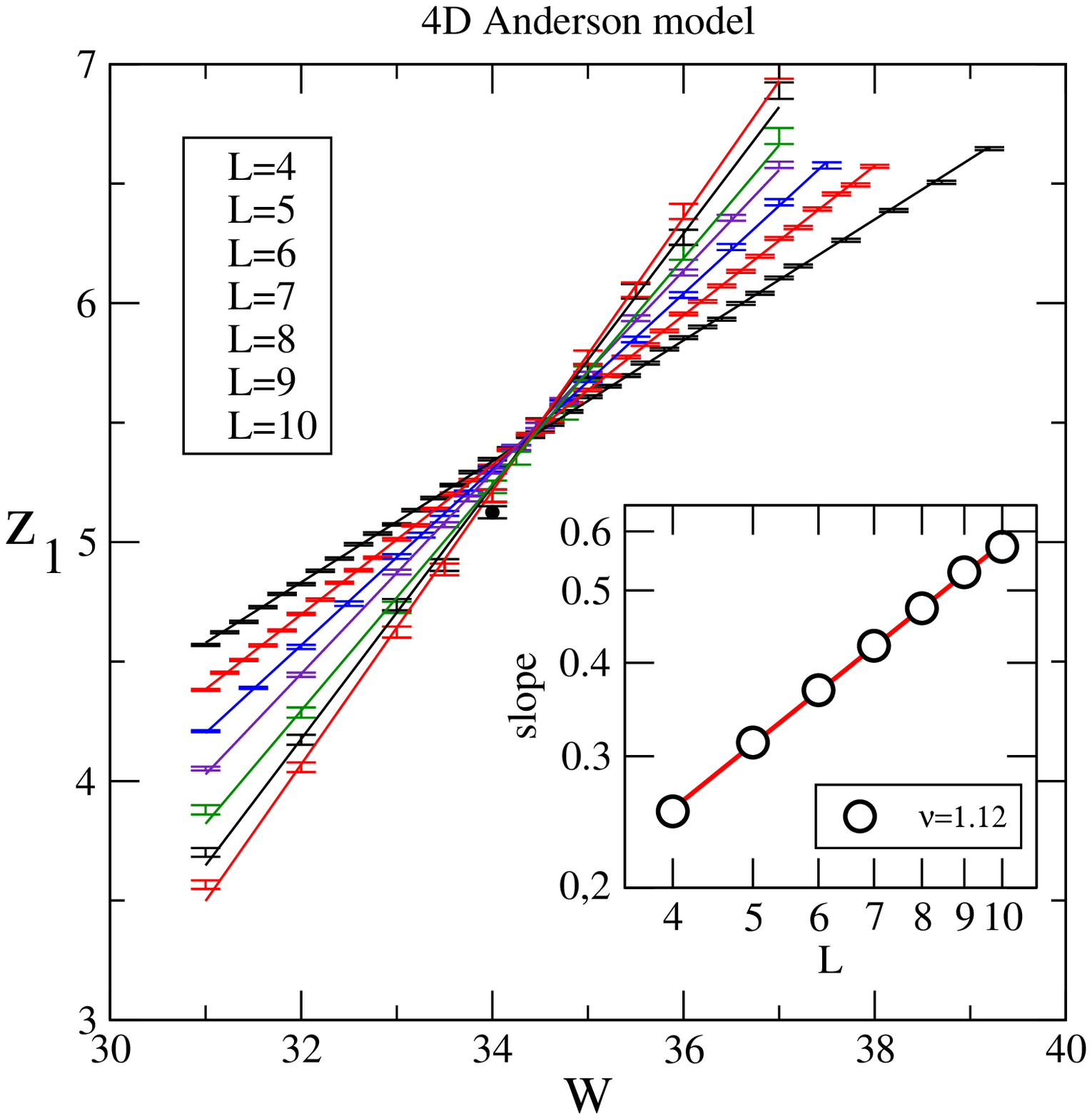}
\includegraphics[clip,width=0.35\textheight]{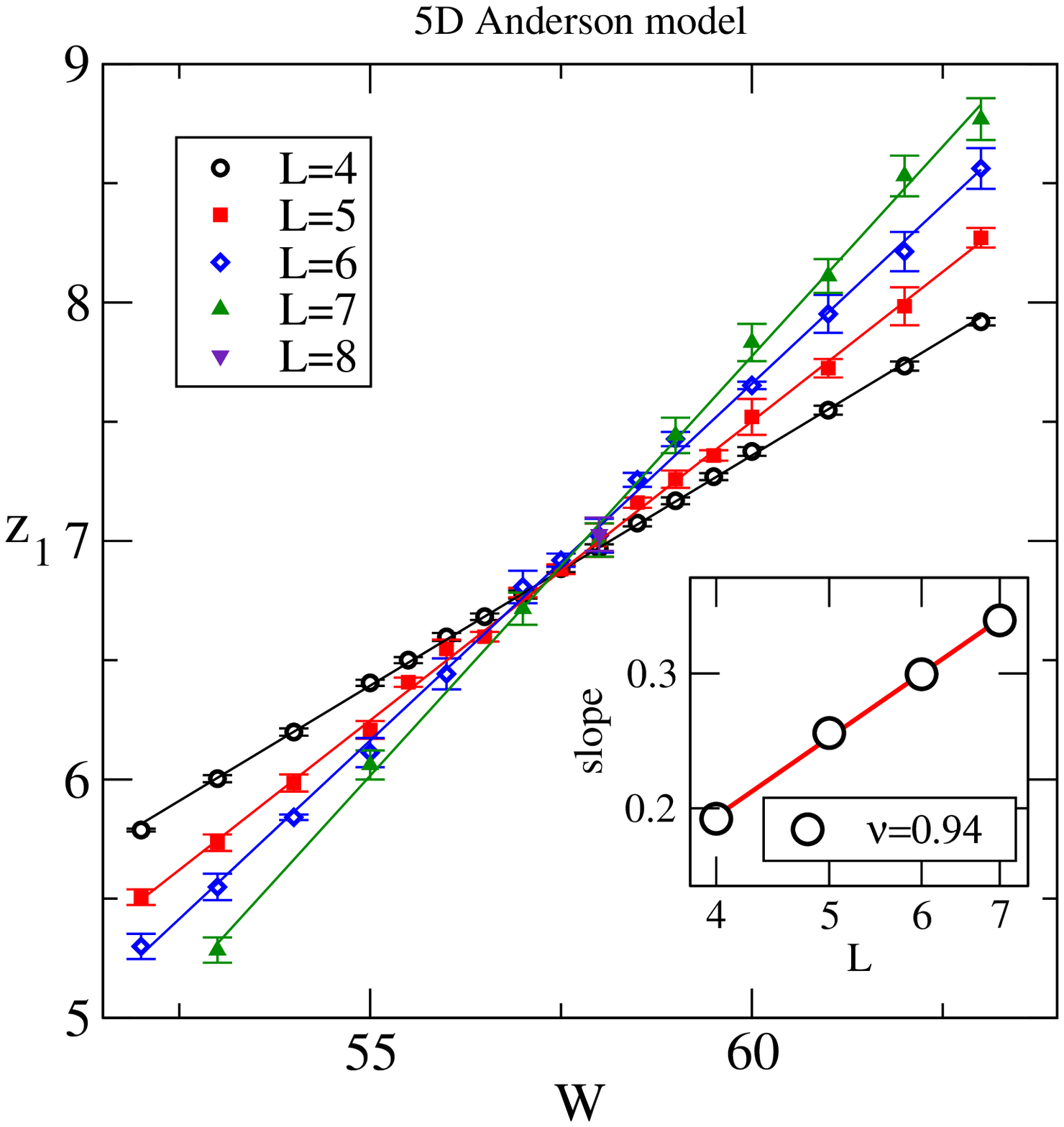}
\end{center}
\caption{The first Lyapunov exponent, $z_1$, 
as a function of disorder, $W$, for various $L$
for the 4D (left)  and 5D (right) Anderson model.
The critical parameters are $W_c=34.3\pm 0.2$, $\nu=1.12\pm 0.05$ for 4D
and  $W_c=57.26\pm 0.2$, $\nu=0.93\pm 0.05$ for 5D.
Inset shows the $L$-dependence of $z_1^{(1)}$, defined by Eq. (\ref{z1-xx}).
Power fit $z_1^{(1)}$ \textsl{vs.} $L$, given by Eq. (\ref{z1-xx2}) 
determines the critical exponent, $\nu$.
}
\label{4D}
\end{figure}

The numerical data for the critical exponent in higher dimensions
does not agree with the self-consistent
theory \cite{Woelfle}
which predicts that
\be\label{peter}
\nu(d)=\left\{
\begin{array}{ll}
(d-2)^{-1}   &  2<d\le 4\\
1/2          &  d>4.
\end{array}
\right.
\ee
For instance, for the 3D systems,  Eq. (\ref{peter}) predicts $\nu=1$,
which clearly disagrees with the numerical result, $\nu=1.57$.

In order to get 
insight into the dimension dependence of the critical exponents, the finite size
scaling analysis 
for $d=4$ was performed in Refs. \cite{Henneke,SG,Travenec},
and for $d=5$  in the present paper. 
Figure  \ref{4D}  presents numerical data for $z_1$  obtained for the quasi-1d systems
$L^3\times L_z$ and $L^4\times L_z$. Although only the data for small $L$ can be 
calculated, obtained results confirm the existence of the critical points
in both systems. The simple scaling analysis, given by Eqs. (\ref{z1-xx}-\ref{z1-xx3})
was used to estimate the position of the critical points and of the critical exponents.
For the 4D Anderson model, we find
\be
\nu_{\rm 4D}=1.12\pm 0.05.
\ee
This result agrees with
previously obtained data \cite{SG,Travenec}. To the best of our knowledge, there
have been  no published data for the critical exponent of the 5D Anderson model yet.
Our  estimation of the critical exponent is 
\be
\nu_{\rm 5D}= 0.94\pm 0.05.
\ee

\begin{figure}[b!]
\begin{center}
\includegraphics[clip,width=0.35\textheight]{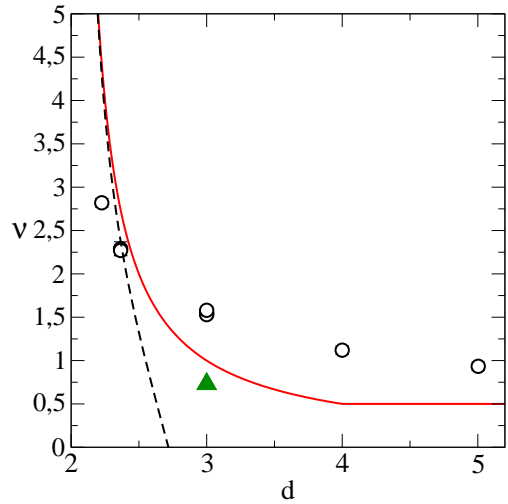}
\end{center}
\caption{Dimension dependence of the  critical exponent. Numerical data differ considerably
from the theoretical predictions, both for $d-2\ll 1$ and for $d\gg 2$.
The dashed line is the $\veps$-expansion, given by Eq. (\ref{nu-eps}). Triangle is
The estimation of Hikami, obtained by the Pade approximation of the $\veps$ 
expansion of the $\beta$ function,
$\nu_{\rm Hikami}\approx 0.73$. The solid line is the analytical prediction (\ref{peter})
}
\label{fig-nid}
\end{figure}

Figure \ref{fig-nid} summarizes all obtained 
numerical data for the critical exponent of the orthogonal Anderson model in $d$
dimensions and compares them with predictions
of analytical calculations. 
We conclude that   there is no agreement of numerical data with the theory,
neither  for small dimension, nor for $d>3$. 

\subsection{Theory \textsl{vs} (numerical) experiment}

Disagreement between the theoretical predictions and numerical data might lead
to the conclusion that the numerical scaling analysis is insufficient or even wrong
\cite{Suslov,Suslov-2,Kuzovkov}. It is not our aim to  discuss these objections here
(see, for instance, the comment to Ref. \cite{Kuzovkov}, published
in Ref. \cite{Wey}). We only concentrate on the discussion of
 advantages and disadvantages
of  numerical analysis.

The main objection against the numerical methods is that they are always restricted to
the systems of finite size. This limitation can be partially avoided by the scaling
analysis, Eq. (\ref{z1-s3}), and by including the irrelevant scaling variables, Eq. (\ref{z1-s34}).
Also, the accuracy   of obtained critical parameters can be estimated by
elimination of numerical data for small system size,
as discussed in Sect. \ref{sect:fse}.
Fortunately, the finite size effects seem to play a negligible role
in higher dimension.  The first estimation
of the critical exponent for the 3D Anderson model, $\nu\approx 1.5$,  obtained 
in pioneering work of MacKinnon and Kramer, \cite{McKK}, differs only in a few per cent
from the best today's estimation, $\nu=1.57$ \cite{SMO-1}.
Also, 
for  the 4D Anderson model,  the estimation 
$\nu=1.1\pm 0.1$ was obtained already in Ref. \cite{SG}, where really small
lattices, typically of the size $3^3\times L_z$ were analyzed. Nevertheless,
the analysis of larger samples, up to $7^3\times L_z$, \cite{Travenec}
brought no corrections to the critical  exponent.
As is shown in Fig. \ref{4D}, 
the additional  data for even large systems
have no influence to the critical exponent, originally estimated for  smaller systems.
We conclude therefore that the numerical simulation provides us with reliable data
in higher dimension. 

Our belief that the numerical data for the critical exponent are correct,
is supported also by results of the scaling analysis of various groups.
In the 3D  systems, 
various models were studied, 
isotropic and 
anisotropic \cite{zambetaki}, with 
diagonal or  off-diagonal disorder \cite{cain,schreiber}.
Scaling of various parameters was analyzed, inclusive
the smallest Lyapunov exponent, higher Lyapunov exponents,
conductance \cite{SMO-1,SMO-2}, 
conductivity \cite{croy}, level statistics \cite{ZK-e,tomi}
and inverse participation ratio \cite{brndiar}. All these works report the
critical exponent close to 1.5, with the accuracy which definitely excludes
the possibility that $\nu=1$.

As discussed in Sect.  \ref{sect:2d},
finite size effects become stronger in lower dimension, 
 The reason is that the mean free path, $\ell$ is larger (Fig. \ref{fig-mfp}). 
A typical problem caused by the finite size of the system
is shown  in Fig. \ref{2D-gW} which presents
the latest numerical data for the mean conductance, $\langle g\rangle$ of the 2D
Anderson model. Although we accept that there is no Anderson transition in $d=2$,
the numerical data seem to mimic the metallic behavior. For $W=1$, the
mean conductance $\langle g\rangle$  increases
when $L$ increases. One might argue that  there is a critical disorder,
$W_c\approx 2$. 
However, as discussed already in Sect. \ref{sect:beyond},
the above conclusion is not correct. 
To determine the character of the transport regime, the size dependence of the
conductance must be carefully analyzed.
The increase of the mean conductance
for $W=1$ is the manifestation of
the \textsl{ballistic} regime. 
Indeed,
the mean free path $\ell\approx 17$ for $W=1$  is comparable with the
size of the system when $L\le 100$. Also, 
the variance, $\textrm{var}~g$, is much larger than expected universal value,
typical for the metallic regime
(Fig. \ref{2d-ucf}).  

Also, the decrease of the mean conductance  with the system size for
disorder $W=3$ and $W=4$
\textsl{cannot} be associated with localization. Indeed, the same data,
plotted in Fig. \ref{2D_w3_g} in the logarithmic scale,  show that the decrease
of the conductance is not exponential, but logarithmic, and is
caused by  the weak localization. To see the exponential localization, one needs
much larger system size.

\smallskip

The big advantage of the  numerical simulation is that it can relatively easy  
analyze statistical properties of any quantity of interest.
No averaging is necessary in the course of calculations. All mean values can be
calculated ``from first principles''. This cannot be done analytically.
The analytical theory must solve the problem how to perform the 
average over the disorder.
Wrong averaging might lead immediately to wrong results \cite{Kuzovkov}.
In our opinion,  the discrepancy between the numerical data and results of analytical
theories is due to the inability of  analytical theories  to analyze completely
the statistical fluctuations in the critical regime.


\section{Two dimensional critical regimes}\label{sect:2d}

The critical regime in the 2D models  deserves a special attention. As discussed above, only
systems with symplectic symmetry exhibits  the metal insulator transition in  dimension $d=2$.
The 2D systems
with unitary symmetry posses, in the presence of strong magnetic field,
the  critical energies, $E_c$ where the localization length diverges.

On first sight it seems that the 2D systems  can be easier simulated numerically since 
the lower dimension of the system allows one to calculate the conductance for much larger samples.
However, this advantage is ``compensated'' by much stronger finite size effects. 

Besides the calculation of the critical exponents and  critical conductance 
distribution, the 2D critical regime is suitable for the verification of  the general relation between the
conductance and conductivity,
\be\label{rovna}
\langle g\rangle =\sigma,
\ee
given by Eq. (\ref{g-vodivost}).
Contrary to the orthogonal 2D systems, where Eq. (\ref{rovna}) holds
only in the diffusive regime, i.e. when the size of the system is smaller than the
localization length, $L\ll \lambda$, Eq. (\ref{rovna}) holds also in the limit of
infinite system size at the critical point of the unitary and symplectic  models.

Note, equation (\ref{rovna})  compares two different  quantities. The conductance, $g$, is given,
by definition, by the transmission properties of the disordered sample at zero temperature.
The value of the conductance depends on the actual distribution of disorder inside the sample.
Owing to the quantum character of electron propagation,  the conductance is not a self-averaged
quantity. 

On the other hand, the \textsl{conductivity}, $\sigma$, is a material parameter. It characterizes the
transport properties of an \textsl{infinite} system. When calculated for the system of \textsl{finite size},
$L$,~
\footnote{Numerical algorithm for the calculation of the 
conductance is described in Ref. \cite{AMcK-2,croy} and in Ref. \cite{Schweitzer} for the case of
critical quantum Hall regime.}, $\sigma$ fluctuates around the mean value, but the fluctuations decrease
when $L$ increases.  Contrary to the conductance, the  conductivity, $\sigma$ \textsl{is} a self-averaged 
quantity. 

For the critical 2D regimes,
Falko and Efetov \cite{FE} derived 
the following relation
between the fractal dimensions of the critical wave function,  $d_q$, and the critical 
conductivity, $\sigma$, derived the relation
\be\label{falko}
d_q=2-\ds{\frac{q}{\beta\pi\sigma (h/e^2)}}.
\ee
Here, $\beta=1$, 2 and 4 determines the symmetry 
of the system\footnote{Note, that $\beta=1/2$, 1 and 2 is
used in original papers, which adds an additional factor of 2 in the
 denominator on the r.h.s. of Eq. (\ref{falko}).
Also, note that factor of 2 for two orientation of electron spin is not 
included in the definition of the conductivity,  $\sigma$.}.
Eq. (\ref{falko}) holds for small $q$, when terms proportional to higher powers of $q$ can be neglected.
Since both the wave functions and the conductance can be calculated numerically in the critical point,
we can verify the relation (\ref{falko}) by direct numerical simulations. 
It will be shown in next two Sections
that indeed 
both  Eqs. (\ref{rovna}) and (\ref{falko}) are satisfied within the accuracy of numerical data.

\subsection{Symplectic models}\label{sect:ando}

The difficulty of the analysis of the 2D symplectic models  is manifested by a wide variety of values 
of the critical exponent, $2\le \nu\le 2.88$,  reported in the literature within the last 15 years 
\cite{Fastenrath,ludwig-isa,EZ,ASD}.  These discrepancies are due to the strong finite size effects.
Recently,  the finite size scaling of the smallest  Lyapunov exponent 
on the SU(2) model, \cite{ASD} provided the following estimate of the critical exponent,
\be
 \nu=2.75\pm 0.01.
\ee
This value can be considered as the  most accurate estimation of the critical exponent.
The analysis of the scaling
of the mean conductance for the 2D Ando model, \cite{Ludwig-1} led to the 
similar value, 
\be
\nu\approx 2.80\pm 0.04.
\ee
The scaling behavior of the mean conductance, $\langle g\rangle$,
is shown in Fig. \ref{fig-ando-scaling}.
The critical conductance distribution for the 2D Ando model is shown in Fig. \ref{2D-ando}.
Since the mean conductance, 
\be
\langle g\rangle_c\approx 0.71,
\ee
is close to 1, the distribution
possesses all characteristic properties of the critical distribution for the 3D Anderson model
(Fig. \ref{pcg-ando}).
Comparison of the critical distribution, calculated for the periodic and  hard wall boundary
conditions is shown in Fig. \ref{pcg-boundary}.

\begin{figure}[b!]
\bc
\includegraphics[clip,height=0.25\textheight]{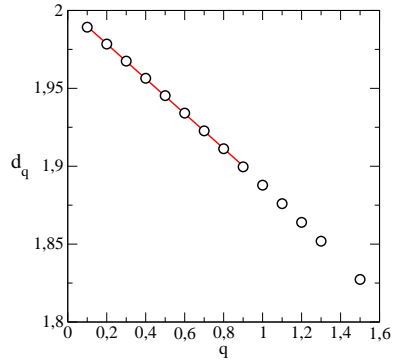}
\ec
\caption{The fractal dimension, $d_q$, as a function of $q$, calculated for the 2D Ando model.
For $q\le 1$,
the fractal dimension, $d_q$, is a linear function of
$q$, in agreement with Eq.  (\ref{falko}). The solid line is the linear fit with the slope  
 $0.11202$ which determines  the critical conductance, $\sigma_c=0.71$, and which agrees with the
estimation of the critical conductivity, $\langle g\rangle_c=0.71$ \cite{Ludwig-2}.
}
\label{dqqq}
\end{figure}

The numerical data for the conductance, together with 
Eq. (\ref{rovna}) enables us  to verify the relation (\ref{falko}) 
between the conductance and the fractal dimensions, $d_q$. To do so, 
the wave function of the 2D Ando model was calculated numerically for the disordered
 sample of size $260\times 260$. Then, the sample was divided into the small squares $\Omega_a$
of size $L_0\times L_0$, and the quantity 
\be\label{ls-1}
p_q(L_0)=\sum_a\left\{\sum_{\vec{r}\in\Omega_a} |\Psi_n(\vec{r})|^2\right\}^q
\ee
was  calculated for the $\ns=10$ different realizations of the disorder.
The wave function $\Psi_n$ is the eigenfunction of the disordered Ando Hamiltonian,
corresponding to the eigenvalue $E_n$, closest to the band center, $E=0$.
Since the wave function is expected to posses the multifractal spatial structure,
the $L_0$-dependence of $p_q(L_0)$ is determined  by the  fractal dimensions, $d_q$,
\be\label{ls-2}
p_q(L_0)\sim L_0^{-(q-1)d_q}.
\ee
Numerically calculated fractal dimensions, $d_q$ are plotted in Fig. \ref{dqqq} as a 
function of $q$. The results confirm that $d_q$ decreases linearly with $q$, for $q\le 1$.
The slope, given by Eq. (\ref{falko}), determines the critical conductivity, $\sigma_c=0.71 e^2/h$,
which is exactly equal to the critical conductance, $\langle g\rangle_c$.

\subsection{Critical quantum Hall regime}\label{sect:QHE}

The 2D disordered system in a strong magnetic field
possesses, inside each Landau band, the critical energy, $E_c$
shown schematically in the left panel of Fig. \ref{qhe-land}. 
When the Fermi energy crosses the energy $E_c$, the transmission 
from one Hall plateau to  another one appears
\cite{Hall,hansen}. 
The existence of the critical energy, $E_c$, was numerically proved in Refs. \cite{Huck}.
The correlation length, $\xi$, diverges on both sides of the critical energy as
\be
\xi(E)\propto |E-E_c|^{-\nu},
\ee
with the   critical index, $\nu\approx 2.33$ \cite{Huck,KOK}.

The critical quantum Hall regime
can be studied numerically within  the Hamiltonian,  (\ref{ham-peierls}) with the
Peierls phase, $\phi=B(ea)^2/\hbar$. Since we are interested only in the
critical parameters, the  periodic boundary conditions are  used in the direction
perpendicular to the propagation. Then,
the Peierls phase, $\phi$, must fulfill the relation $\phi L=2\pi n$, 
where $L$ is the size of the lattice
in the transversal direction, and $n$ is an integer. 

The numerical analysis of the critical regime is difficult, since the disorder
$W$ must be small in order to keep the Landau levels separated from each other. Then,
however,
the mean free path, $\ell$ is large. This problem can be avoided 
with he use of the spatially
correlated disorder \cite{janssen}.

\begin{figure}[t!]
\bc
\includegraphics[clip,height=0.1\textheight,width=0.2\textheight]{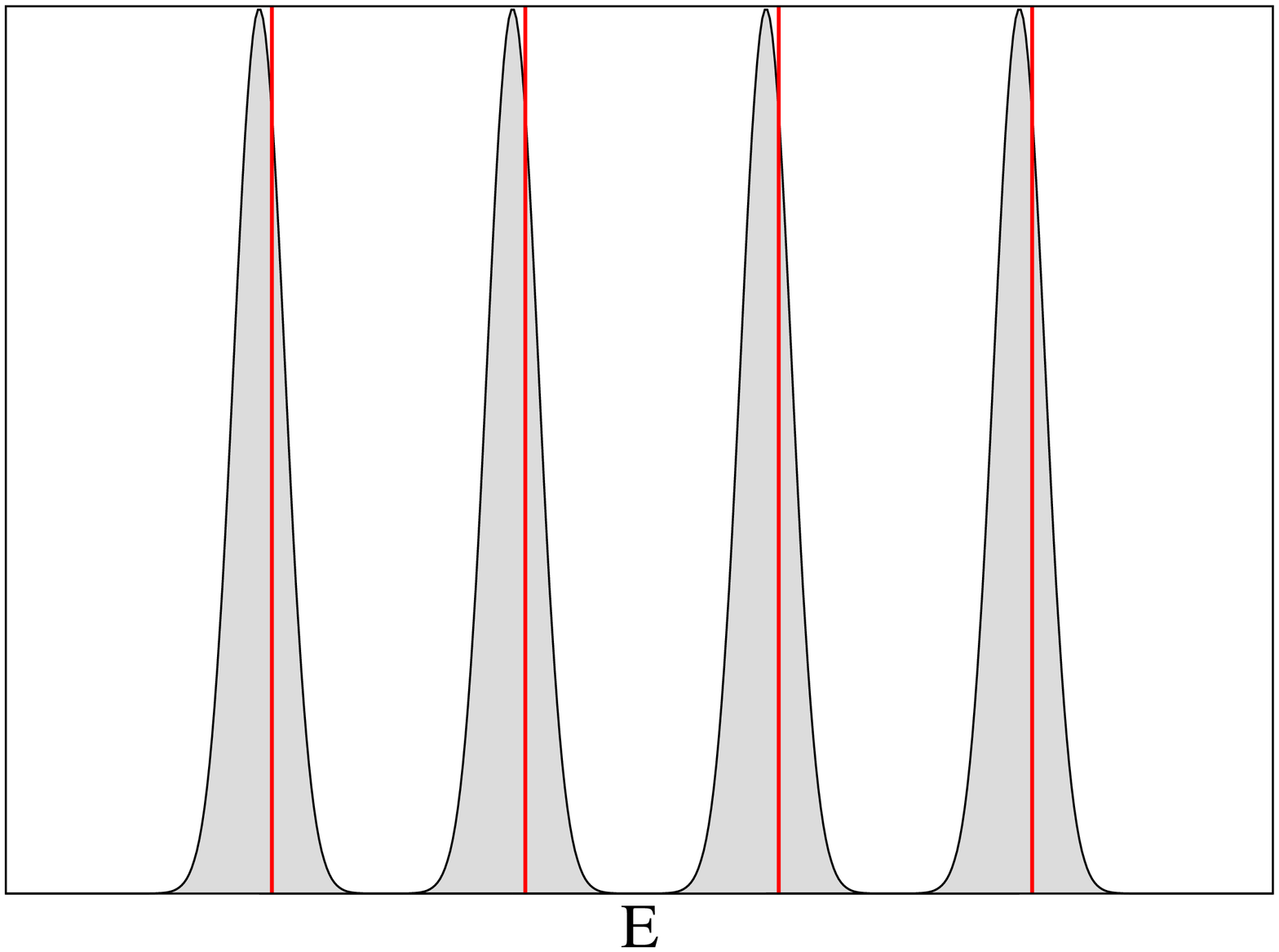}~~~
\includegraphics[clip,height=0.25\textheight]{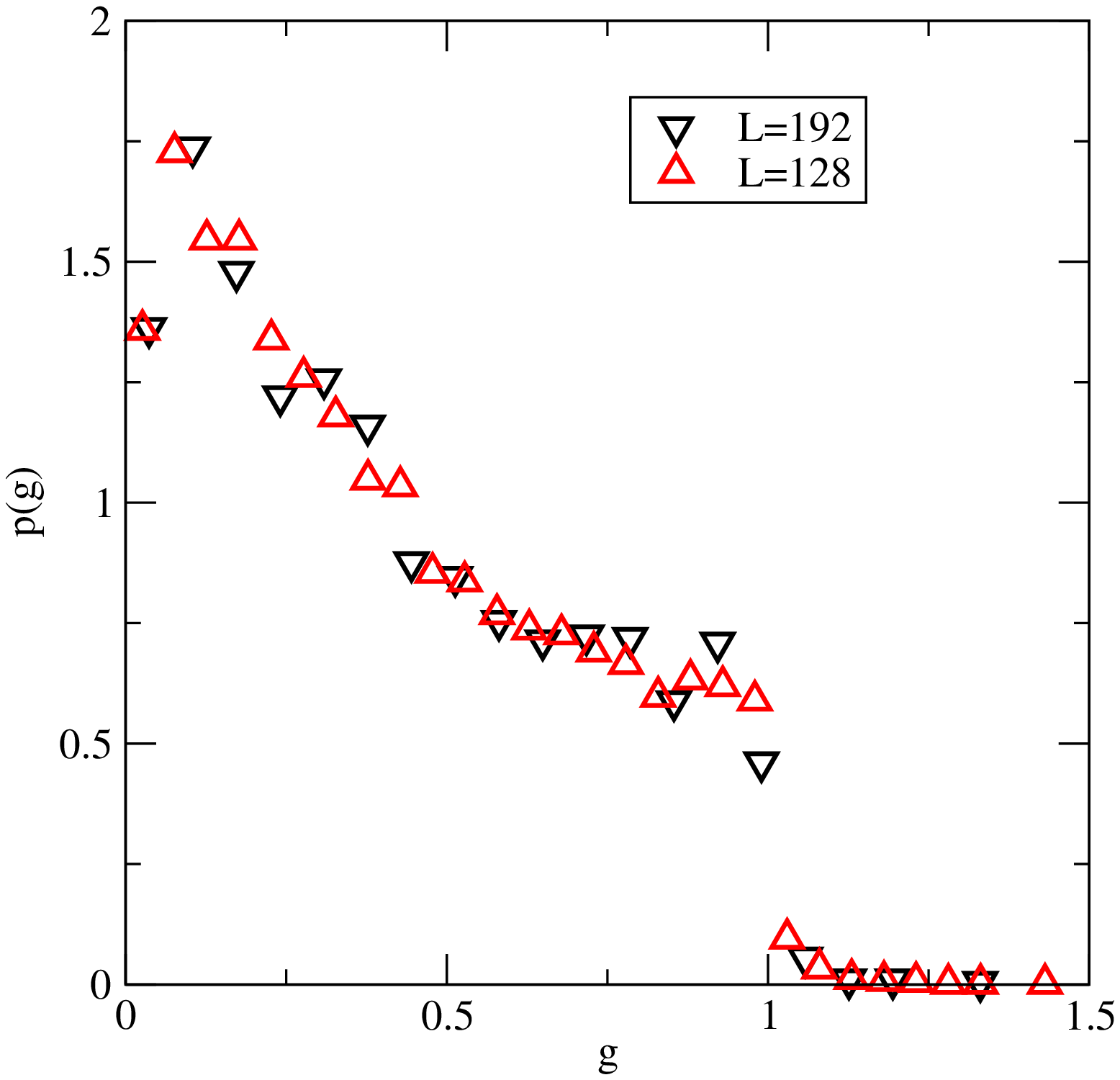}
\ec
\caption{Left: The density of states of the weakly disordered 2D system in a strong magnetic field.
The energy spectrum consists of the Landau bands. Inside each Landau band there
is a critical energy, $E_c$, where the localization length diverges.
Note, the position of the critical energy depends on the strength of the disorder.
Right:
The critical conductance distribution for the first Landau band for $\phi=2\pi/8a$.
The disorder is $W=1.4$ and the critical energy, $E_c=-3.29$.
}
\label{qhe-land}
\end{figure}

In Ref. \cite{Ludwig-2}, the sample averaged conductance, $\langle g\rangle$, and the conductivity,
$\sigma$, were calculated for the first and the second Landau band. 
To eliminate the   finite size effects, the  random disorder was spatially correlated.
Numerical data was fitted to the formulas
\be\label{s-y1}
\langle g(L)\rangle =g_c-g_0\left(\frac{L_0}{L}\right)^y
\ee
and
\be\label{s-y2}
\sigma(L)=\sigma_c-\sigma_0\left(\frac{L_0}{L}\right)^{y'}.
\ee
where $y\approx y'$ are irrelevant scaling exponents.

Numerical analysis, performed for various  strength of the disorder proved that
both the mean conductivity and conductance 
converge to the same  values listed in Table \ref{tab-qhe}.
This data proves the relation (\ref{rovna}) for the critical quantum Hall regime
and they also shows that the critical  conductance (or conductivity) is universal, independent
of the microscopic model and on the Landau level.

Critical values listed in Table \ref{tab-qhe} 
are significantly larger than the commonly accepted value, $0.5 e^2/h$, found in
Ref. \cite{kivelson}, and confirmed in previous  numerical simulations \cite{wang}.
However, they are  in agreement with numerical data obtained on the 
Chalker Coddington model \cite{2D,OSK}. 
Critical conductance, $\sigma_c\approx 0.61 e^2/h$,
is also consistent with the recently calculated
fractal dimension, $d_1=1.739$, and with the relation  (\ref{falko}),
\be
\sigma=\ds{\frac{1}{2\pi(2-d_1)}\frac{e^2}{h}}.
\ee
The difference between the theoretical and numerical data is probably due to the
spatial inhomogeneity of the electron distribution, which was  neglected in the 
analytical calculations.

\begin{table}[b]
\bc
\begin{tabular}{|l|l|l|}
\hline
critical conductivity &  $\sigma_c$   &  $0.58\pm 0.02$  \\
critical conductance &       &  \\
1st. Landau band     &   $g_{c1}$    &  $0.60\pm 0.02$ \\
2nd Landau band      &   $g_{c2}$    &  $0.61\pm 0.03$\\
\hline
\end{tabular}
\ec
\caption{The critical conductivity, $\sigma_c$,  and the critical conductance,
$\langle g\rangle$, (in units of $e^2/h$) for the two lowest Landau bands 
\cite{Ludwig-2}. The data was obtained by fit of the numerical data to Eqs. (\ref{s-y1})
and (\ref{s-y2}) with $y\approx y'\approx -0.4$
}
\label{tab-qhe}
\end{table}


\section{Possible theoretical description of the localized regime}\label{sect:poss}

In this Section, we present two possible descriptions of the localized regime. Both of them are based
on the generalization of the theoretical methods developed for the analysis of the transport in diffusive regime.
In Sect. \ref{gen}, we discuss the possibility to generalize the DMPK equation,
\cite{MK}
and in Sect. \ref{sect:rmf}, a simple generalization of random matrix theory is proposed \cite{M-1995}.

\subsection{Generalized  DMPK equation}\label{gen}

In Appendix \ref{app-dmpk} we present the DMPK equation, which describes 
successfully the transport properties of weakly disordered quasi-1d systems. 
The theory contains only one parameter - the mean free path $\ell$, which measures the strength
of the disorder.

The DMPK equation was derived under the assumption of the homogeneity of
the matrices $u$ and $v$ which parametrize the transfer matrix (see Appendix \ref{sect:param}
for details). 
Physically, homogeneity of matrices $u$ and $v$ means homogeneous distribution of the
electron on the opposite side of the sample. This is possible only if the electron has  many
paths to travel from one side of the sample to  another side.
Mathematically,
this requirement is  expressed by Eq. (\ref{dmpk-a2}), which can be written 
in the form
\be\label{gen-kab}
K_{ab}=\left\langle \sum_c |u_{ac}|^2|u_{bc}|^2\right\rangle=\ds{\frac{1+\delta_{ab}}{N+1}}.
\ee
Here, $N$ is the number of channels.
Clearly, this assumption is fulfilled only in the limit of weak disorder, when the electron
can choose, on its travel through the sample, 
many equivalent  paths. This is not true when the disorder is strong
and the electron hardly finds a  single trajectory propagating from one side  of the 
sample to opposite   one (Fig. \ref{fig-dmpk}).

Recently, Muttalib and Klauder \cite{MK}
proposed the generalization of the DMPK equation.   Without any restriction to
the value of  the matrix elements
$K_{ab}$, they  generalized the DMPK equation into the form
\begin{equation}\label{three}
\frac{\partial p_{L_z}(\lambda)}{\partial (L_z/l)}
=\frac{1}{{J}}\sum_a^N
\frac{\partial}{\partial\lambda_a}\left[\lambda_a(1+\lambda_a){K_{aa}}
{J}\frac{\partial p}{\partial \lambda_a}\right]
\end{equation}
with the Jacobian
\begin{equation}\label{gen-j}
{J}\equiv\prod_{a<b}^N|\lambda_a-\lambda_b|^{{\gamma_{ab}}},~~~~~~
{\gamma_{ab}}\equiv\frac{{2K_{ab}}}{{K_{aa}}}.
\end{equation}
In Eq. (\ref{three}), the symmetry parameter $\beta=1$.

For the weak  disorder, one can  substitute for $K_{ab}$ from 
Eq. (\ref{gen-kab}) and obtain
that Eq. (\ref{three}) reduces to the ``classical'' DMPK equation.
For strong disorder, the parameters 
$K_{aa}$ and $\gamma_{ab}$  represent the   additional free parameters
which must be estimated from numerical experiment.

To estimate values of $\gamma_{ab}$ 
in the localized regime (regime with strong disorder),
we remind the reader that the probability distribution
of the difference, $\delta_{a}=x_{a+1}-x_a$ is similar to the Poisson
distribution in the localized regime (Fig. \ref{pd}). 
Therefore, it is natural to assume that
the repulsion between the two levels in the Jacobian (\ref{gen-j}) is weaker than
in the diffusive regime. Consequently, 
we  assume that $\gamma_{ab}\to 0$ in the case of strong 
disorder. A similar conclusion was derived also in Ref. \cite{M-1995}.

\begin{figure}[t!]
\begin{center}
\psfrag{LK}{$\!\!\!\!\!\!\! LK_{11}$}
\includegraphics[clip,width=0.32\textheight]{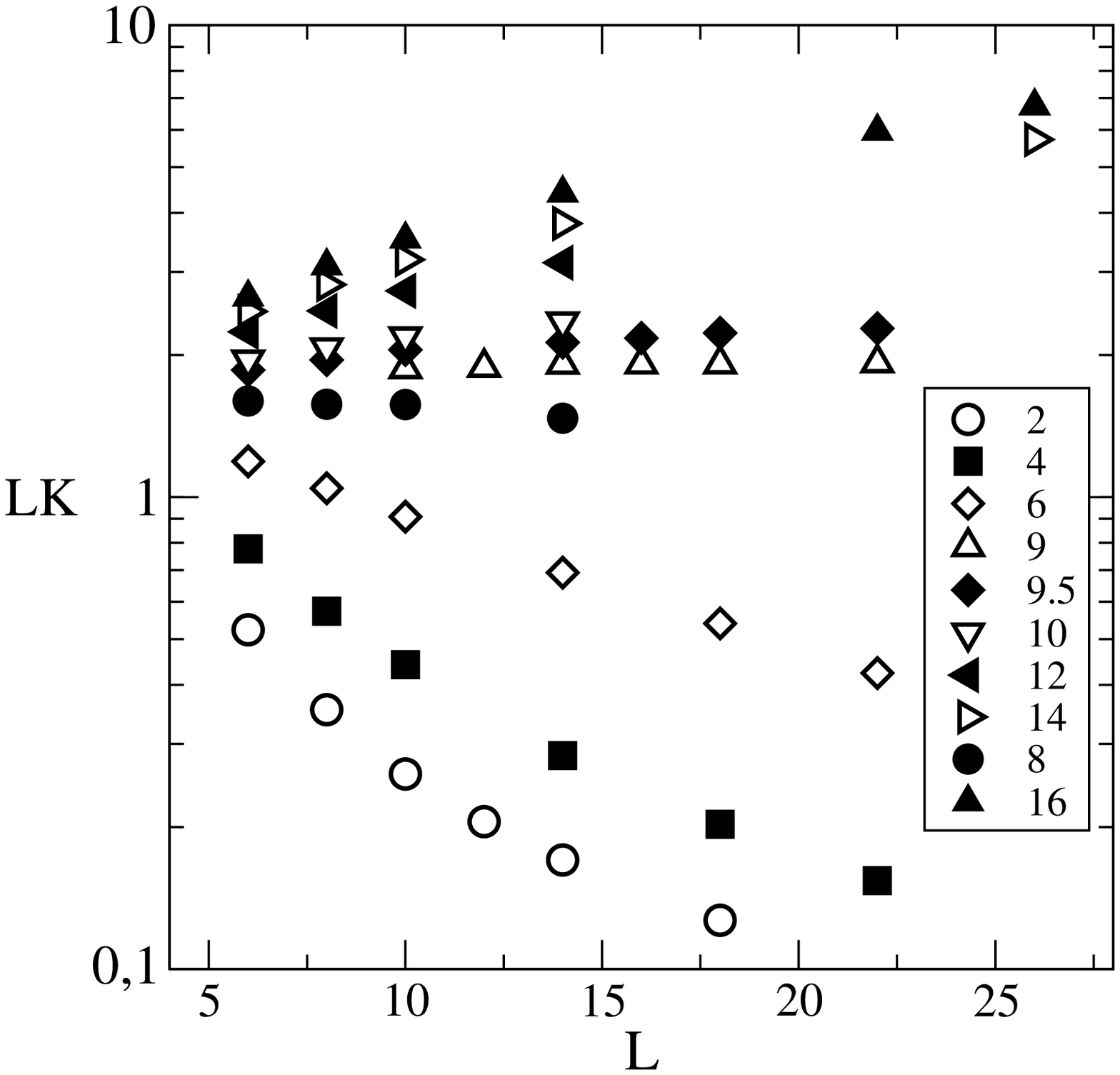}~~
\includegraphics[clip,width=0.32\textheight]{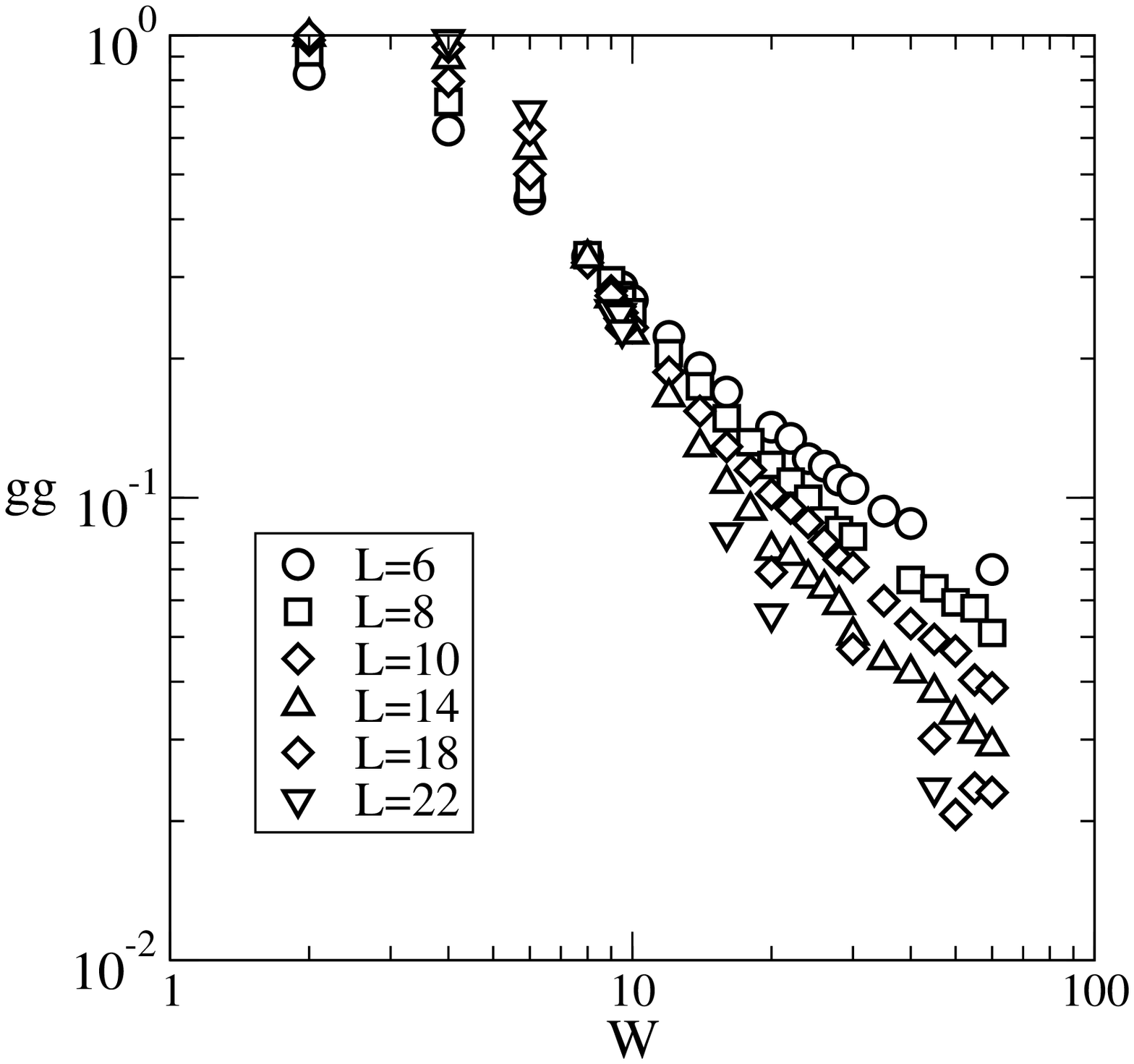}
\ec
\caption{Left: The disorder dependence of $LK_{11}$ for the 3D anisotropic
Anderson model, given by Eq. (\ref{k11}) with $t=0.4$.
The symbols correspond to the different disorder. The data confirm that
$LK_{11}$  decreases (increases) with $L$ for $W<W_c$ ($W>W_c$), respectively.
Here, the critical disorder $W_c\approx 9.5$. Note, $LK_{11}$ does not  depend on the
system size when $W=W_c$.
The right panel   shows that $\gamma_{12}\to 1$ when $W<W_c$ and $L\to\infty$,
but $\gamma_{12}$ decreases with $L$ when disorder $W>W_c$ (insulating regime).
At the critical point, $\gamma_{12}\approx 0.25$ does not depend on the system size
\cite{MMW}.
}
\label{K3}
\ef

\begin{table}[b!]
\bc
\begin{tabular}{|l|c|c|}
\hline
	       &     $L\times {K_{11}}$    &   {$\gamma_{12}$}\\
\hline
$W\ll {W_c}$        &        $L^{-1}$ &               1   \\
${W_c}$          &     {const}    &   {const} \\ 
$W\gg {W_c}$  & $\sim L/\xi$          &     $L^{-1}$\\ 
\hline
\end{tabular}
\end{center}
\caption{The size dependence  of parameters $LK_{11}$ and $\gamma_{12}$ 
in the metallic, localized and critical regime.
}
\label{table-K}
\end{table}

The second assumption made in Ref. \cite{MK} was  that 
$K_{aa}\gg K_{ab}$ for $ a\ne b$. 
In particular, in the localized regime, one can assume that
\be\label{gen-aa}
K_{aa}\propto L^0.
\ee
To understand the physical meaning of (\ref{gen-aa}), note that
\be\label{k11}
K_{aa}=\left\langle \sum_c |u_{ac}|^4\right\rangle.
\ee
is nothing but the inverse participation ratio for the transversal wave function 
on the opposite side of the sample.  In analogy with the properties
of $I_q$, discussed in Sect. \ref{wavefunction}, we conclude that the
condition (\ref{gen-aa}) reflects the non-homogeneity of the distribution 
of the electron on the opposite side of the sample, which follows directly from the
existence of only the single possible path through the sample (Fig. \ref{fig-dmpk}).

Conjecture (\ref{gen-aa}) was numerically tested in Refs. \cite{MMWK,MMW}
by the direct numerical calculation of the matrix $K_{ab}$.
The numerical method of calculation of the conductance, described in Appendix \ref{app-b2},
can be easily generalized for calculation of the eigenvectors $u$.
To avoid the problem with evanescent modes, the calculations were performed for
the anisotropic 3D Anderson model, defined by Eq. (\ref{hama}) with $t=0.4$.
This model exhibits the metal-insulator transition for critical disorder $W_c\approx 9.5$.

The numerical analysis confirmed that 
the parameters $K$ indeed depend on the disorder. In Fig.  \ref{K3}
we see that the size dependence of parameters $LK_{11}$
and $\gamma_{12}$ differs depending on whether the system is
in the metallic, localized or critical regime. Note,
since $\lambda_1<\lambda_2<\dots$,
and the matrix $t^\dag t$  has eigenvalues $(1+\lambda_a)^{-1}$,
the parameter
$K_{11}$  contains information about the spatial structure of the 
first eigenvector, $u_1$,  which corresponds to the largest eigenvalue, 
and $\gamma_{12}$ is a measure of the overlap of eigenvectors related to the
two largest eigenvalues of $t^\dag t$.

In the metallic regime, $LK_{11}\propto L^{-1}$, and  $\gamma_{12}$ converges to 1,
as was assumed in the derivation of the ``classical'' DMPK equation. However,
in the localized regime, $LK_{11}$  increases with the system size,
which indicates that indeed $K_{11}$ is constant, in agreement with Eq. (\ref{gen-aa}). 
Also, $\gamma_{12}$ decreases when $W>W_c$
being $\propto L^{-1}$ in the limit of large system size.

At the critical point, both parameters, $LK_{11}$ and $\gamma_{12}$,
converge to the size independent constants,
indicating that they could be used as the  order parameters for the calculation of the
critical parameters of the model by scaling theory.
The size dependence of $LK_{11}$ and $\gamma_{12}$ is summarized in Table \ref{table-K}.

\begin{figure}[t!]
\begin{center}
\includegraphics[clip,width=0.32\textheight]{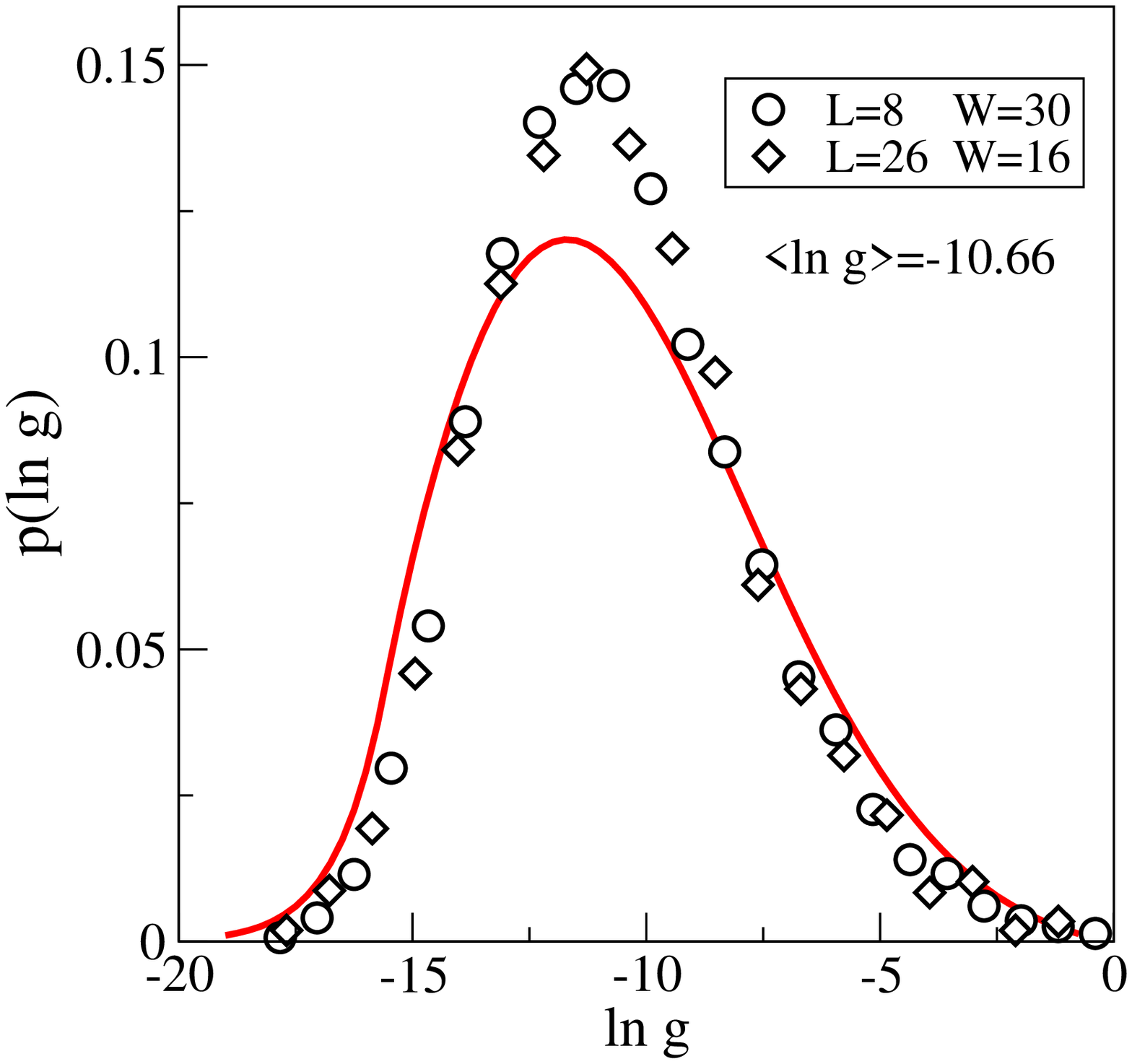}~~
\end{center}
\caption{%
Probability distribution of a strongly disordered three dimensional
system.   Symbols show data form numerical simulations, solid line
is the  solution of the generalized DMPK equation which gives the same
value of $\langle\ln g\rangle$ and with $\gamma_{12}=\Gamma/2$. Dotted line
is the distribution obtained for the same value of $\Gamma$ but with 
$\gamma_{12}=1$.
}
\label{mmw-24}
\ef

With known size dependence of $K_{11}$ and $\gamma_{12}$, we need  to estimate also
other parameters, $K_{ab}$. Detailed numerical analysis, performed in Ref. \cite{MMW}
confirmed that  these quantities can be expressed as a simple functions
of indices $a$ and $b$. It is also reasonable to assume that $K_{aa}=K_{11}$
and $\gamma_{ab}\approx \gamma_{12}$. This reduces the number of free parameters
to two. The first one,
\be
\Gamma=\ds{\frac{\ell}{L_zK_{11}}}
\ee
measures the strength of the disorder, and the second one,
$\gamma_{12}$, 
influences the mutual correlation of channels in the generalized DMPK equation.
In the limit of
\be
\gamma_{12}\sim \ds{\frac{\lambda}{L}}\ll 1,
\ee
the probability distribution $p(g)$ can be 
obtained analytically  by solving the  generalized DMPK equation, given by Eq. (\ref{three}),
by methods developed in Refs. \cite{MuttGW,MuttGW1}
The obtained conductance distribution is shown in Fig. \ref{mmw-24}
and   compared with the numerically calculated distribution  $p(\ln g)$.
Parameter $\Gamma$  was chosen such that
$\langle\ln g\rangle$, equals to the numerically obtained value. 
We see that the analytical
model reproduces qualitatively correctly 
the numerical data. The quantitative difference is probably due
to oversimplification of the model, which neglects the differences between $K_{ab}$
for higher channels.

The  generalized DMPK equation represents the most promising step toward the 
analytical description of the localized regime. 
However, the main  assumption which allows the analytical solution is  that
$\gamma_{12}$ is very small. This is not true at the critical point, as can be seen
in Fig. \ref{K3}.
Therefore, it is not known at present, how accurately the generalized DMPK equation 
describes the critical regime.


\subsection{Random matrix model of the Anderson transition}\label{sect:rmf}

We have shown  in Appendix \ref{app:rmt} that in the diffusive regime,
the probability distribution for the eigenvalues $\lambda$ of the 
transmission matrix can be relatively easily obtained  from the assumption that
the matrix $(t^\dag t)^{-1}$ belongs to the orthogonal class of random matrices.
The form of the distribution was determined
by the additional constrain that the density $\sigma(x)$ of parameters
$x$,
\be
\sigma(x)=\langle \sum_a \delta(x-x_a)\rangle,
\ee
is constant for $x<L_z/\ell$.   This constraint follows directly from the linear
dependence, $\langle x_1\rangle\propto a$, shown in Fig. \ref{fig-xa}. 

Since the spectrum of $x$ exhibits the universal properties also at the critical point
and in the localized regime, it is tempting to try to generalize the random matrix
analysis also for transport beyond the diffusive regime. This expectations are 
inspired by the numerical data for $\sigma(x)$, shown in Fig. \ref{rmt}.

In Sects. \ref{sect:crit}  we found that contrary to the diffusive regime, 
the density $\sigma(x)$ is not constant in the critical regime. 
Since $\langle x_a\rangle^2\propto a$ at the critical point (Fig. \ref{3D-zi}),
the density  must be linear,
\be
\sigma_{\rm crit}(x)\propto x,
\ee
at least in the lower part of the spectra.
 Note, only this part of the spectra 
is relevant  for the description of the transport. The linearity of
the density is confirmed numerically in the left panel of Fig. \ref{rmt}.

\begin{figure}[t!]
\begin{center}
\psfrag{sx}{$\sigma(x)$}
\includegraphics[clip,width=0.32\textheight]{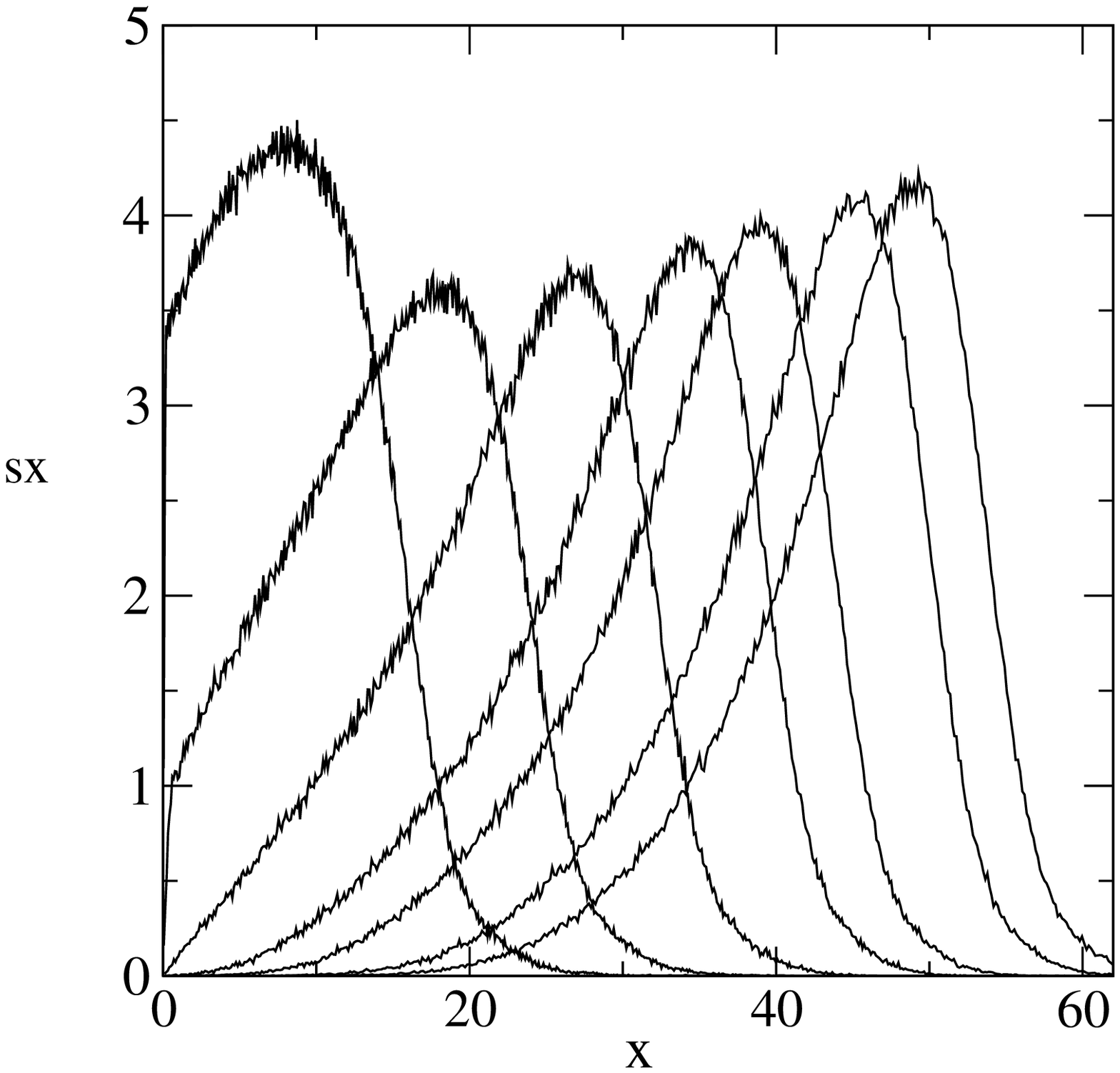}
\includegraphics[clip,width=0.32\textheight]{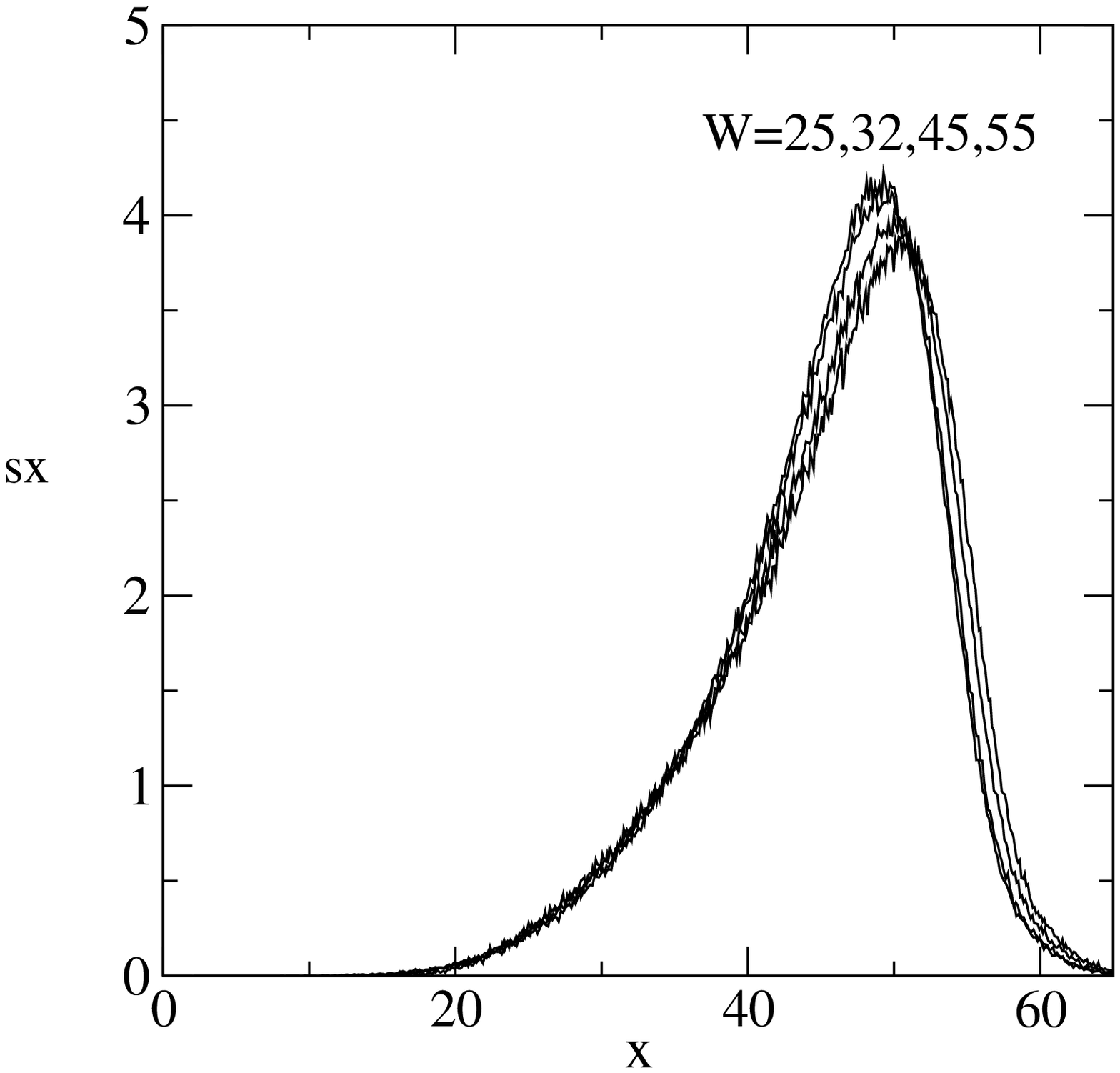}
\end{center}
\caption{%
Left: The density $\sigma(x)$ calculated numerically for the 3D Anderson model with disorder 
$W=6,~10$, $W=W_c=16.5$, and $W=25$ and 32 (from left to the right). Note, 
at the critical point, the density is linear for $x\le 10$.
Right: $\sigma(x)$  for the strongly localized regime with  disorder
$W=25$, 32, 45 and 55. To show the universality of the density,
the data are shifted on the horizontal axis by the difference
$\langle x_1(W)\rangle- \langle x_1(W=55)\rangle$ \cite{C-2}.
}
\label{rmt}
\end{figure}

\begin{figure}[b!]
\begin{center}
\psfrag{sx}{$\sigma(x)$}
\includegraphics[clip,width=0.35\textheight]{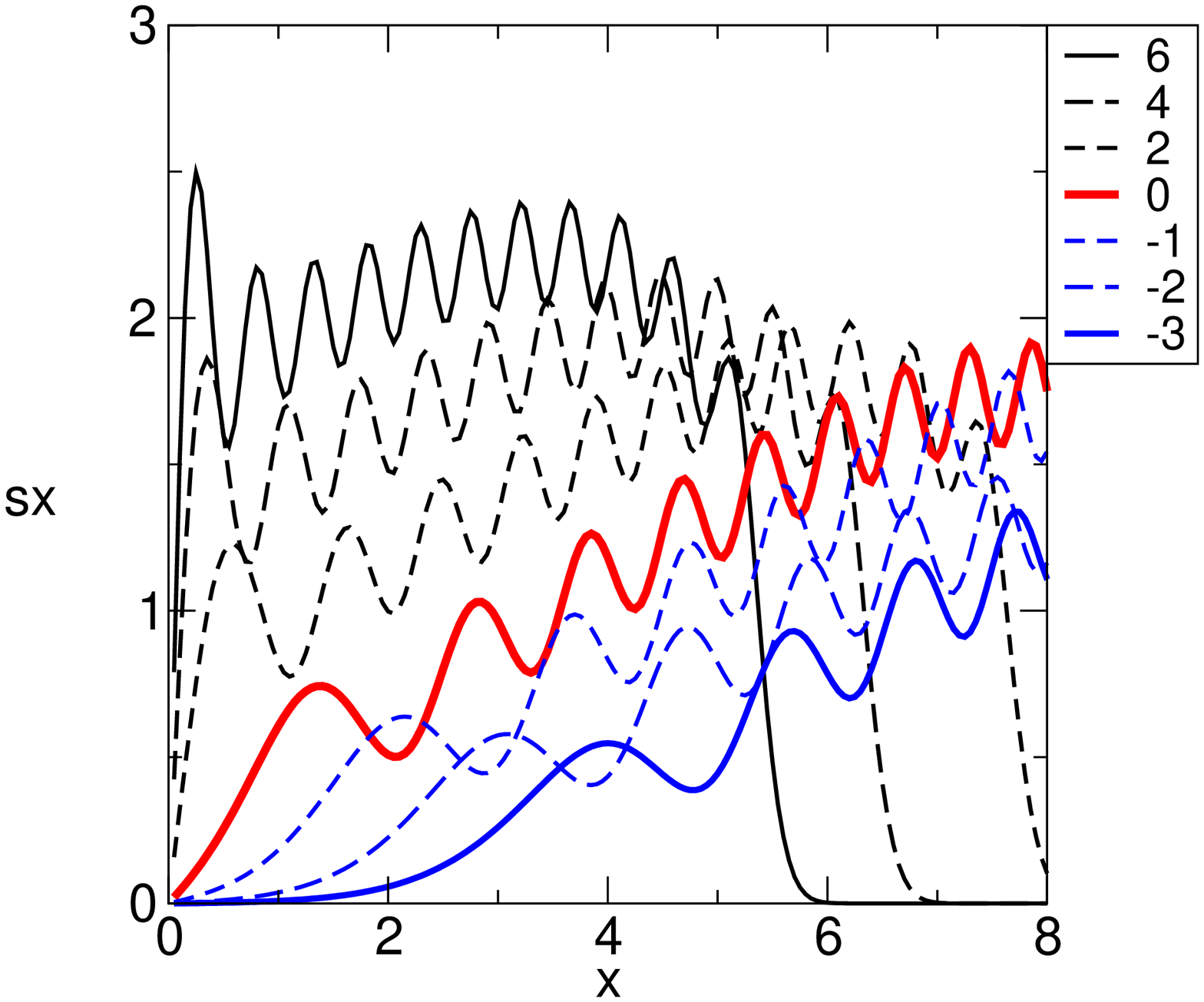}
\end{center}
\caption{The density, $\sigma(x)$, calculated from the random matrix model, given by Eq. (\ref{phen-1}).
Various  values of parameters $a$ are given in the legend \cite{C-2}.
}
\label{rmtt}
\end{figure}

The right panel of Fig. \ref{rmt} shows that
in the localized regime, the density $\sigma(x)$ is a function of the difference,
$x-2L/\lambda$. This follows directly from Eq. (\ref{DDD}).

It is therefore natural to 
postulate that  the density $\sigma(x)$ behaves as
\be\label{phen-1}
\sigma(x)= c(W)\times \left[x+a(W)\right],
\ee
where the function  $a(W,L)$ changes the sign at the critical point
\cite{M-1995}. 

The disorder and size dependence of functions $a$ and $c$ are not known.
The analysis of the size dependence of 
of parameters $x$ in the critical region \cite{M-1997}   confirmed  that
\be\label{phen-2}
 a(W,L)\sim \left\{
 \begin{array}{ll}
 +2L/\xi(W)  &  W\ll W_c\\
 0           &  W=W_c\\
 -2L/\xi(W)  &  W\gg W_c
 \end{array}
 \right.
 \ee
and that $c(W,L)$ is a size independent constant for $W=W_c$.
In Eq. (\ref{phen-2}), $\xi(W)$ is the correlation length introduced in Sect. \ref{scal-LE}.
In strongly localized regime, $\xi(W)$ equals to the localization length $\lambda$.

We can  verify that the  density $\sigma(x)$, given by Eqs. (\ref{phen-1})
and (\ref{phen-2}) reproduces numerical data for parameters $x$.
Indeed, inserting in Eq. (\ref{phen-1}) we find that in the metallic regime,
$\sigma(x)\sim \textrm{const}$ since $x\ll 2L/\xi(W)$ (we remind the reader
that $\langle x_a\rangle = L/(N\ell)a$ in the metallic regime). 
Since $\sigma= c x$ at the critical point, 
we immediately recover that $\langle x_a\rangle^2\propto a$. 
In the localized regime, $\sigma = c(x-2L/\xi)$, so that
the entire spectrum of parameters $x$ shifts by  $2L/\xi(W)$  for higher values of the
disorder $W$.

Using the  method explained in Appendix \ref{app:apl}, it was  found in Ref. \cite{M-1995} that 
the distribution of parameters $x$ at the critical point is given by the
universal distribution
Eq. (\ref{rmf-p1}) and (\ref{rmf-p3}), but with a cubic one particle
potential,
\be\label{phen-3x}
V_{\rm crit}=cx^3.
\ee
Then,  applying the methods of orthogonal polynomials, \cite{Mehta},
 the density $\sigma(x)$ can be calculated within the random matrix model.
 Results, shown in Fig. \ref{rmtt}, agree qualitatively with results of
 the numerical simulations, shown in the left panel of Fig. \ref{rmt}.

An interesting consequence of the expression (\ref{phen-1}) is the 
following scaling relation for higher Lyapunov exponents $z_i$:
\be\label{phen-8}
[z_i-a(W,L)]^2-[z_j-a(W,L)]^2=\frac{1}{c}(i-j),~~~~i,j\le L
\ee
which enables us to find the critical parameters of the  3D Anderson model
from numerical data for only one system size \cite{M-1997}. It also provides us with
numerical evidence
that the model (\ref{phen-2}), based on the change of the density at the critical point,
 is correct.

The simple form of the potential (\ref{phen-3x}) indicates that the statistical description
of the critical regime might be as simple as the analysis of the diffusive regime. However,
we are not aware of any microscopic model  which confirms the above phenomenological
model.


\section{Conclusion}

We have discussed in this paper the main transport properties of the  single electron 
disordered electronic systems. The quantum character of the electron propagation
is responsible for the wide variety of interesting transport phenomena, which we
demonstrated numerically.

In the limit of weak disorder, the weak localization  and anti-localization
corrections to  the conductance and universal conductance fluctuations were observed
and the conductance was found to exhibit the universal conductance fluctuations.
In the localized regime,   the exponential decrease of the
wave functions and the conductance with the system length has been  demonstrated and discussed.

One of the most important phenomena in the localization theory is the absence of the self averaging
of the conductance. In the metallic regime, the electron wave function is spread over the entire sample,
and is very sensitive to any change of the disorder configuration, as well as to the change of the 
boundary conditions. This remarkable property survives also in the limit of
an infinite system size. The sensitivity of the energy spectra 
to the change of the boundary conditions can be
 used as a measure of the electron localization, and defines the main parameter of the localization
theory, the conductance $g$.

The statistical properties of the conductance in the strongly localized regime are even more complicated. 
The transport is possible only by tunneling between localized centers, and the conductance decreases 
exponentially with the size of the sample. Still, a few samples exist in the statistical ensemble
which possess rather high conductance. These samples determine the mean conductance, which might be
in orders of magnitude larger than the typical (the most probable) conductance of the ensemble.
For a given sample, the conductance is also very sensitive to the distribution of
random impurities. Equivalently, a small change of the Fermi energy might cause a huge change in the conductance.

The critical regime of the metal-insulator transition has been
discussed in detail. First, we have analyzed the statistical
properties of the conductance, and we have 
presented numerical data for the critical conductance distribution.
Then, the scaling theory of localization is introduced and verified by numerical scaling analysis
of Lyapunov exponents and conductance distribution. 
In order to compare the  numerical data for the critical parameters
with predictions of the  analytical theories, the critical electronic transport on 
lattices with dimension close to the lower critical dimension, $d_c=2$ as well as with dimension
$d=4$ and $d=5$ 
was simulated.

We have tried to convince the reader that the numerical data, collected within the  last 25 years  
supports the validity of  the one parameter scaling theory  of localization.
We  believe that the numerical data 
provides us with the most reliable estimation of the  critical exponents in various 
disordered models. We hope that the disagreement between the  results of numerical simulations
and theory will motivate further theoretical research \cite{dobro}.

Finally,  we want to mention some other aspects of the single electron localization.
We have not discussed the transport in systems with correlated disorder.
The mobility edge in the 1D system with \textsl{correlated} disorder
has been found in Ref. \cite{krokhin} (see Ref. \cite{izrailev} for review). 
Special attention is deserved also for disordered systems with chiral symmetries
\cite{brouwer}. Also,
application of the concepts of the electron localization to the propagation
of the electromagnetic \cite{SEGC,Genack,asatryan,PRB2005}, and acoustic  \cite{seizmic} waves
in disordered  media opens a new field for studies of the localization and promise further
development of the localization theory.

\section{Acknowledgments}

I thank Martin Mo\v{s}ko for many valuable discussions which inspire writing of this
paper. This work was supported by grant VEGA  2/6069/26  and project APVV-51-003505.

\appendix


\section{Properties of the transfer matrix}\label{app-a}

We summarize here the most important properties of the transfer matrix. 
For details, we refer the reader to Refs. \cite{SlevinNagao,book}.

First, we derive the expression for the transfer matrix, given by Eq. (\ref{TM}).
From Eq. (\ref{one}) we obtain 
\be\label{app-tm-1}
\begin{array}{lll}
C &=& t^+A+r^-D\\
B&=&r^+A+t^-D
\end{array}
\ee
If we have $N$ incoming channels, then the amplitudes $A$, $B$, $C$ and  $D$ are the (complex) vectors 
of length $N$, and the transmission and reflection amplitudes are the matrices
of size $N\times N$.

We express $D$ from the second equation (\ref{app-tm-1}) as
\be
D=(t^-)^{-1}B-(t^-)^{-1}r^+A
\ee
and insert it in the first equation (\ref{app-a}).  We obtain that
the transfer matrix is given by Eq. (\ref{TM}),
\be\label{TM-q}
\textbf{T}=\left(\begin{array}{ll}
t^+-r^-(t^-)^{-1}r^+&
 r^-(t^-)^{-1} \\
 -(t^-)^{-1}r^+  & 
(t^-)^{-1} 
\end{array}\right).
\ee
Note, $t^-$ and $r^-$ are the transmission and reflection amplitudes of the wave 
coming from the \textsl{right} side of the sample, and
 $t^+$ and $r^+$ are the transmission and reflection amplitudes of the wave 
coming from the \textsl{left} side of the sample.

\subsection{The composition law}\label{app:comp}

\begin{figure}[t!]
\begin{center}
\psfrag{t12}{$t^+_{12}$}
\psfrag{r12}{$r^+_{12}$}
\psfrag{tm}{$t^-_{12}$}
\psfrag{rm}{$r^-_{12}$}
\includegraphics[clip,width=0.25\textheight]{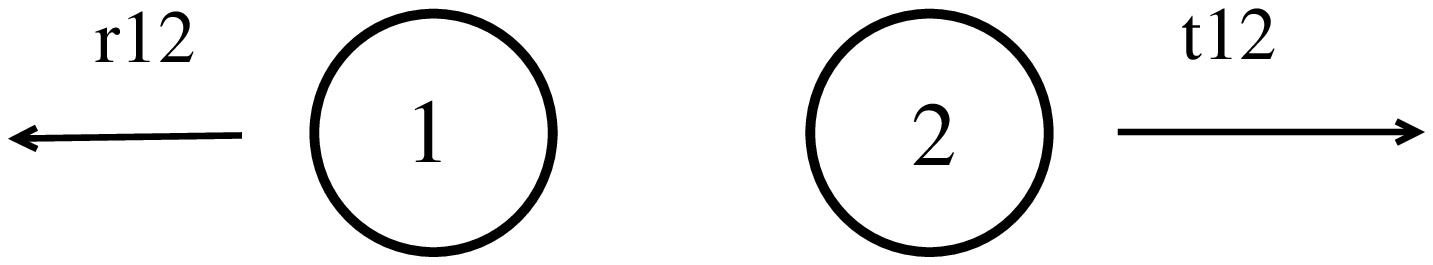}
\end{center}
\caption{The composition law for the transfer matrix.}
\label{uk}
\end{figure}

The transfer matrix given by Eq. (\ref{TM-q})
fulfills the composition law. If the sample consists of two
subsystems, then the transfer matrix $\textbf{T}_{12}$ of the whole sample can 
be calculated  from the transfer matrices of the two  subsystems as
\be\label{cl}
\textbf{T}_{12}=\textbf{T}_2\textbf{T}_1.
\ee
The resulting transfer matrix  $\textbf{T}_{12}$ has again the form (\ref{TM}). This composition law
enables us to calculate the transmission of a complicated structure from the transfer
matrices of its parts. In particular, the transmission through 
two scattering centers is given by the relation
\be\label{twobar}
t_{12}^-=t_1^-\left[1-r_2^+r_1^-\right]^{-1}t_2^-.
\ee
Similarly for the reflection we obtain
\be\label{1-r}
r_{12}^+=r_1^++t_1^+\left[1-r_2^+r_1^-\right]^{-1}r_2^+t_1^-.
\ee

\subsection{Symmetries}

The elements of the transfer matrix are not independent.  Various physical symmetries
provide important  relations between the transmission and reflection amplitudes. Here, we investigate
the consequences of the flux conservation and of the symmetry with respect
to the inversion of time.

\smallskip

The \textsl{flux conservation} requires that the flux on the right-hand side of the sample
equals to the flux on the left-hand side. This gives the following constrain
to the transfer matrix \cite{book},
\be\label{intro-fc}
\textbf{T}^\dag\m2{1}{0}{0}{-1}\textbf{T}=\m2{1}{0}{0}{-1}.
\ee
Inserting 
\be\label{TM-pom}
\textbf{T}=\m2{T_{11}}{T_{12}}{T_{21}}{T_{22}}
\ee
we obtain from Eq. (\ref{intro-fc}) that
\be\label{TM-q1}
T_{22}^\dag T_{22}-T_{12}^\dag T_{12}=1
\ee
and
\be\label{TM-q2}
T_{11}^\dag T_{11}-T_{21}^\dag T_{21}=1.
\ee
We also have 
\be\label{TM-kvakva}
\begin{array}{lcl}
T^\dag_{11}T_{12}-T^\dag_{21}T_{22}&=&0\\
T_{12}^\dag T_{11}-T_{22}^\dag T_{21}&=&0.
\end{array}
\ee
With the use of the explicit form of the transfer matrix, given by Eq. (\ref{TM-q}), we 
obtain from Eq. (\ref{TM-q1}) the relation
\be\label{TM-q3}
(t^-)^\dag t^-+(r^-)^\dag r^-=1.
\ee
The second equation (\ref{TM-kvakva}) gives
\be\label{rplus}
r^+=-(t^-)^{-1\dag}(r^-)^\dag t^+,
\ee
which can be written in the form
\be\label{rpplus}
r^-(t^-)^{-1}=-(t^+)^{-1\dag}(r^+)^\dag.
\ee
Inserting (\ref{rplus}) into Eq. (\ref{TM-q2}) gives, after some algebra, the relation
\be\label{TM-q4}
(t^+)^\dag t^++(r^+)^\dag r^+=1.
\ee
The equations (\ref{TM-q3}) and (\ref{TM-q4}) are a direct consequence of the
flux conservation. They tell us that the electron either
transmits through the sample or is reflected back.

The relation (\ref{rpplus}) enables us also to simplify the expression for $T_{11}$,
\be\label{TM-t11}
\begin{array}{lcl}
T_{11}&=&t^+-r^-(t^-)^{-1}r^+=t^++(t^+)^{-1\dag}(r^+)^\dag r^+\\
   &=& (t^+)^{-1\dag}\left[(t^+)^\dag t^+ + (r^+)^\dag r^+\right]\\
   &=& (t^+)^{-1\dag}.
\end{array}
\ee

The \textsl{time reversal symmetry} implies that the transfer matrix fulfills the relation
\be\label{intro-ts}
\m2{0}{1}{1}{0}\textbf{T}\m2{0}{1}{1}{0}=\textbf{T}^*.
\ee

From Eq. (\ref{intro-ts}) we obtain that for systems with
time reversal symmetry the transfer matrix can be written in the form
\be\label{timer}
\textbf{T}=\m2{T_{11}}{T_{12}}{T_{12}^*}{T_{11}^*},
\ee
where $T_{11}$ and $T_{12}$  are the $N\times N$ complex matrices.
Also, from Eq. (\ref{intro-fc}) we see that $|\textrm{det}~\textbf{T}|=1$ 
and from Eq. (\ref{intro-ts})
we get
$(\textrm{det}~\textbf{T})^*= 
\textrm{det}~\textbf{T}$.  Consequently,
$\textrm{det}~\textbf{T}=1$ when the time reversal symmetry is preserved.

From Eq. (\ref{timer})  we also find that in the case of time reversal symmetry the
 matrices $t^+$ and $t^-$  are related by the relation
\be
t^+=(t^-)^T.
\ee

\subsection{Parametrization of the transfer matrix}\label{sect:param}

Consider now the system with time reversal symmetry. Then, the transfer matrix, $T$, is determined
by two complex matrices, $T_{11}$ and $T_{12}$, which are completely determined  by $4N^2$ \textsl{real}
parameters. However, these parameters are not independent of each other, since
the matrices $T_{11}$ and $T_{12}$ must fulfill the relations of flux conservation.

We insert the expression (\ref{timer}) into Eq. (\ref{intro-fc})
and from Eq. (\ref{TM-q2})  we obtain the relation
\be\label{mp-1}
T_{11}^\dag T_{11}-T_{12}^T T_{12}^*=1, 
\ee
which implies $N^2$ additional relations between the elements of the matrices
$T_{11}$ and $T_{12}$. Other constraints are given by the first of Eq. (\ref{TM-kvakva}),
\be
T_{11}^\dag T_{12} - T_{12}^T T_{11}^* =0,
\ee
which can be written in the form 
\be\label{mp-11}
A^T=A,~~~{\rm where}~~~~A=T_{12}^TT_{11}^*.
\ee
Equation (\ref{mp-11}) requires that the complex matrix $A$ is symmetric. This requirement
implies $\no(\no-1)$ relations between the elements of the matrices $T_{11}$ and $T_{12}$.
Finally, we have that the matrices $T_{11}$ and $T_{12}$ are fully determined
by  $N(2N+1)$ real parameters.

\medskip

Following Ref. \cite{MP} let us look for the solution of Eqs. (\ref{mp-1}) in the form 
\be\label{mp-2}
T_{11}=uLv~~~~\textrm{and}~~~~~T_{12}=u'L'v',
\ee
where $u$, $u'$, $v$, and $v'$ are the unitary matrices and $L$, $L'$ are the diagonal 
\textsl{real} matrices. 
Inserting into Eq. (\ref{mp-1}) we obtain that
\be\label{mp-3}
\begin{array}{lcl}
v^\dag L^2 v - (v')^T (L')^2(v')^* &=& 1,\\
~~ & ~~ & ~~\\
v^\dag L u^\dag u' L' v' - (v')^T L' (u')^T u^* L v^* &=& 0,
\end{array}
\ee
which can be solved with
\be\label{mp-solution}
u'=u,~~~~~~~v'=v^*,~~~~~~~~~(L')^2 = L^2-1.
\ee
Since the complex unitary matrices $u$ and $v$ are determined together by $2N^2$ real parameters
and the real diagonal matrix $L$ needs $N$ real parameters, we see that the solution
(\ref{mp-solution})
is consistent with our estimation of the number of free parameters of the transfer matrix.

Taking $L=\sqrt{1+\lambda}$ we obtain  that 
the transfer matrix can be parametrized in the form
\begin{equation}\label{app-one}
\textbf{T}=\left(\matrix{ u & 0  \cr 0 & {u}^* \cr }\right) \left(\matrix{
\sqrt{1+\lambda} & \sqrt{\lambda}   \cr \sqrt{\lambda}   &
\sqrt{1+\lambda} \cr }\right)
\left(\matrix{ v & 0  \cr 0 & {v}^* \cr }\right),
\end{equation}
where $u,v$ are the  $\no \times \no$ {unitary}  matrices. We also use the following
parametrization of the diagonal elements $\lambda_a$:
\be\label{app-ones}
\lambda_a=\frac{1}{2}\left[\cosh x_a-1\right].
\ee

It was shown in Ref. \cite{MP} that parametrization (\ref{app-one}) is the most general
one. Any transfer matrix of the orthogonal
system can be expressed in the form given by Eq. (\ref{app-one}).
A similar  parametrization can be derived for the transfer matrices with 
unitary and symplectic symmetry.
For unitary symmetry, the transfer matrix has the form
\begin{equation}\label{app-two}
\textbf{T}=\left(\matrix{ u_1 & 0  \cr 0 & u_2 \cr }\right) \left(\matrix{
\sqrt{1+\lambda} & \sqrt{\lambda}   \cr \sqrt{\lambda}   &
\sqrt{1+\lambda} \cr }\right)
\left(\matrix{ v_1 & 0  \cr 0 & v_2 \cr }\right),
\end{equation}
i. e., it is determined by four unitary matrices $u_1$, $u_2$, $v_1$ and $v_2$
 and by the diagonal matrix $\lambda$.
A detailed derivation of relations (\ref{app-one},\ref{app-two}) can be found in Ref. \cite{MP}.

\subsection{Transfer matrix \textsl{vs} conductance}

From the  flux conservation  (\ref{intro-fc}) we obtain the following
relations for the transfer matrix:
\be
\left[\textbf{T}^\dag\right]^{-1}=\m2{1}{0}{0}{-1}\textbf{T}\m2{1}{0}{0}{-1},
\ee
and 
\be
\textbf{T}^{-1}=\m2{1}{0}{0}{-1}\textbf{T}^\dag\m2{1}{0}{0}{-1},
\ee
With the use of these equations we obtain
\be\label{eq208}
\left[\textbf{T}^\dag \textbf{T}\right]^{-1}=\m2{1}{0}{0}{-1}\textbf{T}^\dag \textbf{T}\m2{1}{0}{0}{-1}.
\ee
Using Eqs. (\ref{eq208})  and 
(\ref{TM-kvakva}) we obtain
\be\label{tm-gg}
\left[\textbf{T}^\dag \textbf{T}\right]^{-1}+\textbf{T}^\dag \textbf{T}=\m2{4T_{11}^\dag T_{11}-2}{0}{0}{4T_{22}^\dag T_{22}-2}.
\ee
Inserting into Eq. (\ref{tm-gg})  the  matrix elements $T_{11}=(t^+)^{-1\dag}$,
given by Eq. (\ref{TM-t11}), and $T_{22}=(t^-)^{-1}$
we  obtain the formula of Pichard
\cite{Pnato} 
\begin{equation}\label{aq-pichard}
\left(
\begin{array}{ll}
t^+(t^+)^\dag & 0\\
0        & (t^-)^\dag t^-
\end{array}\right)
=\frac{1}{4}\left[\textbf{T}^\dag \textbf{T} + (\textbf{T}^\dag \textbf{T})^{-1}+2\right]^{-1}.
\end{equation}
Note that relation (\ref{aq-pichard}) follows directly from the requirement of
flux conservation.

We can also use the parametric form  
of the transfer matrix, given by Eq. (\ref{app-two}), and we express the inverse of the
r.h.s.  of Eq. (\ref{aq-pichard}) in the form
\be\label{aq-p1}
\frac{1}{4}\left[\textbf{T}^\dag \textbf{T} + (\textbf{T}^\dag \textbf{T})^{-1}+2\right]=
\m2{v_1^\dag}{0}{0}{v_2^\dag}\m2{1+\lambda}{0}{0}{1+\lambda}\m2{v_1}{0}{0}{v_2}.
\ee
Comparing the l.h.s. of Eq. (\ref{aq-pichard}) with Eq.  (\ref{aq-p1}),
we obtain that
\be\label{tm-w1}
t^+(t^+)^\dag  = v_1^\dag (1+\lambda)^{-1} v_1
\ee
and
\be\label{tm-w2}
(t^-)^\dag t^- = v_2^\dag (1+\lambda)^{-1} v_2.
\ee
Both equations are equivalent if time reversal  symmetry is preserved. Indeed,
$(t^+)^\dag=((t^+)^T)^*=(t^-)^*$ and $v_2=v_1^*$.

The  matrices $v_1$ and $v_2$,
obey the relation of unitarity, $v^{-1}\equiv v^\dag$. Using this relation
 we can express the Hermitian matrix 
$(t^-)^\dag t^-$ in the form
\begin{equation}
(t^-)^\dag t^-=v_2^{-1}\frac{1}{1+\lambda}v_2.
\end{equation}
Thus, the eigenvalues of the matrix  $(t^-)^\dag t^-$  are
$(1+\lambda_a)^{-1}$ 
and  the unitary matrix $v_2^{-1}$  
contains in its columns the corresponding  eigenvectors.

Using the eigenvalues of the matrices $(t^-)^\dag t^-$ and $(t^+)^\dag t^+$ we can express the transmission,
\begin{equation}
g=\frac{e^2}{h}~\textrm{Tr}~(t^-)^\dag t^- = \frac{e^2}{h}\sum_a \frac{1}{1+\lambda_a}.
\end{equation}

\subsection{Open channels and evanescent waves}\label{app-open}

In previous Sections, we assumed that there are only  propagating states in the leads.
This might not be true in real systems. 
For instance, for the  3D model, we see from the dispersion relation, given by Eq.
(\ref{density-1}), that
the propagation along the leads, (which is the $z$ direction in our notation),
is determined by the $z$-component of the wave vector, $k_z$, given by the relation
\be\label{aq-1}
2\cos k_z = E-2\cos k_x-2\cos k_y
\ee
where the  transversal wave vector $k_x$ is   given by
\be
k_{n_x}=\ds{\frac{2\pi}{L_x}}n_x,~~~~n_x=0,1,\dots ,L_x-1
\ee
for the periodic boundary conditions and
\be
k_{n_x}=\ds{\frac{\pi}{L_x+1}}n_x,~~~~n_x=1,\dots ,L_x
\ee
for the hard wall boundaries. Similar relations hold for $k_y$.

Using the eigenfunctions $\Phi_{n_x}(x)$ and $\Phi_{n_y}(y)$, related to the eigenvalues of
$k_x$ and $k_y$, we find that the wave function in the leads is
\be\label{wavefucntion}
\Psi(\vec{r})=\sum_{n_xn_y} \phi_{n_x}(x)\phi_{n_y}(y)\ds{\frac{e^{ik_z}}{\sqrt{2i\sin k_z}}}.
\ee
The wave vector $k_z=k_z(n_x,n_y)$ possesses $N=L_xL_y$ values
given by Eq. (\ref{aq-1})  (not necessarily different
from each other).  The propagating solutions exist only for such 
values of $k_z$, for which
\be\label{aq-cond}
|2\cos k_z(n_x,n_y)|<2,
\ee
i. e. when $k_z$ is real.
Otherwise, $k_z$ is purely imaginary. and determines the exponential
decrease of the wave function with  distance form the left (right) boundary of the sample.
Therefore, we have to distinguish between $\no$ and $N$, where  $\no$  is the number of
propagating channels with $k_z$ real, and $N$ is the total number of channels.

Since we assume that both leads are semi-infinite, 
it is clear that the evanescent waves, emitted from the reservoirs,  are not able to reach the sample.
Therefore, the size of the transmission and reflection matrices is $\no\times\no$.
The necessity to distinguish between  $N$ and $\no$
complicates the  calculation of the conductance. 
Indeed, the size of the transmission matrix $t^-$  is $\no$ but the size of the matrix
$T_{22}$  defined by Eq. (\ref{TM-pom}) is $N\times N$.  If we order the eigenvectors in the matrix
$R$ in such way that the eigenvectors with index $1\le a\le\no$ correspond to
the propagating modes, and the remaining eigenvectors correspond to the evanescent modes,
then the transmission is given only by  the $\no\times\no$
sub-matrix $[T_{22}]_{ab}$, with $a,b\le \no$:
\be\label{taaaa}
T=\sum_{ab=1}^{\no}\Big|\left[T_{22}^{-1}\right]_{ab}\Big|^2.
\ee
Other matrix elements of $T_{22}$ correspond to the scattering of the electron into
the evanescent channels.
Note, the composition laws for the transmission and reflection of the system 
 given by Eqs. (\ref{twobar},\ref{1-r}) are not valid in this case
because evanescent waves between two subsystems can also contribute to the transport.

It is important to underline that the \textsl{analytical theories
discussed in next two Appendices assume implicitly that there are no evanescent wave
in the leads}.  

In numerical simulations, we can avoid the evanescent waves for the band center
($E=0$) by using anisotropic models, given, for instance, by Eq. (\ref{hama}).
However, the evanescent waves cannot be completely excluded 
if we are interested in the transmission of the electron with energy close to the band edge. 
Indeed,  from the $E$ dependence of
$k_z$, given by Eq. (\ref{aq-1}), we see that 
$\no$ is maximal for the band center and
decreases to zero when the energy $E$ approaches the band edge  
of the unperturbed system. Therefore, we have to consider evanescent waves 
in the  numerical calculation of the conductance. The  calculation of
the conductance is described in
Appendix \ref{app-b}.

Note, there are no propagating channels for the energy $E$ outside of the unperturbed
band.  Therefore, we are not able to calculate the conductance for such energies. 
This constrain complicates the scaling analysis of the conductance along the critical line
shown in Fig. \ref{fig-schema-pom}.


\section{Dorokhov-Mello-Pereyra-Kumar (DMPK) Equation.}\label{app-dmpk}

The transfer matrix of the  disordered system is a statistical variable. One can ask
if there is a possibility to calculate the probability distribution 
$p(\textbf{T})$, which is a joint probability distribution of all matrix elements
of the transfer matrix, \textbf{T}.
This problem was solved by Mello \textsl{et al.} \cite{DMPK}.
For the transfer matrix, given by expression (\ref{app-one}), they derived
the equation for the joint probability distribution of
the parameters $\lambda$, which parametrize  the transfer matrix
in Eq. (\ref{app-one}). Since the derivation of the DMPK equation 
is rather difficult, we present here only
 main ideas and refer the reader to the original work, \cite{DMPK}.

Consider a weakly disordered system of length $L_z$, connected on both sides to the semi-infinite
leads as shown in Fig. \ref{fig-kanal}. 
There are  $N$ \textsl{open} 
channels in the  leads, with $N$  big enough to neglect all corrections of
order  $N^{-1}$. The channels are equivalent to each other. In the leads, each channel
is characterized by the same wave vector $k_z$. The disorder is introduced 
inside the sample by
random hopping terms between the channels. 

\begin{figure}[t!]
\bc
\psfrag{Lz}{$L_z$}
\psfrag{dLz}{$\delta L_z$}
\includegraphics[clip,height=0.2\textheight,width=.2\textheight]{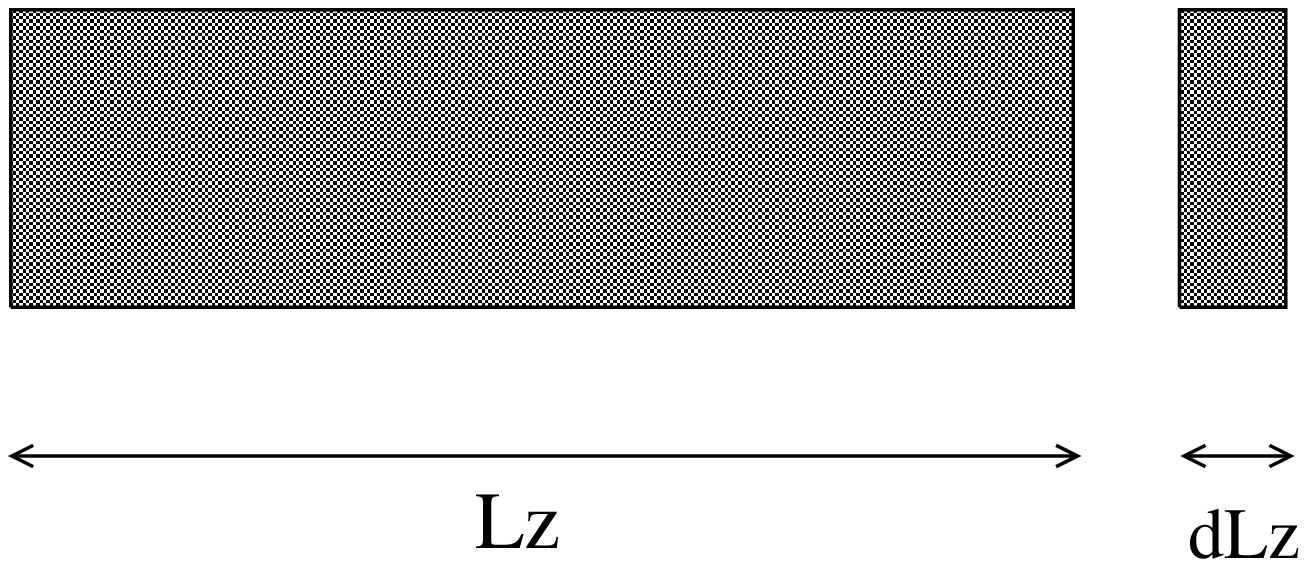}
\ec
\caption{Small segment of length $\delta L_z$ is added to the sample
of length $L_z$.}
\label{dmpk-obr}
\end{figure}

Suppose we have a sample of length $L_z$ and
we know the probability distribution $p_{L_z}(\textbf{T}')d\textbf{T}'$.
Now, we add to the sample of length $L_z$ an additional 
random segment of  length $\delta L$ (Fig. \ref{dmpk-obr}).
The resulting sample will have length $L_z+\delta L_z$ and is characterized by 
the transfer matrix
\be
\textbf{T}=\textbf{T}_{\delta}\textbf{T}'.
\ee 
We suppose that the  length $\delta L_z$
of the additional  segment is sufficiently small  so that
the reflection coefficient of the segment  is linear in $\delta L_z$,
\be\label{dmpk-r}
R^+_\delta=\textrm{Tr}~(r^+)^\dag r^+ \propto a\frac{\delta L_z}{\ell}.
\ee
In Eq. (\ref{dmpk-r}), 
the length $\ell$ is the mean free path
and  we put $a=1$.  We also neglect all
terms proportional to higher orders of $\delta L_z$.  Equivalently, 
we can require that the trace,
\be\label{dmpk-trace}
\textrm{Tr}~\textbf{T}_{\delta}^\dag \textbf{T}_{\delta}=2N\left[1+\frac{\delta L_z}{\ell}\right],
\ee
is linear in $\delta L_z$\footnote{%
Note, in the case of zero reflection,   the matrices 
$(t^-)^\dag t^-$ and $t^+(t^+)^\dag$ are diagonal and
$\textrm{Tr}~T^\dag T \equiv 2N$.}.


We also  need  to know  the probability distribution $p_{\delta L}(\textbf{T}_\delta)$
of matrix elements of the transfer matrix  $T_\delta$. This
distribution represents the main building block of the theory.
In Ref. \cite{DMPK}, the Ansatz
\be\label{dmpk-2}
p_{\delta {L_z}}(\textbf{T}_\delta)\propto\exp -\ds{\frac{(N+1)\ell}{2\delta {L_z}}} \textrm{Tr}~\lambda_{\delta}
\ee
was proposed. It represents the ``most random'' distribution
of matrix elements of the transfer matrix $T_\delta$.
We remind the reader that parameters $\lambda_\delta$  for the transfer matrix $T_\delta$ are defined by
Eq. (\ref{app-one}).
Clearly, $\textrm{Tr}~\langle\lambda_\delta\rangle=N\delta L_z/\ell$.

The probability distribution
$p_{L_z+\delta L_z}(\textbf{T})$ fulfills the Smoluchovsky equation,
\be\label{dmpk-1}
p_{L_z+\delta L_z}(\textbf{T})=\int p_{L_z}(\textbf{TT}_{\delta}^{-1})p_{\delta L_z}(\textbf{T}_\delta) d\textbf{T}_\delta.
\ee

Inserting Eq. (\ref{dmpk-2}) into  Eq. (\ref{dmpk-1}) leads,
after rather complicated mathematical operations, to the equation for the
joint probability distribution of eigenvalues $\lambda$:
\begin{equation}
\label{dmpk}
\frac{\partial p_{L_z}(\{\lambda\})}{\partial (L_z/\ell)}
=\frac{2}{\beta N+2-\beta}~~\frac{1}{{J}}\sum_a^N
\frac{\partial}{\partial\lambda_a}\left[\lambda_a(1+\lambda_a)
{J}\frac{\partial p_L(\{\lambda\})}{\partial \lambda_a}\right]
\end{equation}
Here, $J$ is the Jacobian, given by
\begin{equation}\label{jacobi}
{J}\equiv\prod_{a<b}^N|\lambda_a-\lambda_b|^{\beta}
\end{equation}
and $\beta$ is the   symmetry parameter, $\beta=1,2$ or 4 for
orthogonal, unitary, or  symplectic systems, respectively.
Equation (\ref{dmpk}) is known in the literature as the DMPK equation.
Note,  the mean free path, $\ell$, is the only parameter which enters the theory.  

\begin{figure}[t!]
\bc
\includegraphics[clip,height=0.2\textheight,width=.2\textheight]{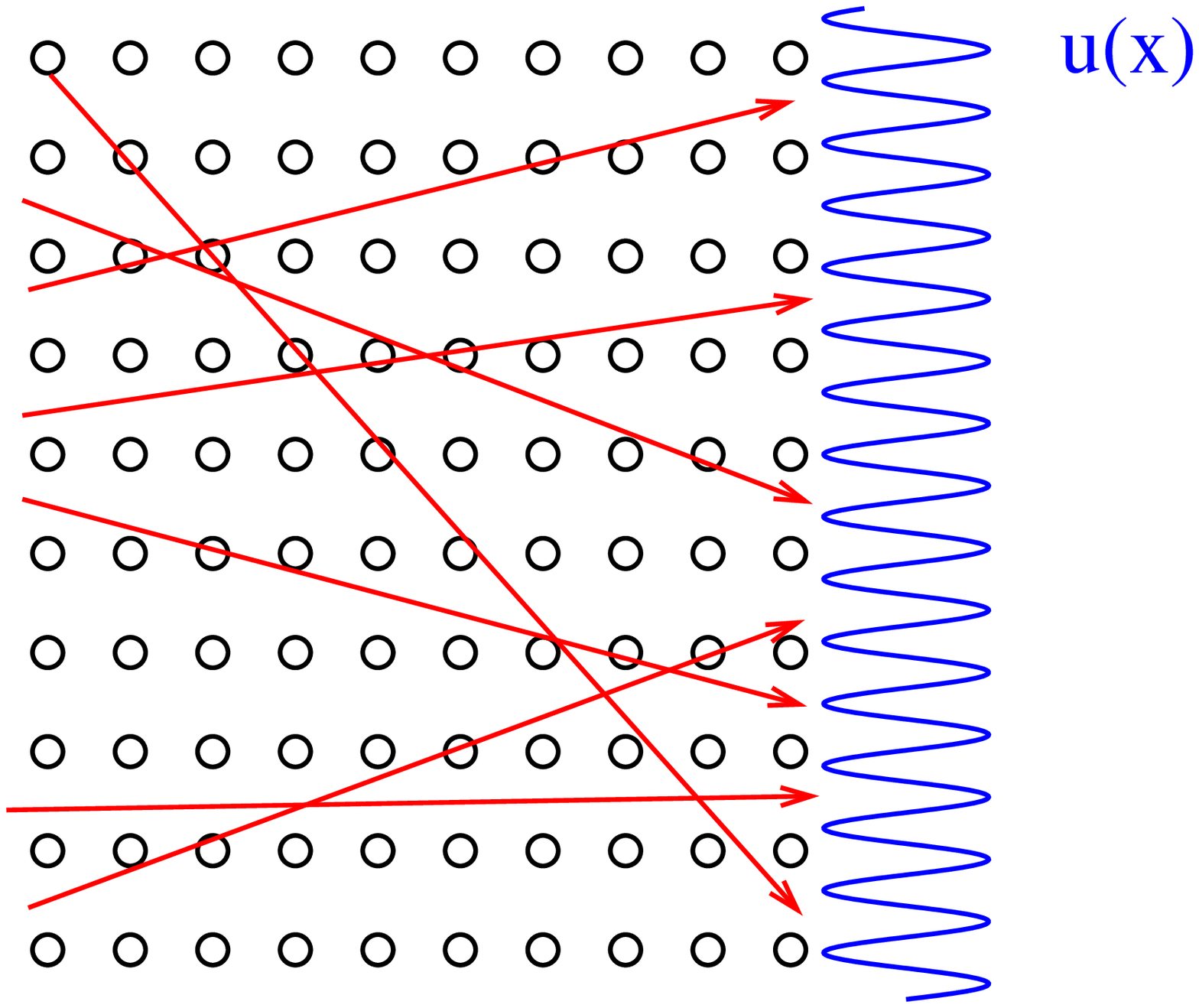}
~~~~~~~~~\includegraphics[clip,height=0.2\textheight,width=.2\textheight]{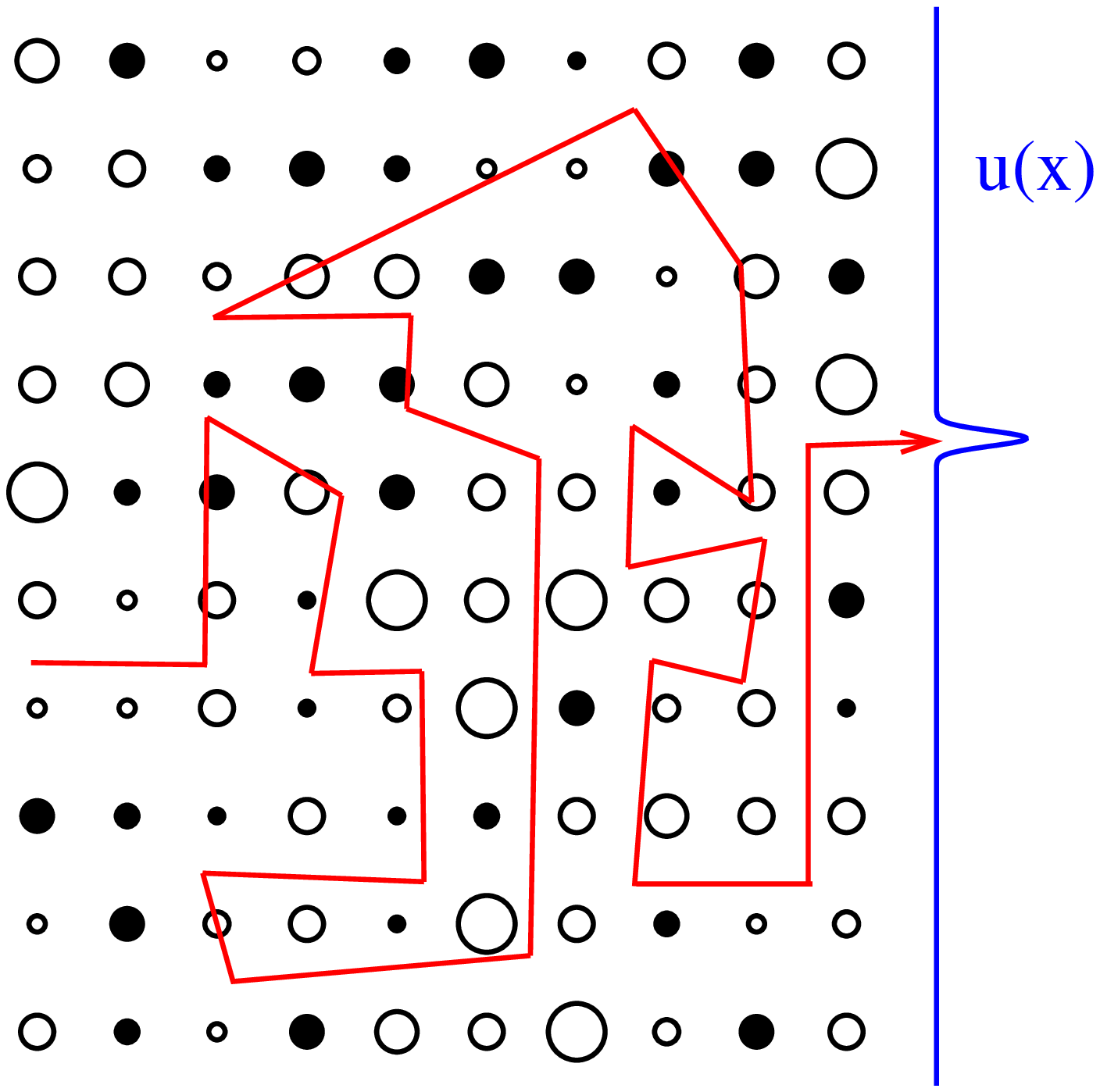}
\ec
\caption{The propagation of the electron through the disordered sample. Left: If the disorder
is weak, then there are many  paths through the structure. We expect that
the wave function of the electron on the opposite side of the sample is
spatially homogeneous. This corresponds to the assumption (\ref{dmpk-a1},\ref{dmpk-a2}).
Right: If the disorder is strong, there is in the best case only one  path
through the structure. Then, the wave function
possesses sharp maxima  at the opposite side of the sample;
the positions of these maxima depend on the realization of the disorder in a given sample.
It is clear that assumption (\ref{dmpk-a2}) is not longer valid. Therefore, we do not expect
that the DMPK equation,
given by Eq. (\ref{dmpk}),  describes the transport in strongly disordered systems.
}
\label{fig-dmpk}
\ef

The DMPK equation, given by Eq, (\ref{dmpk}) does not contain any information
about the matrices $u$ and $v$. These variables were ``integrated out''
in the process of deriving Eq. (\ref{dmpk}). The main
assumption made in this process was  that
$u$, $v$ and $\lambda$  are statistically independent. Next,
it was assumed that the elements of the matrix  $v$  fulfill the 
relations
\be\label{dmpk-a1}
\langle v^*_{ab}v_{cd}\rangle=\ds{\frac{1}{N}}\delta_{ab}\delta_{cd}
\ee
and
\be\label{dmpk-a2}
\langle v^*_{ca}v^*_{cb}v_{da}v_{db}\rangle=\ds{\frac{1+\delta_{ab}}{N(N+1)}}\delta_{cd},
\ee
and similar relations hold also for the  matrix $u$. These assumptions  limit the validity of
the DMPK equation to the systems with weak disorder. 
It is implicitly  assumed that there are  many paths for the electron to propagate through the sample. Then, the
density of the electron on the opposite side of the sample is homogeneous, as shown in the left
panel of  Fig. \ref{fig-dmpk}.

\begin{figure}[b!]
\bc
\psfrag{LL}{$N=L^{2}$}
\psfrag{kax}{$(N+1)K_{11}/2$}
\psfrag{kaa}{$(N+1)K_{aa}/2$}
\psfrag{a}{$a$}
\includegraphics[clip,width=.3\textheight]{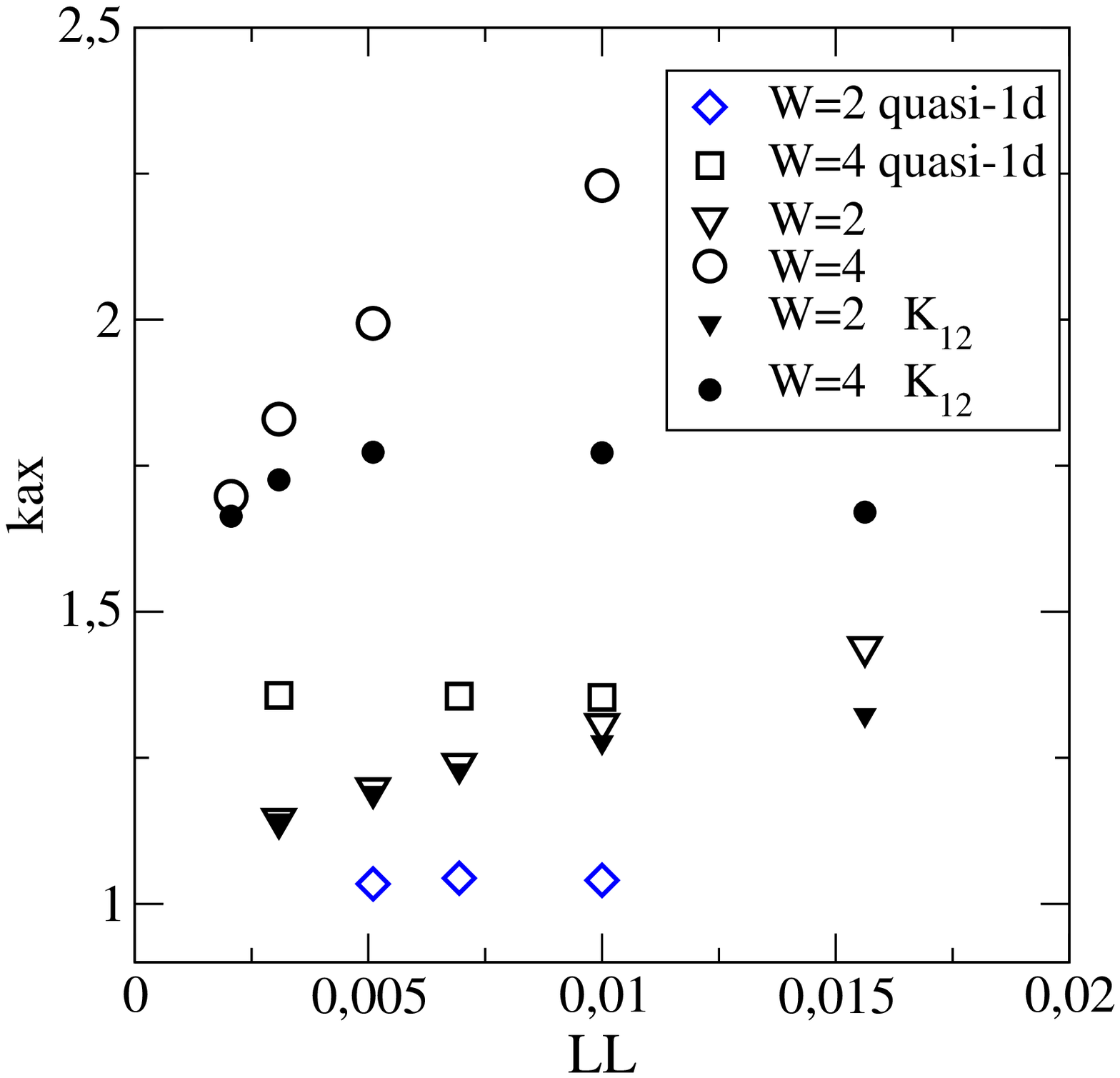}
~~~~~~~~~\includegraphics[clip,width=.3\textheight]{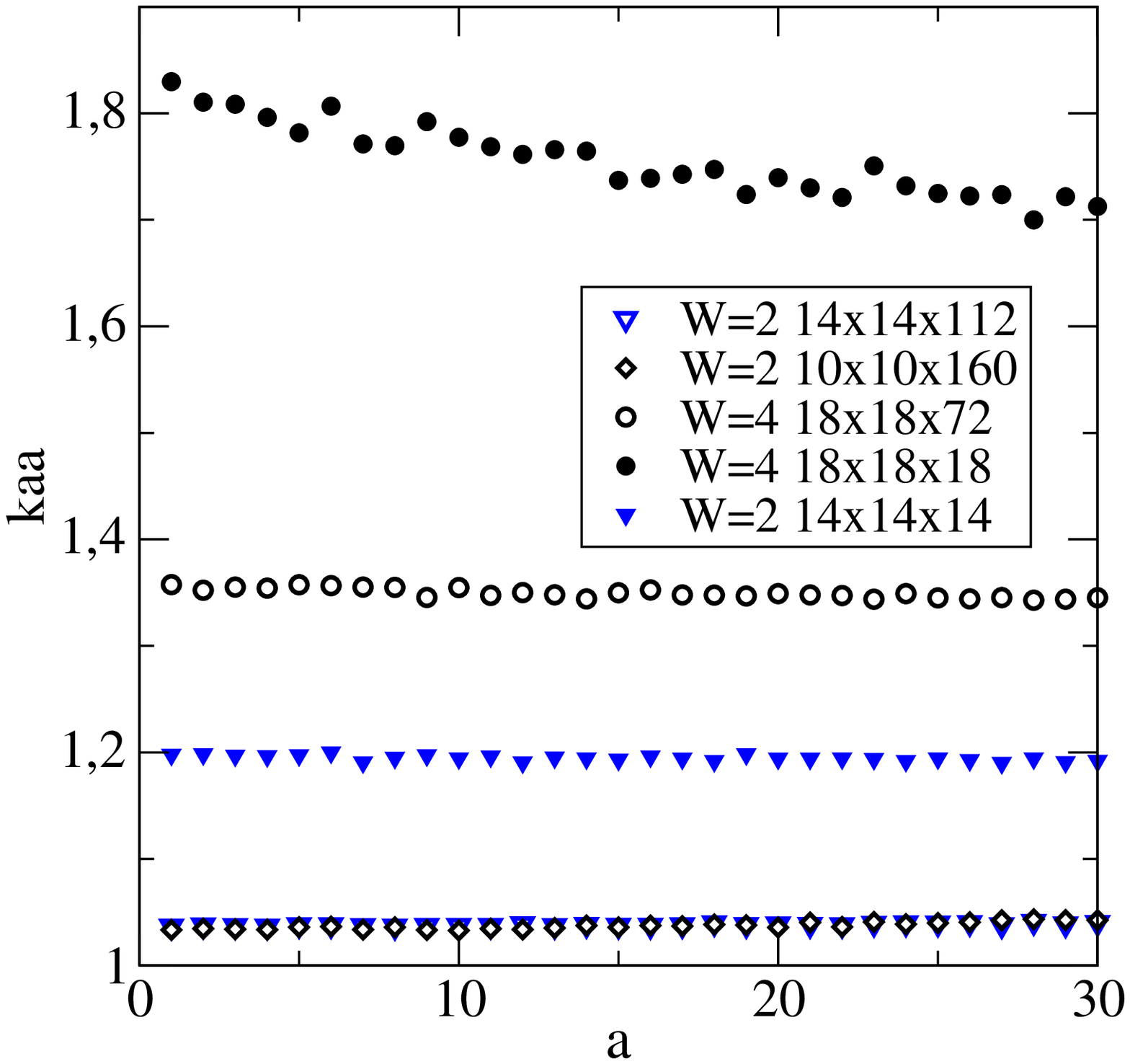}
\ec
\caption{Numerical verification of the validity of the DMPK equation.
Left: The $N=L^{2}$ dependence of the  $(N+1)K_{11}/2$
for the quasi-1d systems $L\times L\times 4L$ and for the cubes $L^3$. In the last case,
also $(N+1)K_{12}$ is shown.  The DMPK is applicable if
these quantities  converge to 1 in the limit
of $L\to\infty$. 
Right: The $a$ dependence of $K_{aa}$ for the quasi-1d systems and for the cubes.
Data were obtained  for the
anisotropic Anderson model with $t=0.4$
with disorder $W=2$ and $W=4$, which corresponds to the mean free path
$\ell=9$ and $\ell=1.8$, respectively
\cite{MMW}.
}
\label{dmpk-k11}

\end{figure}

Another constraint of the DMPK equation is given by the assumption of the equivalence
of  all incoming and outgoing channels. In the real world, the  channels are never 
equivalent, since they  are determined
 by the incident angle (or by the transversal momentum) inside the leads,
as shown in Fig. \ref{fig-kanal}. 
The assumption that all channels are equivalent removes from the
theory any information about the topology of the leads. 
Further, in the derivation of the DMPK it is assumed that channels are equivalent also \textsl{inside}
the sample, since  disorder was introduced by the random hopping  between any two channels.
Thus, the DMPK equation represents the simplest ``mean field theory''. In spite of the above constrains,
the DMPK equation is surprisingly successful in the description of the electronic transport
in weakly disordered quasi-1d systems. 
The  DMPK equation can be easily generalized to description of  more realistic systems. For instance,
instead of the Ansatz (\ref{dmpk-2}) we can use more complicated probability distribution,
which reflects the properties of the system, inclusive anisotropy and
topology \cite{MT}. However, such generalization does not  bring any novel physical information.
It only complicates the resulting formula for the transmission.

\subsection{Numerical verification of validity of the DMPK equation}

In Fig. \ref{dmpk-k11} we verify  numerically the validity of  the 
relation (\ref{dmpk-a2}).  We consider the anisotropic Anderson model,
given by Eq. (\ref{hama}) with $t=0.4$ and 
calculate the matrix
\be
K_{ab}=\left\langle\sum_c |u_{ac}|^2 |u_{bc}|^2\right\rangle,
\ee
where, as usual,  $\langle\dots\rangle$ means averaging over the statistical ensemble.
Requirement (\ref{dmpk-a2}) is equivalent to
\be\label{dmpk-kab}
K_{ab}=\ds{\frac{1+\delta_{ab}}{N+1}}
\ee
with $N=L^2$.

Our data show, that although the channels in the Anderson model are not equivalent to each
other, the requirement (\ref{dmpk-kab}) is perfectly fulfilled for the weakly disordered 
quasi-1d systems, as long as  the mean free path, $\ell$, is comparable to  the width, $L$, of the system
Agreement is worse for 3D systems. This is natural, since the 3D systems are expected to possess
slightly different conductance statistics than the quasi-1d  systems (see Table \ref{table-ucf}
which compares  data for the universal conductance fluctuations, $\textrm{var}~g$.
in the quasi-1d, 2D and 3D systems).
When disorder increases, the mean free path decreases and the transversal structure of
the sample becomes important.  This could be  included into the theory if a more detailed
model for the distribution $p(\textbf{T}_\delta)$ is considered.

\subsection{Conductance}\label{app-dmpk-cond}

The DMPK equation can  be directly used to calculate the first two moments of the conductance
of the weakly disordered quasi-1d system.
The method is described in Ref. \cite{Stone}. When we are interested in the  mean value of the  function
$F(\lambda)$, we multiply both sides of the DMPK equation, given by Eq. (\ref{dmpk}) by $F(\lambda)$
and integrate over $\lambda$. On the r.h.s.,
we obtain 
mean values of some  other functions, $F_i(\lambda)$; repeating this procedure for $F_i$, we obtain
a system of coupled equations for $\langle F\rangle$ and $\langle F_i\rangle$, which can be solved
in the limit of $N\to\infty$. In this way, the mean conductance (in units of $e^2/h$) 
was calculated as
\be\label{dmpk-x1}
\langle g\rangle=\left\langle\sum_a\ds{\frac{1}{1+\lambda_a}}\right\rangle=
\ds{\frac{N\ell}{L_z}-N\frac{\ell^2}{L_z^2}-\frac{1}{3}\left[1-3\frac{\ell}{L_z}\right]+\frac{L_z}{45N\ell}}+\dots
\ee
The first term, $N\ell/L_z$ can be identified with the  conductivity, $\sigma$. 
Other terms represent 
the quantum  corrections.
Of particular interest is the diffusive regime, which is defined by the following relations
between $L_z$ and $\ell$:
\be\label{dmpk-diff}
\ell\ll L_z\ll N\ell.
\ee
If the conditions (\ref{dmpk-diff}) are fulfilled, 
then we can neglect the last term in Eq. (\ref{dmpk-x1}).
Assuming also  that the system is sufficiently long,  so that
\be\label{dmpk-bal}
N\ds{\frac{\ell^2}{L_z^2}}\ll 1,
\ee
we can neglect also the 2nd term on the r.h.s.
and we  finally obtain the following estimation of the mean conductance:
\be\label{dmpk-g}
\langle g\rangle=\ds{\frac{N\ell}{L_z}-\frac{1}{3}}.
\ee
The correction, $\delta g=-1/3$, is the universal weak localization correction to the conductance
of the quasi-1d system. This is a fully  universal value,
independent of any details of the system
provided that the conditions  (\ref{dmpk-diff},\ref{dmpk-bal}) are fulfilled.
The numerical verification of the relation (\ref{dmpk-g}) is shown in Fig. \ref{fig-mfp}.

In a similar way, Mello and Stone calculated in Ref. \cite{Stone} the variance,
\be\label{dmpk-varg}
\textrm{var}~g=\langle g^2\rangle - \langle g\rangle^2=\ds{\frac{2}{15\beta}}.
\ee
The variance of the conductance in the quasi-1d system is universal, given only by the 
physical symmetry (by the value of the parameter $\beta$). The same result was 
obtained in Ref. \cite{LSF} by perturbation Green's function analysis.

The  universal relations (\ref{dmpk-g}) and (\ref{dmpk-varg}) hold only when the system
is long enough (as required by Eq. (\ref{dmpk-bal})). In this case, the electron is scattered many times
inside the sample so that  its wave function and phase are sufficiently randomized. 
As required by Eq. (\ref{dmpk-diff}),
the length of the sample should not  be too long, 
otherwise the effects of localization become important.
This is analyzed in the next Section.

\subsection{Limit $L_z\gg N\ell$}\label{dmpk-ll}

In the limit of  $L_z\gg N\ell$, the localization appears. As  shown in Appendix \ref{app-le},
the wave function decreases exponentially
in the limit of infinitely long system. 
This must be true  also for  the solution of the  DMPK equation. 
Therefore, parameters $\lambda_a=(\cosh x_a-1)/2$  increase exponentially
with the system length.
If we order the parameters $\lambda$,
\be
\lambda_1\ll\lambda_2,\ll\dots ,\ll\lambda_N,
\ee
then the Jacobian, given by Eq. (\ref{jacobi}), reduces to
\be
{J}=\prod_{a=1}\lambda_a^{a-1}
\ee
and the DMPK equation splits into the $N$ independent equations for the 
probability distributions $p(\lambda_a)$,
\begin{equation}
\label{dmpk-loc}
\frac{\partial p_{L}(\lambda_a)}{\partial (L/\ell)}
=\frac{2}{\beta N+2-\beta}~~\lambda_a^{1-a}
\frac{\partial}{\partial\lambda_a}\left[\lambda_a^{a+1}
\frac{\partial p_L(\lambda_a)}{\partial \lambda_a}\right].
\end{equation}
In the derivation of Eq. (\ref{dmpk-loc}
we used that $\lambda_a\gg 1$ so that $\lambda_a(\lambda_a+1)\approx \lambda_a^2$.
Substituting  $\lambda_a=\exp x_a$, 
and  using  
$\partial p(\lambda_a)/\partial\lambda_a=\lambda_a^{-1}\partial p(x_a)/\partial x_a$, we obtain 
\be\label{dmpk-d1}
\frac{\partial p_{L_z}(x_a)}{\partial (L_z/\ell)}
\ds{=\frac{2}{\beta N+2-\beta}~~\left[
a\frac{\partial p_{L_z}(x_a)}{\partial x_a}+
\frac{\partial^2 p_{L_z}(x_a)}{\partial x_a^2}
\right]},
\ee
which is the diffusion equation. The solution of Eq. (\ref{dmpk-d1}) is
\be\label{dm-1}
p_{L_z}(x_a)=\ds{\frac{1}{\sqrt{2\pi\sigma}}\exp -\frac{(x_a-\langle x_a\rangle)^2}{2\sigma}}
\ee
with the mean value,
\be\label{dm-2}
\langle x_a\rangle =2\ds{\frac{1+\beta(a-1)}{\beta N+2-\beta}\frac{L_z}{\ell}},
\ee
and  variance,
\be\label{dm-3}
\sigma= 2\ds{\frac{2}{\beta N+2-\beta}\frac{L_z}{\ell}}.
\ee
Thus, in the limit of $L_z/\ell\gg 1$, the parameters $x_a$ become statistically independent. They
possess Gaussian probability distributions with the same variance, $\sigma$,
and with mean values which increase linearly as a function of index $a$:
\be
\langle x_a\rangle=\langle x_1\rangle
+
2\ds{\frac{\beta a}{\beta N+2-\beta}\frac{L_z}{\ell}}.
\ee
The numerical data for the probability distributions $p_{L_z}(x_a)$ are shown in Fig. \ref{fig-z1}.

\begin{figure}[t!]
\begin{center}
\includegraphics[clip,width=0.25\textheight]{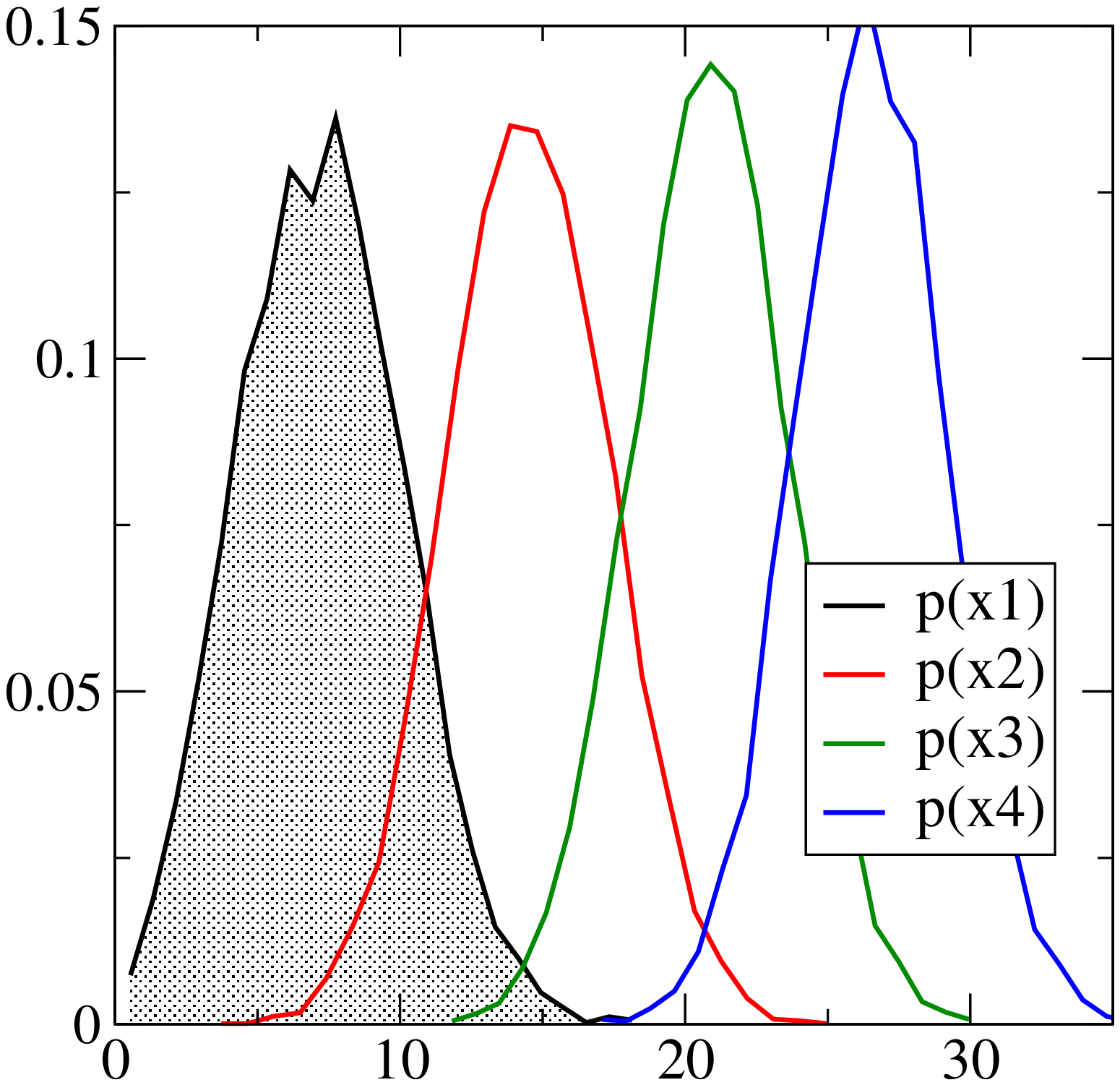}~~~~
\end{center}
\caption{%
The statistical distributions of the parameters $x_a$ for the weakly disordered quasi-1d system
with the box
disorder $W=6$. The  size of the system is $10\times 10\times 200$. Since  disorder
is weak, (the critical disorder $W_c=16.5$ for the box distribution of random energies), we have
$\langle x_a\rangle = a\langle x_1\rangle$, in agreement with the DMPK equation. 
The length of the system, $L_z=200$, is sufficient to localize all electronic states.
All parameters
$x_a$ are large and posses the Gaussian distribution, as predicted by Eqs. (\ref{dm-1}-\ref{dm-3}).
}
\label{fig-z1}
\end{figure}

When calculating the conductance, we can neglect all  contributions from the higher channels
We obtain
\be
g\approx 4 e^{-x_1}
\ee
which confirms that the conductance decreases exponentially when the system length increases.
We also obtain that the \textsl{logarithm} of the conductance, $\ln g = -x_1 +\ln 4$,
possesses the Gaussian probability distribution with the mean
\be
\langle \ln g\rangle = 
- 2\ds{\frac{1}{\beta N+2-\beta}\frac{L_z}{\ell}}+\ln 4
\ee
and with the variance,
\be 
\textrm{var}~g= -2\langle\ln g\rangle - \ln 4.
\ee
The DMPK equation predicts the universal statistical properties of the conductance also in the 
localized regime.

However, it is important  to mention that the localization described by the DMPK equation does not
correspond to the  true insulating regime. Note,  the disorder is still weak in the system.
The localization appears only due to the increase of the length of the system, $L_z$.
We discuss in Sect. \ref{sect:loc}) that the localization in the 
weakly disordered quasi-1d systems differs qualitatively from the
localization in the   \textsl{strongly}
disordered 3D systems.


\section{Random matrix theory}\label{app:rmt}

Random matrix theory was developed 50 years 
ago for the description of the statistical properties of 
large  matrices, which appear, for instance,  in studies of the energy spectrum 
of big nuclei. The experimental data showed that the energy 
spectra possess common universal features -
for instance, the repulsion of the neighboring eigenvalues, similar to that shown
in the left panel of Fig. \ref{fig-spectrum}. It was suggested that the Hamiltonian of a big nucleus
can be approximated by a random matrix. Then, the qualitative properties of the
energy spectra can be obtained from the general properties of the random matrices.
Recently, the random matrix theory for the transfer matrix was developed to
explain some universal properties of the electron transmission in the diffusive regime.

We present here the formula for the joint probability distribution $p(\Lambda)$
of the eigenvalues of the random matrix. More details about the theory 
can be found in Refs. \cite{Mehta,Pnato}.

Consider a real symmetric $N\times N$ matrix \textbf{H} with random elements $H_{ab}$. Let us define the 
joint probability 
distribution of its elements, $P(\textbf{H})d\textbf{H}$. The measure $d\textbf{H}$ is given as
\be\label{rmf-1}
d\textbf{H}=\prod_{i=1}^N dH_{ii}\prod_{i<j} dH_{ij}.
\ee
Our aim is to find the expression for  the joint probability distribution 
of the eigenvalues of the matrix $H$. 
Consider only the orthogonal symmetry class. Then
the matrix $\textbf{H}$ is real and symmetric. It can be diagonalized like
\be\label{rmf-a1}
\textbf{H}=\textbf{Q}\Lambda \textbf{Q}^{-1},
\ee
with the helps of the   orthogonal matrix $\textbf{Q}$, which fulfills
$\textbf{Q}^{-1}=\textbf{Q}^T$. 
It is evident that
\be\label{rmf-a2}
\textbf{QQ}^T=1.
\ee
Then, each element $H_{ij}$ can be expressed as
\be\label{rmf-b1}
H_{ij}=\sum_{k=1}^N \Lambda_k Q_{ki} Q_{kj}.
\ee
Since $\textbf{H}$ is a real symmetric matrix, it is fully determined by
the $N(N+1)/2$ real independent 
parameters, $H_{ij}$, $i\le j$. We want to
express the measure $dH$ in terms of the eigenvalues $\Lambda_i$ and parameters $x_a$ of
the matrix $Q$ (there are exactly $N(N-1)/2$ independent parameters
$x_a$ which determine matrix $Q$).
We need to find the Jacobian, $J$, of the transformation
\be
d\textbf{H}=\prod_{i=1}^N dH_{ii}\prod_{i<j} dH_{ij}=J_N(\{\Lambda\},\{Q_{ij}\})\prod_{i=1}^N d\Lambda_i\prod_{a}x_a.
\ee
The Jacobian matrix can be schematically written in the form
\be\label{rmf-jac}
\textbf{J}=\m2{\ds{\frac{\partial H_{ii}}{\partial\lambda_\alpha}}}
{\ds{\frac{\partial H_{ij}}{\partial\lambda_\alpha}}}
{\ds{\frac{\partial H_{ii}}{\partial x_\mu}}}
{\ds{\frac{\partial H_{ij}}{\partial x_\mu}}},
\ee
where the upper left quadrant is the   $N\times N$ matrix, ($i$ counts columns and $\alpha$
counts rows), and the lower right quadrant  is a square matrix of  size
$N(N-1)/2$ ($i<j$ counts columns and $\mu$ counts rows of the matrix).

With the help of Eq. (\ref{rmf-b1}) we see that the first $N$ rows of the matrix $\textbf{J}$
do not contain $\Lambda$. Also, all elements in the lower  $N(N-1)/2$  rows
are linear in the eigenvalues. Therefore, $J=\textrm{det}~\textbf{J}$
is a polynomial of order of $N(N-1)/2$ in the eigenvalues $\Lambda$.
In order to find the form of this polynomial one has to realize that  if the matrix
$\textbf{H}$ possesses two degenerate eigenvalues, $\lambda_a=\lambda_b$
for $a\ne b$, then  diagonalization of $\textbf{H}$ is not unique
and the matrix $\textbf{J}^{-1}$, which determines the transformation inverse to (\ref{rmf-a1}),
  must be singular. Therefore, $J=\textrm{det}~\textbf{J}=0$
for the degenerate matrix. Then, we easily find that
\be\label{rmf-jj}
J=\prod_{a<b}|\Lambda_a-\Lambda_b|~F(\{x\}),
\ee
where the function $F(x)$ does not contain any information about the eigenvalues $\Lambda$.
We obtain
\be
P(\textbf{H})d\textbf{H}=P(\{\Lambda\})J(\{\Lambda\}d\Lambda P(x)dx.
\ee
Now, we can integrate out all  parameters $x$ to obtain the probability distribution for
eigenvalues, $\Lambda$.

In the case of unitary and symplectic systems, the Jacobian changes to
\be\label{rmf-jjj}
J_\beta=\propto \prod_{a<b}|\Lambda_a-\Lambda_b|^\beta,
\ee
where $\beta=2$ ($\beta=4$) for the unitary (symplectic) symmetry, respectively.

The expression  (\ref{rmf-jjj}) plays an important role in the theory of random matrices.
It tells us that there is a zero probability to find degenerate eigenvalues.
The level repulsion is a typical property of random matrices. 
We have shown already in Fig. \ref{fig-wigner}, that the spectrum of weakly disordered  Hamiltonian 
exhibits level repulsion.

To find an explicit form of the probability distribution  $P(\Lambda)$, we 
look for the 
 probability distribution $P(\Lambda)$ in the form
\be\label{rmf-p}
P(\Lambda)=J_\beta \exp \sum_a F(\Lambda_a).
\ee
Choosing  the expression (\ref{rmf-p}), we implicitly assume that $F(\Lambda)$
is a function of only one eigenvalue.  We will see later
that this assumption might not be fulfilled in some applications of random matrix theory.

In the limit of $N\to\infty$, we introduce the level density, $\sigma(\Lambda)$
and we write the distribution (\ref{rmf-p}) in the  continuum form,
\be
P(\Lambda)=e^{-\beta H}
\ee
where
\be
 H=-\frac{1}{\beta}\int d\Lambda \sigma(\Lambda)F(\Lambda)-\frac{1}{2}\int\int d\Lambda d\Lambda'
\sigma(\Lambda)\sigma(\Lambda')\ln |\Lambda-\Lambda'|.
\ee
To specify the function $F$, 
we must introduce some constrains.
We can, for instance, require a specific
form of the density, $\sigma(\Lambda)$ and look for 
 the ``most random'' distribution $P(\Lambda)$ by minimizing 
the ``Hamiltonian'' $H$. The condition
\be
\ds{\frac{\delta H}{\delta\sigma}}=0
\ee
leads to  the following expression for  $F$:
\be\label{dmpk-F}
F(\Lambda)=-\beta\int d\Lambda' \sigma(\Lambda')\ln|\Lambda-\Lambda'|.
\ee
In the next Section, we apply the above formulas to the transfer matrix.

\subsection{Application to transfer matrix}\label{app:apl}

Assuming that the transmission matrix 
\be
\left[t^\dag t\right]^{-1}=v^\dag(1+\lambda)v
\ee
of the diffusive system belongs to the class of 
random matrices, we immediately obtain that the distribution of parameters 
$\lambda$ has a form
\be\label{rmf-p1}
P(\{\lambda\})=e^{-\beta H}
\ee
with
\be\label{rmf-p2}
H=-\sum_{a<b}\ln|\lambda_a-\lambda_b|+\sum_a\int_0^\infty d\lambda \sigma(\lambda)\ln|\lambda-\lambda_a|.
\ee
The first term is the Jacobian, and the second term is the one particle potential, 
given by Eq. (\ref{dmpk-F}).

It is more useful to express the distribution (\ref{rmf-p1},\ref{rmf-p2})
in terms of  variables $x_a$ instead of $\lambda$. 
Inserting $\lambda_a=(\cosh x_a-1)/2$ into Eq. (\ref{rmf-p2}) we obtain
\be\label{rmf-p3}
H(\{x\})= - \sum_{a<b}^N\ln|\cosh x_a-\cosh x_b|-\frac{1}{\beta}\sum_a^N\ln|\sinh x_a|
+\sum_a V(x_a).
\ee
The second term on the r.h.s. of Eq. (\ref{rmf-p3}) is the Jacobian of the
transformation  $\lambda\to x$, and we have used  
$\sigma(\lambda) d\lambda=\sigma(x)dx$.
The last term is the one particle potential,
\be\label{px}
V(x)=\int_0^\infty dx'~\sigma(x')\ln|\cosh x-\cosh x'|.
\ee
\begin{figure}[t!]
\psfrag{b=1}{$\beta=1$}
\psfrag{b=4}{$\beta=4$}
\psfrag{x1}{$\delta_{12}$}
\psfrag{p(x1)}{$p(\delta_{12})$}
\begin{center}
\includegraphics[clip,width=0.35\textheight]{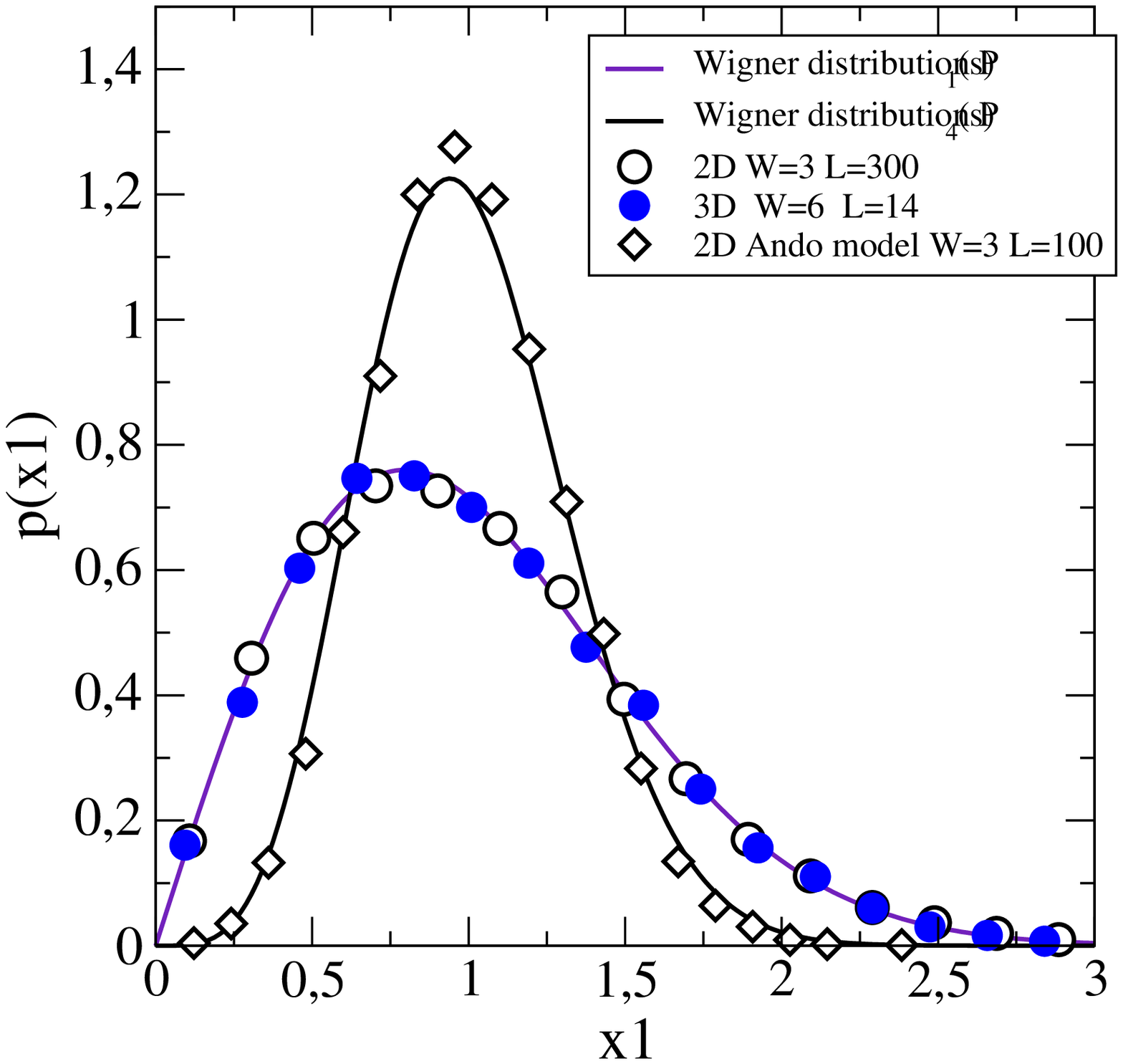}
\end{center}
\caption{The probability distribution $p(\delta_{12})$ of the \textsl{normalized} difference
$x_2-x_1$ for three models in the diffusive regime. 
Two systems with orthogonal symmetry have the same distribution $p_1$, given by
Eq. (\ref{wigner-1}). This confirms that the  important property in the diffusive regime
 is not determined by the  dimension 
of the sample, but by  the randomization of the electronic wave function  due to the multiple scattering.
The third statistical ensemble, 
given by the 2D Ando model,
possesses the symplectic symmetry ($\beta=4$)
The solid lines are the Wigner distributions.
}
\label{fig-pdelta}
\end{figure}

The density $\sigma(x)$ can be estimated from the numerical data. 
Fig. \ref{3D-zi} shows that $\langle x_a\rangle \sim a$. This leads to the constant
density
\be\label{rmt-qq}
\sigma(x)=\left\{
\begin{array}{ll}
\frac{N\ell}{L_z} & x<L_z/\ell  \\
0  & \textrm{otherwise}.
\end{array}
\right.
\ee
We insert expression (\ref{rmt-qq}) into Eq. (\ref{px}), and we  calculate the potential $V(x)$  
following the method given in Ref.  \cite{PZIS}.
The first derivative of Eq. (\ref{px}) reads
\be
\frac{\partial V(x)}{\partial x}=\frac{N\ell}{L_z}\int_0^\infty\frac{\sinh x}{\cosh x-\cosh x'}dx'.
\ee
With the new variable, $y=e^{x'}$, we obtain
\be
\frac{\partial V(x)}{\partial x}=\frac{N\ell}{L_z}\int_1^A
\left[\ds{
\frac{1}{y-e^{-x}}
-\frac{1}{y-e^{+x}}
}
\right]
dy,
\ee
with $A=\exp L_z/\ell\gg 1$.
Since $x>0$, the first integral gives $\ln|A-e^{-x}|-\ln|1-e^{-x}|$.
The function $1/(y-e^x)$ possesses a singularity at $e^x$. Since
the (negative) contribution to the integral from $1$ to $e^x$  cancels with the (positive) contribution
from $e^x$ to $2e^x-1$, we obtain that the second integral gives
$\ln|A-e^x|-\ln|e^x-1|$. 
 For $A>>e^x$ 
the difference of the  two integrals 
is $\ln|e^x-1|-\ln|1-e^{-x}|\equiv x$. We finally obtain
\be
\frac{\partial V(x)}{\partial x}=\frac{N\ell}{L_z}~x
\ee

Thus, we have  the \textsl{quadratic} one particle potential,
\be\label{pxx}
V(x)=\frac{N\ell}{2L}x^2.
\ee

The probability distribution $P(x)$ is similar to the statistical sum
of the one dimensional Coulomb gas. In the Coulomb gas  analogy,
the parameters $x_a$ represent the position  of the $a$th particle confined
in the quadratic potential $V(x)$ and interacting with the other particles
by the two particle logarithmic interaction. The parameter $\beta$ is  the ``temperature''.
A detailed analysis of the probability distribution $p(x)$ and of the
consequences for the  diffusive transport are given in a series of papers of Pichard 
\textsl{et al.} \cite{Pnato,PZIS,ZP,SPM,APM}.

The most simple consequence of the probability distribution $p(x)$
is the ``level repulsion''. From random matrix theory it follows that
the \textsl{normalized} difference
\be
\delta_{a,a+1}=\ds{\frac{x_{a+1}-x_a}{\langle x_{a+1}-x_a\rangle}}
\ee
is distributed with the Wigner distribution $p_\beta$. This was confirmed by
numerical simulations \cite{PZIS,ZP,Kramer-PM}. In Fig. \ref{fig-pdelta}
we plot $p(\delta_{12})$ for three statistical ensembles for the  2D and 3D orthogonal systems,
and for the 2D Ando model.  The data confirm that the distributions  
 $p(\delta_{12})$ are indeed very similar to the Wigner distributions.

\begin{figure}[b!]
\begin{center}
\includegraphics[clip,width=0.35\textheight]{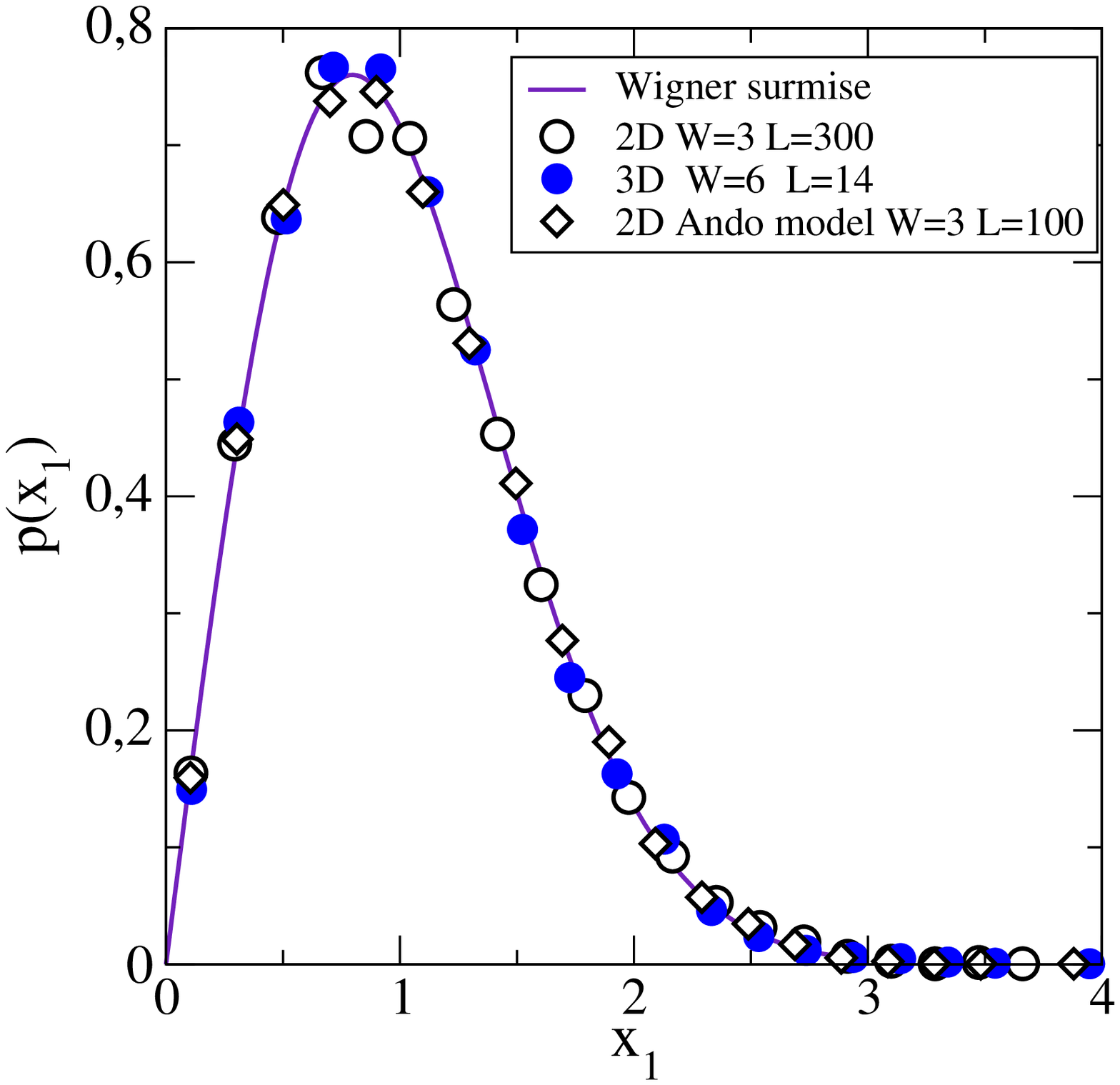}
\end{center}
\caption{The distribution $p(x_1)$ of the \textsl{normalized} parameter $x_1$ for three
different models in the diffusive regime. The solid line is the Wigner distribution, given by Eq. 
(\ref{wigner-1}). Note, the distribution $p(x_1)$ does not depend on the physical symmetry of the model.
}
\label{fig-px1}
\end{figure}

Note, the term $\ln\sinh x$ in the ``Hamiltonian'' (\ref{rmf-p3}) 
can be interpreted as the interaction of the particle located at  $x$ with its ``mirror image'',
i. e. the particle located at $-x$. 
Therefore, the distribution of $p(x_1)$ is  given by the Wigner distribution, too. 
However, since
the ``interaction'', $\ln\sinh x$, does not depend on $\beta$,
the distribution $p(x_1)$ should not depend on the physical symmetry \cite{PZIS}. This is confirmed
in Fig. \ref{fig-px1}.

\begin{figure}[t!]
\begin{center}
\includegraphics[clip,width=0.25\textheight]{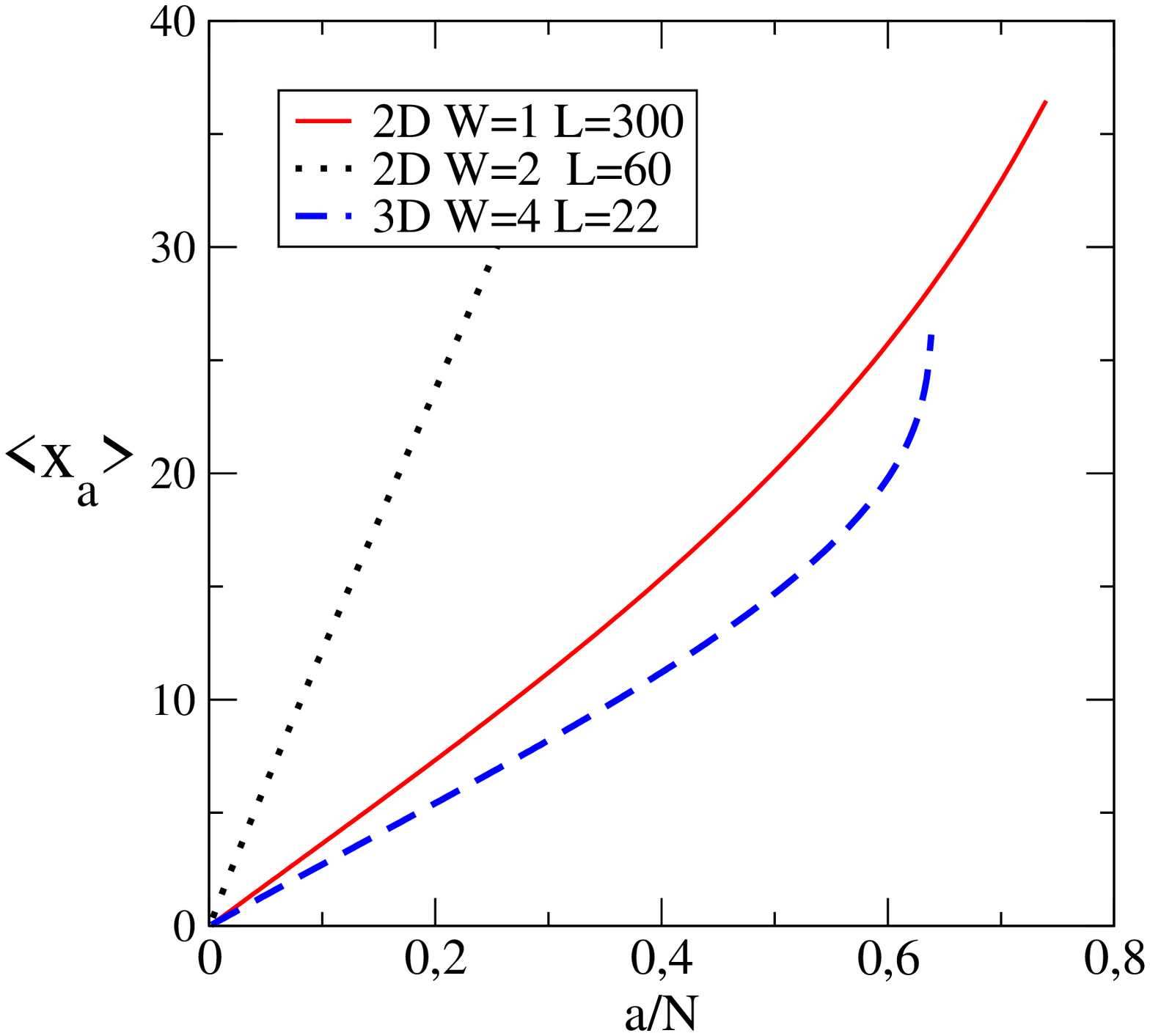}~~~~~
\includegraphics[clip,width=0.25\textheight]{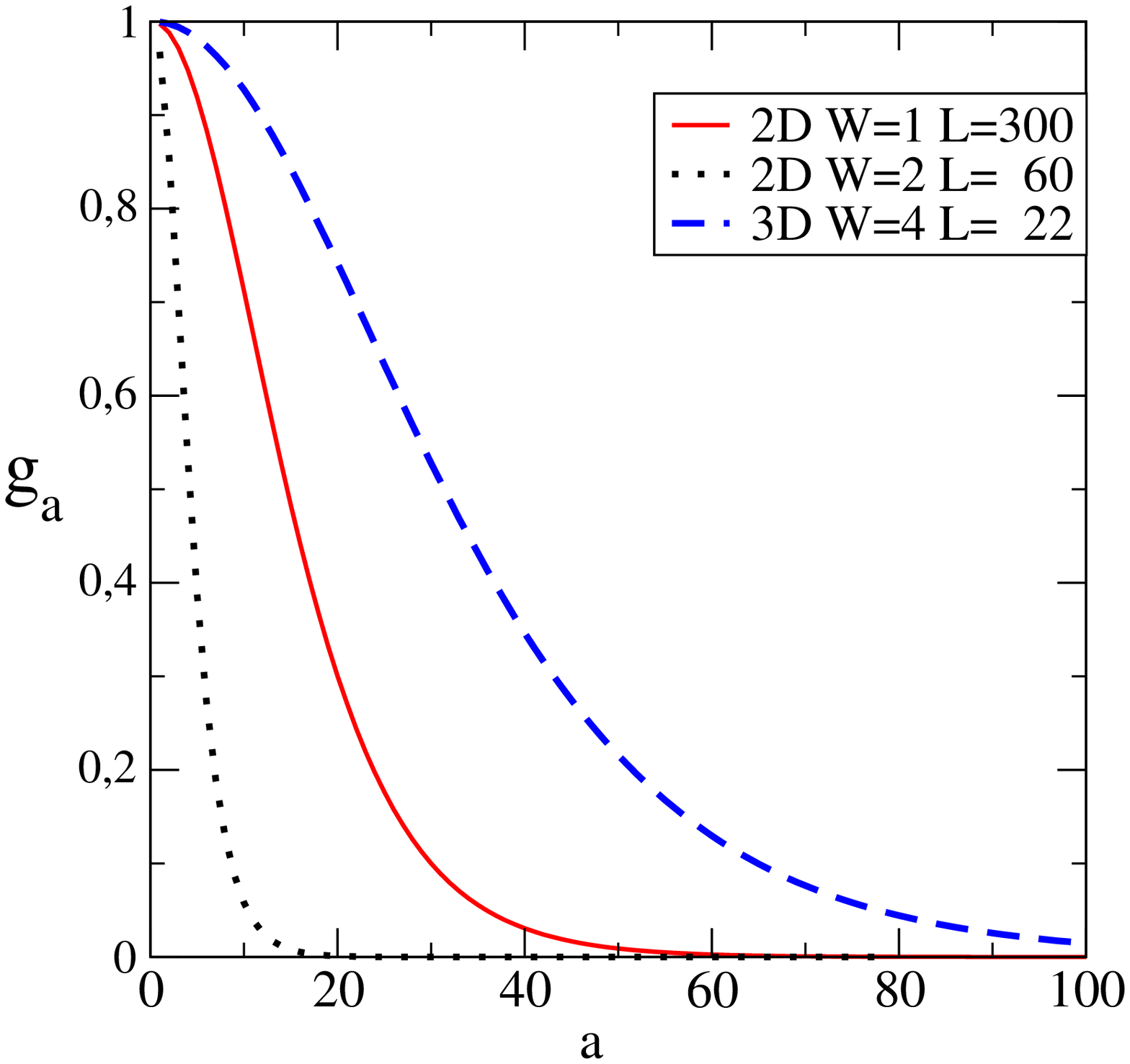}
\end{center}
\caption{The index dependence of the mean $\langle x_a\rangle$ (left) and
$g_a$ (right) for the  2D and 
3D weakly disordered systems.  The data confirm the linearity
given by Eq. (\ref{univ-xa}). To calculate $\langle x_a\rangle$, 
we have to diagonalize the matrix $t^\dag t$ for each sample, extract $x_a$ from
the eigenvalue $(1+\lambda_a)^{-1}$. The mean value, $\langle x_a\rangle$ is obtained by
averaging over the statistical ensemble.  
Note, the data for large $a$ are not
present since $\cosh^2(x_a/2)$ exceeds the numerical accuracy of the computers 
for such large values of $x_a$.
The conductance $\langle g\rangle =
16.1$, 4.3 for the 2D systems with $W=1$ and $W=2$, respectively, and
$\langle g\rangle =42.1$ for the 3D system. In the last model, the anisotropy, $t=0.4$ was
used to avoid the evanescent channels in leads.
}
\label{fig-xa}
\end{figure}

Figure  \ref{fig-xa} confirms that  the spectrum of parameters $x_a$ is linear
in the diffusive  regime, both in the 2D and 3D systems,
\be\label{univ-xa}
\langle x_a\rangle=\left[1+(a-1)\beta\right]\langle x_1\rangle. 
\ee
For $\beta=1$, a more accurate estimation, which agrees also with results of numerical  simulations,
can be derived from random matrix theory,
\be\label{xxa}
x_a\propto \sqrt{X_a}
\ee
where $X_a$ are zeros of Laguerre polynomials $L_N$
\cite{Muttalib,M-1995}, $X_a\approx \left[4(N+1/2)\right]^{-1}j_0^2(a)$,
where $j_0(a)$ is the $a$th zero of the Bessel function $J_0(x)$. In the limit
of $a\gg 1$, $j_0(a)\approx \pi(a-1/4)$, so that  Eq. (\ref{xxa}) gives $x_a\propto a$.

The right Fig. \ref{fig-xa} shows the $a$ dependence of  parameters $g_a$,
\be g_a = \cosh^{-2}x_a/2.
\ee
Clearly, $g_a$ is the contribution of the $a$th channel and
\be
g=\sum_a g_a.
\ee
Because of the rigidity of the spectrum, the fluctuations of $x_a$
are of the order of the mean spacing, which is $\sim L_z/N\ell$. 
We see that $g_a$ are very close to 1 for small $a$
and exponentially  small when $a\to N$. There is $\neff$ channels for which $x_a\le 1$.
Higher channels, with $a>\neff$ give only negligible small contribution
to transport
Then, the conductance $g\approx \neff$.
The slope of the linear dependence, $\langle x_1\rangle$, 
determines the conductance: if $\langle x_{\neff}\rangle=1$,
then we have, from Eq. (\ref{univ-xa}) that
\be
\langle g\rangle =\neff=1+\frac{1}{\beta}\left(\frac{1}{\langle x_1\rangle}-1\right).
\ee
This expression simplifies for $\beta=1$:
\be\label{neff-x1}
\langle g\rangle =\neff=\frac{1}{\langle x_1\rangle}.
\ee
Since $g=N\ell/L_z$, we immediately see that
\be
\langle x_1\rangle=\frac{L_z}{N\ell}.
\ee
Thus, all parameters $x_a$ increase linearly with the length of the system.
The mean value, $\langle x_a\rangle$, as well as 
the spacing between two neighboring parameters, $\langle x_a\rangle$, decreases
as $N^{-1}$. 

The rigidity of the spectrum of parameters $x_a$ was used for the explanation of the
universality of conductance fluctuations in Ref.  \cite{Imry}.

Random matrix theory represents a powerful tool for the analysis of transport in disordered
systems. More details are given  in Refs. \cite{Brody,Pnato,Beenakker}.

\subsection{DMPK equation versus random matrix theory}

Both the DMPK equation and the random matrix theory were successfully applied to the transport
in disordered systems.  Originally it was
believed that the theories are equivalent. This would mean that the probability distribution
$p(\lambda)$, given by random matrix theory, 
solves the DMPK equation.  

However, these two approaches are not equivalent. 
The discrepancy between the two approaches was observed when Beenakker \cite{Been}
calculated the variance of the conductance from random matrix theory. The obtained result,
$\textrm{var}~g=1/8\beta$
differs from the exact value,
$\textrm{var}~g=2/15\beta$, obtained by the diagrammatic expansion \cite{LSF}
and from the DMPK equation \cite{Stone}. This proves that random matrix theory
is not exact. Beenakker and Rejaei \cite{BR} solved the DMPK equation for
$\beta=2$. Both in  the metallic and localized  regimes, the solution reads
\be\label{been-1}
p(x)=\exp -\beta \left[ \sum_a V(x_a)+\sum_{a<b} u(x_a,x_b)\right],
\ee
with the interacting term
\be\label{been-2}
u(x_a,x_b)=-\frac{1}{2}\ln|\cosh x_a-\cosh x_b|-\frac{1}{2}\ln|x_a^2-x_b^2|
\ee
and with the one-particle potential,
\be\label{been-3}
V(x_a)=\frac{N\ell}{2L_z}x_a^2-\frac{1}{4}\ln(x_a\sinh x_a).
\ee
This solution is very similar to the probability distribution derived from random matrix 
theory. However,  note, that it cannot be obtained from the Ansatz (\ref{rmf-p}),
since the two particle interaction, given by Eq. (\ref{been-2}), is not determined
by the Jacobian only. Therefore, the one parameter function $F(\lambda)$, introduced in Eq. (\ref{rmf-p}),
is not sufficient for obtaining the distribution (\ref{been-1}). 


\section{Lyapunov exponent}\label{app-le}

\subsection{One-dimensional case}

The Lyapunov exponent, $\gamma(E)$, of the one dimensional disordered system
is defined by the relation
\be\label{9-one}
\gamma(E)=\lim_{L_z\to\infty}\frac{1}{2L_z}\ln\left(\Psi_{L_z}^2+\Psi_{L_z+1}^2\right),
\ee
Note also, that the wave function 
\be
|\Psi_{L_z}|\propto e^{-\gamma L_z}
\ee
decreases exponentially when $L_z$ increases. From Eq. (\ref{9-one}) we
see that
 the real part of the Lyapunov exponent determines the localization length,
\be
\textrm{Re}~\gamma(E)=\lambda^{-1}.
\ee
Thouless \cite{Thouless} showed that the   \textsl{imaginary part}  
is related to the  density of states 
 by the relation
\be\label{LE-4}
\rho(E)=\frac{2}{\pi}\frac{\partial}{\partial E}~ \textrm{Im}~\gamma(E).
\ee

In the 1D systems, the Lyapunov exponent can be found analytically in the limit of weak disorder.
For the Anderson model, defined by Eq. (\ref{1d-sche}), we find
\be\label{9-result}
\gamma(E) =ik+\ds{\frac{\langle\eeps^2\rangle}{2(4-E^2)}},
\ee
with $E=2\cos k$.
Note that the expression (\ref{9-result}) fails when close to the band edge, $|E|\to 2$.
A more detailed analysis \cite{KW,DG} showed that in this case $\gamma\propto W^{2/3}$.
In general, the 
weak disorder expansion exhibits 
a peculiar behavior in the neighborhood  of energies $E=2\cos \pi p/q$, with
$p$ and $q$ being integers. At the band center, $E=2\cos \pi/2=0$,
the fourth order term of the expansion  diverges and gives rise to the correction of the 2nd order term
\cite{KW,DG}. 
For instance, $\textrm{Re} \gamma =W^2/96$ for $E=0$ and the box disorder, defined by Eq.
(\ref{dis-box}), since  
$\langle\eeps^2\rangle=1/12$. Correct expression, which takes into account
the higher order terms of the expansion, gives 
$\textrm{Re} \gamma =W^2/105.4$, which agrees with numerical data.

The Lyapunov exponent plays an important role in the theory of localization.  In the next
Section, we generalize the 1D case to the quasi-1d systems. 
Note that the Schr\"odinger equation,
\be
\Psi_{n+1}+(\eeps_n-E)\Psi_n+\Psi_{n-1}=0,
\ee
can be written in the matrix form
\be\label{1d-tm}
\mv{\Psi_{n+1}}{\Psi_n}=\m2{E-\eeps_n}{-1}{1}{0}\mv{\Psi_n}{\Psi_{n-1}}.
\ee
Then we can write
\be\label{1d-psi}
\Psi_{L_z+1}^2+\Psi_{L_z}^2=
(\Psi_{L_z+1}, \Psi_{L_z})\mv{\Psi_{L_z+1}}{\Psi_{L_z}}=
(\Psi_{1}, \Psi_0)(\textbf{M}^{(L_z)})^T\textbf{M}^{(L_z)}\mv{\Psi_{1}}{\Psi_0},
\ee
where  we have introduced the transfer matrix, 
\be
\textbf{M}^{(L_z)} =\prod_{n=1}^{L_z} \textbf{M}_n=\prod_{n=1}^{L_z}\m2{E-\eeps_n}{-1}{1}{0}.
\ee
Thus, in the limit of $L_z\to\infty$, $\gamma$ is given by the eigenvalues
of the matrix $\textbf{M}^{(L_z)}$. Since $\textbf{M}^{(L_z)}$ contains the random energies $\eeps_i$,
$\gamma$ is a statistical variable, too. 

Oseledec \cite{Oseledec} proved that the 
probability distribution, $p(\gamma)$ is Gaussian  with the mean value
$\langle\gamma\rangle\propto L_z$ and variance, $\textrm{var}~\gamma\propto \langle\gamma\rangle$.
Therefore, the variable 
\be
z=\frac{2L}{L_z}\times \gamma
\ee
is a self-averaged quantity in the limit 
of long samples.

Self-averaging of the Lyapunov exponent can be intuitively understood
from the expression (\ref{1d-psi}). The Lyapunov exponent represents 
the logarithm of the eigenvalue of the matrix 
$(\textbf{M}^{(L_z)})^T\textbf{M}^{(L_z)}$, which is the  product of the random matrices.  
Naively speaking, $\gamma$ can be considered as a sum of logarithm of
eigenvalues of the random matrices $\textbf{M}_n$, $n=1,2,\dots L_z$. Then, applying  the
central limit theorem we expect, that both the mean value and  variance 
of $\gamma$ are proportional to $L_z$.

\subsection{Quasi-1d case}\label{le-q1d}

Consider the  quasi-1d system of the size $L^{d-1}\times L_z$, and suppose that
$L_z\gg L$. For the purpose of numerical  analysis,
we divide the system into  a set of vertical slices, shown in Fig. \ref{fig-tm} and
we write the Schr\"odinger equation in the form
\be\label{tm-1}
\Psi_{n+1} = (E-{\cal H}_n)\Psi_n -\Psi_{n-1}
\ee
where ${\cal H}_n$ is the Hamiltonian related to  the $n$th slice in Fig. \ref{fig-tm},
and $\Psi_n$ is the vector which contains in its elements the wave function
in sites of the $n$th slice. The length of the vector $\Psi_n$ is $N=L^{d-1}$.

Equation (\ref{tm-1}) can be rewritten in the matrix form,
\be
\mv{\Psi_{n+1}}{\Psi_n}=\textbf{M}_n\mv{\Psi_n}{\Psi_{n-1}}
\ee
where the transfer matrix, $\textbf{M}_n$, is given by the relation
\be\label{tm-m}
\textbf{M}_n=\m2{E-{\cal H}_n}{-1}{1}{0}.
\ee
This relation is formally equivalent to Eq. (\ref{1d-tm}) but now the matrix $\textbf{M}_n$
has the size $2N\times 2N$.
Note, $\textrm{det}~M_n\equiv 1$.
Also, the eigenvalues of $\textbf{M}_n$ appear in pairs, $\lambda$
and $\lambda^{-1}$.
 
\begin{figure}[t!]
\begin{center}
\includegraphics[clip,width=0.23\textheight]{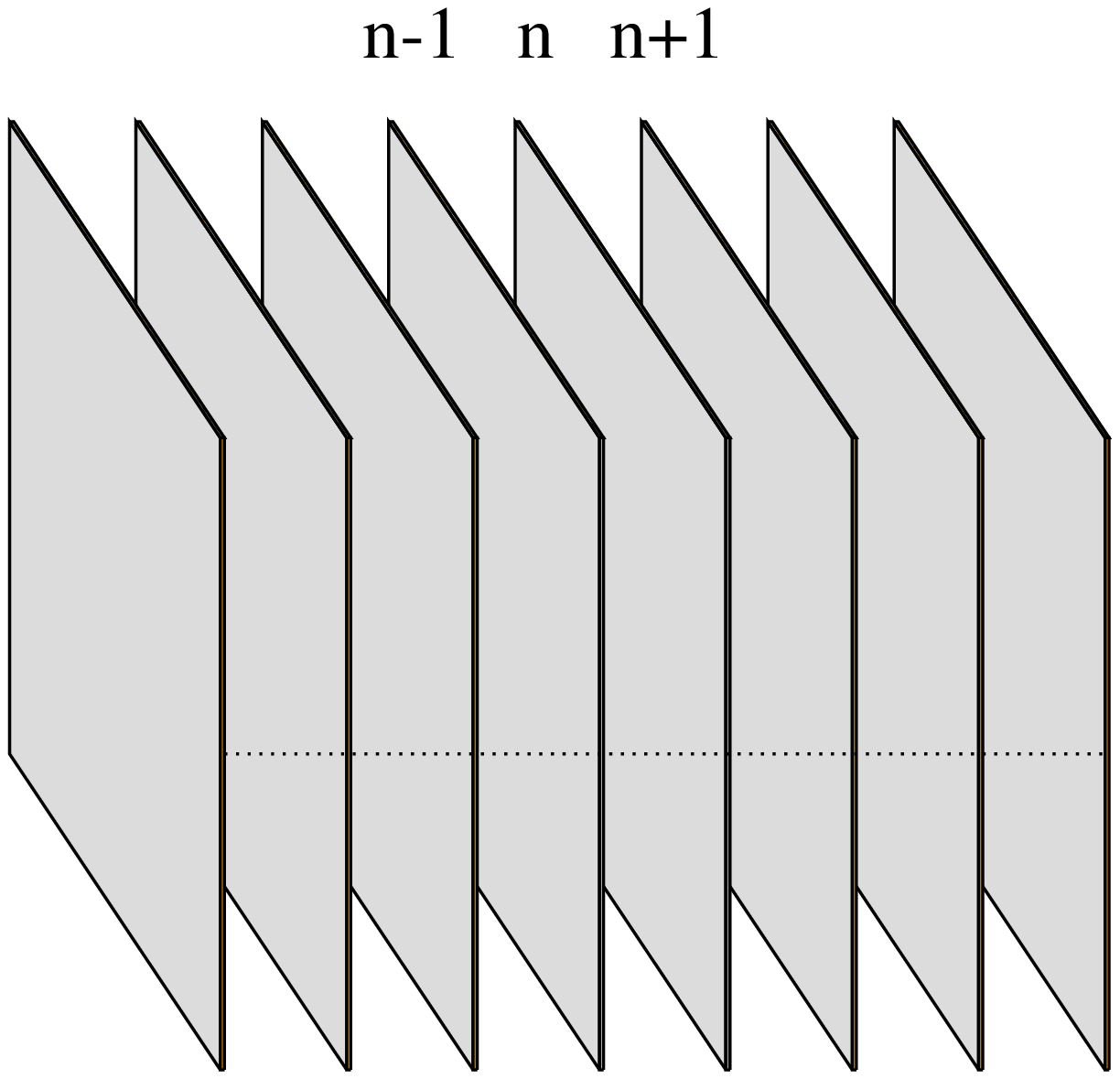}
\end{center}
\caption{We cut the quasi-1d system into the  $L\times L$ vertical slices. The propagation
of the electron in slice
$n$ is given by the Hamiltonian ${\cal H}_n$.
}
\label{fig-tm}
\end{figure}

Now we calculate the $2N\times 2N$ matrix 
\be\label{4614}
\textbf{M}^{(L_z)} =\prod_{n=1}^{L_z} \textbf{M}_n=\prod_{n=1}^{L_z}\m2{E-{\cal H}_n}{-1}{1}{0}.
\ee
The matrix $\textbf{M}^{(L_z)}$ determines the behavior of the wave function at long distances.
Since the system is effectively one dimensional, the wave function must decrease
exponentially when the length of the system, $L_z$, increases. This exponential decrease
is given by  the eigenvalues, $\exp-\gamma_a$, with $\gamma_a>0$.
Clearly, the smallest $\gamma_a$ determines the localization length.

Since the matrices $\textbf{M}_n$ contains the random variables, $\eeps_n$,  $\alpha_a$ are
also statistical
variables, and we need to know their probability distributions. Following 
Oseledec \cite{Oseledec}, we have that 
in the limit of $L_z\to \infty$ all  the eigenvalues of the matrix
\be\label{Oseledec}
[\textbf{M}^{(L_z)}\textbf{M}^{(L_z)}]^{L/L_z}
\ee
converge to $\lambda_a=e^{\zeta_a}$  and $\lambda_{N+a}=e^{-\zeta_a}$
where 
\be
\zeta_a=\frac{2L}{L_z}\times\gamma_a.
\ee
The mean value,   $\langle \zeta_a\rangle$, does not depend on $L_z$,
and the  variance,
\be
\textrm{var}~\zeta_a=\langle \zeta_a^2\rangle - \langle \zeta_a\rangle^2\propto \frac{L}{L_z}.
\ee
Thus, Oseledec theorem states that the  parameters $\zeta_a$ are the self-averaged quantities in the limit
of $L_z\to\infty$.  For a given realization for the disorder, the eigenvalues of the matrix
(\ref{Oseledec}) converge to their mean values. In numerical simulations, the length $L_z$ is finite,
so that the obtained numerical data, $\zeta_a$, differs from the limited mean values, $\langle \zeta_a\rangle$.
Oseledec theorem enables us to estimate the typical difference, 
\be
\zeta_a - \langle\zeta_a\rangle \sim \sqrt{\textrm{var}~\zeta_a}\propto \left(\frac{L}{L_z}\right)^{1/2}.
\ee
Thus, we can  avoid the problem of  statistical fluctuations
in numerical calculations.
It is sufficient to calculate the product
of the transfer matrices, $\textbf{M}^{(L_z)}$, for a sufficiently large  system length,
$L_z$, and then to calculate the eigenvalues, $\zeta_a$.
If $L_z$ is large  enough, then the  obtained 
values $\zeta_a$ lie close to the mean values, $\langle \zeta_a\rangle$.
This procedure is applied in the finite size scaling analysis
of Lyapunov exponents, discussed in Section \ref{scal-LE}.

In what follows we order $\zeta_1>\zeta_2>\dots >\zeta_N$.
That means that 
\be
\zeta_a=z_{N+1-a},
\ee
 where $z_a$ are discussed in   Sect. \ref{section:nsa}.
Clearly, the smallest Lyapunov exponent, $z_1$, equals to the smallest eigenvalue $\zeta_N$.

In the  numerical simulations, we  need to calculate the smallest (in absolute value) parameter
$\zeta_N=z_1$. This can be done by the following numerical algorithm.

\begin{enumerate}
\item Choose the required accuracy $\eeps$ of $\zeta_N$. Then, start with the following variables:
first, we need the $N\times 2N$ matrix $\textbf{V}$ which contains in $N$ columns the mutually orthogonal vectors 
$v_a$ of the length $2N$. We will also need   the  vectors  $d_a$ and $e_a$
of the length $N$ and put their stating values,
$d_a=(0,0,\dots 0)$ and
$e_a=(0,0,\dots 0)$.  Also, put 
$L_z=0$.

\item
Calculate the $n_i$ iterations
\be
\textbf{U}=\prod_{n=1}^{n_i} \textbf{M}_n \textbf{V}
\ee

\item Perform the Schmidt orthogonalization of vectors $u_a$, given  by columns of the
matrix $\textbf{U}$. Obtain new vectors $w_a$, which are already orthogonal to each other.

\item Calculate the norm of vectors $w_a$, 
and add these  norms into vectors $d$ and $e$ as follows:
\be
d_a=d_a+\ln |w_a|
\ee
and 
\be
e_a=e_a+(\ln |w_a|)^2.
\ee

\item Put
\be
u_a=  w_a/|w_a|
\ee
and  the length, 
\be
L_z=L_z+ n_i.
\ee
Calculate  $\zeta_a=d_a/L_z$, 
$\eta_a=e_a/L_z$
and the accuracy
\be
\eeps_a=\ds{\frac{\sqrt{\eta_a-\zeta_a^2}}{\zeta_a}}
\ee

\item If $\eeps_N>\eeps$ then go to the step 2, otherwise stop.

\end{enumerate}

The obtained value of $\zeta_N$ is the required smallest Lyapunov exponent $z_1$.

The length $L_z$, necessary for the calculation of the smallest Lyapunov exponent,
depends on the required accuracy, $\eeps$.
For the 3D Anderson model, the length of the system, $L_z$  was estimated  in Ref. \cite{M-1995}
as
\be
L_z=\frac{1}{2\eeps^2}L.
\ee
In numerical simulations, the accuracy $\eeps=0.001$ is usually required. Then,
$L_z=500.000\times L$. The number of iterations between two successive Schmidt
orthogonalization is $n_i\approx 6-8$ for the critical disorder, and smaller (larger)
when $W\gg W_c (W\ll W_c)$, respectively.


\section{Calculation of the conductance}\label{app-b}

\subsection{1d case}\label{app-b1}

It is convenient to write the Schr\"odinger Equation (\ref{1d-sche}) in the form
\be\label{1d-2}
\mv{\Psi_{n+1}}{\Psi_n}= \textbf{M}_n\mv{\Psi_{n}}{\Psi_{n-1}}
=\m2{E-\eeps_n}{-1}{1}{0}\mv{\Psi_{n}}{\Psi_{n-1}}.
\ee
To calculate  the conductance, consider the  sample of the length $L_z$ sites\footnote{For simplicity,
we use the  lattice constant, $a$  as the unit length}, connected
to two semi-infinite ideal leads. This means that $\eeps_n=0$ for both $n\le 0$ and
for $n>N-1$.

Note that the  transfer matrix  $\textbf{M}$, defined in Eq. (\ref{1d-2})
does not have the structure
of the transfer matrix $\textbf{T}$. Indeed,  $\textbf{T}$ connects the propagating waves 
on the left and right hand side of the sample,  
while $\textbf{M}$ relates the wave functions in the
the site representation.
Both matrices are connected by the transformation,
\be\label{2-qmq}
\textbf{T}_n=\bQ^{-1}\textbf{M}_n\bQ
=\textbf{Q}^{-1}\m2{E-\eeps_n}{-1}{1}{0}~\textbf{Q},
\ee
where
\be
\textbf{Q}=\m2{1}{1}{e^{-ik}}{e^{+ik}}.
\ee

Consider the electron coming from the right hand  side of the system.
Then, on the left hand side of the system, there is only a transmitted wave going
to the left. Its  wave function at the sites $n=0$ and $n=-1$
can be written as
\be\label{8-init}
\begin{array}{lcl}
\Psi_{-1}&=&e^{+ik}\\
\Psi_{0}&=&1.
\end{array}
\ee
The wave function on the  right-hand 
side of the system is given by  superposition
of the incoming and reflected waves. We can use the transfer matrix,
$\textbf{M}$, to express the wave function at sites $L_z$ and $L_z-1$:
\be\label{4-tm-n}
\mv{\Psi_{L_z}}{\Psi_{L_z-1}}=\textbf{M}_{L_z-1}\textbf{M}_{L_z-2}\dots \textbf{M}_1\textbf{M}_0\mv{\Psi_0}{\Psi_{-1}}
=\textbf{M}^{(L_z)}\mv{\Psi_0}{\Psi_{-1}}.
\ee
Now we multiply both sides of the   of Eq.
(\ref{4-tm-n}) by the matrix $\bQ^{-1}$. We get
\be\label{4-tm-nx}
\bQ^{-1}\mv{\Psi_{L_z}}{\Psi_{L_z-1}}=\bQ^{-1}\textbf{M}^{L_z}\bQ\bQ^{-1}\mv{\Psi_0}{\Psi_{-1}}.
\ee
We obtain that
\be\label{4-ab}
\ds{\frac{1}{2i\sin k}}
\mvr{e^{ik}\Psi_{L_z}-\Psi_{L_z-1}}{-e^{-ik}\Psi_{L_z}+\Psi_{L_z-1}}=
\textbf{T}^{(L_z)}
\ds{\frac{1}{2i\sin k}}
\mvr{e^{ik}\Psi_0-\Psi_{-1}}{-e^{-ik}\Psi_0+\Psi_{-1}},
\ee
where we used the relation (\ref{2-qmq}).
between the transfer matrices $\textbf{M}$ and $\textbf{T}$.
Using the explicit form of  $\Psi_0$ and $\Psi_{-1}$, given by Eqs. (\ref{8-init})
we find  that the r.h.s. of Eq. (\ref{4-ab}) equals to 
\be
\textbf{T}^{(L_z)}\mv{0}{1}.
\ee
From Eq. (\ref{4-ab}) we   obtain
\be\label{8-uax}
\ds{\frac{1}{2i\sin k}}
\mvr{e^{ik}\Psi_{L_z}-\Psi_{L_z-1}}{-e^{-ik}\Psi_{L_z}+\Psi_{L_z-1}}=
\mvr{r^+(t^-)^{-1}}{(t^-)^{-1}}.
\ee
So we  obtain  the transmission  coefficient in the form
\be\label{8-1}
T=|t^-|^{-2}=\ds{\frac{4\sin^2 k}{|e^{-ik}\Psi_{L_z}-\Psi_{L_z-1}|^2}}.
\ee
Equation (\ref{8-1}) is very useful for numerical calculations of the transmission $T$.

\subsection{Quasi-1d case}\label{app-b2}

To calculate the conductance in the quasi-1d case, we have to generalize 
the method of the previous Section. 
We again assume that our system is connected to two semi-infinite leads
with zero disorder, $W=0$. The electron is coming from the  right
and is scattered by the sample. The resulting waves either continue
to the left in the left  lead, or  travel back to the right
in the right side lead. 

In numerical simulations, we use the transfer matrix $\textbf{M}$, given by Eq. 
(\ref{4614})
\be\label{iter}
\textbf{M}^{(L_z)} =\prod_{n=1}^{L_z} \textbf{M}_n=\prod_{n=1}^{L_z}\m2{E-{\cal H}_n}{-1}{1}{0}.
\ee
Similarly as in the case of the 1D problem, discussed in the previous section,
 we have to transform this transfer matrix 
into the ``wave'' representation.
In this representation, the transfer matrix in the leads is diagonal.  
Therefore, in the first step we have to  diagonalize  the transfer matrix
\be
\textbf{M}_0=\m2{E-{\cal H}_0}{-1}{1}{0}
\ee
where ${\cal H}_0$ is Hamiltonian of the transversal slice \textsl{without}
disorder.

In general, the  transfer matrix has some eigenvalues with modulus equal to 1, i.e.
$\lambda=\exp ik_z$. The corresponding eigenvectors represent  the
propagating waves. Other eigenvalues  are of the form
$\lambda=\exp \pm\kappa$.  They correspond to the evanescent modes. 
Note, if $\lambda$ is an eigenvalue, then $\lambda^{-1}$ is also an eigenvalue
corresponding to the wave traveling in the opposite direction.

As the transfer matrix $\textbf{M}_0$  is not  Hermitian, 
we have to calculate both the left and right
eigenvectors.
Then, we  construct four 
matrices: The $N\times 2N$ matrix $\Rleft$ ($\Rright$) which contains in its columns 
 $N$ right eigenvectors, for waves 
traveling to the \textsl{left}, (\textsl{right}),
respectively.
Similarly,
matrices $\Lleft$ and $\Lright$ are $2N\times N$ matrices which contains in $N$
rows the left eigenvectors of the transfer matrix $\textbf{M}_0$
which represent the  waves traveling 
to the left and to the right, respectively.
Then, by definition of the transfer matrix,
\be
\textbf{T}=\m2{T_{11}}{T_{12}}{T_{21}}{T_{22}}=\m2{\Lright \textbf{M}\Rright}{\Lright \textbf{M}\Rleft}{\Lleft \textbf{M}\Rright}{\Lleft\textbf{M}\Rleft}
\ee
so that
\be\label{ltr}
T_{22}=\Lleft \textbf{T} \Rleft.
\ee
At this point we have to distinguish between the propagating and evanescent  modes. 
If there are evanescent modes,  then  $T_{22}\ne [t^-]^{-1}$ because the two  matrices have
different size. 
If we order the eigenvectors in the matrix
$R$ in such a way that the eigenvectors with index $1\le a\le\no$ correspond to
the propagating modes, and the remaining eigenvectors correspond to the evanescent modes,
then the transmission is given by  the $\no\times\no$
sub-matrix $[T_{22}]_{ab}$, with $a,b\le \no$:
\be
T=\sum_{ab=1}^{\no}\Big|\left[T_{22}^{-1}\right]_{ab}\Big|^2.
\ee
Other matrix elements of $T_{22}$ correspond to the scattering of the electron into
the evanescent channels. They do not contribute to the transmission since 
evanescent waves decay to zero in the semi-infinite  leads.

\medskip

It seems that the relation (\ref{ltr})  solves our problem completely.
However, the above algorithm must be modified. The reason is that
the  elements of the matrix $(t^-)^{-1}$ are given by their largest 
eigenvalues. 
We are, however,  interested in the largest eigenvalues of the
matrix $t^-$.   
As the elements  of the transfer matrix increase
exponentially in the iteration procedure given by Eq. (\ref{iter}),
any information about the smallest eigenvalues of $(t^-)^{-1}$ 
will be quickly lost. We have therefore to introduce some re-normalization
procedure. We use the procedure described in
Ref. \cite{PMcKR}.

Relation (\ref{ltr}) can be written as
\be\label{new}
T_{22}=\Lleft r^{(L_z)}
\ee
where we have defined the $N\times 2N$ matrices $r^{(n)}$, $n=0,1,\dots L_z$
as
\be\label{rr}
r^{(n)}=\textbf{M}_nr^{(n-1)},~~~~~{\rm and}~~~~~ r^{(0)}=\Rleft
\ee
Each matrix $r^{(n)}$ can be written as 
\be
r=\left(r_1\atop r_2\right)
\ee
with $r_1$, $r_2$ being the $N\times N$ matrices.
We transform $r$ as
\be\label{trik}
r=r' r_1,\quad\quad r'=\left(1\atop r_2r_1^{-1}\right)
\ee
and define $r^{(n)}=\textbf{M}_n(r')^{(n-1)}$.
In contrast to the matrices $r_1$ and $r_2$, all eigenvalues of the matrix $r_2r_1^{-1}$
are of the  order of unity. The 
relation (\ref{new}) can now be re-written into the form
\be\label{kva}
T_{22}=
\Lleft\left(1\atop r^{(n)}_2\left[r^{n}_1\right]^{-1}\right)
r^{(n)}_1r^{(n-1)}_1\dots r^{(1)}_1r^{(0)}_1
\ee
from which we get 
\be\label{result}
\begin{array}{ll}
\left[T_{22}\right]^{-1}= 
\left[r^{(0)}_1\right]^{-1}
\left[r^{(1)}_1\right]^{-1}\dots
\left[r^{(n)}_1\right]^{-1}
\left[ 
\Lleft\left(1\atop r_2^{(n)}\left[r^{(n)}_1\right]^{-1}\right)
\right]^{-1}.
\end{array}
\ee
All elements of the matrices on the r.h.s. of Eq. (\ref{result}) are 
of the   order of unity.

\addcontentsline{toc}{section}{References}
\renewcommand{\refname}{}
\def\xref#1#2#3#4#5{#1:\ #2 \textbf{ #3}, #4 (#5)}
\def\refer#1#2#3#4#5#6{#1:\  {\rm #3}\ {\bf #4},\ {(#5)}\ #6}
\def\referc#1#2#3#4#5#6#7{#1:\  {\textrm{#3}}\ {\bf #4},\ {(#5)}\ #6,}

\end{document}